\documentclass[10pt]{iopart}
\usepackage{graphicx}
\usepackage{amssymb}
\begin{document}

\review[Depinning and nonequilibrium dynamic phases of particle assemblies]{
  Depinning and nonequilibrium dynamic phases of particle assemblies driven
  over random and ordered substrates: a review
} 
\author{
C Reichhardt and C J Olson Reichhardt 
} 
\address{
Theoretical Division,
Los Alamos National Laboratory, Los Alamos, New Mexico 87545, USA
}
\eads{\mailto{reichhardt@lanl.gov},\mailto{cjrx@lanl.gov}}

\begin{abstract}
We review the depinning and nonequilibrium phases of
collectively interacting particle systems driven over random or periodic substrates.  
This type of system is relevant to vortices in type-II superconductors, 
sliding charge density waves, electron crystals, colloids,
stripe and pattern forming systems, and skyrmions, and could also have connections
to jamming, glassy behaviors, and active matter.
These systems are also ideal for exploring the broader issues of
characterizing transient and steady state
nonequilibrium flow phases as well as        
nonequilibrium phase transitions between distinct
dynamical phases, analogous to phase transitions between different equilibrium states.  
We discuss the differences between elastic and plastic depinning on random substrates
and  the different types of nonequilibrium phases
which are associated 
with specific features in the velocity-force curves, fluctuation spectra,
scaling relations, and local or global particle ordering.
We describe how these quantities can change depending on the dimension, 
anisotropy, disorder strength, and the presence of hysteresis.
Within the moving phase we discuss how there can be 
a transition from a liquid-like state to dynamically ordered
moving crystal, smectic, or nematic states.   
Systems with periodic or quasiperiodic 
substrates 
can have multiple nonequilibrium second or first order transitions in the
moving state between chaotic and coherent phases, and can exhibit hysteresis.
We also discuss systems 
with competing repulsive and attractive interactions,
which undergo dynamical transitions into stripes and other complex morphologies 
when driven over random substrates.
Throughout this work we highlight open issues and 
future directions such as  absorbing phase transitions, nonequilibrium work
relations, 
inertia, the role of non-dissipative dynamics such as Magnus effects, 
and how these results could be extended to the broader issues of plasticity in 
crystals, amorphous solids, and jamming phenomena.  
\end{abstract}
\submitto{\RPP}
\maketitle
\ioptwocol

\section{Introduction}

A wide class of systems can be effectively described as a collection
of interacting point particles that are driven over disordered or ordered substrates.
In equilibrium and in the absence of a substrate,
the interactions between the particles may favor a certain 
type of symmetry, such as a triangular lattice for repulsively interacting
particles in two dimensions.
Other types of crystalline phases are possible, however, and in some cases
the ground states are frustrated or disordered.
When the particles also interact with a substrate,
a wealth
of new types of phases are possible
that depend on whether the substrate is random, quasiperiodic, or
periodic, as well as the strength of the coupling between the particles and
the substrate. 
Individual potential minima in the substrate,
which we call pinning sites, exert a maximum force on individual particles that
we call a pinning strength.
For random or disordered substrates, as the substrate coupling strength increases
a transition can occur from a crystalline to a disordered or
partially disordered state containing grain boundaries or topological defects such
as dislocations, with a defect density that increases as a function of the disorder
strength.
For periodic substrates, various types of commensurate and incommensurate phases
can arise depending on  the ratio of the number of particles to the
number of substrate minima as well as on the pinning strength.
For any type of substrate, if a driving force is applied, a depinning transition occurs
at a critical driving threshold; below this force, the particles remain stationary, while
above this force, some or all of the particles move over the substrate.
Examples of systems in which such effects occur
include vortices in type-II superconductors interacting with 
randomly placed  pinning sites \cite{1} 
or periodic arrays of nanostructured  pinning sites \cite{2},
disordered electron crystals \cite{3},
sliding charge density waves under conditions where plasticity can occur \cite{4}, 
charge transport in metallic dot arrays \cite{5},
colloidal systems and soft matter assemblies on random  substrates \cite{6} or
optically created periodic \cite{7} or quasiperiodic substrates \cite{New1},
driven stripe and pattern forming systems \cite{8}, and
atoms and molecules moving over two-dimensional (2D) substrates \cite{9}.
There are also other examples of the depinning of strictly elastic objects in which
the system can be described as an assembly of elastically coupled
elements or particles that maintain the same neighbors over time, 
including magnetic domain walls \cite{10,11},
individual superconducting vortex lines \cite{12,12N},
 elastic strings \cite{14,15},
 interfaces \cite{16},
 certain models of charge density waves \cite{4}, 
 contact line depinning \cite{17},
 and slider block models for earthquakes \cite{18}.

\begin{figure}
\includegraphics[width=\columnwidth]{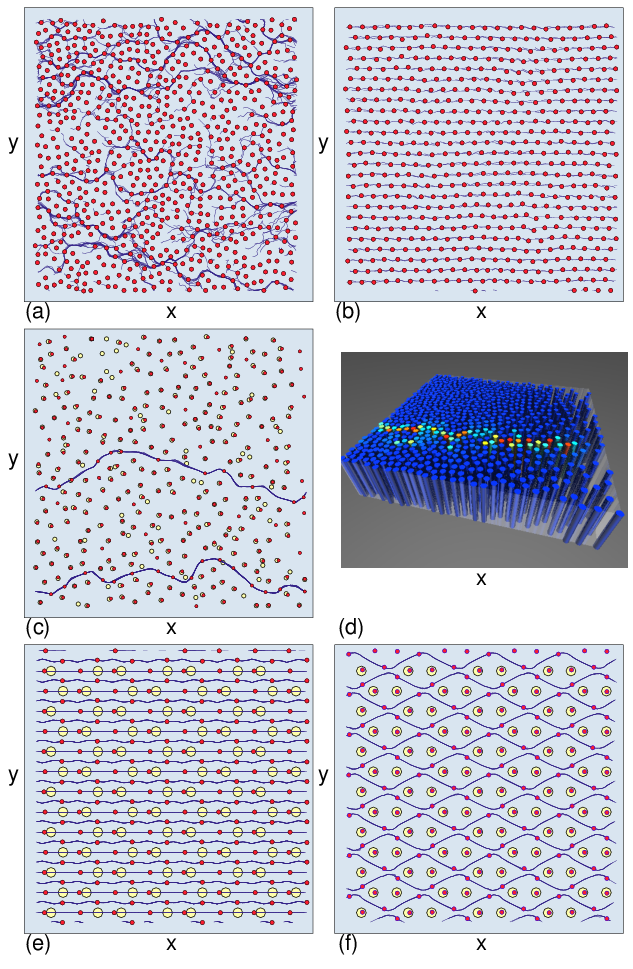}
\caption{
  Different varieties of plastic and elastic flow for
  particles driven in the positive $x$-direction over substrates
  with different pinning strengths.
  Red dots: particle positions; lines: particle trajectories; yellow circles: pinning sites.
  (a,b) Yukawa interacting colloidal particles on a 
  substrate composed of randomly placed pinning sites (not shown)
  just above the depinning threshold.
  (a) For strong coupling to the substrate, given by a
  pinning strength of $F_p=0.25$,
  the depinning transition is plastic and the particles
  change neighbors as they move.
  (b) For a weaker pinning strength of $F_p=0.12$, the depinning is elastic and the
  colloids
  maintain the same neighbors as they move.
  (c-f) Particles representing superconducting vortices.
  (c) An example of finite size
  plastic flow in which the particles move in steady
  state winding channels determined by the substrate disorder.  
  (d) A transient plastic avalanche event in a gradient-driven vortex system,
  with the vortices represented as columns.
  Colors indicate speed of motion, from stationary
  blue particles to rapidly moving red particles.
  (e) A moving crystal in a periodic honeycomb substrate.
  (f) An ordered plastic flow phase in a periodic honeycomb substrate.
  Adapted with permission from:
  (a,b) C. Reichhardt and C.J. Olson,
  Phys. Rev. Lett. {\bf 89}, 078301 (2002).  Copyright 2002 by the American Physical Society;
  (c) C. Reichhardt, C.J. Olson, and F. Nori,
  Phys. Rev. B {\bf 58}, 6534 (1998).  Copyright 1998 by the American Physical Society;
  (e,f) C. Reichhardt and C.J. Olson Reichhardt,
  Phys. Rev. B {\bf 78}, 224511 (2008).  Copyright 2008 by the American Physical Society.
}
\label{fig:1}
\end{figure}

 In this review we focus on particle-based systems which allow
 the possibility of some form of plastic deformation in which
the particles can exchange neighbors.  Under 
certain conditions such as weak disorder, however,
these systems can still display elastic behavior.
For example, in the case of
superconducting vortices or
colloids interacting with a random substrate, when
the coupling between the particles and the substrate
is weak,
the depinning transition is elastic and the particles keep the same neighbors as they move,
as highlighted in figure 1(b) for a 2D assembly of particles with repulsive Yukawa
interactions moving over a weak random substrate.
The image shows that motion just above the depinning threshold is  
coherent, and the particles maintain their triangular ordering. 
For sufficiently strong particle-substrate coupling, 
the nature of the depinning changes and plastic motion emerges in which
a portion of the particles
remain immobile while the other particles can flow around them,
permitting the particles to change neighbors over time.
This is illustrated in figure~1(a) just above the depinning
threshold for the same system in figure~1(b) but with a pinning strength
that is twice as large \cite{19}.
There can also be a dynamical transition at
higher drives between a liquid-like plastic flow state
to a coherently moving or dynamically ordered 
state \cite{20}. 
Sliding friction or tribology is
a related phenomenon, and
various frictional models have been proposed
that take the form of
a collection of particles
siding over an ordered substrate with either fully elastic
motion or with plasticity \cite{9,21}.
Other descriptions of atomic friction involve realistic three-dimensional
all-atom simulations that allow phonon propagation and even electronic degrees 
of freedom to be taken into account \cite{9,22}.
The field of tribology is very large and is not the focus of this review;
we refer the interested reader to several excellent reviews of this field \cite{9,21}.

One of the reasons why the driven dynamics of particles on ordered and disordered 
substrates is such a fascinating field is that it  
has implications well beyond the specific systems
that have been studied.
It provides an ideal testing ground for understanding general issues of 
nonequilibrium phases and nonequilibrium phase transitions;
thus, a review unifying various aspects of the field is particularly useful.
In equilibrium systems, numerous well-developed concepts and procedures
exist for  
characterizing equilibrium states of matter and the
transitions between these states \cite{23,24}.
The different states can be identified by symmetries in the system,
while transitions between 
these states can be characterized  by the breaking of these symmetries or 
the measurement of an order parameter that
becomes finite or goes to zero at the transition point.
Such transitions can be discontinuous or  first order,
with hysteresis across the transition, or
they can be second order, where 
the transition is continuous.
It is also possible to have various mixed first order and second order transitions,
weak first order transitions, 
or simply crossover behaviors. 
Continuous transitions can fall into different universality classes 
which have different  scaling behaviors 
near the transition point.
A particularly powerful result of universality is that even if
different systems have very different microscopic details,
the essential behavior at the transition remains the same
if they are in the same universality class.   
One of the most outstanding open questions in physics and materials
science is whether systems
that are out of equilibrium can also exhibit specific types of
nonequilibrium states, whether there exist well-defined nonequilibrium phase 
transitions between such states, and whether there are
universal properties of these transitions that remain the same
between different systems \cite{25,26}. 
If a unifying framework can be established for
understanding nonequilibrium systems, it would  have
profound implications for statistical physics, condensed matter,
materials science, biological science, information  science, and even the social sciences.

There are many difficulties  in understanding nonequilibrium phases, such as  
which measurable properties of the system could play the role of an order parameter,
what quantities would be diverging, and where exactly the transition occurs. 
Only recently have some systems been experimentally realized
which clearly exhibit nonequilibrium  phase transitions, such 
as transitions between different turbulent states in
a liquid crystal \cite{27} and reversible to irreversible transitions
in periodically sheared dilute colloidal suspensions \cite{28}.
It would be   
highly desirable
to identify additional model systems where such questions could be
readily addressed or in which the
system parameters could be easily controlled.  
Systems that can be represented as individual particles moving over
random and periodic substrates
have both of these properties, making them ideally suited for
testing the concepts of nonequilibrium phases.
As
advances in microfabrication, nanotechnology, optics, and synthesis techniques
continue, more precise control over many aspects of
these systems such as the strength and type of substrate and
the nature of the particle-particle interactions becomes feasible. 
For example, in superconducting systems it is now possible to create tailored
substrates with specific geometries, to control
the ratio of the number of vortices to the number of pinning sites
with an applied magnetic field, and to control the driving force
with an applied current \cite{2}.
In colloidal systems, optical traps can be used to create a periodic substrate in which
the depth of 
the traps  can be readily controlled, and in
many cases the interactions between the colloids can also be tuned \cite{7}.

In this review, we specifically
highlight several different 
systems in which depinning and nonequilibrium phases have been studied,
including vortices in  superconductors interacting with random and periodic substrates in
two and three dimensions,
colloids driven over ordered and disordered substrates,
sliding Wigner crystals, and charge hopping in dot arrays.
We show 
how the different dynamical phases have been identified,
describe the possible order parameters, and examine the roles
of dimensionality, ergodicity breaking, 
transient behaviors, and transport properties.
We also discuss pattern forming systems in which the interactions between
the particles have both attractive and repulsive components, so that
when driven over a substrate the particles form stripe, bubble, glassy, 
and emulsion states.
We examine the recent developments in nonequilibrium phases in
skyrmion systems where the non-dissipative effects from the Magnus term
can dominate the dynamics.
Finally, we briefly discuss other systems
which could potentially be driven over random and periodic substrates,
such as active matter or self-driven particles, and mention
possible connections to jamming phenomena and quantum effects.

\begin{figure}
\includegraphics[width=\columnwidth]{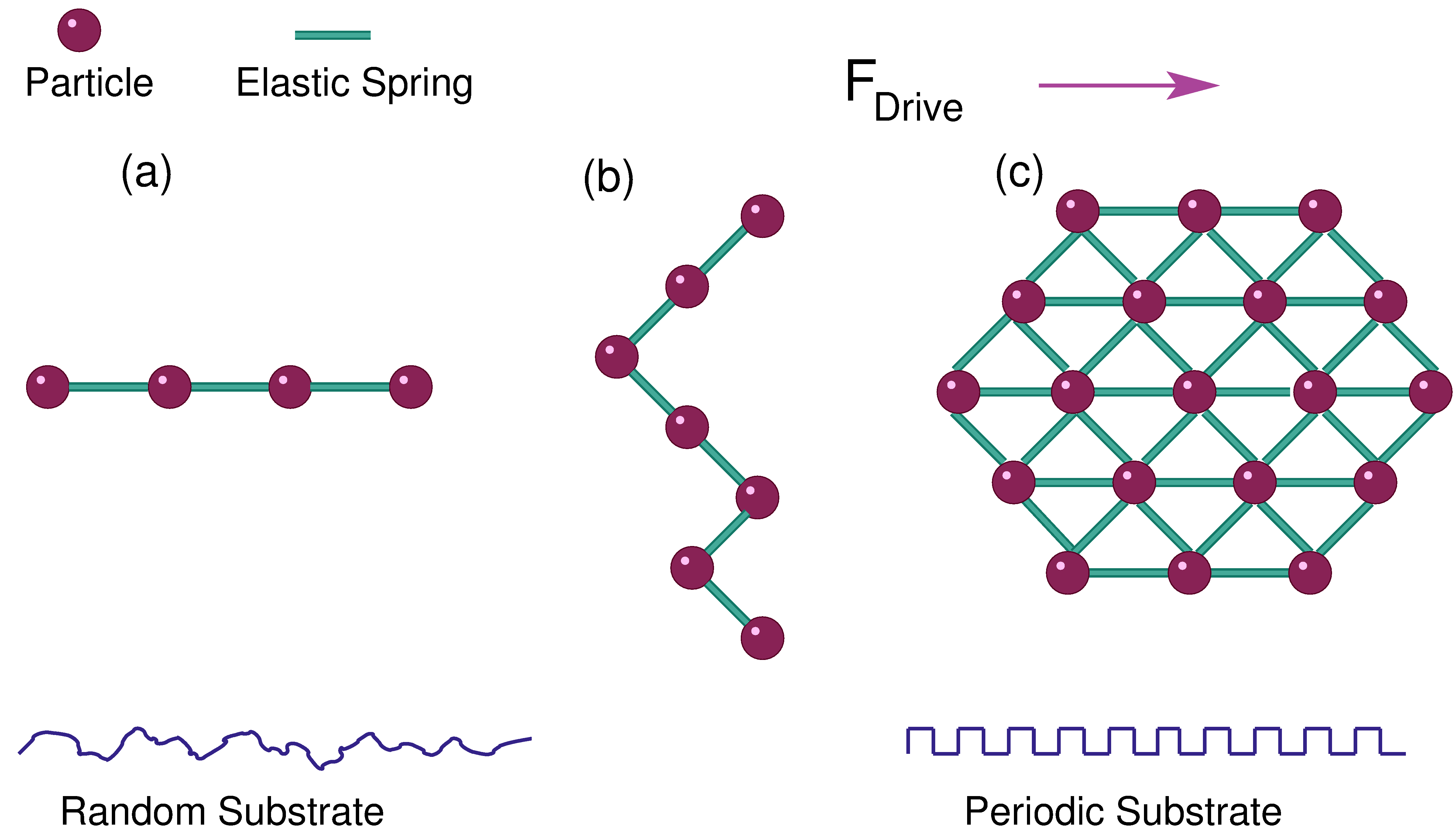}
\caption{ Schematic representations of elastic systems. Lines  
  represent  unbreakable elastic springs, circles are particles,
  and an external driving force $F_{\rm Drive}$ is applied to the particles
  in the positive $x$-direction.  In the remainder of this review we write
  $F_{\rm Drive}$ as $F_D$.
  The particles can interact with a random or disordered substrate (lower left)
  or a periodic substrate (lower right).
  (a) A 1D system with a 1D chain of particles.
  (b) A 2D system with a 1D string driven perpendicularly to its length
  to model a domain wall or interface.
  (c) A 2D system with a 2D elastic lattice.
  Generalizations can also be made to higher dimensions.
  These systems exhibit an elastic depinning transition from a pinned state
  to a moving state in which the particles maintain their same nearest neighbors.   
}
\label{fig:2}
\end{figure}

This review is organized as follows.  In Section 2 we provide an overview
of depinning transitions and dynamic phase transitions, as well as the
classes of systems that exhibit such phenomena.
In Section 3 we describe experimental, theoretical, and 2D simulation
studies of plastic and elastic depinning and
dynamical ordering transitions for vortices in type-II
superconductors.
In Section 4 we discuss depinning
and first order dynamical phase transitions
in 3D models of superconducting vortex systems,
as well as phase locking effects that occur when
an ac drive is added to the dc driving force.
In Section 5 we describe depinning and first order dynamical
phase transitions in other layered systems
with many or few layers, including
charge density waves (CDWs), mean field models, and coupled 1D channels.
In Section 6 we cover depinning and dynamic phases for
superconducting vortices and colloids
moving over periodic substrates, including both egg carton and
muffin tin substrate types.
In Section 7 we discuss dynamic phases that arise for
particles driven over quasiperiodic substrates.
Section 8 covers depinning and dynamics in charge transport for 2D Wigner
crystal systems and Coulomb-coupled metallic dot arrays.
In Section 9 we describe the depinning and dynamics of
magnetic skyrmions, where Magnus effects play an important
role in the behavior.
In Section 10 we connect depinning phenomena with jamming transitions
observed in 2D packings of hard disks.
Section 11 describes the depinning and dynamics in systems
with competing attractive and repulsive particle-particle interactions,
where pattern formation occurs, including pairwise and
non-pairwise competing interactions, phase field models, and driven
binary systems.
In Section 12 we discuss the connection between depinning transitions
and nonequilibrium absorbing phase transitions.
In Section 13 we briefly describe other systems in which
depinning and dynamical phases have been studied, and indicate some
future directions for study.
There is no uniform terminology in the literature for driven systems and depinning
phenomena.  Throughout this review we use the symbol $F_D$ to represent a
driving force, $V$ to represent the average velocity of the particles,
$F_p$ to represent the pinning strength, and $F_c$ to represent the
critical depinning threshold.
In each
section and figure caption we indicate how these symbols map onto the physical
driving forces and velocities in a given system, and to the symbols used in a
given figure.

\section{Types of Systems and  Depinning Phenomena}

\begin{figure}
\includegraphics[width=\columnwidth]{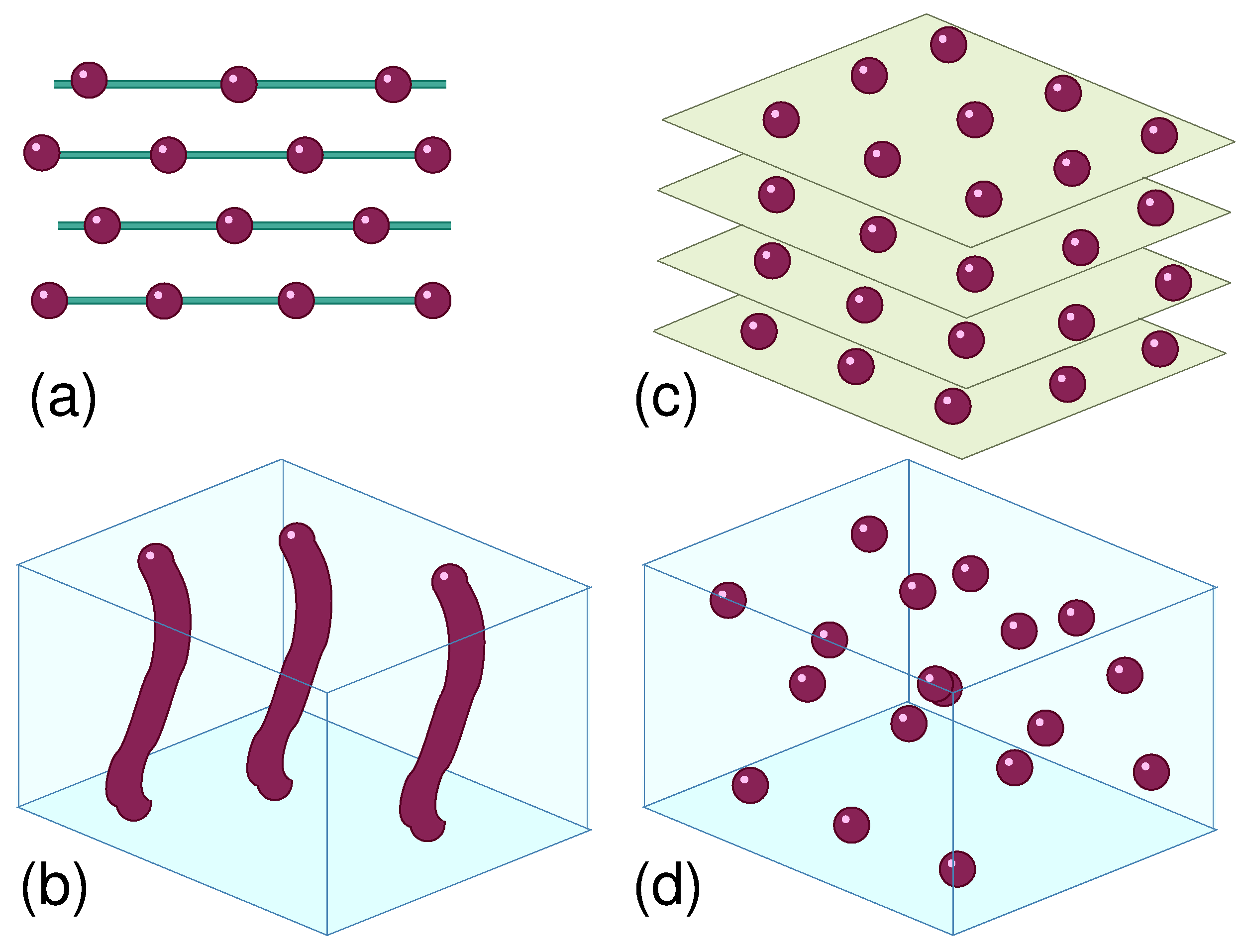}
\caption{ Schematic representations of systems that allow for plasticity or exchange of
  particle neighbors.
  Lines represent unbreakable elastic springs and circles are particles.
  (a) A 2D system of 1D layers.  The particles keep the same
  neighbors in each layer but adjacent layers can decouple.
  A 2D system in which the particles can exchange all their neighbors is illustrated
  in figure~\ref{fig:1}(a).
  (b) A 3D system of elastic lines with endpoints in a pair of parallel planes. Plastic motion
  with exchange of neighbors can occur parallel to the planes, but the lines remain
  elastic along their length.
  (c) A layered 3D system of particles, in which particles can exchange neighbors within
  a given plane but remain confined by that plane.
(d) An isotropic 3D system where particles can exchange neighbors in all directions.  
}
\label{fig:3}
\end{figure}

\subsection{Elastic depinning}

At a depinning transition, a system changes from a state in which all of the particles
are stationary to one in which some or all of the particles are moving
under the influence of a driving force.
Depending on the nature of the coupling between the particles, their geometry,
and the nature of the substrate, as illustrated in figures~\ref{fig:2} and \ref{fig:3},
different types of depinning transitions can occur.
We first consider elastic depinning, where the moving state above the depinning
transition is perfectly elastic, with all the particles moving and no plastic flow 
so that all particles maintain the same neighbors as a function of time.
Figures~\ref{fig:1}(b) and (e) illustrate systems that have
depinned elastically.
For models of
harmonically coupled particles driven over random or periodic substrates,
which could be in one, two, or three dimensions, 
plasticity does not occur, and the depinning transition is elastic. 
In figure~\ref{fig:2} we show schematic examples of elastically coupled systems
in which the particles 
interact via unbreakable  elastic springs and are driven by a force
$F_{\rm Drive}$ over some form of substrate.   We call the driving force
$F_D$ in the remainder of this review.
Individual potential minima in the substrate,
which we call pinning sites, exert a maximum force on
individual particles that we refer to as a pinning strength $F_p$.
Figure~\ref{fig:2}(a) illustrates a
one-dimensional (1D) system with a chain of coupled particles,
figure~\ref{fig:2}(b) shows a 2D system containing a 1D string driven perpendicular
to its length to model a domain wall or moving interface, and
figure~\ref{fig:2}(c) shows a 2D system with a 2D array of elastically coupled particles.
Similar features appear in
three-dimensional (3D) systems (not shown), which can contain 1D strings, 2D
membrane-like arrays, or 3D lattices of elastically coupled particles.
Inclusion of a random or periodic substrate introduces a depinning threshold $F_c$;
for drives $F_{D}<F_c$ the particles exhibit no steady state motion.
As the driving force is increased from zero, 
the string, membrane, or lattice becomes increasingly rough as the
depinning threshold is approached.
At $F_c$, depinning occurs and there is a transition to
an elastic sliding phase.
In most cases, the roughness of the sliding particle structure
decreases for sufficiently high drives when the perturbations from the pinning sites
become negligible.
Since the particles are elastically coupled,
they keep their same neighbors over time,
so that the system can be viewed as an unbreakable rubber sheet 
pulled over a carpet.

\begin{figure}
\includegraphics[width=\columnwidth]{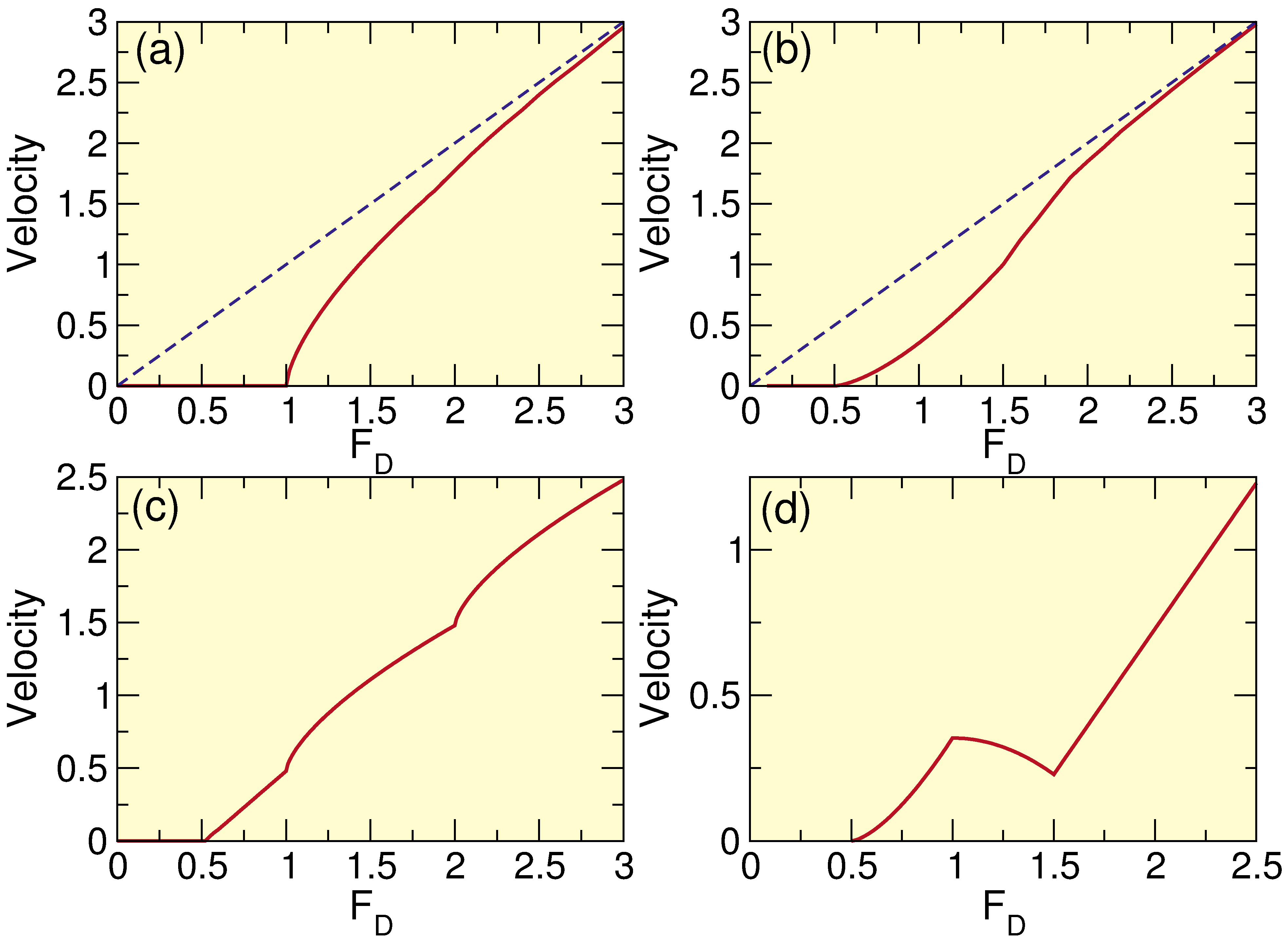}
\caption{ Schematics of typical velocity-force curve behaviors
  for particle assemblies driven over random or periodic substrates. $V$,
  written as Velocity on the figure panels, is the
  average velocity of the particles,
  $F_D$ is the driving force, and $F_c$ is the depinning threshold.
  (a) For elastic depinning, near the depinning threshold
  $V \propto (F_D- F_{c})^\beta$ with $\beta = 2/3$. The dashed
  line indicates the substrate-free velocity-force response with $V\propto F_D$.
  (b) For a system similar to that shown in figure~\ref{fig:1}(a), where
  plastic depinning occurs,
  near the depinning threshold
  $V \propto (F_D- F_{c})^{\beta}$ with $\beta = 1.5$.
  The velocity-force curve has an overall S-shape, with
  a transition to a dynamically reordered state at high drives.
  (c) A system with multiple dynamical transitions, marked by the kinks in
  the velocity-force curve.
  (d) A system in which a change in the dynamical flow occurs that causes the average
  particle velocity to decrease with increasing $F_D$ over a certain range of $F_D$,
  producing negative differential conductivity.
}
\label{fig:4}
\end{figure}

The velocity $V$ at a given value of $F_D$ 
is taken to be the steady state time-averaged velocity of all particles
in the system, $V=\langle N^{-1}\sum_i^N ({\bf v}_i\cdot {\bf \hat x})\rangle$
for a sample containing $N$ particles
which each have velocity ${\bf v}_i$ that are subjected to a driving force applied
along the $x$ direction.
The curve $V$ versus $F_D$ is referred to as a velocity-force curve.
The velocity-force curves for elastic depinning transitions typically exhibit 
features similar to those shown schematically in figure~\ref{fig:4}(a).
There is a critical
value $F_c$ of the driving force corresponding to the highest drive that can be applied
for which the system remains pinned with no steady-state motion.
This is termed the critical depinning threshold.
For overdamped systems, the average particle velocity $V$ in the absence of a substrate
increases linearly with $F_{D}$ according to
$V = F_{D}/\eta$, where $\eta$ is the damping term.
In the presence of a substrate, 
once the particles are moving for drives above depinning $F_D>F_c$, 
$V$ is normally smaller than the clean limit value, particularly
near the depinning threshold $F_c$.
In figure~\ref{fig:4}(a) the dashed line indicates
the clean or substrate-free velocity-force curve
for a system with $\eta = 1.0$.
As $F_{D}$ increases,
the value of $V$ for a system with a substrate gradually approaches the clean limit
value. 

Fisher \cite{29} was one of the first to show that 
the pinned to sliding transition in an elastic system has
similarities to a second order equilibrium phase transition,
and exhibits critical phenomena in which certain quantities obey
power law scaling close to the critical point $F_{c}$.
Here, the velocity-force curve takes the form
\begin{equation}
V \propto (F_D - F_{c})^\beta
\end{equation}
where $\beta$ is a critical exponent \cite{29},
while at high drives,
\begin{equation}
  V \propto F_D .
\end{equation}
The power of the approach introduced by Fisher is that it suggests that
the class of
nonequilibrium systems that exhibit elastic depinning can be understood using an
approach similar to that applied to equilibrium critical phenomena,
and that if critical exponents and university classes can be identified,
the same critical exponents could arise in many different types of systems 
even though the microscopic details and size scales of these systems
could be vastly different.
If $F_{c}$ is a critical point, other quantities such as a correlation length $\xi$ would
also diverge near depinning,
$\xi \propto (F_D- F_{c})^{-\nu}$, with critical exponent $\nu$.
Since this initial proposal, there has 
been extensive work in identifying
the critical  exponents for
depinning in
CDW models,
such as the thorough study  by 
Myers and  Sethna who find $\beta = 0.45$ in dimension $d = 1$,
$\beta=0.65$ and $\nu=0.5$ in  $d = 2$, and $\beta=0.8$ in $d = 3$ 
\cite{30}. 
In addition to the motion in the sliding phase, transient particle
rearrangements can occur below the depinning threshold if the external drive
is increased by some increment and the particles adjust their positions
in order to balance the driving and pinning forces.
The transient time over which this motion occurs also diverges as the depinning
threshold is approached.
The rearrangements below $F_{c}$ can be viewed as avalanches of motion
in which correlated groups of particles move for a time before becoming
repinned, and the spatial size of the jumps of these particles
can provide a measure of the correlation
length \cite{30}.

There are many complications to understanding 
critical behavior in driven particle systems.
One example is the identification of the critical depinning threshold $F_{c}$.
In experiments, the driving force is increased at a certain rate; however,
if there are transient times associated with criticality
at depinning, then at some point these transient 
times become larger than the inverse rate at which the driving force is being swept,
so the system no longer has enough time to settle back into the pinned phase and 
appears to depin at a drive lower than the true critical point.
Additionally, if inertial effects are included in the dynamics of the system,
the driving rate dependence may be even stronger, or 
new types of dynamics could appear that change the criticality
of the system or eliminate it altogether. 
Other possible effects include retardation in modes of dissipation,
which could cause local buildups of heat or strain, as well as nonuniformity
in the substrate potential itself.
Many real materials can have large inhomogeneities in the effective pinning strength and 
in the spatial distribution of the pinning sites,
which means that as the depinning threshold is approached, one part
of the sample may reach the critical threshold before the rest of the sample,
thus making a scaling analysis difficult to perform.
Instead of approaching the depinning threshold from below, it is possible to decrease
the driving force from high values in order to approach the threshold from above;
however,
it is possible that the scaling exponents could be different on
each side of the transition \cite{30}.
Many  models for
elastic depinning are in the strongly overdamped limit and include the assumption
that the damping constant is fixed;
however, in experiments it is possible for the damping
constant itself to have a drive dependence 
if additional damping modes become excited in the sliding state.  

There are still many open questions in elastic depinning,
particularly regarding the role of inertia and hysteresis effects.
There
are a growing number of biological systems such as membranes and
networks that exhibit highly nonlinear elasticity, and in many cases these
biological objects are moving over some type of substrate,
so another direction for future research is to study  depinning in systems
with nonlinear elasticity  \cite{31,32}.
Recent studies considered wrinkling transitions and crumpling in elastic sheets \cite{New2}, 
and it would be interesting to see whether the ideas of wrinkling transitions
could be applied to depinning of elastic sheets and membranes or 
to elastic depinning in curved geometries.

\subsection{Plastic depinning}

This brings us to the next level in
complexity in depinning, where  some form of plasticity can occur
either by the breaking and reforming of physical bonds between neighboring
particles or by a change of neighbors for particles without explicit bonds.
In the elastic models illustrated in figure~\ref{fig:2}, if the pinning
is very strong, it can induce such large elastic distortions in the particle assembly
that a portion of the particles can move past one
another a distance larger than the equilibrium lattice constant $a$,
so that the elastic springs connecting the particles would likely break at
one or more locations.
In experimental systems, where the number of particles can be very large, 
it is likely that
even for weak random pinning there can be rare regions of strong disorder  that
can produce some plasticity.
Thus, it is of paramount importance to
understand plastic deformation at depinning, and
it has even been argued that plasticity is relevant in most CDW systems \cite{33}.

In 2D, the simplest model in which it is realistic 
to include plastic effects consists of unbreakable
1D elastic chains aligned parallel to each
other and coupled by breakable bonds, as illustrated in figure~\ref{fig:3}(a).
Here  different depinning processes as well as a variety of
dynamical phases can occur.
If the pinning is weak, local variations in pinning strength are
unimportant and the entire system can
behave elastically and depin in a single step, as shown schematically in
figure~\ref{fig:4}(a).  For strong pinning, local variations in pinning strength cause
individual 1D chains to have different depinning thresholds $F_c^i$ so that
the depinning is now plastic, and only
the subset of chains for which $F_D>F_c^i$ are able to move.
The moving chains slide past the remaining pinned chains,
resulting in a coexistence of pinned and flowing chains.
The depinning threshold of the entire system
is equal to the smallest $F_c^i$ in the system, making plastic depinning an
example of an extreme value event.
As the driving force is increased, more and more chains begin to move.
When $F_D>{\rm max}(F_c^i)$ and the depinning threshold of every chain
has been exceeded,
the sample enters a state in which all the 1D chains are 
in motion
but travel with different average velocities, so that 
they continue to slide past one another.
The difference in velocities arises since at a given $F_D$, different chains are
different distances $F_D-F_c^i$ above their depinning threshold.
At high enough drives, $F_D-F_c^i$ is large for all chains and the effective pinning
forces experienced by the moving chains become
weak enough that 
the chains can dynamically recouple and form a moving elastic solid.

The existence of
different dynamical phases
in the unbreakable 1D elastic chain system
can produce signatures in the velocity-force curves
as shown schematically in figure~\ref{fig:4}(c), where only a portion of the particles are
moving just above the depinning threshold, and cusps appear
as additional chains of particles begin
to move or as neighboring chains dynamically recouple.
The individual particle velocity distribution function $P(v_i)$ just above the onset of
sliding can be bimodal if there 
is a coexistence of pinned and moving channels or
if the channels form distinct groups that slide past each other,
while $P(v_i)$ becomes sharply peaked above the dynamical recoupling transition when the
motion throughout the sample becomes coherent.
The features in  
the velocity-force curves and velocity distribution functions
can depend strongly on dimension, disorder, pinning strength, and
inter-particle coupling.
The system size is also important.
For example, if plastic events
occur at a specific size scale, then in systems of infinite size,
the elastic response at small scales is washed out,
while in small systems the
behavior might appear purely elastic, 
and in intermediate
sized systems, elastic and plastic behaviors can compete but the plastic events would
be a dominant feature, so that obtaining a meaningful average velocity would be
very problematic.
Additionally, as a function of driving force,
the length scales at which plastic events occur can change.

At the next level of complexity in 2D, the particles are free to plastically deform in
any direction, such as in the samples illustrated in figure~\ref{fig:1}.
A system of this type could have isotropic or anisotropic coupling between the particles.
Figure~\ref{fig:3}(b) shows that
another class of systems that allow 2D plastic deformation is
an assembly of elastic line objects with their endpoints confined to two parallel planes.
Plastic motion with exchange of neighbors can occur parallel to the planes, but the lines
remain elastic along their length.
In figure~\ref{fig:3}(c) we illustrate a 3D layered system where the particles are
confined within a given layer but can exchange neighbors within that layer.
Finally, figure~\ref{fig:3}(d) illustrates an isotropic  system of 
point particles where plasticity can occur in all directions equally.
In addition to the different geometries for the particle arrangements,
the substrate disorder can also have different spatial features
which depend on dimensionality, such as 3D columnar, splayed, or planar defects.

Although all the systems highlighted in figures~\ref{fig:1} and~\ref{fig:3}
exhibit plasticity, it is an open question
whether there are distinct types of plastic flow phases,
whether transitions between different types of plastic flow can occur,
and how such phases could be characterized.
For example, figure~\ref{fig:1}(a) illustrates a strongly fluctuating plastic
flow, figure~\ref{fig:1}(c) shows a steady state plastic flow through fixed
winding channels, figure~\ref{fig:1}(d) shows transient plastic flow in an
avalanche event for a very slowly driven system, and figure~\ref{fig:1}(f)
illustrates an ordered plastic flow state.  All of these flows are plastic,
but they may have features that would allow them to be subdivided
into separate types of plastic flow.
The problem of plasticity is generally considered one of the most
outstanding problems in materials 
science, in part because systems that exhibit plasticity 
can have both solidlike and liquidlike properties at the same time.

One well-defined question in systems with pinning  
is whether, under certain conditions, 
plastic depinning is associated with dynamical critical phenomena similar to that
found for elastic depinning. 
For example, in the strong pinning limit where the particles are highly disordered
in the pinned state, the flow just above depinning follows
channels or riverlike features
of the type illustrated in figure~\ref{fig:1}(a).
Such plastic channel flow
behavior has been observed in vortex systems \cite{19,34,35,36,37}, 
Wigner crystals  \cite{38,40},  
and colloidal systems \cite{6,19}.  
For plastic depinning in 2D systems in the strongly amorphous limit,
the velocity-force curves also appear to exhibit a scaling of 
the form $V = (F_D-F_{c})^\beta$ with $\beta > 1.0$.
This is in contrast to elastic systems in 2D and higher dimensions
where $\beta < 1.0$.
As a result, velocity-force curves associated with plastic depinning have a concavity
opposite to that found for elastic depinning, as shown schematically
in figure~\ref{fig:4}(a,b) which contrasts
a representative example of 
a typical velocity-force curve for elastic depinning with $\beta=2/3$ with 
a plastic depinning velocity-force curve with $\beta = 1.5$.
These results suggest that if plastic depinning has critical properties, it falls into a different
university class than elastic depinning;
however, the existence of such criticality and its exact nature has not been verified,
as measured exponents span a wide range of values from
$\beta=1.25$  to $\beta=2.5$  \cite{19,20,35,40,39,41,42}.
The variation in the exponents could result from
strong finite size effects, transient effects,
or differences in the nature of the particle-particle interactions,
or it could indicate that multiple 
types of criticality occur at plastic depinning due to the different
dominant degrees of freedom,
such as glide and climb of dislocations, kinetic barriers,
and the distinction between long range interactions of the particles and long range  
interactions between the emergent topological defects.
This also suggests that new approaches to understanding plastic depinning will be useful
to understand whether plastic depinning possesses truly universal features.

Many of the same
challenges that are faced in general studies of
plasticity in crystalline and amorphous solids also
arise in the plastic depinning context.
Just one example of a challenge in materials plasticity is
the fact that stress-strain curves in  amorphous systems 
often have different properties depending on how the material is prepared,
such as a deep quench compared to a slow quench.
It is likely that similar effects occur for plastic depinning; however, little is known
about how this would arise. 
Another challenge in plastic depinning  
is that there can be differences between the response of systems with
intermediate-range order and those in the strongly  amorphous limit.
In materials science, crystalline materials can exhibit plastic behavior
called crystal plasticity where
a well defined
number of topological defects such as dislocations or grain boundaries
are present and
where critical behavior appears near yield that is similar to the behavior observed  near
depinning \cite{43,44}.    This crystal plasticity is distinct from the plasticity that
arises in amorphous systems where topological defects are not well defined.
For the case of systems that exhibit plastic depinning  transitions,
it is not known whether a similar distinction can be drawn between depinning that
has properties similar to those associated with
crystal plasticity and depinning that is amorphous in nature.

\subsection{Dynamical transitions}

Despite the challenges in studying plasticity,
it appears that plastic depinning of particle systems 
in the strongly amorphous limit
produces some generic features in the transport and dynamics 
even for very different types of interactions between the particles.
Figure~\ref{fig:4}(b) shows a schematic of a typical
velocity-force curve illustrating a pinned phase at low $F_D$,
plastic depinning with $V \propto (F_D - F_{c})^\beta$ and $\beta = 1.5$,
and an overall S-shape produced by the transition to Ohmic or linear velocity at
high drive when the substrate becomes ineffective, associated with a
peak in $dV/dF_{D}$ just below the Ohmic regime. 
The presence of a peak in $dV/dF_{D}$ often indicates that
there is a transition within the moving phase from a plastic or 
liquid state into a moving solid where the particles are moving so rapidly
over the underlying pinning landscape that the particle-particle interactions
dominate over the particle-pinning interactions, causing the particles to adopt a
structure similar to that found in the substrate-free limit, which could be a triangular
lattice in 2D.
This suggests that there can be two generic transitions associated with
plastic depinning: the depinning transition itself,
and the dynamical ordering transition at higher drive. 
Such pairs of transitions have been observed in experiments and simulations
of vortices in type-II superconductors \cite{20,36,41,45},
driven Wigner crystals \cite{40}, colloids \cite{46},
frictional studies \cite{47},
skyrmions \cite{48}, and pattern forming systems driven over random substrates \cite{8}. 
The dynamically
ordered moving state can still be affected by the
pinning even in the high drive limit so that the reordered structure 
is generally not an isotropic crystal but rather an anisotropic moving crystal \cite{49},
moving smectic \cite{50,51,52,53},
or even anisotropic liquid state \cite{54} depending on the strength of the disorder, 
the drive, and the dimensionality.
An open issue is whether the dynamical ordering transition can be viewed 
as a nonequilibrium phase transition and, if so, whether the type of nonequilibrium
transition can be identified.
For example, the ordering transition might have the same properties as
the criticality observed for elastic depinning \cite{29}
or an absorbing phase transition such as directed percolation \cite{25},
or it might be possible to effectively map it
to some class of equilibrium phase transition by replacing the
dynamical fluctuations with  an effective temperature \cite{23,24}.

Plastic depinning can also occur for
particles driven over periodic substrates, as illustrated in
figure~\ref{fig:1}(f), 
and the type of dynamics that occur depend on whether the particle arrangement
is 
commensurate or incommensurate
with the substrate \cite{2,7,55,56,57,58}.
There are number of examples of periodic substrate systems
in which the  depinning and sliding dynamics
appear to exhibit very well-defined transitions between distinct plastic flow phases,
suggesting that such systems could be ideal for testing methods for
characterizing different nonequilibrium
flow states \cite{58,59,60,61,62}.
Figure~\ref{fig:4}(c) shows an example schematic
velocity-force curve for a system
that depins plastically and then exhibits transitions
between different flow states at higher drives, which are marked by
kinks in the curve.
The jump up in $V$ at the transitions can arise either because more particles have
become mobile or because there has been a change from disordered flow to a more
coherent flow.
Figure~\ref{fig:4}(d) shows schematically that in some cases
the average particle velocity can actually decrease with increasing drive due
to a change in the flow pattern from a more coherently moving state to one 
with stronger fluctuations, or to a state in which more particles
are pinned.

\begin{figure}
  \includegraphics[width=\columnwidth]{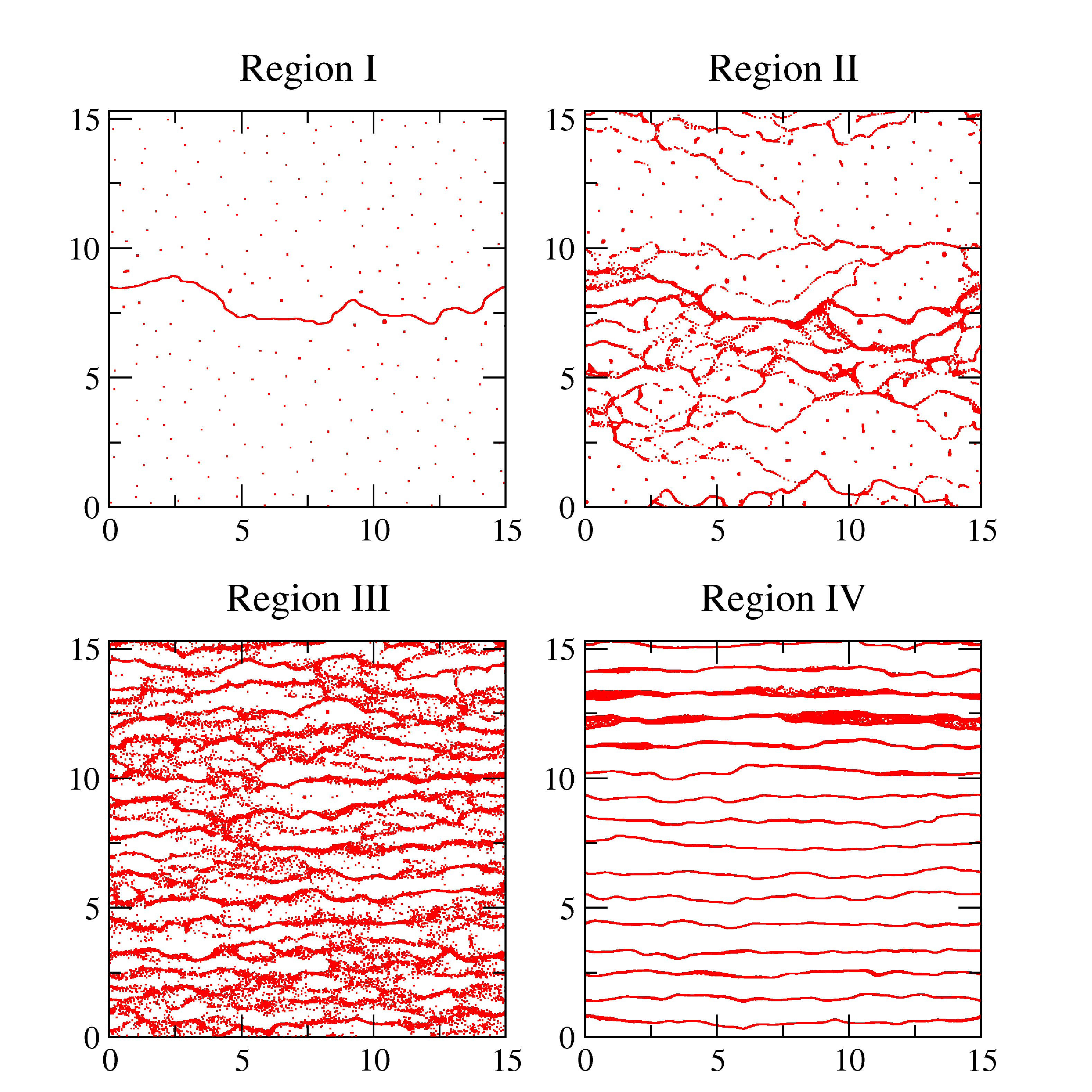}
\caption{
  Trajectories of moving
  superconducting vortices from a 2D simulation in which there is a random
  substrate and a driving force is applied in the positive $x$ direction,
  to the right in the figure.
  (a) Region I just above depinning.
  (b) Region II at high drives where disordered channels form.
  (c) Region III, where there is some additional nematic or smectic type
  ordering of the trajectories.
(d) Region IV, where the system is in a dynamically reordered state.  
Reprinted with permission from Y. Fily, E. Olive, N. Di Scala, and J.C. Soret,
Phys. Rev. B {\bf 82}, 134519 (2010).  Copyright 2010 by the American Physical Society.
}
\label{fig:Fily6}
\end{figure}

\begin{figure}
  \includegraphics[width=\columnwidth]{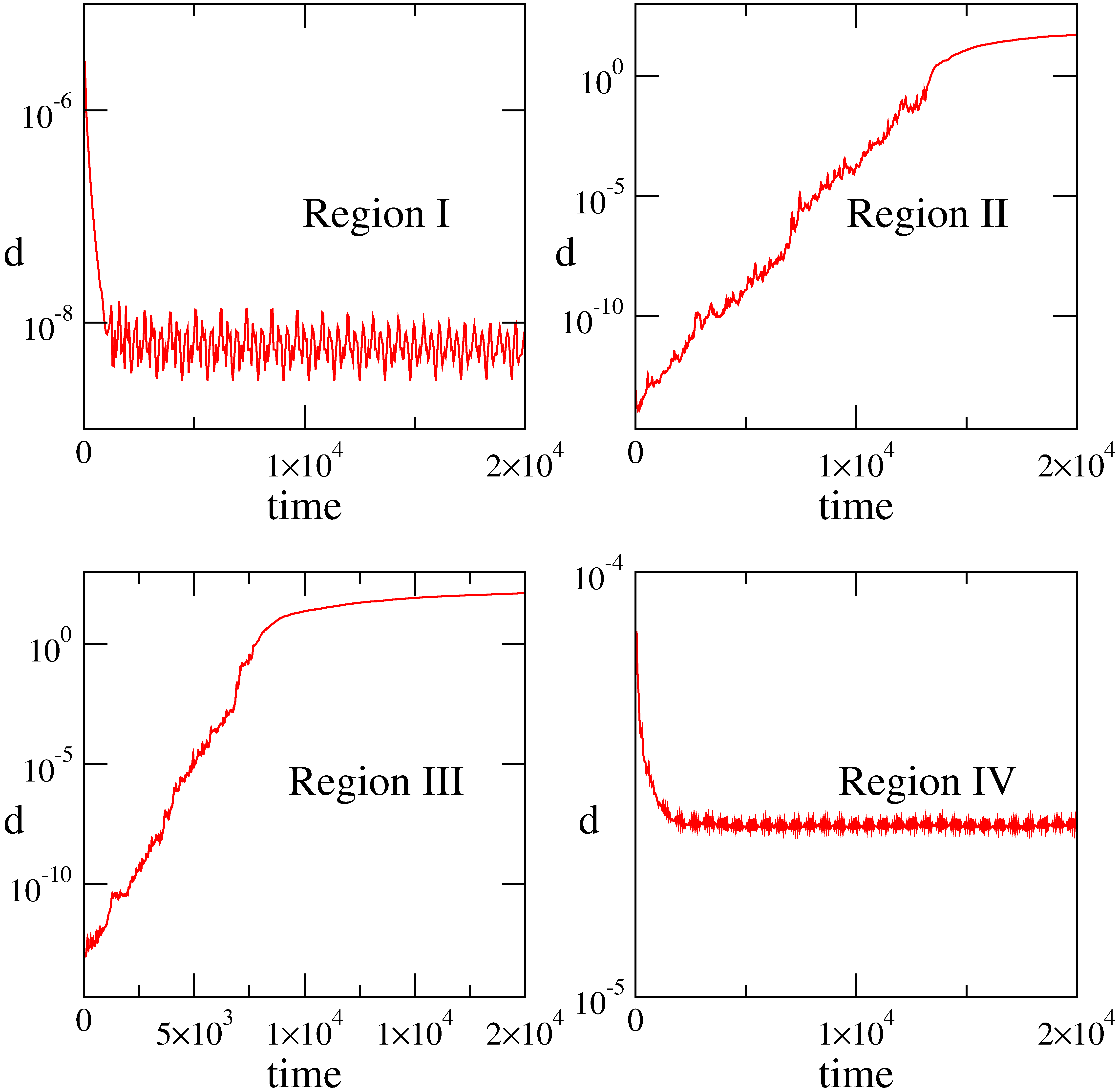}
  \caption{
    For the superconducting vortex system in figure~\ref{fig:Fily6}, the
    distance $d$  between two neighboring particle flow trajectories
    as a function of time in
  Region I,  which is non-chaotic, Regions II and III, which are chaotic,
  and Region IV, which is non-chaotic. 
Reprinted with permission from Y. Fily, E. Olive, N. Di Scala, and J.C. Soret,
Phys. Rev. B {\bf 82}, 134519 (2010).  Copyright 2010 by the American Physical Society.
}
\label{fig:Fily7}
\end{figure}

An example of some of the complications in analyzing plastic depinning 
and the possibility of different kinds of plastic flow states 
appears in the work of Fily {\it et al.} \cite{41} who study
the depinning of 2D superconducting vortices from a random substrate.
They observe combinations of elastic non-chaotic flow and
plastic chaotic flow, making it
difficult to establish the true value of the critical depinning threshold $F_{c}$.
Figure~\ref{fig:Fily6}
shows the vortex trajectories in this system.
The  depinning transition is associated with
the opening of isolated channels of motion, labeled Region I in figure~\ref{fig:Fily6}(a).
At higher drives, as shown in figure~\ref{fig:Fily6}(b) and labeled Region II,
multiple channels appear that change as a function of time.
 Although both Region I and Region II are plastic flows since they
both exhibit a coexistence of moving and pinned particles, the two
regions have very different dynamics in phase space,   
as illustrated in figure~\ref{fig:Fily7} which shows the value of
\begin{equation}
d(t) = \sqrt{ \sum^{N}_{i=1} |{\bf r}^{1}_{i}(t) - {\bf r}^{2}_{i}(t)|^2} .
\end{equation}
This quantity measures the difference in
phase space of  two  neighboring trajectories $[ {\bf r}^1_{1}(t),...,{\bf r}^{1}_{N}(t)]$ 
and $[{\bf r}^{2}_{1}(t),....,{\bf r}^{2}_{N}(t)]$.
In Region I, figure~\ref{fig:Fily7}(a)
shows that $d$ rapidly drops to  a small constant value,
indicating that the system has entered a closed orbit in phase space and is
non-chaotic,
while for Region II in figure~\ref{fig:Fily7}(b),
$d$ increases exponentially with time, indicating chaotic behavior with
a positive Lyapunov exponent and diverging trajectories.
The saturation in $d$ in the chaotic region
occurs at the point at which $d(t)$ becomes of the same order
as the size of the chaotic attractor.  
This result indicates that the flowing state above depinning can have
characteristics of both chaotic plasticity and non-chaotic plasticity.
In Region I the vortices move in an isolated channel, so this region can also
be described as an effectively 1D system of elastically coupled particles that
undergoes an elastic depinning transition while the rest of the system remains
frozen.
As more channels begin to flow with increasing drive,
they initially also behave elastically, but
at some point the channels begin to intermingle, producing
a transition to a chaotic or plastic flow phase.
This raises
the question of how to accurately determine
the critical depinning threshold value $F_{c}$ since there are now 
two effective  depinning transitions,
one at $F_{c1}$ for the non-chaotic Region I flow, and one at
$F_{c2}$  for the onset of chaotic plastic flow.
Another issue is whether the difference between
$F_{c1}$ and $F_{c2}$ changes with system size
and if, in the infinite size limit, there is only a single transition with a unique value
of $F_c$.
Figure~\ref{fig:Fily6}(c) shows the trajectories at even higher drives in Region III,
where there are more channels and the
flow has a more nematic or smectic type character.
Although this flow differs in appearance from the Region II flow,
figure~\ref{fig:Fily7}(c) shows that $d(t)$ grows exponentially,
indicating that the flow is still chaotic.
It is not 
clear whether Region III is truly a separate flow phase or just a crossover from
the Region II flow. 
At higher drives, the flow transitions into the static 1D flow channels of
Region IV, illustrated in 
figure~\ref{fig:Fily6}(d),
and
$d(t)$ saturates at a low value as shown
in figure~\ref{fig:Fily7}(d), indicating that Region IV is non-chaotic.

\begin{figure}
  \includegraphics[width=\columnwidth]{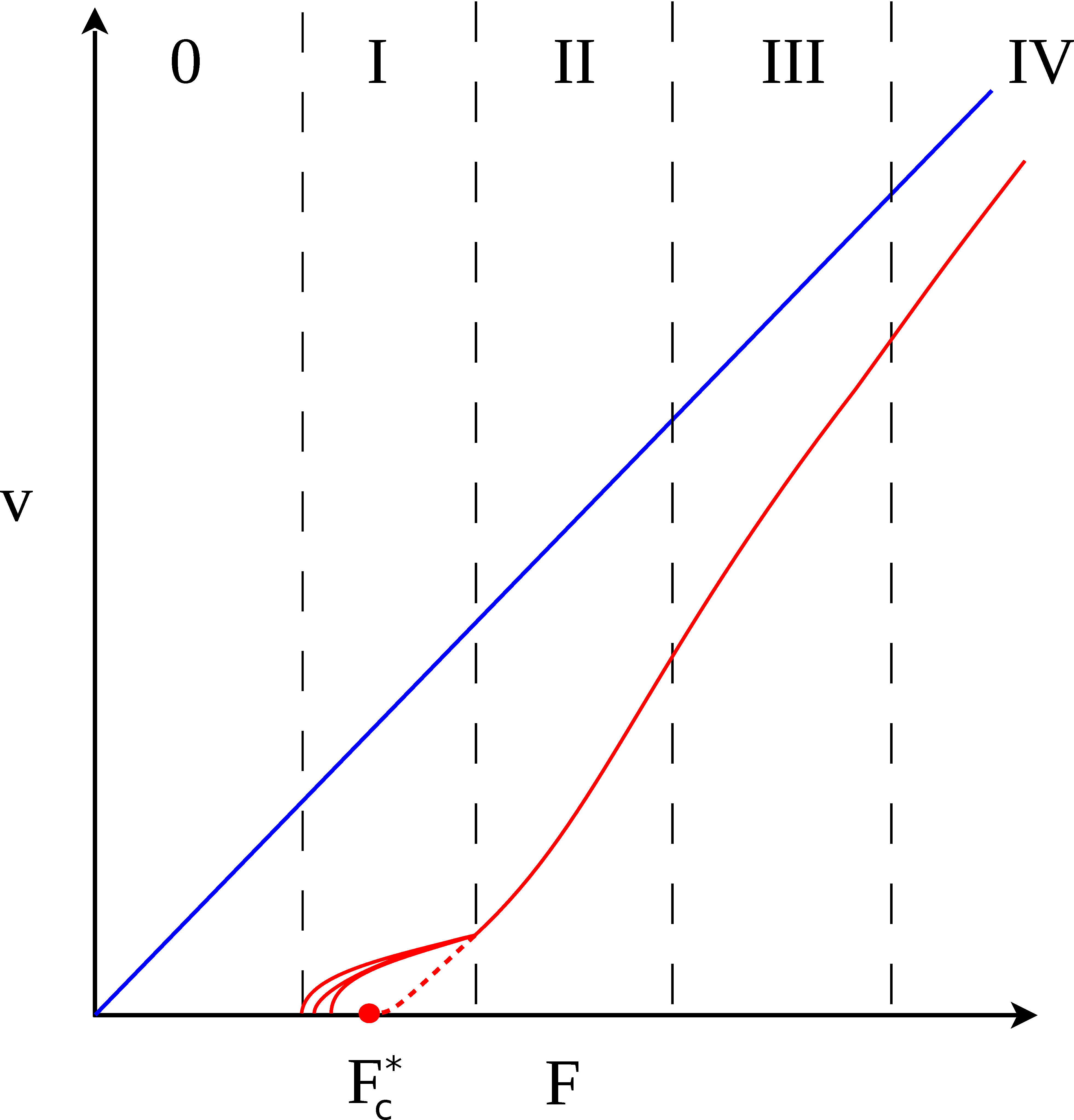}
\caption{ 
  A schematic of velocity-force curves (red lines) from the
  superconducting vortex
  system in figures~\ref{fig:Fily6} and \ref{fig:Fily7} highlighting where
  the transitions from Regions 0 through IV occur, where Region 0 is the
  pinned phase. The blue line is the substrate-free velocity-force response.
  Here $V$ is velocity and $F$ corresponds to the driving force $F_D$, while
  $F^{*}_{c}$ denotes the point at which the extrapolated onset of the
  chaotic flow phase would occur.
  The actual onset of Region I occurs below $F^{*}_c$, indicating
  that there is a regime in which the
  velocity-force curves have elastic rather than plastic features.
Reprinted with permission from Y. Fily, E. Olive, N. Di Scala, and J.C. Soret,
Phys. Rev. B {\bf 82}, 134519 (2010).  Copyright 2010 by the American Physical Society.
}
\label{fig:Fily8}
\end{figure}

Figure~\ref{fig:Fily8} shows a schematic of the
velocity-force curve for this system, where the 
upper line denotes the velocity-force response in a clean system without
pinning and where the locations of the different regions are indicated.  Region 0
corresponds to the pinned phase.
The dashed line shows the point $F^*_c$ at which the onset of the chaotic
plastic flow phase would be expected to occur based on an extrapolation
of the velocity-force curve;
however, the true depinning threshold $F_c<F^*_c$, and since the motion in
Region I is effectively elastic, the velocity-force curve initially
exhibits an elastic scaling exponent of
$\beta < 1.0$
before crossing over to a shape consistent with $\beta > 1.0$ at the
transition to the chaotic plastic flow regime.  
Using various scaling techniques, Fily {\it et al.} determined that
$\beta=1.3$ at the onset of the chaotic plastic flow region.
Some questions that remain is whether the transition
between Region I and Region II is a dynamical phase transition, or if it instead
obeys the period-doubling  behavior expected for a transition to chaos
in low dimensional systems \cite{ott1}.
Also unknown is
the nature of the transitions between the other regions,
and how general the behavior at these transitions is to
other systems.

\begin{figure}
  \includegraphics[width=\columnwidth]{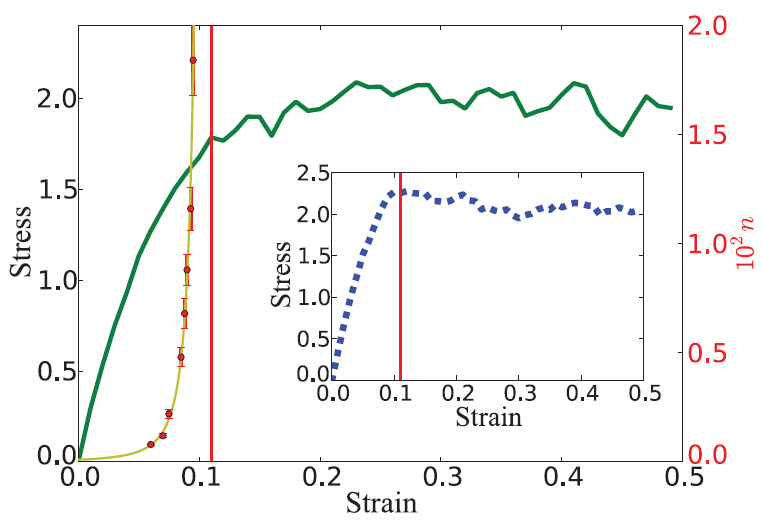}
\caption{
Stress-strain  curve  (green) from  molecular
dynamics simulations of 16384 particles under quasistatic shear.
Red dots represent the number of shear cycles $n$
required to reach periodic
behavior under oscillatory shear
(the scale is on the right side of the figure in red).
The vertical red line is the strain amplitude at which
the time to reach reversible behavior diverges. Inset: stress-strain
behavior for the same parameters as the solid green curve but with
different initial particle configurations.  The vertical red line is the
same as in the main figure.
Reprinted with permission from I. Regev, T. Lookman, and C. Reichhardt,
Phys. Rev. E {\bf 88}, 062401 (2013). Copyright 2013 by the American Physical Society.
}
\label{fig:Ido}
\end{figure}

Although this review concentrates on depinning phenomena,
we note that
there may be deeper connections between depinning and the onset of
yield and  plasticity in pin-free materials.
In recent work on 
non-chaotic to chaotic transitions near yield
in amorphous systems, similar effects arise
where reversible or non-chaotic plastic flow appears below yield,
followed by a transition to irreversible plastic flow.
This would imply that yielding could be understood as a transition
to a chaotic state, similar to the transition to chaotic plastic depinning,
and raises the question
of whether these two different systems fall in the same universality class
\cite{ott2,ott3,ott4,ott5}.          
Figure~\ref{fig:Ido} shows two examples of stress-strain 
curves from simulations of sheared amorphous materials,
where the solid vertical line represents the 
yield point above which the system starts to flow.
Below yield, the stress increases 
with strain in what could be considered a pinned regime, while the yielding point could
be associated with a depinning transition, and the higher strain regime could be similar
to a flowing regime.
Recent work has shown that below yield,
there can be plastic rearrangements of the particles similar to the rearrangements
that occur in pinned systems subjected to drives that are below the depinning threshold.
The yellow line in figure~\ref{fig:Ido} indicates that the
transient time required for these plastic rearrangements to die out
diverges as a power law as the yielding transition is approached, suggesting that
yield has
critical properties  \cite{ott3}
similar to those observed at critical depinning transitions.
In Section 12
we discuss how plastic depinning and the onset of irreversible
plastic deformations in materials
may be governed by the same type of critical behavior.

\section{Depinning and Dynamic Phases in Superconducting Vortex Systems}

One of the best examples of a system that exhibits
different kinds of depinning and dynamical sliding behavior is 
vortices in type-II superconductors \cite{1}.
Under application of a magnetic field, the flux does not enter a type-II
superconducting sample in a uniform manner,
but rather forms quantized vortices that each carry an  elemental unit of magnetic
flux $\phi_0=h/2e$.
The field is most intense at the non-superconducting core of the vortex
and then falls off at further distances.
The existence of a gradient in this field indicates that there is
a supercurrent ${\bf J} \propto \nabla {\bf B}$
circulating around the core, in analogy with the fluid circulating around the core of
a superfluid vortex.
The superconducting
vortices interact repulsively with each other through a potential that has
the form of a Bessel function $K_{1}(r_{ij}/\lambda)$ in a superconducting slab,
where $\lambda$ is the London penetration depth of
the magnetic field away from the core.
This interaction behaves as $\ln(r)$ at short range and falls
off exponentially at larger distances.
The Bessel function vortex-vortex interaction can be
used for 3D systems in which the vortices behave as stiff lines that form a
2D triangular lattice in the plane perpendicular to the magnetic field,
and the overall dynamics can be modeled as effectively 2D.
Models for vortex lines that are not stiff are described in Section 4.
In thin film samples the  vortices have a
longer range interaction of the form $\ln(r)$ or $1/r$ depending on
the sample thickness and
the behavior of the magnetic field at the surface.
An externally applied current
${\bf J}$ interacts with the magnetic field of each
vortex and induces a Lorentz driving force
$F_{D} \propto {\bf B} \times {\bf J}$ that causes
the vortex to move.
Since the core of the vortex is non-superconducting,
the electrons located in the core behave dissipatively,
resulting in overdamped vortex motion with
a damping term $\eta$.
The vortex velocity $V$
can be measured as a finite voltage response in the sample
produced by the dissipation.

In the absence of pinning produced by a substrate or by intrinsic
inhomogeneities in the sample, the vortex velocity increases
linearly with the driving force, $V = F_D/\eta$.
Most samples, however, contain intrinsic
disorder in the form of sites where the order parameter of the superconducting
condensate is suppressed.
These locations act as pinning sites for the vortices,
and allow the vortex lattice to remain immobile even when a driving current is applied
to the sample.
The critical current can be defined as the applied current at which a finite voltage response
first appears, and corresponds to a depinning threshold $F_c$.
Under a driving current, the vortex velocity-force curves are directly
proportional to the voltage-current curves, so that features in the voltage-current curves 
or the time dependent fluctuations of the voltage reflect how the
vortices are moving and their velocity fluctuations.
In the interest of generality, we refer to voltage-current curves as velocity-force curves
in our discussion.
Many applications of type-II superconductors
require the material to remain superconducting while transporting high currents,
so there has been extensive work on understanding how to
optimize the pinning and maximize the critical current
in these systems using techniques such as
ion irradiation, chemical synthesis \cite{1},
and nanostructuring of artificial pinning arrays \cite{2}.

At magnetic fields $H$ well below the critical magnetic field $H_{c2}$ at which
the sample ceases to superconduct,
the vortex
core undergoes only small distortions when the vortex moves or interacts with
other vortices or pinning sites,
so the vortex dynamics can be described by
a particle-based equation of motion which for a single vortex $i$ can be written as
\begin{equation}
\eta \frac{ d {\bf R}_{i}}{dt}  = {\bf F}_{i}^{vv} + {\bf F}^{p}_{i} + {\bf F}_{D} + {\bf F}^{T}.
\end{equation}
Here ${\bf R}_{i}$ is the position of vortex $i$
and the vortex-vortex interaction force is
${\bf F}^{vv}_{i} = \sum^{N_{v}}_{j=1}F_{0}K_{1}(R_{i}/\lambda){\bf \hat R_{ij}}$,
where $F_{0} = \phi_{0}^{2}/2\pi\mu_{0}\lambda^3$,
$\phi_{0}$ is the elementary flux quantum,
$\mu_0$ is the permittivity of free space,
$K_{1}$ is the modified Bessel function, 
$R_{ij} = |{\bf R}_{i} - {\bf R}_{j}|$,
and ${\hat {\bf R}_{ij}} = ({\bf R}_{i} - {\bf R}_{j})/R_{ij}$.
The density of vortices in the system is proportional to
the applied magnetic field.
The interaction with the pinning sites is represented by
${\bf F}^{p}_{i}$ , which can be modeled
in various ways.  For example, parabolic traps
cut off at a radius $R_{p}$ or localized Gaussian-shaped traps
can be used to represent
pointlike pinning sites, while different potentials can represent
smoother substrate landscapes or linelike defects such as twin boundaries.
In general,
a single pinning site or substrate feature exerts a maximum pinning force
of $F_{p}$ on an individual vortex.
The external driving force is given by ${\bf F}_{D}$, and represents
a dc and/or ac applied current
which generates a force of magnitude $F_D$.
On periodic substrates the drive can be oriented
along different angles to the
substrate symmetry directions.
In typical velocity-force simulations, $F_{D}$ is increased in small increments $\delta F_{D}$
and the velocity of the vortices is measured during a fixed
simulation time period to obtain an average value of $V$ for each value of $F_D$.
It is important to wait long enough for the system to settle into
a steady state, particularly near the depinning threshold.
The thermal
fluctuations  ${\bf F}^{T}$ are modeled as random
Langevin kicks with the properties $\langle {F}^{T}(t)\rangle = 0$
and
$\langle F^{T}_{i}(t)F_j^T(t^{\prime})\rangle = 2\eta k_{B}T\delta_{ij}
\delta(t - t^{\prime})$,
where $k_{B}$ is the Boltzmann constant.
In many superconducting vortex systems, thermal forces are not relevant; however, 
high-temperature superconducting materials can exhibit regimes
in which the thermal effects are strong enough to produce liquid-like behavior
of the vortices, resulting in strong creep effects in the velocity-force curves.
In this review we focus on regimes where thermal fluctuations
do not dominate the behavior.
In simulations, the vortex ground states can be obtained using
various energy minimization schemes or by simulated annealing
where the system is started at a high temperature and gradually cooled to $T = 0$.
For systems with a Bessel function vortex-vortex interaction, the
mutual repulsion of the vortices causes them to naturally
avoid the short-distance divergence of the interaction force
provided that
the vortex density does not become too large,
while the interaction can be truncated
at sufficiently large distances where it becomes negligible without affecting the results.
For $\ln(r)$ or $1/r$ vortex interactions, no long-range cutoff is possible and
techniques such as Euler or Lekner summations must be employed in dynamical
simulations.

The equation of motion for superconducting vortices can also contain
additional non-dissipative terms, such as
a Magnus force \cite{63,64} generated by the interaction of the driving current with
the circulating vortex supercurrent.  This 
produces an additional force on the vortex that is transverse to the Lorentz force.
In superconductors, Magnus force effects are usually very  small \cite{1},
but
in skyrmion systems, the Magnus term is large and affects the particle
dynamics, as we discuss in section 9.
In principle the vortex motion can also include an inertial contribution
$ M \ddot {\bf R}_i$, but the mass $M$ of a vortex is very small so such
effects can normally be ignored.

Features of the vortex system that make it ideally suited for
studying depinning include the numerous readily experimentally tunable 
parameters, such as magnetic field
which can be used to sweep the system from
the single vortex limit at low
magnetic fields to the strongly interacting vortex limit at high magnetic fields,
as well as the applied current, which can sweep the system
from low to high driving forces.
Due to the small size of an individual vortex, experimental studies
can readily access
the long-time dynamics of large assemblies of vortices,
so that statistically averaged measures of dynamical steady
states can be obtained.
It is also possible to tune the vortex-vortex interactions
using temperature since $\lambda$ 
diverges near $T_{c}$ and $H_{c2}$, which denote the temperature and
magnetic field at which superconductivity is destroyed.
By operating
close to $T_c$, the vortex-vortex interactions can be weakened
relative to other energy scales in the system.

\subsection{Elastic and plastic depinning transitions}

Ideally, the best samples for studying depinning and dynamic flows have low
intrinsic pinning strength, permitting access to the flux flow regime over a wide
range of currents without significant generation of local heating effects by the
moving normal vortex cores.
In such samples, the basic assumptions used in constructing a
particle-based vortex equation of motion remain valid.
Many weak pinning  samples exhibit what is called a ``peak effect''
in which the critical depinning threshold $F_c$
initially decreases with increasing
magnetic field before suddenly increasing to a peak value
and then dropping to zero
at the superconducting-to-normal transition \cite{20,65}.
The initial decrease in $F_c$ with increasing magnetic field
is expected since the
vortex-vortex
interactions become stronger as the vortices get closer together,
reducing the effectiveness of the intrinsic pinning,
but the peak in $F_c$ has not been satisfactorily explained.
The peak coincides with pronounced changes
in the current-voltage curves \cite{20,66} and  noise fluctuations \cite{67,68},
suggesting that the peak effect
is associated with changes in the vortex lattice structure in the pinned phase
as well as changes in the depinning transition and the vortex motion above depinning.

\begin{figure}
  \includegraphics[width=\columnwidth]{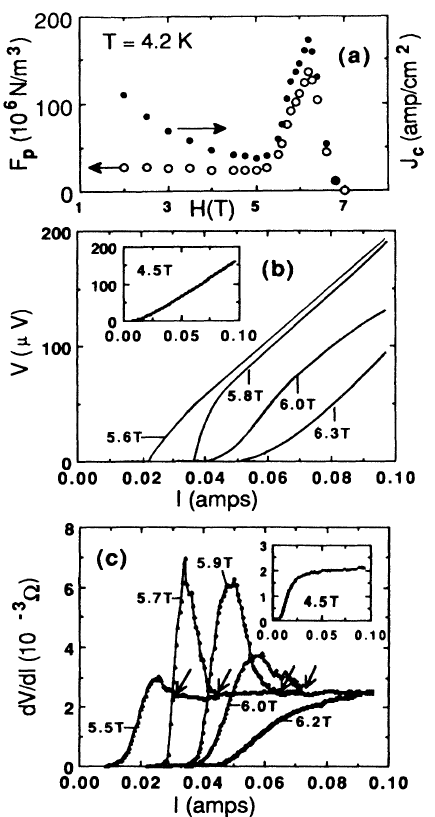}
\caption{
  (a) The experimentally measured
  depinning force
  $F_p$ (open circles)
  and critical current $J_c$ (filled circles) versus magnetic field $H$ for vortices in a
  superconducting sample of 2H-NbSe$_{2}$.
There is a peak in the pinning force as a function of field.
  Here $J_c$ corresponds to the critical depinning threshold $F_c$, while the vortex
  density increases with increasing $H$.
(b) Voltage $V$ vs current $I$ curves 
obtained at different magnetic field values (labeled)
show a change in the shape of the curves
across the peak effect.
The voltage corresponds to the vortex velocity $V$, while the current 
corresponds to the driving force $F_D$.
(c) The corresponding $dV/dI$ curves, which are proportional to $dV/dF_D$ curves, 
obtained at different magnetic field values (labeled)
each exhibit a spike due to the S-shape of the
velocity-force curves. This spike reaches its largest magnitude near the apex of the
peak effect.
The inset shows that at lower fields below the peak effect, the $dV/dI$ curve
does not contain a spike.
Reprinted with permission from S. Bhattacharya and M.J. Higgins,
Phys. Rev. Lett. {\bf 70}, 2617 (1993).  Copyright 1993 by the American Physical Society.
}
\label{fig:5}
\end{figure}

Higgins and Bhattacharya 
experimentally measured the effective
vortex
velocity-force
curves
across the peak effect in 
a low pinning sample of 2H-NbSe$_2$ \cite{20,66}.
Based on the features of these curves, they deduced
that below the peak effect the vortices depin elastically,
while at and above the peak effect the
vortices depin plastically
and undergo a transition at higher drives to
a dynamically ordered moving vortex state.
Figure~\ref{fig:5}(a) shows the measured effective
pinning strength $F_p$, determined using the relation
$F_{p} = |{\bf J}_c \times {\bf B}|$, vs $H$, indicating that there is a
peak in the pinning force near $H=6$ Tesla.
Here $J_c$ corresponds to the critical depinning threshold $F_c$, while
the vortex density increases with increasing $H$.
The
velocity-force curves in figure~\ref{fig:5}(b)
show changes in concavity when
measured at different magnetic fields across the peak effect,
and close to the peak effect
the curves are S-shaped.
The changes in the curves are more clearly visible in the
$dV/dF_D$ plots in figure~\ref{fig:5}(c).
At fields below
the peak effect regime, as shown in the inset of
figure~\ref{fig:5}(c), there is only a plateau in $dV/dF_D$, while
near the peak effect $dV/dF_D$ develops a spike feature, shown in the
main panel of figure~\ref{fig:5}(c), that reflects the S-shape of the
velocity-force curves.
For fields above the peak effect, the spike in the $dV/dF_D$ curves diminishes and
disappears.

\begin{figure}
  \includegraphics[width=\columnwidth]{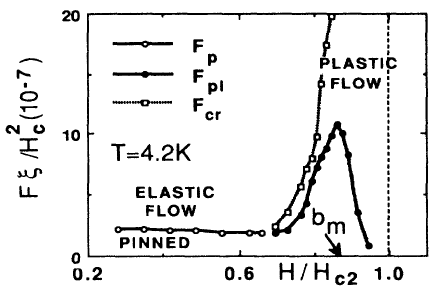}
  \caption{
    Experimentally measured
    dynamic phase diagram as a function of
    the effective applied driving force $F \xi/H_c^2$, which corresponds to $F_D$,
    vs magnetic field $H/H_{c2}$, which corresponds to the vortex density,
    from the 2H-NbSe$_2$ sample in figure~\ref{fig:5}.
Below the peak effect there is
an elastic depinning transition $F_p$ from a pinned state to elastic flow (open circles).
As the peak effect is approached, the depinning becomes plastic, as indicated by
the $F_{pl}$ line (filled circles), and there is a higher drive transition from
plastic flow to a dynamically reordered elastic flow state, indicated by the $F_{cr}$
line (open squares).
Both $F_p$ and $F_{pl}$ correspond to the depinning threshold $F_c$.
Reprinted with permission from S. Bhattacharya and M.J. Higgins,
Phys. Rev. Lett. {\bf 70}, 2617 (1993).  Copyright 1993 by the American Physical Society.
}
\label{fig:6}
\end{figure}

The velocity-force curves in figure~\ref{fig:5}(b)
can be scaled to fit $V \propto (F_D - F_c)^\beta$, with $\beta=1.75$ above
the peak of the peak effect where plastic flow is believed to occur.
It was proposed that when the vortex lattice depins elastically,
$dV/dF_D$ saturates to a constant value and exhibits no spike feature,
whereas when the vortex lattice
depins plastically, the vortex structure is initially strongly disordered
but dynamically orders at higher drives
into a defect free structure, producing
a spike in the $dV/dF_D$ curve \cite{20}.
According to this interpretation, 
the change in the velocity-force curves reflects a transition from elastic to
plastic depinning and also provides evidence for
a dynamical transition from plastic flow  to a moving
ordered structure at higher drives.
Figure~\ref{fig:6} shows the proposed dynamic phase diagram for
this scenario.
At magnetic fields below the peak effect, the vortex lattice is ordered and depins elastically
directly to a moving ordered state, marked in the figure as elastic flow.
The peak effect coincides with a transition from a pinned ordered lattice to a disordered
pinned structure that depins plastically and dynamically reorders
at higher drives.
As $H_{c2}$ is approached, the drive at which the dynamical reordering occurs
diverges.

A change from elastic to plastic depinning at the peak effect is consistent with
the experimental observations, but the reason for this change must be explained.
One argument invokes the increase in the size of the London penetration depth
$\lambda$ as $H_{c2}$ is approached.  Since the vortex-vortex interaction strength
is controlled by gradients in the local magnetic field, as $\lambda$ increases these
gradients become smaller and the vortex-vortex interactions become weaker.
When the vortex-vortex interactions are weak enough, dislocations
can proliferate in the vortex lattice, destroy the triangular ordering, and
cause a transition to plastic depinning.
In the elastic regime, when the stiff triangular lattice formed by the vortices interacts
with randomly placed pinning sites, it is too energetically costly for the vortices to
deform their triangular lattice in order to take advantage of the pinning energy, so many
of the pinning sites remain unoccupied and therefore ineffective.  Once dislocations
appear in the vortex lattice, however, the softened or amorphous lattice can deform
easily in order to allow vortices to sit in the disordered pinning sites, so that a much
higher fraction of the pinning sites are occupied and the effectiveness of the pinning
increases dramatically.
In other words, the stiffer triangular vortex lattice couples to the substrate more weakly
than the softer amorphous
vortex structure.
One analogy that has been used to describe why elastic depinning occurs at a
lower driving force than plastic depinning
is that the elastic depinning process is akin to the easy motion of
a stiff solid such as an ice cube over a carpet, while the plastic depinning
process is like attempting to move a soft solid such as a gelatin cube over a carpet.
The soft solid easily adjusts to fit in the asperities of the carpet and can be moved
only with difficulty \cite{69}.

The idea that softer systems are more strongly coupled to the pinning substrate
applies
only when the number of pinning sites
is greater than or equal to the number of particles
in the system.
For example, consider a substrate composed of
a flat surface decorated by a
small number of well-defined pinning sites that can each capture no more than
a single vortex.
When there are more vortices than pinning sites, some of the vortices have
no pinning site available in which to sit and are confined only by the vortex-vortex
interaction force from their neighbors.  These interstitial vortices can still be
effectively pinned if enough of their neighbors are directly pinned; however, the strength
of this interstitial pinning
depends on the shear modulus of the vortex lattice and not merely on
the strength of the pinning sites.
In a stiff elastic vortex lattice,
the strong shear modulus 
makes it difficult or impossible to depin the interstitial vortices separately
from the vortices in the pinning sites,
leading to an elastic depinning transition.
If the vortex-vortex interactions are weakened
substantially, then the interstitial vortices can be depinned at lower drives than the
vortices in the pinning sites, resulting in a plastic depinning transition.
In simulations of 2D systems containing a small number of pinning sites, when
the vortex interactions are modified in order to tune the shear modulus from a high
value to a low value,
a very high shear modulus results in a low depinning threshold as the stiff vortex
lattice depins elastically, 
while for a reduced shear modulus the depinning threshold
increases and reaches a maximum at the onset of plastic depinning \cite{70}. 
For even smaller shear modulus values, the depinning
threshold decreases again when the vortices at the pinning sites are no longer able
to exert a confining force on the interstitial vortices \cite{70}.
It is not clear whether this model can
be applied to the experimentally  observed peak effect in superconducting systems since 
the number of intrinsic pinning sites is generally at least as large as if not greater than
the number of vortices, which would imply that the pinning force should be maximal
for low or zero shear modulus \cite{71}.

\subsection{Dynamical ordering transitions}

It was first noted
in neutron scattering experiments by Thorel {\it et al.}
that vortices can exhibit increased
ordering in the moving state \cite{71N},
while 2D vortex simulations by Shi and Berlinsky
showed that the number of topological defects in
the vortex lattice decreases with increasing
drive \cite{72}.
Koshelev and Vinokur (KV) performed
2D simulations of vortices driven over random pinning and mapped
out a dynamic phase diagram containing a pinned phase,
a plastic flow phase characterized by a highly
defected vortex lattice, and a dynamically ordered phase at high drives where the
vortices form a sixfold ordered structure \cite{45}.
These simulations also included a thermal fluctuation force, and showed that
in the absence of pinning, there is
a bulk melting temperature $T_{m}$ above which the vortices
lose their sixfold ordering.
In the presence of pinning, as the temperature $T$ increases toward $T_m$,
the depinning threshold $F_c$ decreases while the driving force at which dynamical
reordering occurs shifts to higher drives, with a divergence near $T_m$.
The divergence of the reordering force at the point at which
the flow becomes liquidlike or plastic
is similar to the divergence
observed in the experimentally determined dynamic phase diagram
of Bhattacharya and Higgins shown in figure~\ref{fig:6} \cite{20};
however, the transition observed by KV \cite{45}
occurs as a
function of temperature rather than as a function of magnetic field.
Additionally, there is no peak in the depinning threshold
in the simulation studies
since the thermal fluctuations
monotonically reduce $F_c$.
This work does, however, provide further
evidence that there are at least three generic dynamic phases in driven vortex systems
with disorder: pinned, plastic flow, and elastic flow, with transitions between these states.
KV argued that vortices moving over a random landscape
experience dynamical fluctuations induced by the pinning that
can be treated as
an effective shaking temperature $T_{sh}$, so that when the
combination of the bulk temperature $T$ and $T_{sh}$ are greater than or equal 
to $T_{m}$, the system is in a disordered plastic flow phase.
As $F_D$ increases, $T_{sh}$ decreases since the vortices 
are moving too rapidly to 
respond to the
perturbations induced by the pinning sites,
and using analytical arguments KV find 
$T_{sh} \propto 1/F_{D}$.
The vortices dynamically order when
$T + T_{sh} < T_m$, which explains the divergence in
the driving force at which dynamic reordering occurs
as $T$ approaches $T_{m}$.
Several further experiments and simulations
also find similar evidence for three dynamical phases \cite{50,54,New4,73,74,75,76,77}.

\begin{figure}
  \includegraphics[width=\columnwidth]{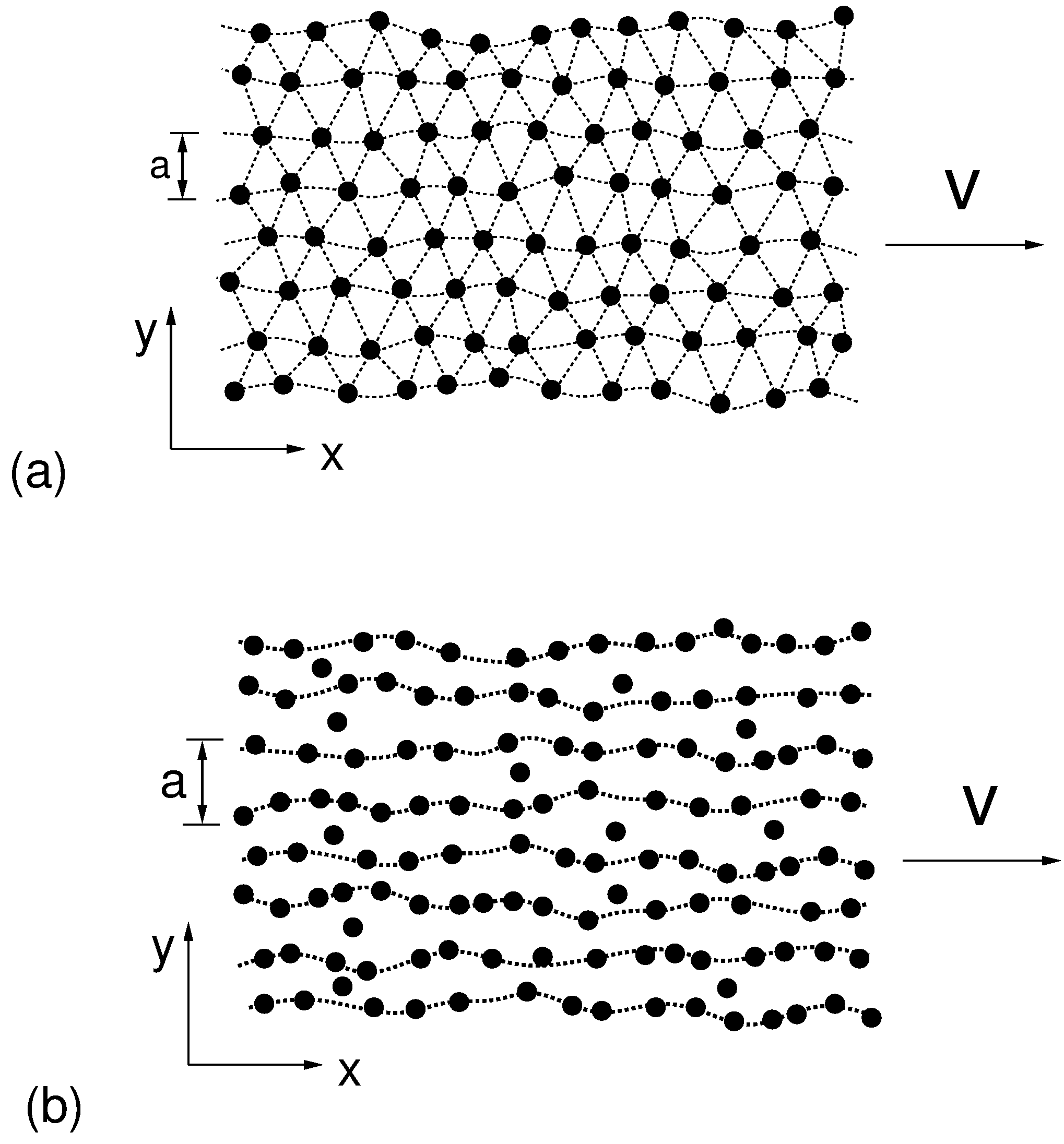}
\caption{
  Schematic illustrations of two possible moving states
  for a lattice of particles driven along the $x$ direction. (a) The moving Bragg
  glass, which contains no dislocations.  (b)  A moving smectic state, which
  is periodic transverse to the driving direction but has liquid order along the
  driving direction.
Reprinted with permission from L. Balents, M.C. Marchetti, and L. Radzihovsky,
Phys. Rev. B {\bf 57}, 7705 (1998).  Copyright 1998 by the American Physical Society.
}
\label{fig:M}
\end{figure}

\begin{figure}
  \includegraphics[width=\columnwidth]{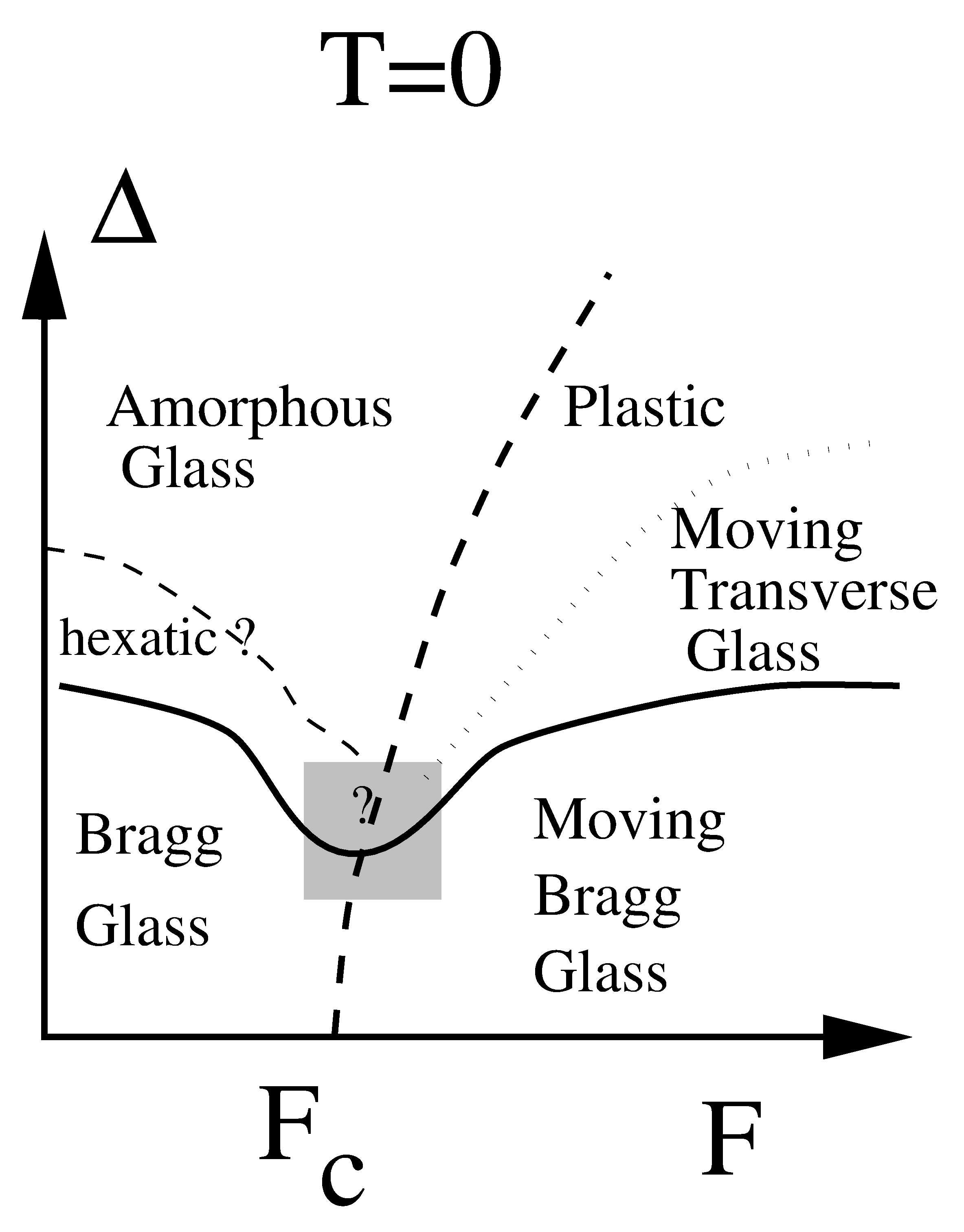}
\caption{
Schematic dynamic phase diagram as a function of
disorder strength $\Delta$ versus driving force $F$ for $d=3$ dimensions.
$\Delta$ corresponds to the pinning force $F_p$, while $F$
corresponds to the driving force $F_D$.  The heavy dashed line is a depinning transition.
 The  behavior  in  the shaded  square  region  is  unclear.  An
interesting possibility would be a direct depinning of a hexatic into
the moving transverse glass, but other scenarios are possible.
  Reprinted with permission from P. Le Doussal and T. Giamarchi,
Phys. Rev. B {\bf 57}, 11356 (1998). Copyright 1998 by the American Physical Society.
}
\label{fig:ledoussal}
\end{figure}

It is interesting to ask whether there are
different kinds of plastic flow or different dynamically ordered moving phases.
In some ways the dynamically ordered
phase is easier to address, since if it has some crystalline properties, then the
presence or absence of certain symmetries
can be used to distinguish different ordered moving phases.
Giamarchi and Le Doussal argued that in the moving phase,
the pinning-induced shaking temperature is anisotropic rather than isotropic,
so that the dynamically ordered state is not 
a moving crystal but a moving anisotropic Bragg glass
\cite{49}.
A Bragg glass is an equilibrium glass state that arises for systems with
quenched disorder in which there are no dislocations present \cite{78}.
Within the moving Bragg glass phase, the flow consists of
vortices traveling in coupled channels, and
if an additional driving force is applied perpendicular to the original driving force,
there is a finite
transverse depinning threshold \cite{49}.
In the moving frame, the transverse depinning threshold can be viewed as an effective
pinning force in the direction perpendicular to the channel direction that is
experienced by the moving
vortex channels, each of which acts like an elastic object.
In further theoretical studies, it was
argued that the anisotropy
of the shaking temperature is strong enough that
it produces dislocations in the vortex lattice,
destroying the Bragg glass state,
but that these dislocations are aligned by the driving force,
producing a moving smectic phase illustrated schematically
in figure~\ref{fig:M}
that has long range or quasi-long range order transverse to the driving direction
but short range order parallel to the driving direction \cite{51,52}.
Such a smectic phase
would be generic to 2D systems.
Giamarchi and Le Doussal then proposed 
that the
moving anisotropic Bragg glass
and moving smectic phase
are different regimes of the same system \cite{53}, as shown in the
schematic phase diagram of figure~\ref{fig:ledoussal}
as a function of disorder strength versus driving force.
At low disorder strengths the system
forms a dislocation-free Bragg glass that depins elastically  into a moving Bragg glass. 
At higher disorder strength there is a transition from a Bragg
glass to an amorphous glass phase that
depins plastically and then dynamically reorders into a moving  smectic,
also termed a moving transverse glass. 
It is possible that a pinned hexatic phase could exist
between the pinned Bragg glass and
the pinned amorphous glass \cite{78}.

\begin{figure}
  \includegraphics[width=\columnwidth]{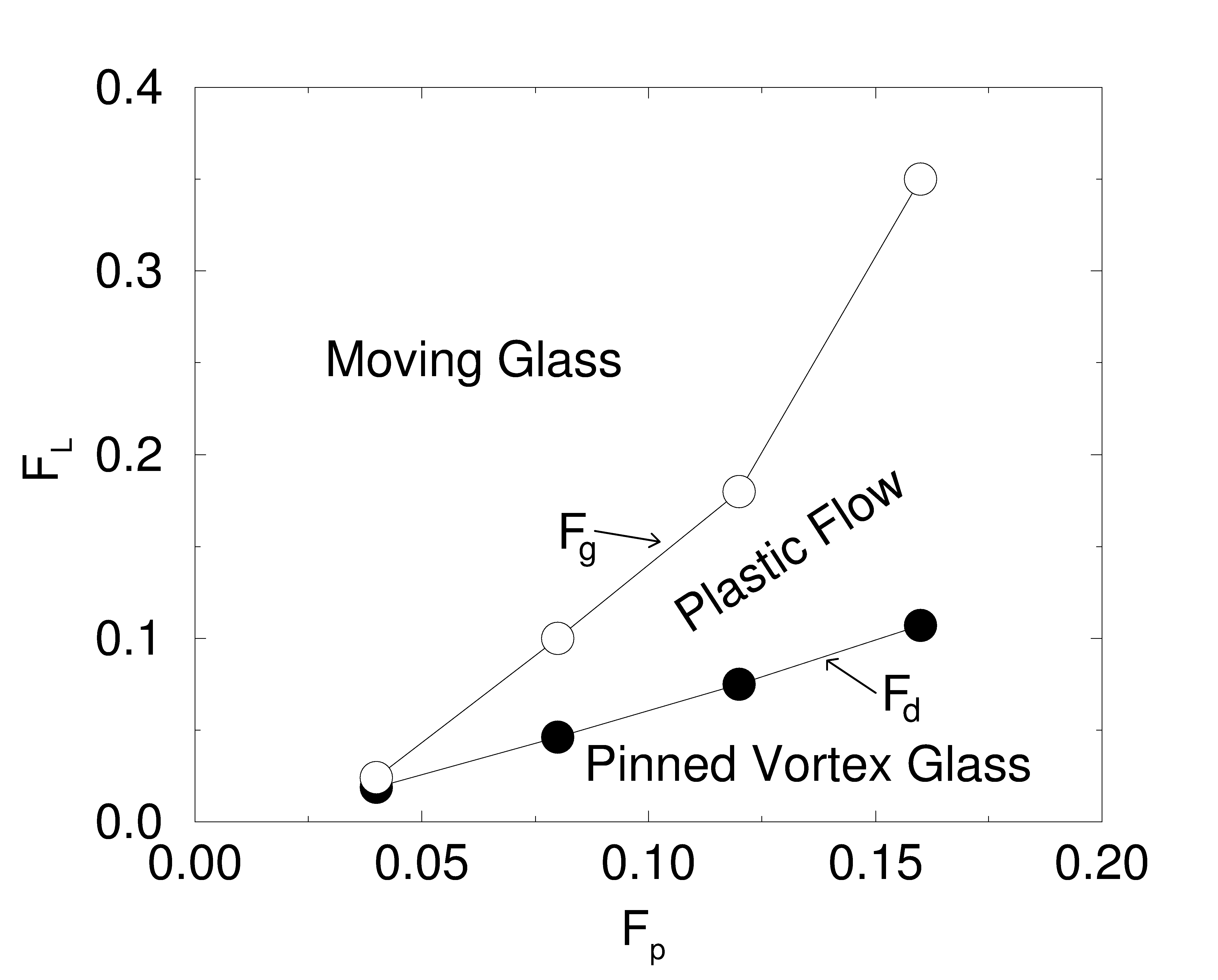}
\caption{
  The dynamic phase diagram from
  simulations of a 2D vortex system for driving force $F_{L}$
vs pinning force $F_{p}$ highlighting a pinned phase, plastic flow phase, and
a dynamical partially ordered moving glass phase.
Filled circles are the depinning line $F_d$, corresponding to $F_c$, while
the open circles indicate the dynamical reordering transition.
Here $F_L$ corresponds to the driving force $F_D$.
Reprinted with permission from K. Moon, R.T. Scalettar, and G.T. Zim{\' a}nyi,
Phys. Rev. Lett. {\bf 77}, 2778 (1996).  Copyright 1996 by the American Physical Society.
}
\label{fig:7}
\end{figure}

\begin{figure}
  \includegraphics[width=\columnwidth]{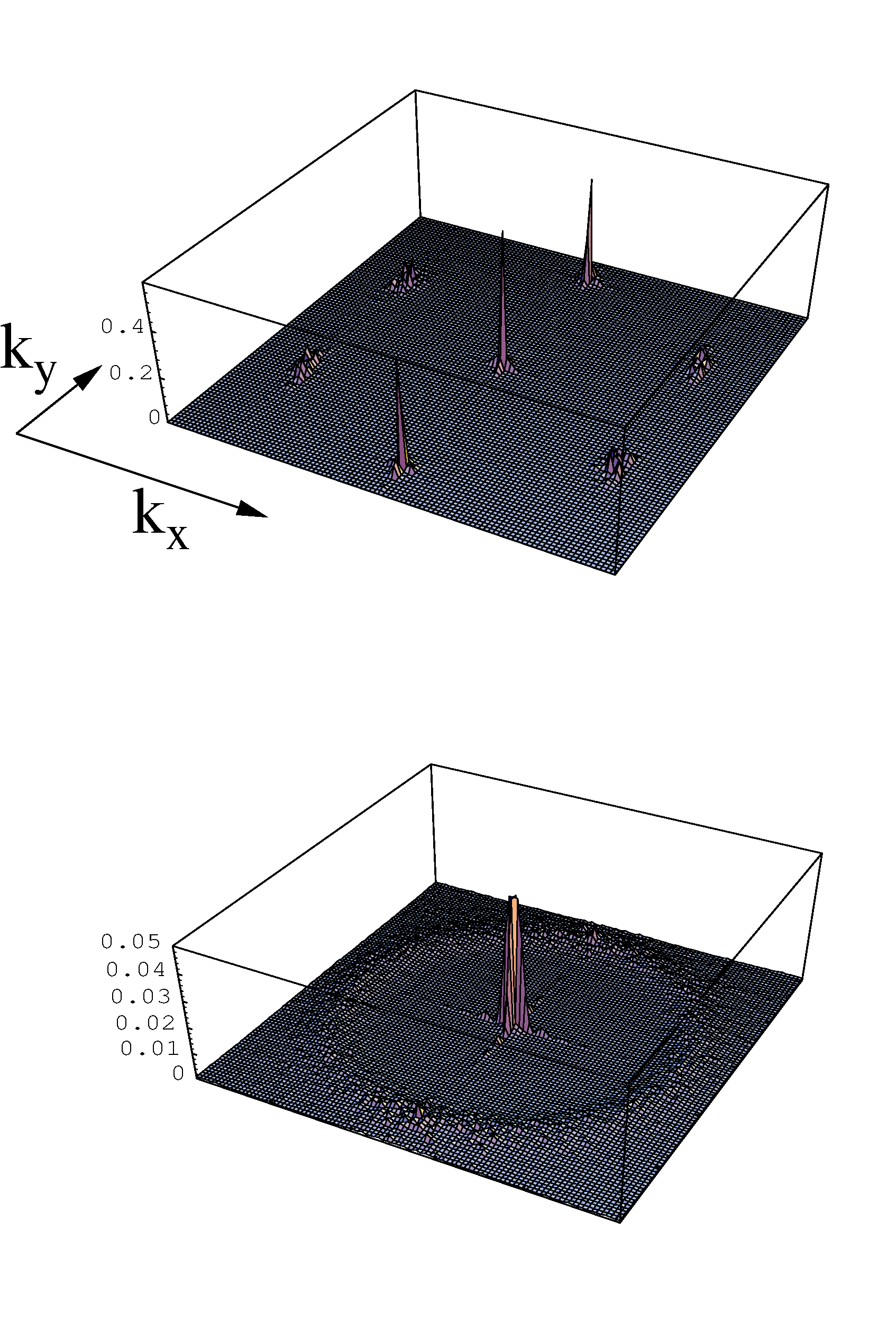}
\caption{
  The structure factor $S({\bf k})$ from the system in figure~\ref{fig:7}
  at a pinning strength of $F_{p}=0.16$.
  Upper panel:  At a driving force of $F_{L} = 0.6$ in the dynamically
  ordered phase, there are two peaks indicative of a
  smectic structure.
  Lower panel: At $F_{L} = 0.2$ in the plastic flow phase, there is a ring shape
  indicative of a liquid or amorphous structure.
  Here $F_L$ corresponds to the driving force $F_D$.
Reprinted with permission from K. Moon, R.T. Scalettar, and G.T. Zim{\' a}nyi,
Phys. Rev. Lett. {\bf 77}, 2778 (1996).  Copyright 1996 by the American Physical Society.
}
\label{fig:8}
\end{figure}

\begin{figure}
  \includegraphics[width=0.9\columnwidth]{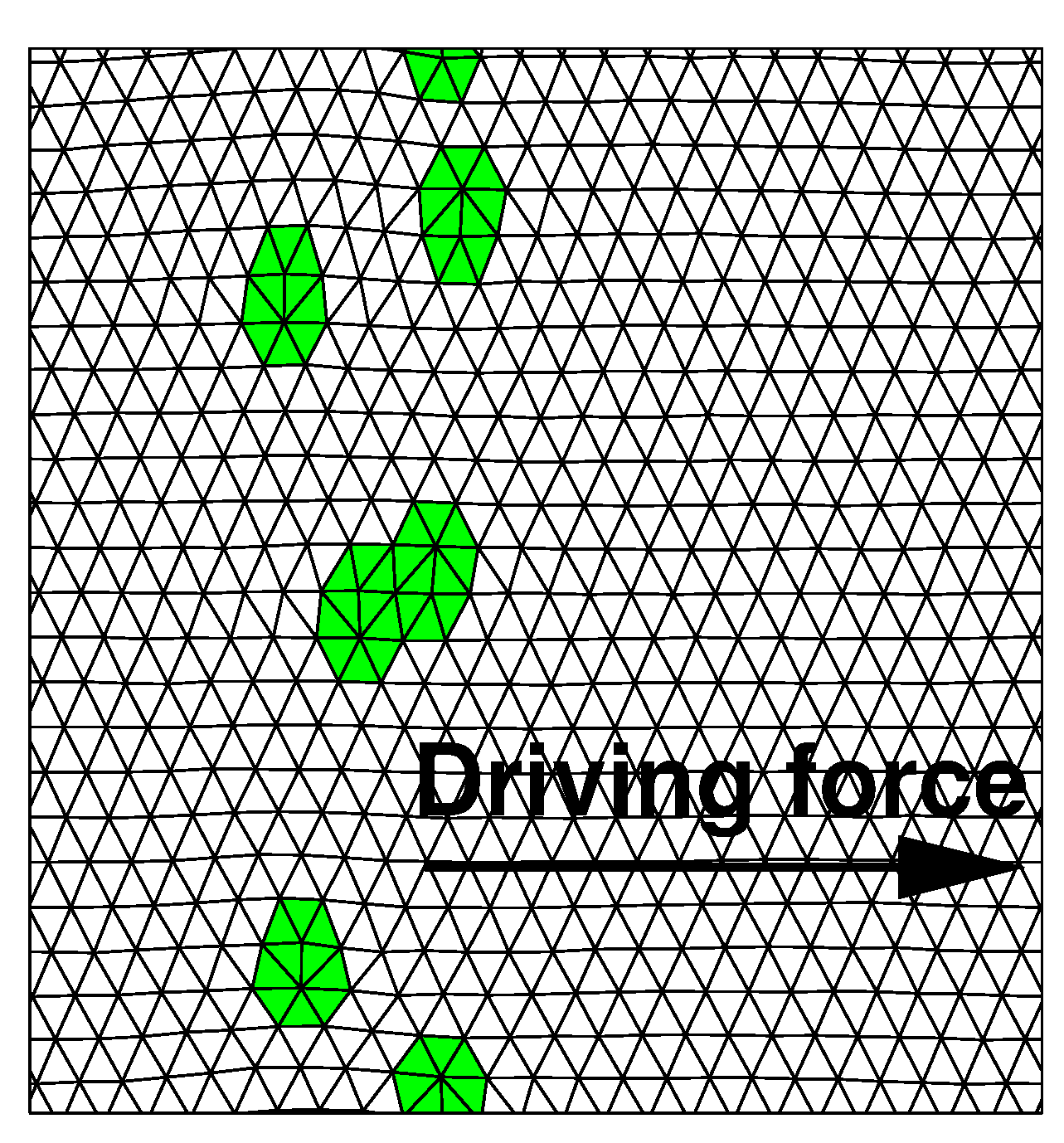}
\caption{
  Alignment of dislocations in a coherently moving smectic phase from a 2D
  vortex simulation.
White regions contain sixfold coordinated vortices
while green regions contain fivefold or sevenfold coordinated
vortices.
Reprinted with permission from H. Fangohr, S.J. Cox, and P.A.J. de Groot,
Phys. Rev. B {\bf 64}, 064505 (2001).  Copyright 2001 by the American Physical Society.
}
\label{fig:New1}
\end{figure}

Using Bitter decoration experiments,
Pardo {\it et al.} found
direct evidence for a moving smectic phase as well as a
weaker pinning regime in which only an anisotropic crystal
appears  \cite{79}, while
other imaging experiments
show ordered vortices moving along coupled 1D channels \cite{80}.
Additional theoretical works and simulations also provide evidence
of moving crystal phases and  smectic phases
\cite{50,74,36,82,85a,83,84,85}.
In other studies,
a transverse depinning threshold is observed in
the moving crystal or smectic phase, but is
absent in the plastic flow phase when
the system acts like a liquid \cite{50,74,36,81a,82a,83a}.
In figure~\ref{fig:7}, the dynamical phase diagram of driving force versus
pinning strength from 2D simulations of Moon {\it et al.} \cite{50}
contains a pinned vortex phase, a plastic flow phase, and a moving ordered phase or
moving glass.
The lines on the phase diagram are determined from velocity-force curve measurements
as well as
changes in the structure factor
\begin{equation}
  S({\bf k})=\frac{1}{N}\left|\sum_i^N \exp(-i{\bf k}\cdot {\bf R}_i)\right|^2 .
\end{equation}
In the plastic phase, $S({\bf k})$
has a ringlike structure indicative  of glass or liquid ordering,
as shown in the lower panel of
figure~\ref{fig:8},
while in the dynamically ordered phase, the upper panel of figure~\ref{fig:8} shows that
the ring structure
is replaced by two peaks
indicative of a moving smectic phase.
If a true moving crystal had formed, $S({\bf k})$ would contain six peaks
of equal weight.
An example of aligned dislocation structures in the moving smectic phase
from 2D vortex simulations
appears in figure~\ref{fig:New1}.
Fivefold and sevenfold
coordinated dislocations in the sixfold coordinated lattice pair up with
each other and preferentially align so that they can glide along the driving
force direction \cite{75}.
Moon {\it et al.} also identified several aligned dislocations in the smectic state
\cite{50}.

\begin{figure}
  \includegraphics[width=\columnwidth]{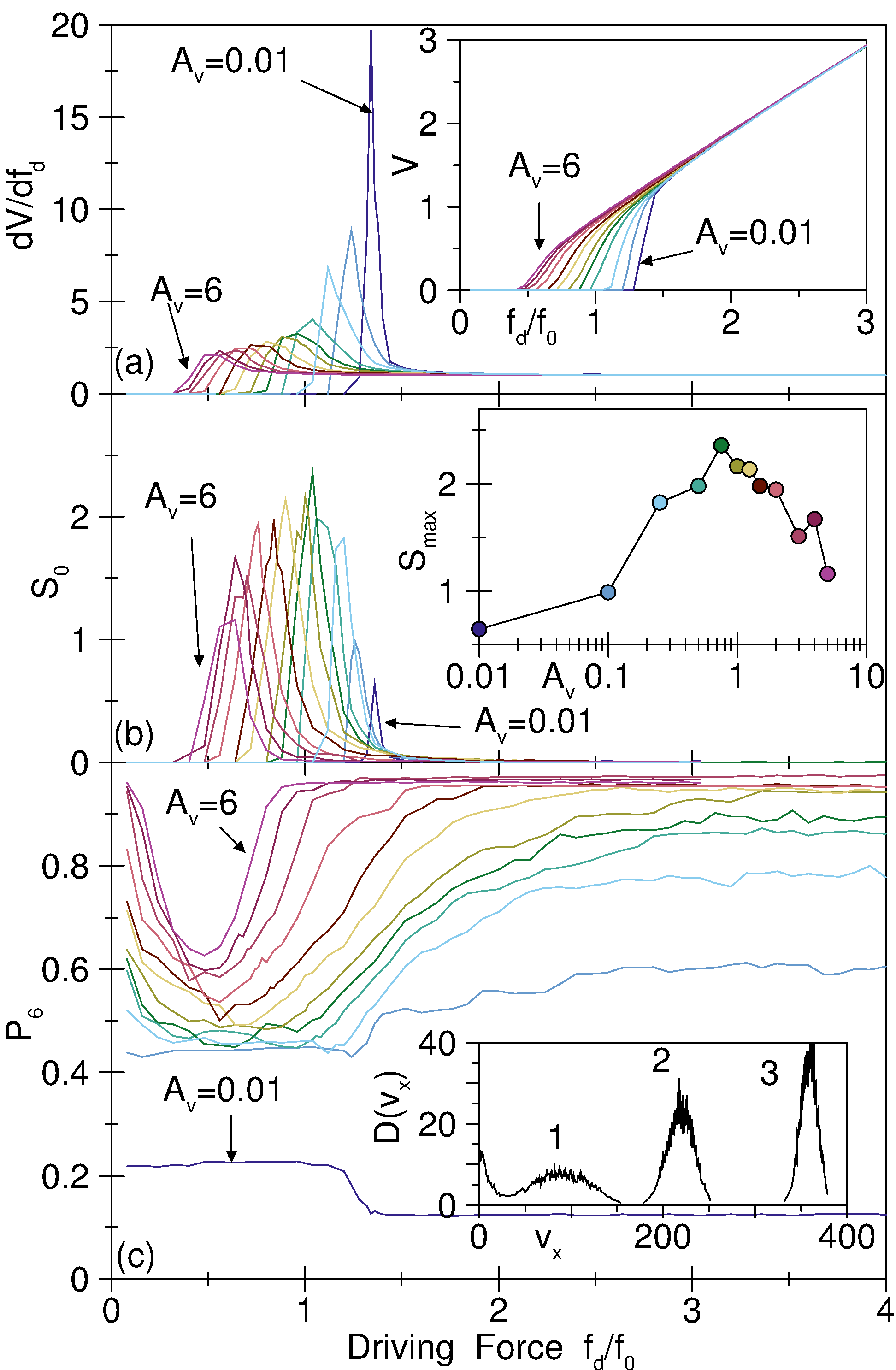}
\caption{
  (a) Inset: Velocity $V$ vs driving force $f_d/f_0$ curves from 2D simulations
  of vortices moving over a disordered substrate as the vortex-vortex
  interaction strength $A_v$, or effective shear modulus, is varied from $A_v=6$ to
  $A_v=0.01$.
  Main panel: $dV/df_{d}$ vs driving force $f_{d}/f_{0}$.
  (b) The corresponding velocity noise power $S_{0}$ vs $f_{d}/f_{0}$.
  The inset shows the maximum value $S_{\rm max}$ of $S_0$ for each value of $A_v$.
  (c) The fraction of six-fold coordinated vortices $P_{6}$ vs $f_{d}/f_0$
  for varied $A_{v}$, showing that above depinning,
  $P_6$ increases with increasing $f_d/f_0$.
  Here $f_d/f_0$ corresponds to the driving force $F_D$
  and $dV/df_d$ corresponds to $dV/dF_D$.
Reprinted with permission from C.J. Olson, C. Reichhardt, and F. Nori,
Phys. Rev. Lett. {\bf 81}, 3757 (1998).  Copyright 1998 by the American Physical Society.
}
\label{fig:9}
\end{figure}

The exact nature of the moving phase, and whether it forms a true smectic
structure or a moving dislocation-free Bragg glass,
depends on the dimensionality of the system and the strength of the disorder.
Olson {\it et al.} performed 2D simulations of vortices interacting
with random pinning arrangements
in which there were more pinning sites than vortices,
and tuned the strength $A_v$ of the vortex-vortex interactions 
or the effective shear modulus from zero to high values \cite{85a}, obtaining
the velocity-force curves shown in
the inset of figure~\ref{fig:9}(a).
As $A_{v}$ increases, the depinning threshold $F_c$ decreases.
The main panel of figure~\ref{fig:9}(a) shows that
the corresponding $dV/dF_D$ curves have a peak at the depinning threshold which
increases in size with decreasing $A_v$, and that for high drives all of the
$dV/dF_D$ curves saturate to the same constant value.
Figure~\ref{fig:9}(c) shows the fraction $P_6$ of sixfold coordinated
vortices versus $F_D$.
To measure $P_6$, the coordination number $z_i$ of each vortex is obtained
by performing a Voronoi construction \cite{voronoi}, and
taking $P_6=N^{-1}\sum_i^{N}\delta(z_i-6)$.
For  a perfect triangular lattice, $P_{6} = 1.0$.
In these simulations, the vortices are initialized in a triangular lattice,
so that as the driving force is increased in the
pinned state, short-lived vortex rearrangements
or avalanches occur that cause $P_6$ to drop with increasing $F_D$
until it reaches a minimum
value at the depinning threshold.
Above depinning,  $P_6$ gradually increases with increasing drive,
reaching a value of nearly $1.0$ for the larger values of $A_v$
where the vortex lattice is fairly stiff, but saturating
to a value significantly lower than $1.0$ for lower values of $A_v$,
indicating that numerous dislocations persist in the dynamically reordered
state.
The power spectrum of the time series of the vortex velocity fluctuations
$V(t)$
is obtained from
\begin{equation}
S(\omega) = \left |\int V(t)e^{-i2\pi \omega}dt \right |^2.
\end{equation}
Figure~\ref{fig:9}(b) shows the
noise power $S_0$ \cite{weissman}, obtained by integrating $S(\omega)$ 
over a fixed frequency
range, $S_{0} =\int^{\omega_{1}}_{\omega_{2}}df S(\omega)$.
In the plastic flow phase, $S(\omega)$ 
has the form $S(\omega) \propto 1/f^\alpha$
with $\alpha = 1.5$ to $2.0$, indicating the existence of large noise power at
low frequencies
due to the slowly changing plastic flow channel structure
similar to that illustrated in figure~\ref{fig:1}(a).
At higher drives the spectrum becomes white
with $\alpha = 0$ or develops a peak at a single frequency,
corresponding to narrow band noise, and this coincides with
a drop in $S_{0}$ and the saturation of $P_6$ to its highest
dynamically ordered value.
The inset of figure~\ref{fig:9}(b) shows the
maximum value $S_{\rm max}$ reached by $S_{0}$ across all driving forces
as a function of $A_v$.
At higher $A_v$, the noise power is small since there are few defects in the system,
while for small $A_v$ when the depinning threshold is highest, the noise power is also
small due to the loss of collective vortex-vortex interaction effects which are needed
in order to produce large-scale velocity fluctuations.

\begin{figure}
  \includegraphics[width=\columnwidth]{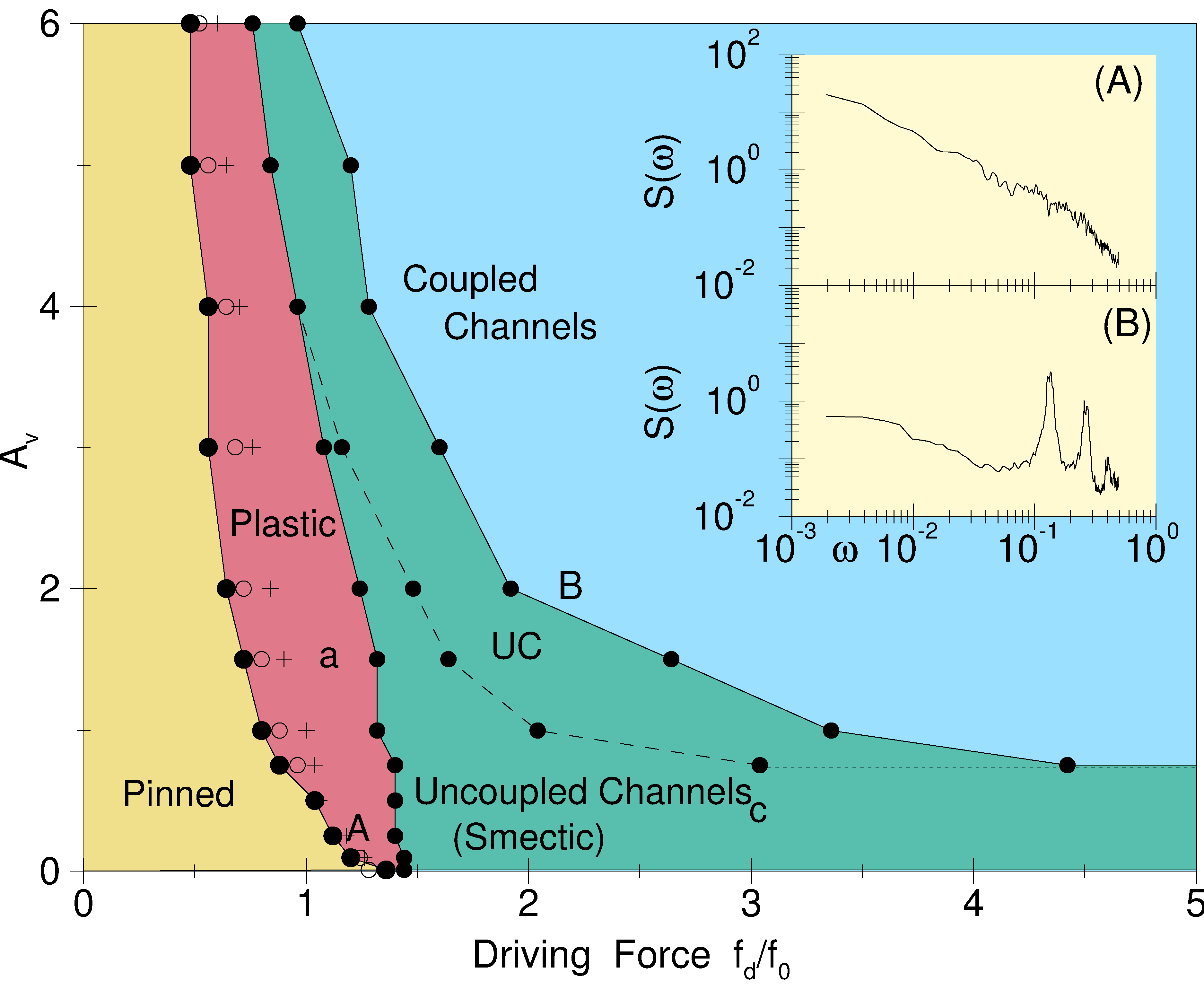}
\caption{
  The dynamic phase diagram for the system in figure~\ref{fig:9} as a function of
  vortex-vortex interaction strength $A_{v}$ vs driving force $f_{d}/f_{0}$,
  showing a pinned phase, plastic phase, uncoupled channel (UC) phase,
  and coupled channel or moving lattice phase.
  Here $f_d/f_0$ corresponds to the driving force $F_D$ and $A_v$ is an effective
  shear modulus.
The open circles and crosses indicate the locations of the peaks in the noise power
and $dV/dF_D$ curves, respectively.
  Inset (A) shows the velocity noise power spectrum $S(\omega)$ obtained in the plastic flow
  phase at the point marked A at $A_{v} = 1.0$, where there is $1/f^\alpha$ noise.
  Inset (B) shows $S(\omega)$ obtained in
the coupled channel phase at the point marked B at $A_{v} = 2.0$ where there is
a narrow band noise signal.
Adapted with permission from C.J. Olson, C. Reichhardt, and F. Nori,
Phys. Rev. Lett. {\bf 81}, 3757 (1998).  Copyright 1998 by the American Physical Society.
}
\label{fig:10}
\end{figure}

A dynamic phase diagram as a function of $A_v$ versus $F_D$
constructed from the simulation measurements appears
in figure~\ref{fig:10}.
The width of the pinned phase decreases as $A_v$ increases since a stiffer system
is not as well pinned by random disorder.
The open circles and crosses indicate the locations of the peaks in the noise power
and $dV/dF_D$ curves, respectively.
For $A_v < 1.0$, the system depins into a plastic flow phase and then
reorders into a moving phase in which the dislocations have their Burgers vectors aligned
perpendicular to the driving direction, forming a smectic structure in which
$S({\bf k})$ contains only two prominent peaks.
For $A_v  > 1.0$, after the vortices depin plastically and 
dynamically reorder into a smectic
state, at higher drives a second dynamical reordering occurs into
a defect-free anisotropic crystal with six peaks in $S({\bf k})$.
The drive at which the vortices dynamically reorder into the crystalline state
diverges as $A_{v}$ decreases.
Inset (A) of figure~\ref{fig:10} shows a representative
velocity noise power spectrum $S(\omega)$ obtained in the plastic flow phase, where
$1/f^{1.5}$ noise appears, while in inset (B) of figure~\ref{fig:10}, 
$S(\omega)$ in the moving crystal phase has a
washboard character with a narrow band noise peak.
These results indicate that noise fluctuations
can serve as another powerful tool for examining
transitions between plastic and dynamically ordered phases.
In Section 4 we
provide a further discussion of noise measurements for nonequilibrium states.

In 2D systems such as these,
finite size effects can arise.  If a sufficiently large system were simulated,
it is possible that some aligned dislocations would appear in what is labeled
a moving crystal state
in figure~\ref{fig:10}, indicating that the moving crystal or coupled channel phase
is actually smectic at large scales.
If this is the case,
the moving smectic would exhibit another length scale in the form of
the average spacing between dislocations.  This length scale 
is controlled
by the ratio of the vortex-vortex interaction strength to the vortex-pinning
interaction strength.
Thus, it is
possible to form
a strong smectic in which each flowing channel of vortices contains numerous dislocations
separated by distances of the order of a few lattice constants,
as well as a weak smectic in which
large groups of channels are coupled and the dislocations are separated
by hundreds or thousands of lattice
constants.
Such finite size effects could impact
the results of
Pardo {\it et al.} \cite{79}, where the observation of a moving crystal phase
rather than a moving smectic phase could indicate that
the length scale at which dislocations appear was larger than the region over
which the experimental images were obtained.

Despite the large number of studies on dynamical reordering in 2D vortex systems,
there are still many open questions.
The distances between the aligned dislocations in the smectic phase may depend
on the drive and on the disorder strength, and might exhibit additional divergences.
The nature of the dynamical reordering transition is unknown.
It might be similar to an equilibrium freezing transition; alternatively, it might
be better understood as a type of absorbing phase transition similar to directed
percolation, as described in Section 12.
There could be a true transition between the moving smectic and moving crystal phases,
and if so, the nature of this transition is also unknown.
Additional transitions between different phases could occur, such as
anisotropic plastic to moving nematic to moving smectic to moving crystal,
and methods would need to be identified to distinguish between these phases if
they are truly distinct phases.
If the disorder is very weak, it is possible that only a moving floating crystal could
occur, with the effects of the pinning vanishing completely;
alternatively,
it is possible that the
pinning effects never completely disappear.
There are also some studies that find that the orientation of the mostly triangular
dynamically ordered moving state can rotate as a function of driving force, but the
exact reason for this rotation has not yet been identified;
it could be due to extra dissipation created by
additional modes of motion of the vortex core \cite{86}.

\section{Depinning and First Order Dynamical Phase Transitions in 3D Systems} 

Most of the results in Section 3
were obtained in
2D or effectively 2D systems, where the depinning and dynamic
reordering transitions are generally continuous or second order in nature.
In this Section we discuss systems in which
some of these transitions are
sharper or first order in character.
Transitions of this type can occur in overdamped 3D vortex systems
and in 2D layered systems with geometries similar to those illustrated
in figure~\ref{fig:3}.
First order behavior can also arise in systems containing inertia or
stress overshoot effects.

The system in which 3D depinning and dynamic effects have been most extensively 
studied is vortices in type-II superconductors. 
The vortices are line-like objects, and in the limit where the lines are stiff, the 
2D approximation of their behavior described in Section 3 works well; however,
there are many superconducting systems that are strongly anisotropic or contain
a series of superconducting layers, and in these systems
the vortex lines can break up 
or decouple along the direction of the applied magnetic field \cite{1}.
Even for isotropic 3D superconductors or only weakly anisotropic
superconductors, near $T_c$ or $H_{c2}$ the vortex lines may 
become soft enough 
to break up along their length in the presence of strong pinning.
Thus, an important question is how the
dynamics
of interacting vortices changes when plasticity is allowed to occur in the 
third direction.

There are numerous features observed in real superconductors, 
particularly near the peak effect, that  
have not been successfully captured in 2D simulations.  
In general, 2D simulations show that as the vortex-vortex interactions weaken relative 
to the vortex-pinning interactions, the critical depinning threshold
increases gradually \cite{50,76,85,87,88N},
while in many experiments, 
the increase in $F_{c}$ is usually much sharper.
Paltiel {\it et al.} \cite{88} showed that the onset of the peak effect in 
2H-NbSe$_{2}$ is very sharp when edge contamination 
effects are removed, and argued that the vortex order-to-disorder transition associated
with the  
peak effect is  a first order transition \cite{89}.
Near the peak effect,
strong hysteresis in the velocity-force curves
is commonly observed \cite{90,91,92} along with a variety of  
memory effects \cite{93,94,95,96,97},
indicating the presence of metastability which is consistent with
an underlying first order phase transition.
Imaging experiments reveal the
coexistence of 
ordered and disordered pinned phases separated by sharp,
well defined boundaries, lending support to the first-order nature
of the order-disorder transition
\cite{98}.
In 3D strongly layered superconducting systems, 
it is generally agreed that 
the 3D to  2D decoupling or
melting transition is first order \cite{99}.   
The fact  that 2D simulations
do not produce many of these features is consistent with the idea
that  
isotropic 2D systems of the type illustrated
in figure~\ref{fig:1}(a,b) generally do not exhibit sharp transitions.
The melting transitions in 
most 2D systems are continuous, taking the form
of either a Kosterlitz-Thouless
transition,
a second order transition, or at
most a weak first order transition,
while when random disorder is
added to 2D systems, it  becomes even less likely that 
first order transitions can occur \cite{100}. 

The question is whether driven 3D systems in the presence of random disorder 
exhibit transitions that are first or second order in nature, and whether
these features are consistent with what is observed in vortex experiments. 
Most theoretical and simulation work
has focused on 3D elastic line models or 3D layered models
of the type illustrated in figure~\ref{fig:3}(b,c).
In 3D elastic line models, the vortices can exhibit plasticity in the
$x-y$ plane perpendicular to their length, but not along the $z$ direction
parallel to their length.
One issue with such elastic models is how to treat
the behavior when the line wandering
along the $z$-direction becomes large enough that crossing or
entanglement of the lines can occur.
The vortex lines can be modeled as unbreakable objects
like directed polymers that cannot pass through each other, or
they can be allowed to cross each other with line breaking 
and reconnection taken into account.
An alternative 3D model consists of coupled planes that each contain a fixed number
of 2D vortices called pancake vortices \cite{clem}.
Each pancake vortex has a repulsive interaction with
vortices in the same layer but experiences an 
attraction to vortices in adjacent layers.
In this model, the 2D vortices can stack up into 3D line-like objects but can also
undergo decoupling transitions in which the line-like object breaks
apart and the vortices move independently in each layer.

\begin{figure}
  \includegraphics[width=\columnwidth]{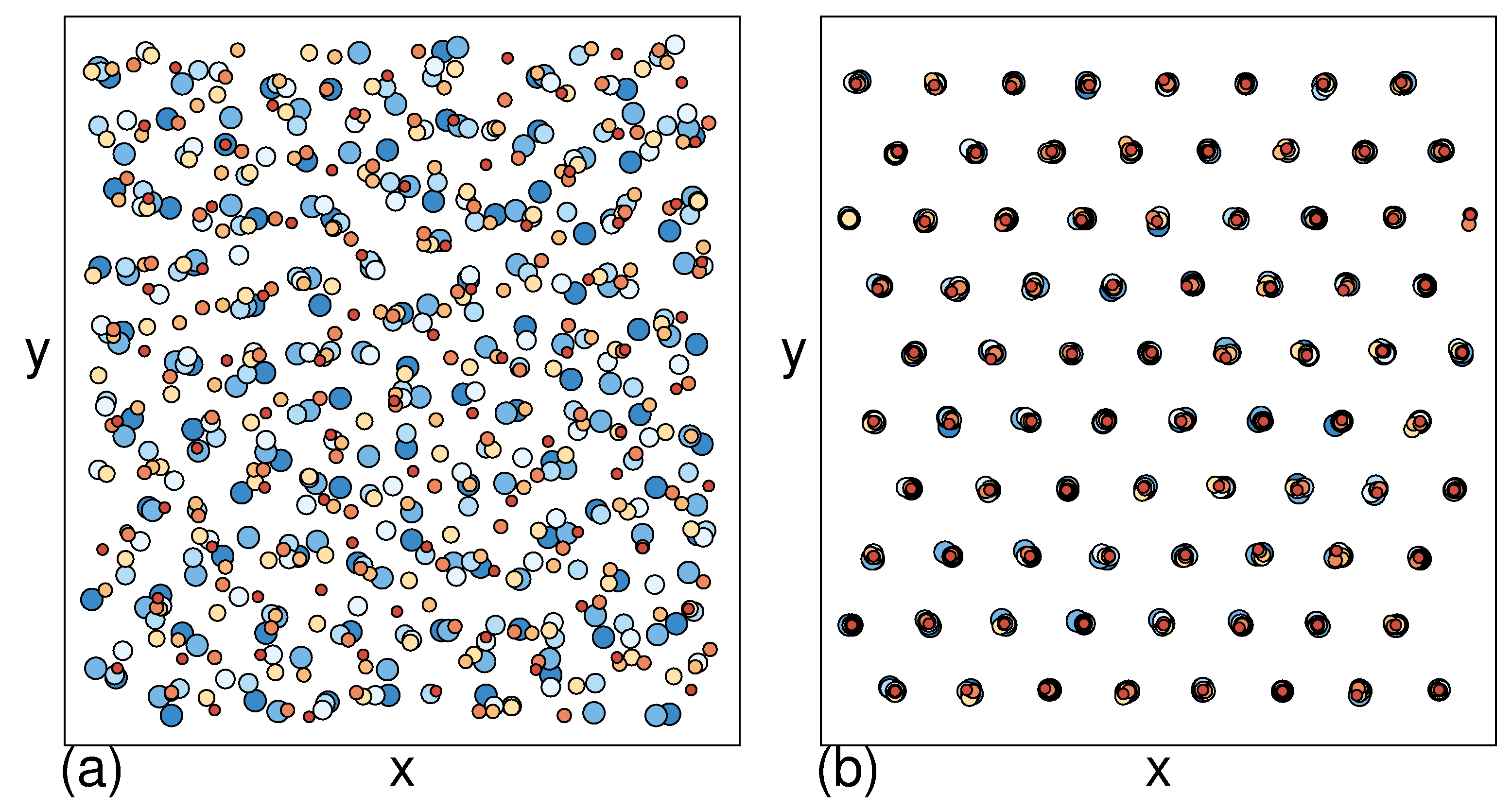}
\caption{ Top view of a simulation of a 3D layered system
  of magnetically interacting vortices in the presence of 
  pinning in the form of random point defects.
  Circles are the vortices in each layer, with different colors and sizes indicating the different
  layers; the pinning sites are not shown.
  (a) A disordered phase in which the vortices are  decoupled and move
  independently  in each layer.
  (b) An ordered state in which the vortices are aligned in the $z$ direction and
  form a triangular lattice in the $x-y$ plane.
Adapted with permission from C.J. Olson, C. Reichhardt, and V.M. Vinokur,
Phys. Rev. B {\bf 64}, 140502(R) (2001).  Copyright 2001 by the American Physical Society.
}
\label{fig:11}
\end{figure}

\begin{figure}
  \includegraphics[width=\columnwidth]{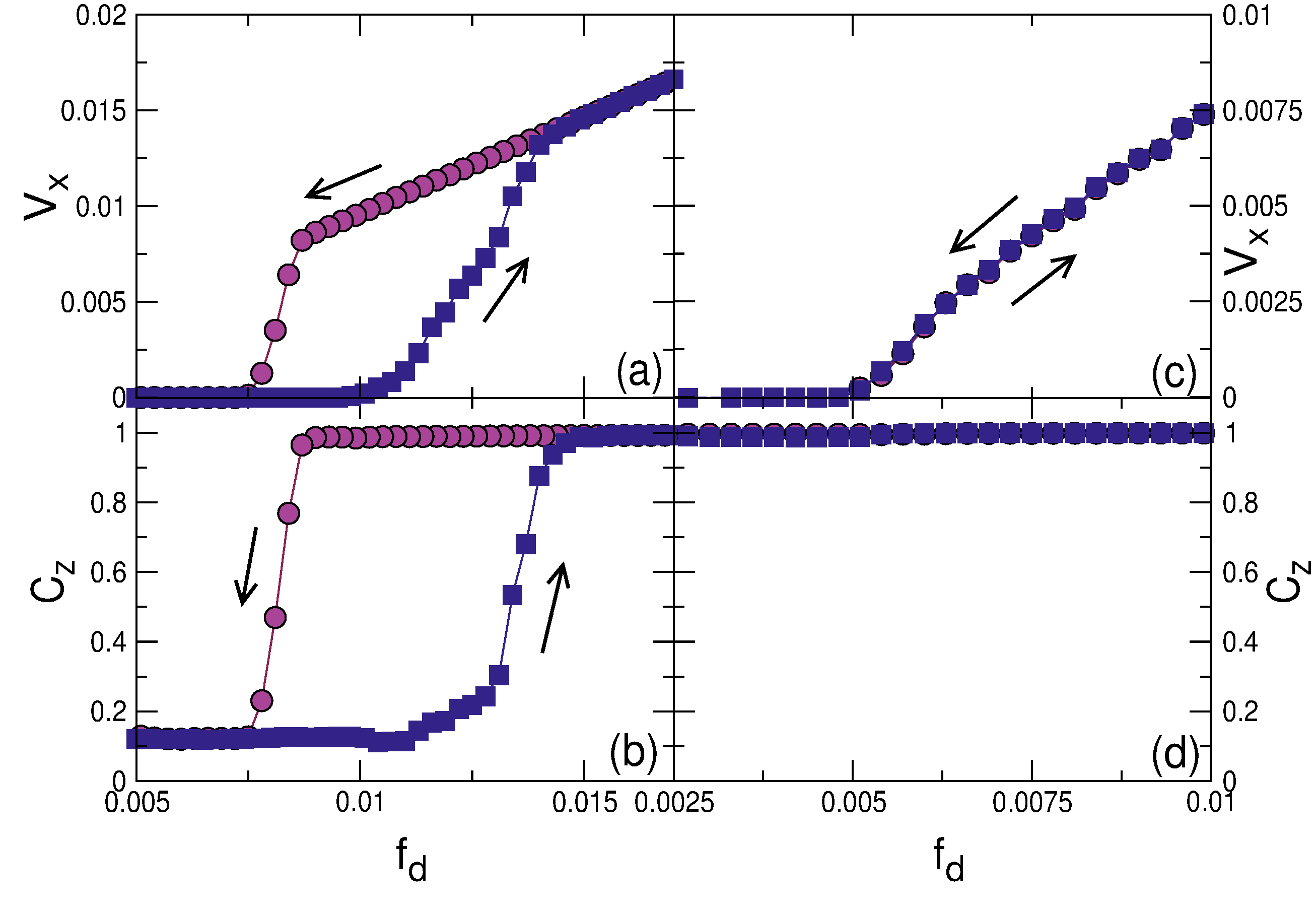}
\caption{
(a) Average vortex velocity $V_x$ in the driving direction vs 
driving force $f_d$ in the disordered decoupled phase for 
simulations of the 3D layered system in figure~\ref{fig:11}(a). 
The vortices depin plastically and dynamically reorder at higher drives.  Strong hysteresis
is present for sweeping the drive up (blue squares) and down (red circles), indicative
of behavior that is first order in nature.
(b) $C_z$ vs $f_d$.
The quantity $C_z$ measures the correlations of the vortex lines along the $z$ direction,
$C_z=1-\langle (2R_z/a_0)\Theta(a_0/2-R_z)\rangle$
where $R_z=|{\bf R}_{i,l}-{\bf R}_{i,l+1}|$, ${\bf R}_{i,l}$ is the location of vortex
$i$ in layer $l$, $a_0$ is the vortex lattice constant, $\Theta$ is the Heaviside
step function, and the average is performed over all vortices in neighboring layers.
Below depinning and for low drives, $C_z$ is small, and at higher drives
there is a dynamical transition into an ordered 3D solid.
(c) $V_x$ vs $f_{d}$ curves for the ordered 3D solid illustrated in figure~\ref{fig:11}(b).
The depinning is elastic and there is no hysteresis in 
in the velocity response. 
(d) $C_{z}$ vs $f_d$ for the ordered 3D solid also shows no hysteresis.
Here $V_x$ corresponds to the velocity $V$ and $f_d$ corresponds to the driving force
$F_D$.
Adapted with permission from C.J. Olson, C. Reichhardt, and V.M. Vinokur,
Phys. Rev. B {\bf 64}, 140502(R) (2001).  Copyright 2001 by the American Physical Society.
}
\label{fig:12}
\end{figure}

\subsection{Depinning and dynamical phase transitions}

Simulations of 3D elastic lattice line models show that dynamical ordering can occur
between a plastic flow state and a moving crystal; however, a
first order or peak effect transition
is not observed \cite{101,102},  suggesting that some form
of 3D plasticity or line breaking that is neglected by the models must be important
in the experimental systems.
Olson {\it et al.} considered a 3D layered model of magnetically interacting vortices as
illustrated in figure~\ref{fig:11}(b), which shows
a top view of the system in which the vortices in $N_L=8$ layers
are aligned with each other along the $z$ axis and form a triangular lattice
in the $x-y$ plane \cite{103}.
In the absence of pinning and as a function of temperature,
this model shows a single first order melting transition
where the lines break
up into decoupled layers similar to those shown in
figure~\ref{fig:11}(a).
At $T = 0$ 
it also shows a first order transition as a function of increasing pinning strength
from a 3D ordered state to a 2D disordered decoupled state. 
For weak pinning under an applied drive $F_{D}$,
the 3D ordered system depins elastically into a
moving  3D line solid, and there is
no hysteresis in the velocity-force curves,
as shown in figure~\ref{fig:12}(c).
In both the pinned and moving phase,
a measure $C_z$ of the correlations of the vortex lines along the $z$ direction
gives $C_z=1.0$,
as shown in 
figure~\ref{fig:12}(d).  
The lack of hysteresis or any discontinuous behavior
upon sweeping the driving current up and down
suggests that elastic depinning in this 3D system is a continuous transition; 
however, a scaling analysis of the velocity-force curves has not been performed.
When the strength of the pinning is increased or the 
effective vortex-vortex coupling between layers is reduced by thermal effects or
other means,
then the pinned state is no longer a 3D ordered state but is instead a disordered
decoupled 2D state of the type illustrated in figure~\ref{fig:11}(a).
This decoupled 2D state depins plastically, as shown in figure~\ref{fig:12}(a),
and at higher drives the system exhibits a sharp transition into a moving 
3D ordered solid,
as indicated by the abrupt increase of $C_z$ from a low value to $C_z=1.0$ within the
moving phase in figure~\ref{fig:12}(b).  
Both the velocity-force curves and $C_z$ are strongly hysteretic upon
sweeping the drive up and down, with the coupled 3D ordered state persisting
down to much lower drives during the decreasing drive sweep.  The net vortex velocity
in the 3D ordered phase is higher than in the 2D decoupled phase, indicating that
the pinning is less effective for the 3D ordered state.

\begin{figure}
  \includegraphics[width=\columnwidth]{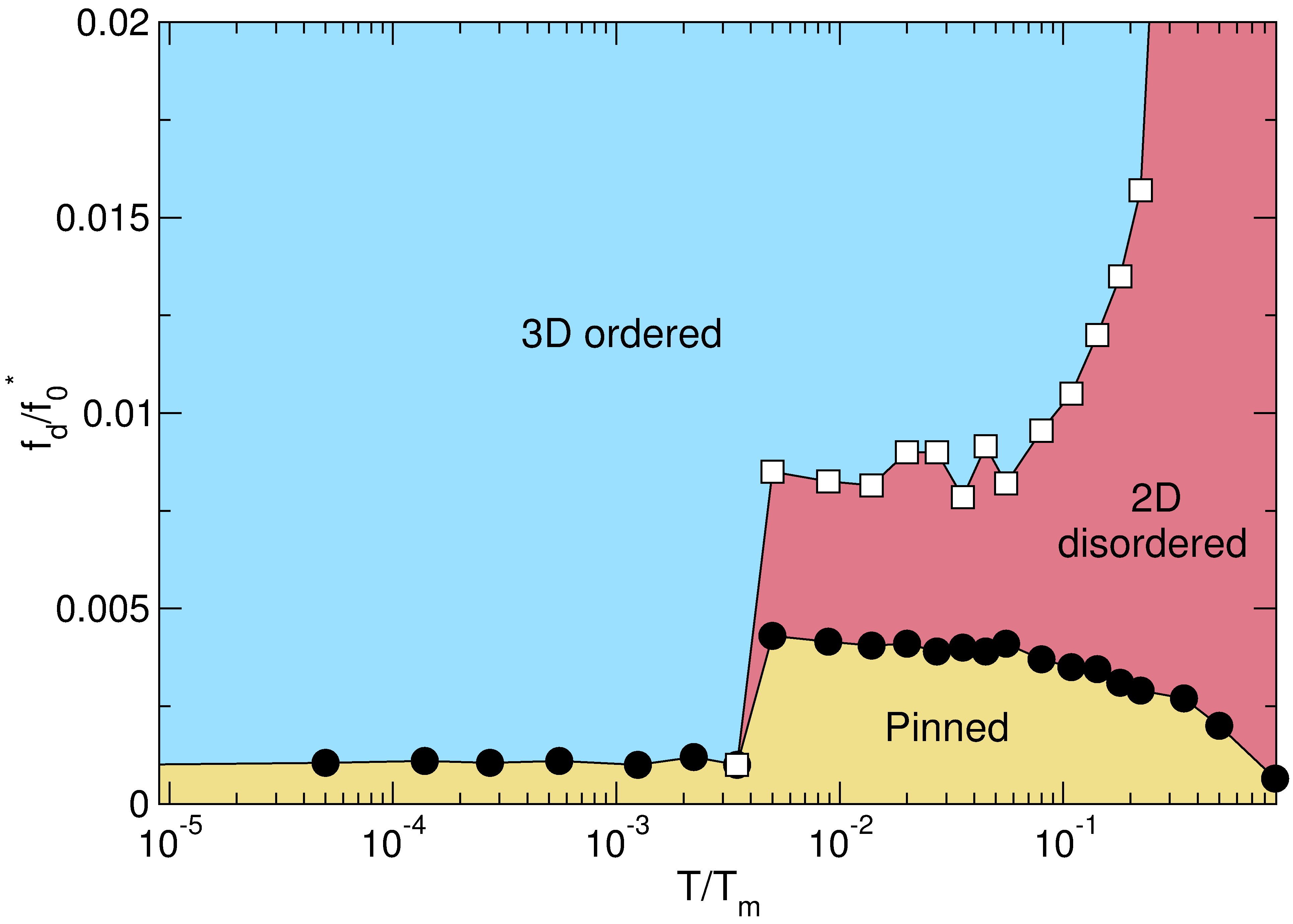}
\caption{
The simulated dynamic phase diagram as a function of $f_d/f_0$ vs temperature $T/T_m$ 
for the  3D magnetically coupled layered vortex system shown in 
figures~\ref{fig:11} and ~\ref{fig:12}
with weak pinning.
$T_m$ is the melting temperature and $f_d/f_0$ corresponds to the driving force $F_D$.
Black circles indicate the depinning threshold $F_c$, while white squares show the
dynamical reordering line.
At  low $T$, the 3D solid depins elastically as shown in figure \ref{fig:12}(c,d), while
at higher $T$ there is a transition to a pinned 2D disordered state  that depins
plastically and undergoes a first order dynamical transition into a moving 3D solid
at higher drives, as shown in figure~\ref{fig:12}(a,b).
There is a sharp increase in the depinning
threshold at the transition from the 3D ordered pinned 
state to the 2D decoupled pinned state.  
Adapted with permission from C.J. Olson, C. Reichhardt, and V.M. Vinokur,
Phys. Rev. B {\bf 64}, 140502(R) (2001).  Copyright 2001 by the American Physical Society.
}
\label{fig:13}
\end{figure}

Figure~\ref{fig:13} shows
the dynamic phase diagram for the
3D layered vortex system from Figures~\ref{fig:11} and \ref{fig:12}
with weak pinning as a function of 
$F_{D}$ versus temperature $T/T_{m}$, where $T_m$ is the melting temperature
of the clean system.  At low $T$ the vortices depin elastically
from a 3D pinned state to a moving 3D solid, while
at higher $T$ there is a first order transition within the pinned state
to a disordered 2D state which is associated with a 
sharp increase in the depinning threshold $F_{c}$. 
From the 2D disordered state, the vortices depin plastically 
and at higher drive undergo a first order dynamical transition into 
an ordered 3D moving solid. 
The driving force at which the dynamical ordering occurs shows
a divergence as $T$ approaches $T_{m}$ from below, 
similar to the behavior observed in the 
2D simulations of Koshelev and Vinokur \cite{45}.
As $T$ is further increased, the depinning threshold decreases
since 2D vortices 
can be easily thermally depinned. These results show that many of the 
features observed experimentally in the peak effect  
regime, including a first-order transition from a pinned elastic solid to a pinned
disordered solid and a discontinuous dynamical ordering transition,
can be captured 
by a 3D model that allows plasticity in the $z$-direction.

\begin{figure}
  \includegraphics[width=\columnwidth]{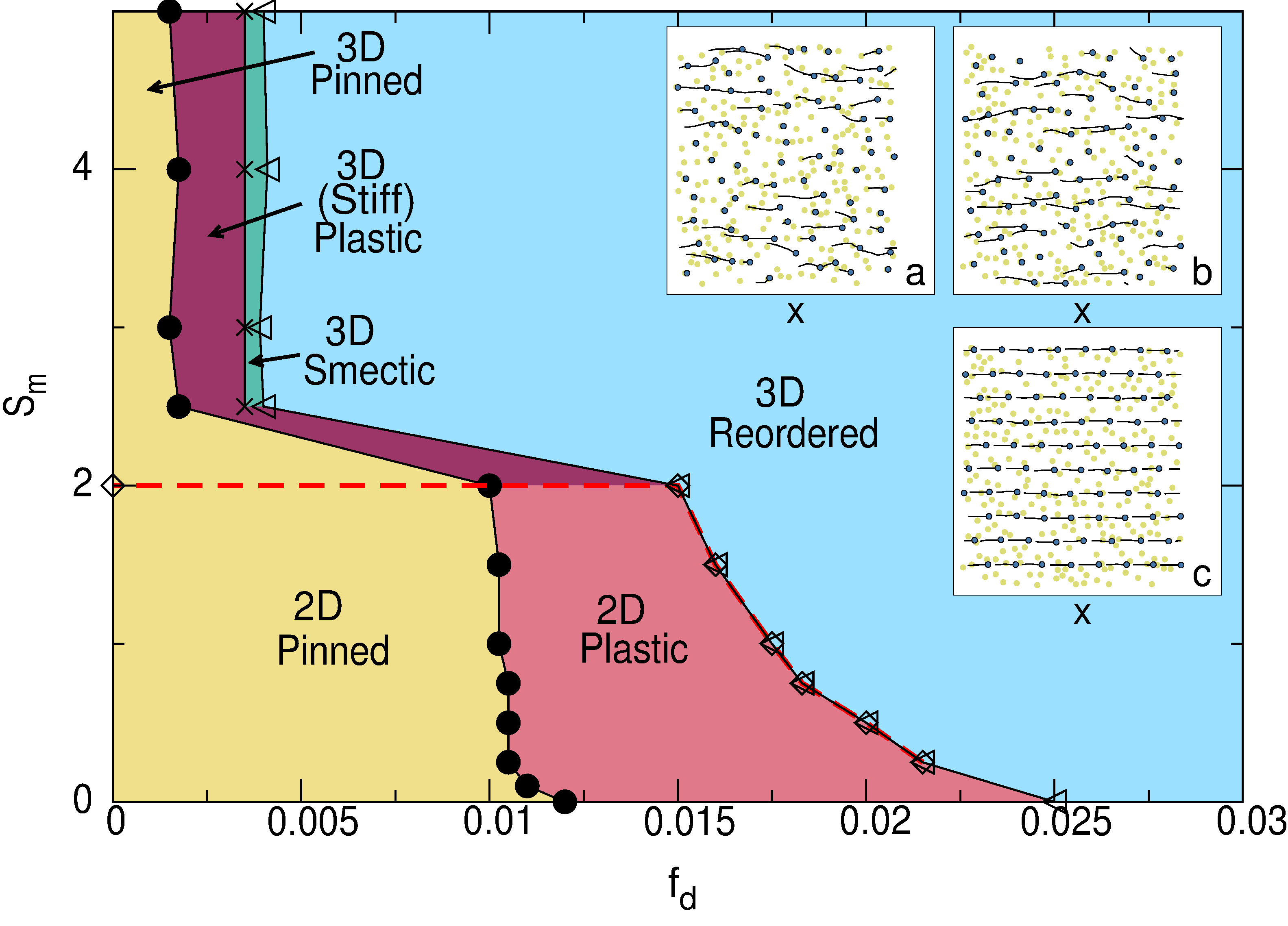}
\caption{
  Simulated dynamic phase diagram as a function of interlayer coupling strength $S_m$ 
  vs driving force $f_{d}$ for $T = 0$ for the 3D magnetically coupled layered vortex system
  shown in figures~\ref{fig:11}  to ~\ref{fig:13} with weak pinning.
  There is a first order phase transition,
  indicated by the dashed red line, from a 3D ordered solid into a 2D disordered 
 decoupled phase.
 The other phases are plastic flow of stiff 3D lines, a 3D smectic phase, a 3D dynamically
 reordered phase, and a 
 2D disordered plastic flow phase.
 Insets (a) and (b) show the top and bottom layer, respectively, of the pinning sites
 (green dots), vortex positions (blue dots) and trajectories (lines)
 in the 2D plastic flow phase, while inset (c) illustrates one layer of the
 3D ordered moving phase.
 Here $f_d$ corresponds to the driving force $F_D$.
Adapted with permission from C.J. Olson, G.T. Zim{\' a}nyi, A.B. Kolton, and
N. Gr{\o}nbech-Jensen,
Phys. Rev. Lett. {\bf 85}, 5416 (2000).  Copyright 2000 by the American Physical Society.
}
\label{fig:14}
\end{figure}

Other numerical studies of the same 3D
layered vortex model at $T = 0$ focused on the effect 
of modifying the strength of the interlayer coupling $S_m$
\cite{104}.
Figure~\ref{fig:14} shows a dynamical phase diagram
for this model as a function
of $S_{m}$ versus $F_{D}$ 
at a fixed pinning strength.  For 
strong interlayer coupling $S_{m} > 2.0$,
the system forms a 3D ordered pinned state that depins into 
a 3D plastic flow state in which the vortices form stiff lines that can flow plastically
around each other in the $x-y$ plane.
At higher drives there is a transition into a 3D smectic
phase followed by a transition to a 3D moving crystal phase.
In this 3D regime the $dV/dF_D$ curve exhibits 
a single peak similar to that found in 2D simulations of completely stiff vortex lines.
As 
$S_{m}$ decreases,
there is a first order transition into a disordered decoupled
2D state that depins plastically and at higher drives
undergoes a first order
transition into a 3D ordered moving line state.
The $dV/dF_{D}$ curves
have distinct double-peak features associated with
this two step depinning and dynamical reordering process.
The top insets in figure~\ref{fig:14}
show that the vortex positions
and trajectories obtained simultaneously in the top and bottom layers of the
sample in the 2D decoupled plastic flow phase
are distinct, while the bottom inset illustrates vortex flow
in the top layer of the sample 
in the 3D dynamically ordered phase
where the flow in all layers is identical.
Additional studies of the same model show dynamical 
ordering  from a 2D decoupled state to a
3D ordered structure \cite{105,106}, and many of the memory and metastability
effects that are observed in experiments can
be captured in the 3D layered vortex model by superheating and
supercooling the sample across the first order transition
\cite{107}.

In 2D systems
the most general moving ordered state at high drives
is a moving smectic phase; however,
simulations of the 3D layered vortex model
show evidence 
of only small regions of 3D moving smectic phases, and it is not clear if
such 3D smectic states are robust or whether they
are a result of finite size effects.
Other models for 3D vortex systems include 3D anisotropic frustrated
X-Y systems of the type used by Chen and Hu \cite{108} 
to examine the  weak pinning regime
in which the system forms an ordered assembly of 3D lines without topological defects 
or a 3D Bragg glass state.
In this work, 
a first order transition separates the moving
3D ordered state from a moving smectic state, which transitions continuously into a liquid 
state as a function of temperature.
Hern{\' a}ndez and Dom{\' \i}nguez studied a similar model and observed
that a first order transition from a 3D ordered pinned state to
a 2D disordered state is associated with a large increase
in the depinning threshold \cite{109}.

\subsection{Phase locking effects}

In the moving ordered Bragg glass state,
Chen and Hu observed 
a pronounced washboard signal in the velocity noise spectrum, with
a narrow band noise peak called a washboard peak at a characteristic frequency   
of $f_{0} = V/a_0$ \cite{108}, where $V$ is the velocity along the driving
direction and $a_0$ is the vortex lattice constant.
Similar washboard noise spectra have been observed
in the ordered state in 2D systems  \cite{85,101};
however, the washboard peak is particularly sharp in the 3D system \cite{108}, 
indicating that the moving 3D ordered state
undergoes fewer random fluctuations than the moving 2D
ordered state.  Washboard noise and
broad band noise features
have also  been observed
experimentally in layered superconductors \cite{111,112}.
When narrow band noise is present,
interference effects can arise if an ac drive is applied to the sample in addition
to the dc drive.
When the ac drive frequency matches the washboard frequency or
its harmonics, the two frequencies lock and
a series of Shapiro steps appear in the velocity-force curves.
Such effects were first observed by Fiory
\cite{113} 
for the moving vortex lattice, while the washboard effect for vortices moving over random
pinning was explained by  Schmid and Hauger \cite{114}.
Phase locking, including enhanced ordering effects,
of vortices moving over random pinning
has been studied in experiments \cite{112,115,116,117} 
and simulations \cite{110},
while related studies have been performed for
colloids driven over a periodic substrate \cite{118}. 
Narrow band noise and phase locking steps have 
also been studied
in experiments on sliding CDW systems \cite{4}.
In this case, the scaling 
of the shape of the velocity-force curve as it enters and exits individual
phase locked steps is similar to the scaling observed at an elastic
depinning transition, with
$V \propto (F_{D} - F_{s})^\beta$, where $F_{s}$ is the drive    
at which the phase locking occurs and $\beta < 1.0$ \cite{119}.
It is not known whether vortices obey scaling of this type as they enter and
exit phase locking steps,
nor is it known whether the character of the
Shapiro steps is different in 2D and 3D vortex systems, or whether different phase
locking signatures would appear depending on whether the system is in a moving
smectic state or a moving crystal state.

\begin{figure}
  \includegraphics[width=\columnwidth]{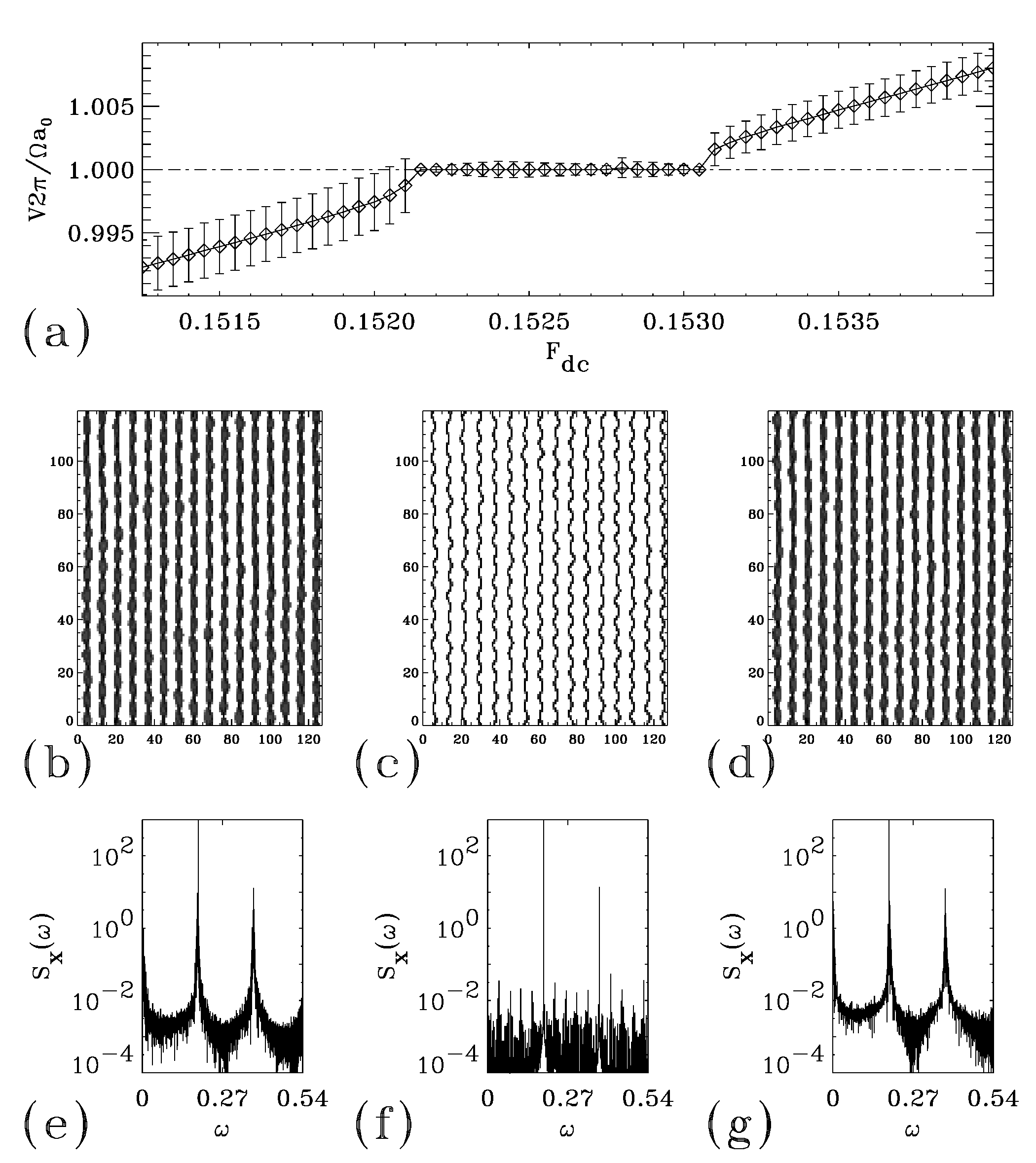}
\caption{
  (a) Simulations of velocity $V$ vs dc driving force $F_{dc}$
  in the window around the main transverse phase locking step in
  a system with a fixed ac drive applied transverse to the dc drive for vortices moving
  over random disorder in the dynamically ordered phase.
(b,c,d) Typical  time-averaged
  coarse-grained  vortex densities.  Here the
  dc driving force is applied along the $y$
  axis and the transverse ac drive is applied along the $x$ axis.
  (e,f,g) Typical voltage power spectra $S_x(\omega)$ of the transverse vortex velocity.
  (b,e)  A  phase-unlocked  state  below the locking step.
  (c,f) A phase-locked state on the locking step.
  (d,g) A phase-unlocked state  above  the  locking step.
  Reprinted with permission from A.B. Kolton, D. Dom{\' \i}nguez, and N. Gr{\o}nbech-Jensen,
Phys. Rev. B {\bf 65}, 184508 (2002). Copyright 2002 by the American Physical Society.
}
\label{fig:kolton}
\end{figure}

If a longitudinal drive is applied to the vortices in order to place them in a dynamically
ordered state, and then a transverse ac drive is added,
the vortices can become even more ordered and can exhibit transverse phase
locking.
Even though the
dc drive causes the vortices to move in the longitudinal
direction, the narrow band noise signal
of the reordered state produces a periodic modulation of the vortex
motion in the transverse direction as well, making it possible for the transverse ac drive
to lock to the narrow band frequency.
Kolton {\it et al.} \cite{110a} examined
transverse phase locking 
in the moving ordered phase for vortices driven over random disorder.
Figure~\ref{fig:kolton}(a) shows that there is a clear phase locking
step in the longitudinal vortex
velocity as a function of the longitudinal dc drive
$F_{D}$ when a transverse ac drive is present.
The vortex trajectories in figure~\ref{fig:kolton}(b) show
that below the locking step,
the vortices are moving in channels
in the longitudinal direction ($y$ direction in the figure) but there is some spread in the
trajectories along the transverse direction.
Above the locking step, the trajectories are similar, as illustrated in
figure~\ref{fig:kolton}(d), but on the step a highly ordered sinusoidal pattern appears,
as shown in figure~\ref{fig:kolton}(c).
In the power spectra of the velocity time series, shown in figure~\ref{fig:kolton}(e-g),
there is a sharp resonance on the locking step due to the highly ordered motion.
Although longitudinal phase locking effects have been observed
experimentally \cite{117}, there has not yet been an experimental observation
of transverse phase locking.
 
\begin{figure}
  \includegraphics[width=\columnwidth]{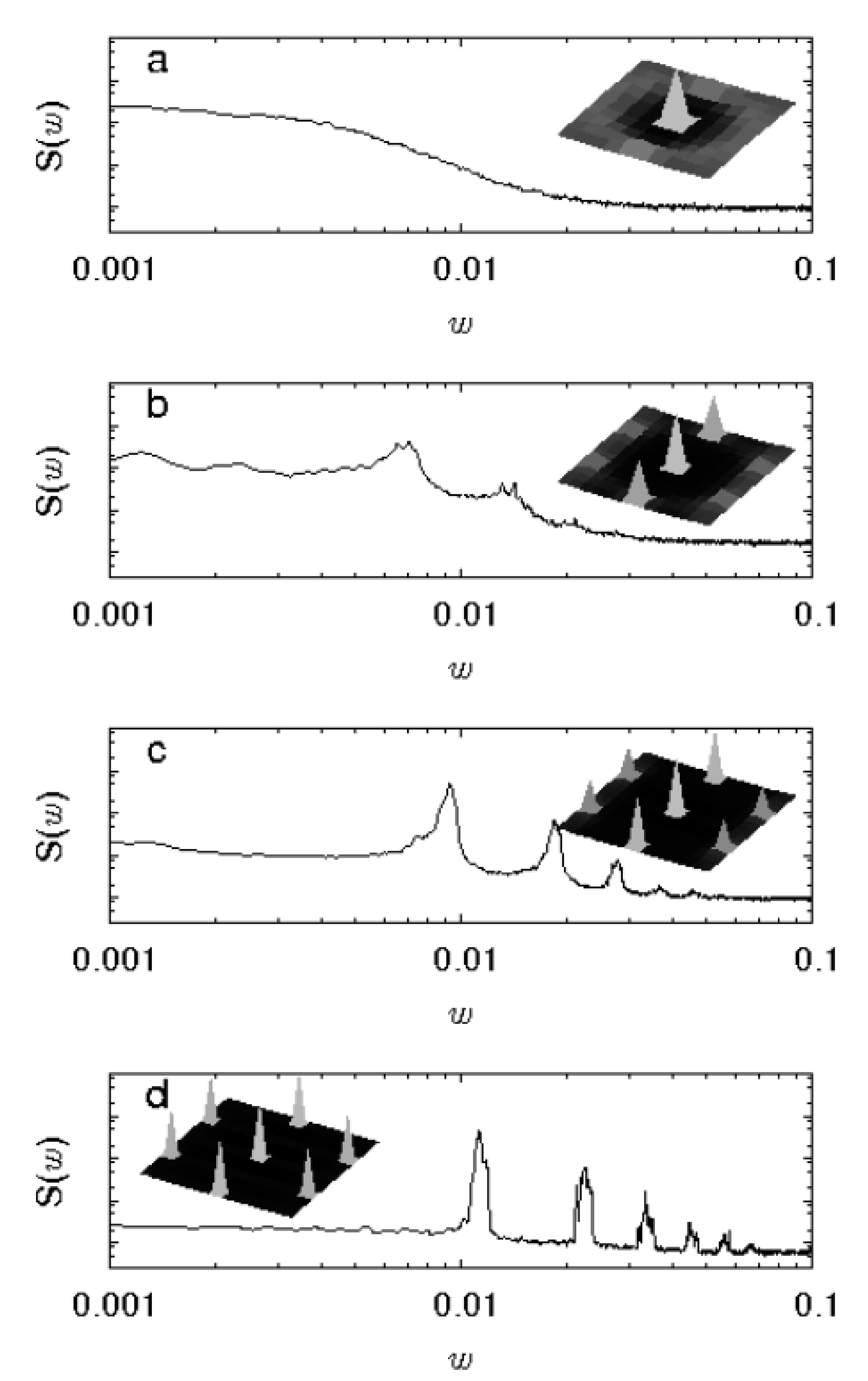}
  \caption{ The vortex velocity power spectra
    $S(\omega)$ from
    simulations of a 3D elastic line vortex model in the moving phase 
for increasing vortex density, from top to bottom. The insets 
show the corresponding structure factor $S({\bf k})$.
(a) Plastic flow phase.
(b) Smectic flow phase.
(c) Anisotropic crystal phase.
(d) Moving crystal phase. 
Reprinted
with permission of Springer
from T.J. Bullard, J. Das, G.L. Daquila and U.C. T{\" a}uber,
Eur. Phys. J. B {\bf 65}, 469 (2008).
Copyright 2008 by the The European Physical Journal.
}
\label{fig:15}
\end{figure}

Voltage noise measurements that can be used to detect the
existence of a washboard frequency provide a useful
experimental method of determining
where the different dynamical phases occur as well as the amount of order in
the moving structure. 
An illustration of how the power spectra and
noise fluctuations are correlated with the amount of order in the 
vortex system appears in figure~\ref{fig:15}, which shows
the evolution of  the velocity noise spectrum
$S(\omega)$ in the moving state for increasing
vortex density for the  3D elastic
line model studied by  Bullard et al \cite{101}.
The inset of each panel shows the corresponding structure factor
$S({\bf k})$ of the moving state. 
For the plastic flow phase in figure~\ref{fig:15}(a), the noise is broad band
and the structure factor has a ring feature
indicative of an amorphous state. 
Figure~\ref{fig:15}(b) shows that at a higher density the system
forms a moving smectic as indicated by the appearance of two peaks
in $S({\bf k})$ at finite ${\bf k}$, while
$S(\omega)$ begins to develop a series of smeared peaks as a washboard
frequency emerges.
In figure~\ref{fig:15}(c) the system is even more ordered and
there are now six peaks in $S({\bf k})$; however, the ordering is still anisotropic as
two of the peaks are much more pronounced than the others.
The washboard frequency peak in $S(\omega)$ is more prominent,
reflecting the stronger ordering in the moving system.  
In figure~\ref{fig:15}(d) a defect-free moving lattice state forms in which
$S({\bf k})$ contains 
six peaks that are nearly equal in weight,
while $S(\omega)$ shows sharp spikes at the
washboard frequency and its
higher harmonics.
The washboard frequency increases with increasing vortex density as 
the vortex lattice constant decreases.     

There are still many questions to ask
about the dynamics of 3D depinning, and very little
is known beyond the behavior of directed line models.
A 3D isotropic lattice of particles could arise in systems 
such as a 3D Wigner crystal \cite{120,121}, 
3D colloidal Yukawa systems \cite{122,123}  and    
3D cubic lattices of skyrmions \cite{124,125}. 
Systems of this type could exhibit
a rich variety of plastic flow and dynamically ordered phases in which
the particles could dynamically order into
1D columns or 2D slabs aligned with the drive or into other anisotropic ordered
structures.
It would be interesting to determine whether such reordering transitions are
continuous or first order.
Other systems in which 3D reordering transitions could be studied include
3D liquid crystals moving in random disorder as well as
pattern forming systems or Lennard-Jones particles, where additional effects
could arise such as the formation of moving labyrinth phases.
Finally, there have been many studies of packing and glass transitions
in dimensions higher than 3 to determine where mean-field behavior emerges.
It may be interesting to examine depinning and dynamical ordering
transitions in systems with 4 or more dimensions to explore whether
a similar mean-field limit can be reached.

\section{Depinning and First Order Dynamical Phase Transitions in Other
  Layered Systems}

\begin{figure}
  \includegraphics[width=\columnwidth]{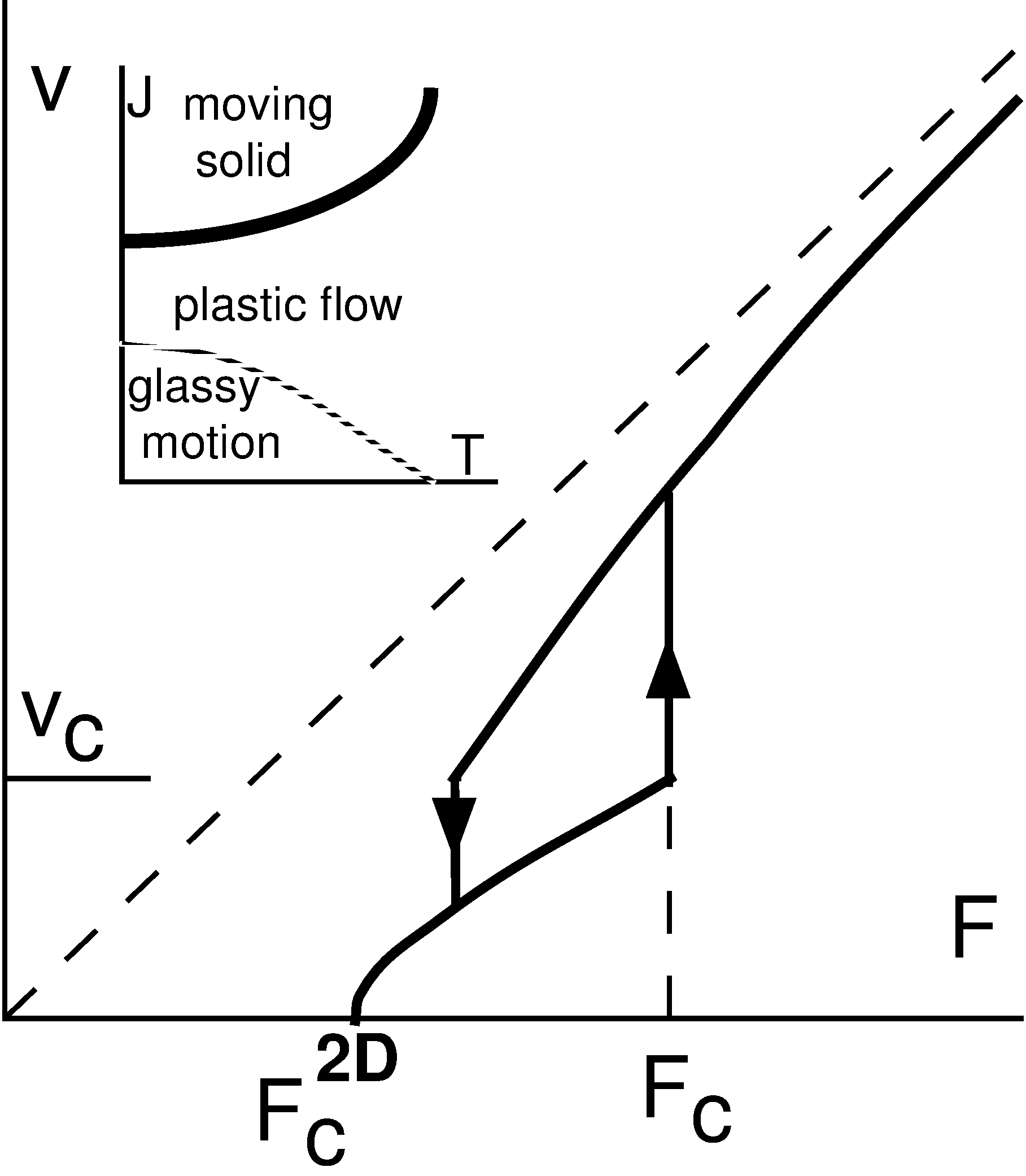}
\caption{ 
  The theoretical velocity $V$ versus driving force $F$
  response for a 3D layered sliding CDW system.
A continuous transition to plastic flow at $F^{2D}_{c}$ is followed by a sharp
hysteretic transition into a coherent moving 3D state at $F_{c}$. 
The inset shows the dynamic phase diagram for driving force $J$
vs temperature $T$, highlighting
the evolution of the two depinning transitions a function of temperature.
Here, $F$ and $J$ correspond to the driving force $F_D$, $F_c^{2D}$ corresponds
to the critical depinning threshold $F_c$, while $F_c$ in the figure
corresponds to the dynamical ordering transition.
Reprinted with permission from V.M. Vinokur and T. Nattermann,
Phys. Rev. Lett. {\bf 79}, 3471 (1997). Copyright 1997 by the American Physical Society.
}
\label{fig:16}
\end{figure}

\subsection{Systems with many layers}

Continuous and first order depinning phenomena can occur 
in other 3D and 2D layered systems beyond 
superconducting vortices, such as in
sliding CDWs \cite{4},
layered Wigner crystals \cite{126},
and particles moving along the easy direction
through periodic asymmetric 1D constrictions 
\cite{127}.
For example,
layering effects can produce first order 
dynamical depinning transitions in 2D or 3D sliding CDWs.
Although many CDW systems are thought to 
exhibit purely elastic behavior,
experiments have revealed a variety of hysteretic,
switching, and discontinuous behaviors,
which have been argued to indicate that a purely overdamped
elastic description of CDWs often breaks down \cite{4,33,128,129}. 
Vinokur and 
Nattermann \cite{130} theoretically examined the depinning transition
for a layered CDW system and
found that the velocity-force curves
exhibit a two-step depinning processes, with an initial
continuous depinning into a plastic flow phase of decoupled layers followed by a second
sharp hysteretic transition at higher drives into a coherent moving structure.
For weaker disorder,
the system undergoes a single depinning transition into a coherently sliding 3D solid.
The velocity-force curves in
figure~\ref{fig:16} show
a depinning transition
at $F_c$
followed by a sharp hysteretic jump in $V$ at
the dynamical reordering transition into a coherently moving 3D solid.
The inset of figure~\ref{fig:16} shows a dynamical phase diagram
as a function of driving force versus temperature
in the strong pinning limit where
pinned, plastic flow, and moving solid phases appear. 
The overall behavior of this system is very similar to that
observed in simulations of a 3D layered vortex system, described in Section 4.

\begin{figure}
  \includegraphics[width=\columnwidth]{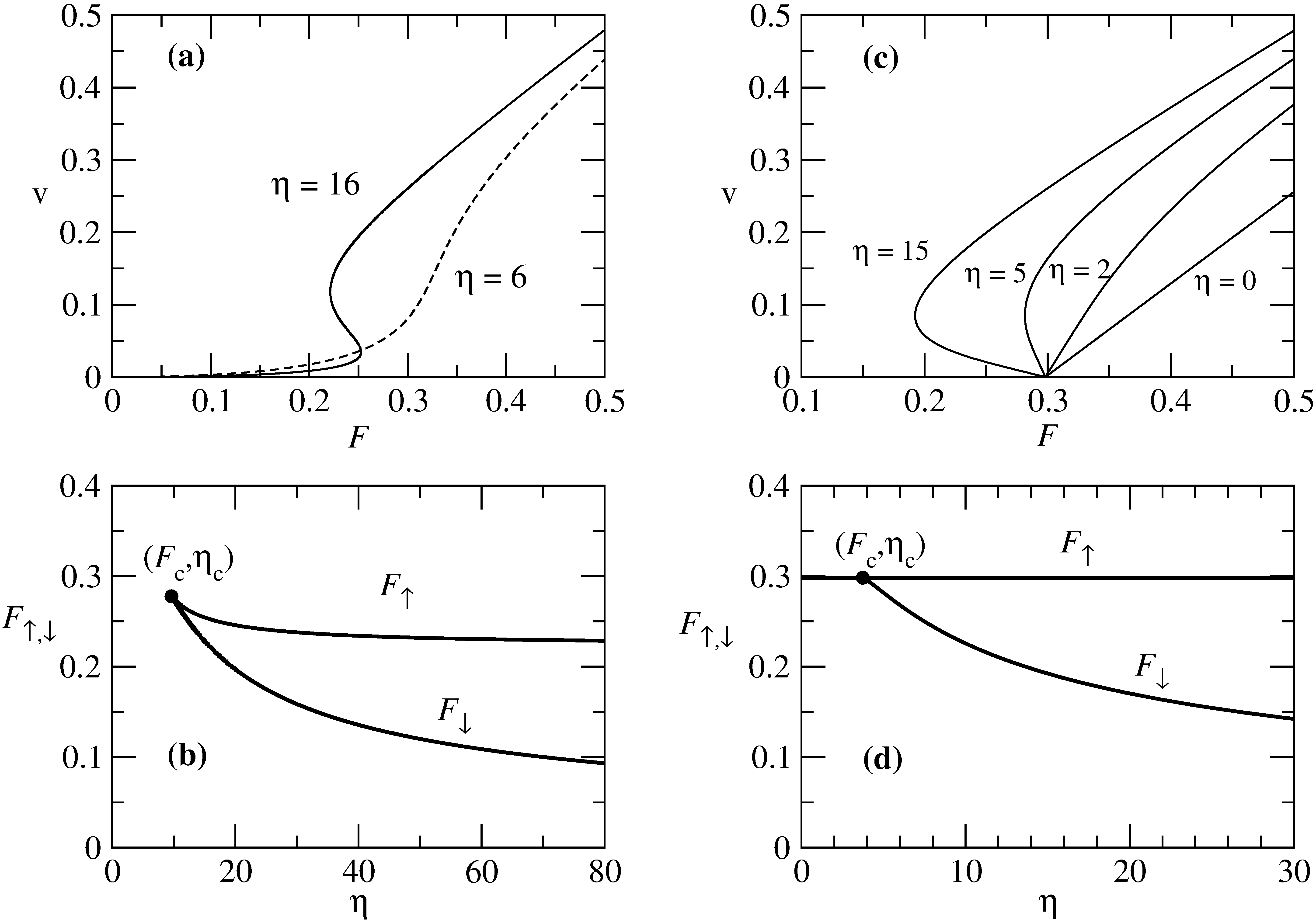}
\caption{ 
  (a) The velocity-force response for an anisotropic
  model with a random substrate, elastic coupling in the direction of 
  drive, and viscous coupling in the transverse direction, where $\eta$ is the
  strength of the viscous coupling term.
  Here $F$ corresponds to the driving force $F_D$.
In the purely viscous case shown here, there is a critical viscous coupling 
$\eta_c$ above which the velocity-force curves become hysteretic. (b) The corresponding
phase diagram showing as a function of $\eta$
the driving forces $F_{\uparrow}$ and $F_{\downarrow}$
at which the hysteretic transitions to the coupled sliding state occur for
sweeping the drive up and down, respectively.
(c) The velocity-force curves for a system with finite elasticity,
where in the absence of transverse viscous
coupling $(\eta = 0)$ the depinning is continuous and non-hysteretic. 
(d)
The corresponding driving force vs $\eta$ phase diagram shows that
there is again a critical viscous coupling above which hysteretic
behavior occurs.
Reprinted with permission from M.C. Marchetti, A.A. Middleton, K. Saunders, and
J.M. Schwarz,
Phys. Rev. Lett. {\bf 91}, 107002 (2003). Copyright 2003 by the American Physical Society.
}
\label{fig:17}
\end{figure}

Marchetti {\it et al.} considered an overdamped anisotropic mean field model 
of a system of particles that are elastically coupled along the driving direction
and viscoelastically coupled transverse to the driving direction, allowing the
possibility for slip of neighboring rows of particles \cite{131}.
The equation of motion is
\begin{equation} 
  \dot \phi^{i}_{l}(t) = K\sum_{\langle j\rangle} (\phi^j_{l} - \phi^{i}_{l}) + F +
  F_{p} + \sigma^{\alpha}_{l,i} 
\end{equation}
where the first term gives the elastic coupling to the neighboring particles
along the direction of drive with elastic constant $K$,
$F$ is the external driving
force corresponding to $F_D$,
$F_{p}$ is the pinning force,
and $\sigma$ is the viscoelastic interaction term
characterized by a  coupling parameter $\eta$.
Various forms can be used
for the viscoelastic time response, and in \cite{131} a linear response was applied,
permitting the system to be
tuned from purely elastic to purely viscous by varying the coupling constants.
Figure~\ref{fig:17}(a) illustrates
the velocity-force response of the system in the purely viscous limit of $K=0$
for different values of the viscous coupling $\eta$, while
the corresponding phase diagram of driving force
versus $\eta$ in figure~\ref{fig:17}(b) 
shows that there is a critical viscous coupling $\eta_c$ 
above which the system develops a hysteretic 
response.  
When a finite elasticity is introduced by setting $K=1$,
the velocity-force curves for different values of $\eta$ take the shapes
illustrated in
figure~\ref{fig:17}(c).
For $\eta = 0$ the depinning transition
is continuous, while for higher values of the viscoelastic coupling
the depinning becomes sharp
and is hysteretic, as shown in the dynamic phase diagram in
figure~\ref{fig:17}(d).
Followup studies and variations of this model that allow plasticity to arise from 
nonconcave interaction potentials also show
that coherent and incoherent flows can coexist in parameter space \cite{131N}.

\begin{figure}
  \includegraphics[width=\columnwidth]{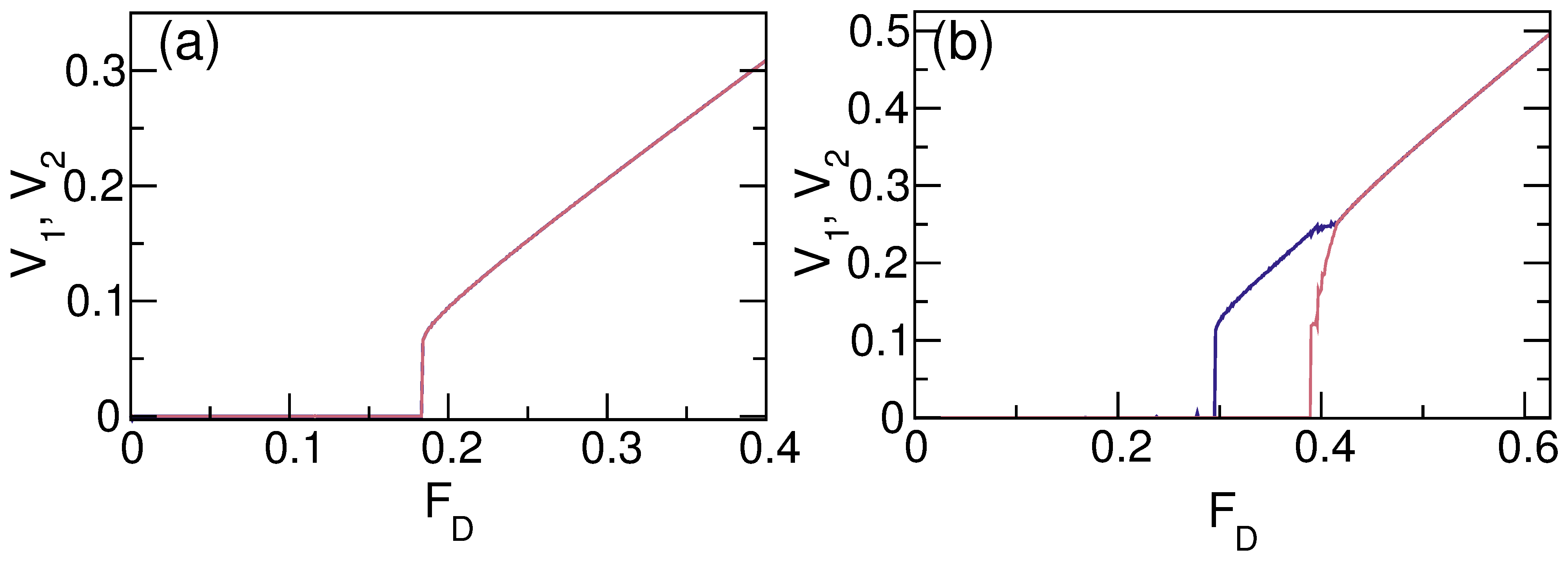}
\caption{
  The simulated velocity-force curves for two coupled 1D channels of
  Yukawa particles moving over random pinning of strength $F_p=6.0$.
The channels are separated by a distance $d$ and the velocities in the two channels
are denoted by $V_{1}$ (blue) and $V_{2}$ (red).
There are four times as many particles as pinning sites, and both particles and pinning
sites are evenly divided between the two channels.
(a) At $d = 1.13$, the depinning is elastic and there is no slip between the channels. 
(b) At $d = 1.47$, the channels depin at different drives
but show a dynamical locking transition
at higher drives near $F_{D} = 0.4$.
Adapted with permission from C. Reichhardt and C. J. Olson Reichhardt,
Phys. Rev. B {\bf 84}, 174208 (2011). Copyright 2011 by the American Physical Society.
}
\label{fig:18}
\end{figure}

\subsection{Systems with few layers}

\begin{figure}
  \includegraphics[width=0.8\columnwidth]{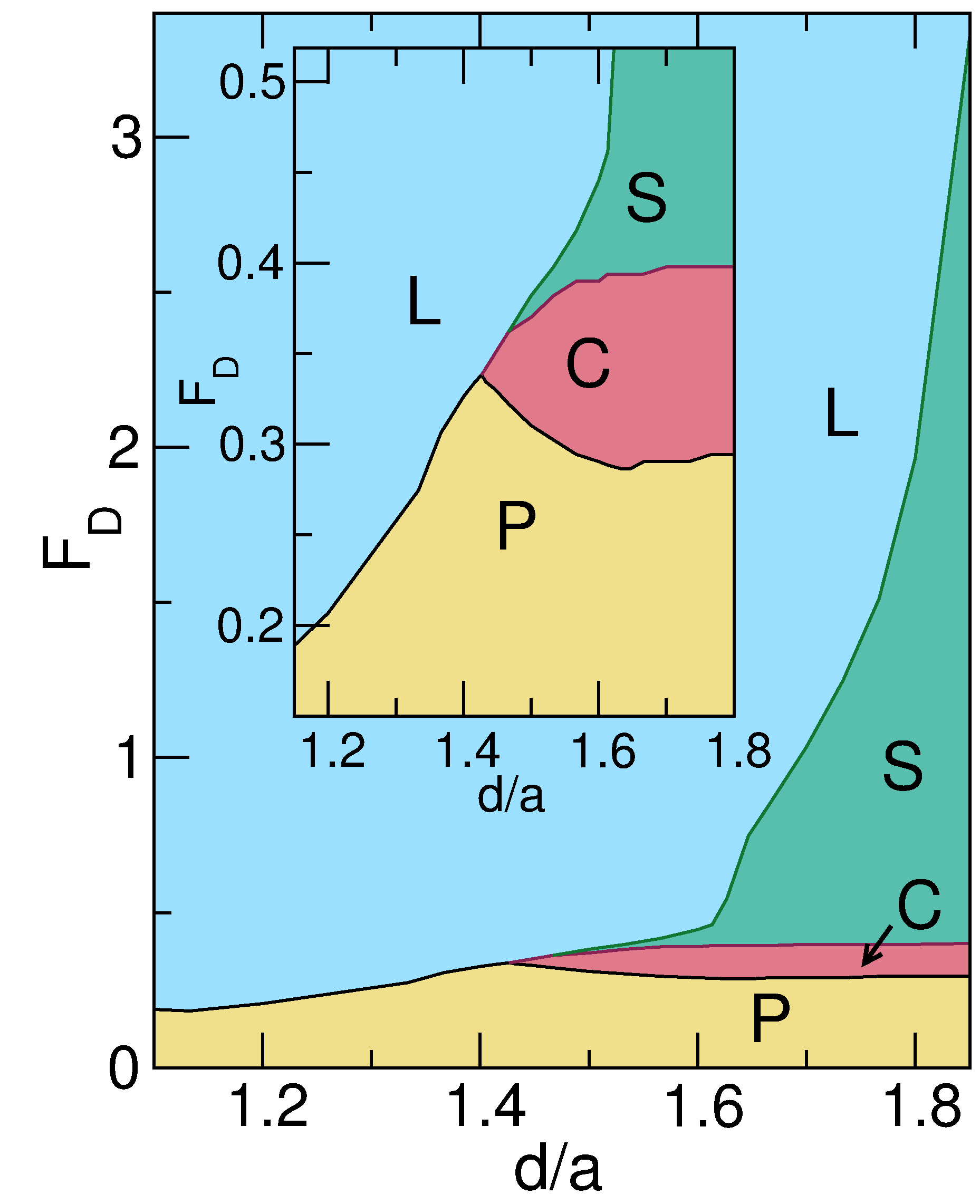}
\caption{
The simulated dynamic phase diagram for the two coupled 1D channels of Yukawa particles 
from Fig.~18 for driving force $F_{D}$ vs interchannel distance $d/a$.
P: pinned phase.  C: coexistence between pinned and moving channels.
S: sliding phase in which both channels
are moving at different velocities and slip past one another.
L: locked channel phase.
Inset:  A detail of the main panel shows that
near the elastic to plastic depinning transition
at $d/a=1.4$ there is peak in the critical depinning threshold. 
Adapted with permission from C. Reichhardt and C. J. Olson Reichhardt,
Phys. Rev. B {\bf 84}, 174208 (2011). Copyright 2011 by the American Physical Society.
}
\label{fig:19}
\end{figure}

It is also possible to observe dynamical phase transitions in systems with only a
small number of layers.
Le Doussal {\it et al.}
showed that in a $d$-dimensional two-layer model, both continuous and sharp
dynamical transitions can occur if plasticity between 
the two layers is allowed \cite{132}.
Experimental realizations of systems of interacting particles confined
in one to three layers can be achieved for
vortices in mesoscopic channels \cite{133},
coupled 1D Wigner crystal wires \cite{134}, colloidal systems \cite{135},
dusty plasmas \cite{136},
and ions in line traps \cite{137}.
Reichhardt and Reichhardt \cite{138} considered Yukawa
repulsively interacting particles in two coupled
1D channels separated by an interchannel distance $d$
in the presence of random pinning.
The particles form zig-zag structures to minimize their interaction energy.
An interesting feature of this layered system is that 
the pinning strength can be adjusted separately in each layer,
producing a different effective damping in the
moving phase for each layer.
Despite its simplicity, this model exhibits a remarkably rich variety of
dynamic phases and  hysteretic
responses.
For example, when the interchannel distance is small and there are
an equal number of randomly placed pinning sites of equal strength in each
layer, the velocity-force curves for each channel in figure~\ref{fig:18}(a) show
that the particles depin elastically and are locked together in the two layers.
Here $V_1$ is the velocity of channel 1 and $V_2$ is the velocity of channel 2.
When the interchannel distance is increased in order to reduce the interchannel
coupling, figure~\ref{fig:18}(b) shows
that each channel depins at a different  value of $F_D$.
At low drives, only one channel is moving; at intermediate drives, the second channel
depins and the channels slip past each other; and at higher drives, 
there is a transition to a  locked phase where the particles in the two channels
move together.
This simple system thus produces four different phases, illustrated
in the dynamic phase diagram as a function of driving force versus
interchannel distance $d$ in figure~\ref{fig:19}: phase P, where both
channels are pinned; phase C, where one channel is pinned and the other
is flowing; phase S, a sliding phase where both channels are flowing but are not
locked; and phase L, a moving locked phase.
The depinning transition is elastic for $d\leq 1.4$ and plastic for $d>1.4$, while
the drive at which the system dynamically orders
from a sliding state to a locked state diverges with increasing $d$. 
The inset of figure~\ref{fig:19} shows a detail of the crossover from elastic to plastic
depinning near $d = 1.4$, where a peak in the critical 
depinning threshold appears at the transition.
This phase diagram shows behavior remarkably similar to that observed in the
complex peak effect phenomenon
for 3D vortex systems described in Section 3,
where a peak in the critical depinning threshold is associated with a transition from
elastic to plastic depinning, and the drive needed to dynamically order the system
diverges as the coupling between neighboring vortices decreases.

\begin{figure}
  \includegraphics[width=\columnwidth]{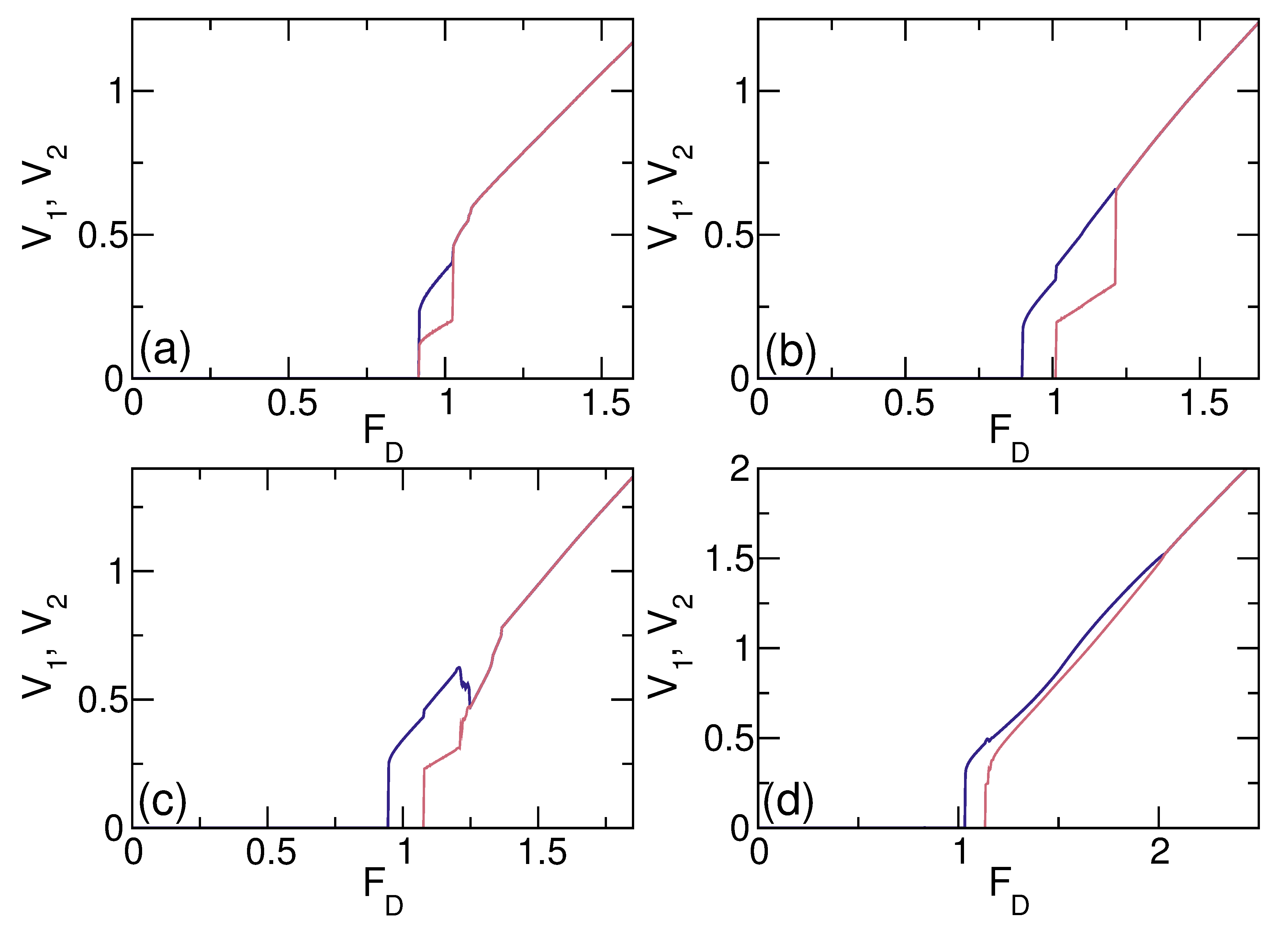}
\caption{
  The simulated velocity-force curves $V_{1}$ (blue) and $V_{2}$(red) vs $F_D$ for
  two coupled 1D channels of Yukawa interacting particles
  moving over random pinning of strength $F_p=15.0$,
  where the channels
  are separated by a distance $d$.
  The pinning strength
  has been increased compared to figures~\ref{fig:18} and ~\ref{fig:19}.
  (a) At $d = 1.147$, both channels depin at the same time to
  a sliding state with $V_{1} > V_{2}$,
  and the channels lock at higher drives.
  (b) At $d = 1.2$, the channels depin at different drives,
  and there is a jump up in $V_{1}$ when $V_{2}$ becomes finite. 
  (c) At $d = 1.27$,
  at the onset of the dynamical locking phase $V_{1}$ drops  while $V_{2}$ increases.   
(d) At $d = 1.47$, the channels do not lock until  a much higher drive is applied.  
Adapted with permission from C. Reichhardt and C. J. Olson Reichhardt,
Phys. Rev. B {\bf 84}, 174208 (2011). Copyright 2011 by the American Physical Society.
}
\label{fig:20}
\end{figure}

If the pinning strength is increased, additional dynamical features appear
in the model of two
coupled 1D channels.
Figure~\ref{fig:20} shows the velocity-force curves for each channel in
a sample with pinning that is strong enough to permit the formation of moving
solitons that can coexist with pinned particles in a single channel.
Figure~\ref{fig:20}(a) shows that for $d = 1.2$, both channels
depin at the same time, but there is a regime
in which $V_{1}>V_{2}$, and
when the moving  channels become locked,
there is a sharp jump in both $V_{1}$ and $V_{2}$. 
In figure~\ref{fig:20}(b) for $d = 1.27$, each channel depins at a
different value of $F_{D}$.  The depinning transition in channel 2 is sharp and
coincides with a jump up in $V_{1}$, while 
a dynamical coupling transition occurs
near $F_{D} = 1.25$.
Figure~\ref{fig:20}(c) 
shows that at $d = 1.27$, the channels depin
separately, while at the dynamic locking 
transition, a jump up in $V_{2}$ is associated with
a jump down in $V_{1}$, producing
a negative differential conductivity in channel 1.
For $d=1.47$ in figure~\ref{fig:20}(d), the
channels do not couple until a much higher drive is applied.
The general shape of the dynamical 
phase diagram in figure~\ref{fig:19} is preserved for models containing up to 8
interacting channels.
In an 8 channel system, within the sliding phase a series of locking and unlocking
transitions of different groups of layers occurs, and only at higher drives do
all of the channels finally lock together.

\begin{figure}
  \includegraphics[width=\columnwidth]{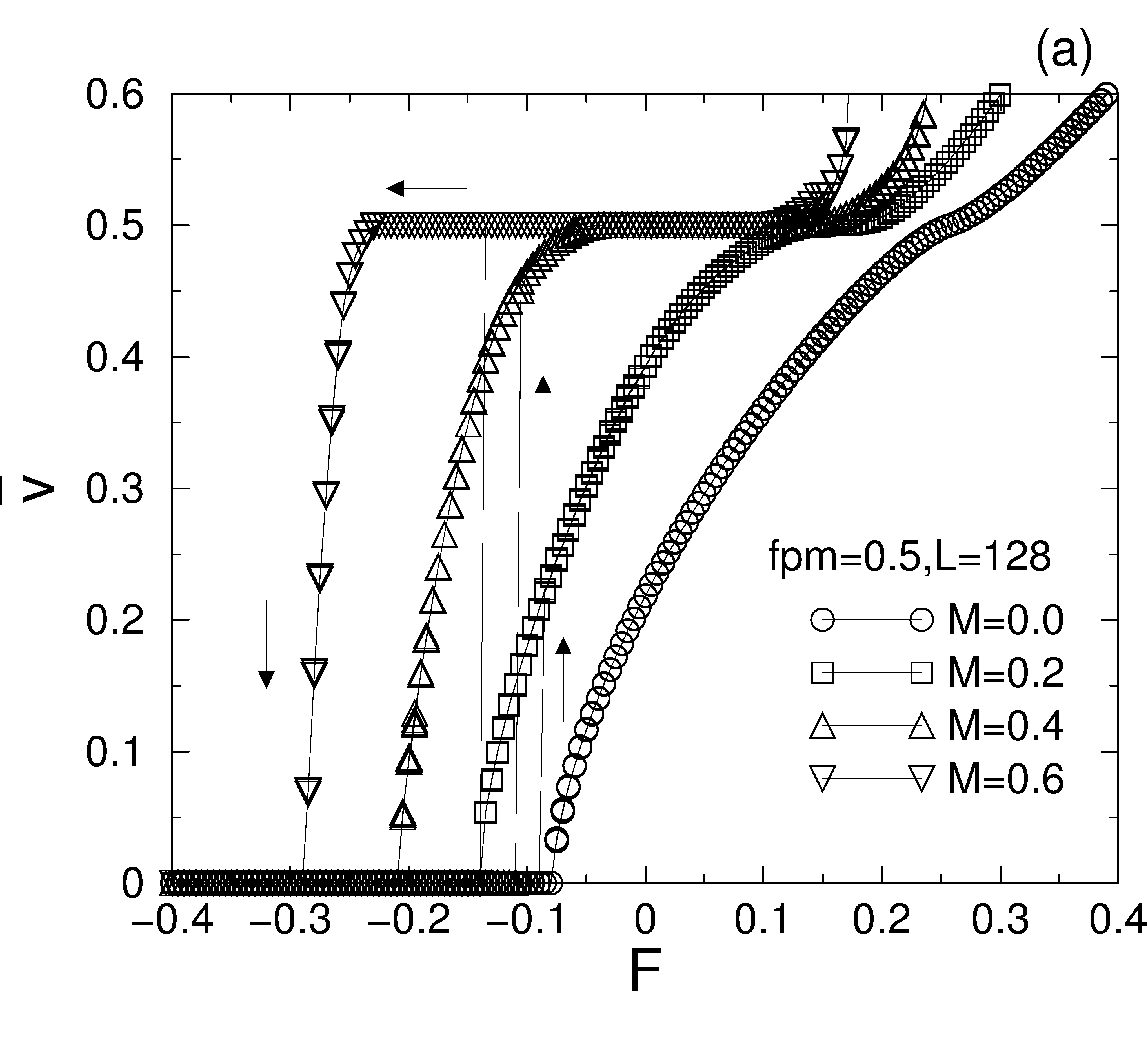}
\caption{
  Simulated velocity-force curves for a mean field elastic model in the presence of random
  disorder, where the strength of inertial overshoot effects is determined by the
  parameter $M$.  At $M=0$ the system depins elastically, while at finite $M$
  the system exhibits a hysteretic response.
  Here $\bar v$ corresponds to the velocity $V$ and $F$ corresponds to
  the driving force $F_D$.
Reprinted with permission from J.M. Schwarz and D.S. Fisher,
Phys. Rev. E {\bf 67}, 021603 (2003).  Copyright 2003 by the American Physical Society.
}
\label{fig:New2}
\end{figure}

Other mechanisms can transform a non-hysteretic 
depinning transition into a hysteretic one.
Schwarz and Fisher examined an 
elastic manifold driven over random disorder, which exhibits
a non-hysteretic depinning transition in the overdamped limit \cite{139}.
Stress overshoots, in which a moving segment of
the manifold generates stress in the non-moving segments, can be introduced by
adding inertial and elastic wave effects, and when stress overshoots are present,
the critical
depinning threshold is reduced and the depinning transition becomes
increasingly hysteretic.  This is illustrated in figure~\ref{fig:New2}, which
shows the velocity-force curves 
from simulations of a mean field model
where $M$ is the inertial or stress overshoot parameter.
There is no hysteresis in the depinning transition
at $M=0$, and the hysteresis grows with increasing $M$. 
Hysteretic depinning transitions
can also arise in underdamped
Frenkel-Kontorova models \cite{47,140},
where there can be a coexistence of large regions of moving particles with areas of
pinned particles.
Less is known about how inclusion of inertia could
affect systems that exhibit plastic depinning. 

\subsection{Kinetic precursors and spinodals}
 
Numerous aspects of first order phase transitions remain current topics 
of active research, particularly the kinetics of such transitions and
the presence of precursor dynamics as the transition
is approached \cite{141,142,143,144}.
For example, below the transition there can be 
fluctuating droplets containing the new phase.
Since the size of these fluctuating droplets increases as the first order
phase transition is approached,
it is possible for a droplet to reach the critical droplet size that nucleates the new
phase
at a point below the actual phase transition,
making it possible for the system to exhibit critical features in spite of the fact
that the transition is first order \cite{142}.
Features of this type could
be strongly enhanced by the inhomogeneities introduced by a pinning landscape.
Additionally, the ideas of spinodal points in mean field
models and pseudo-spinodal points in systems
with finite range interactions that describe the extent of metastability
could be applied to
nonequilibrium first order dynamical transitions  \cite{143,144}.
For example, if a system
undergoing phase separation or coarsening as a function of time
were simultaneously driven over random disorder,
the fluctuating forces induced by the pinning could stabilize the coarsening process
at some limit, or they could enhance the speed of the coarsening or destroy the
coarsening process altogether.

\section{Depinning and Dynamic Phases on Periodic Pinning Arrays}

Another class of systems that can exhibit depinning transitions and various
types of nonequilibrium phase transitions is 
particles driven over
periodic substrates.
As mentioned in the introduction, there
has already been substantial work
on a variety of frictional systems, which in some cases
can be modeled as atoms or molecules moving over a periodic substrate.  Such work
is beyond the scope of this review, and we refer the
reader to an excellent recent review of this field \cite{9}.
There are, however, examples of systems that can
be effectively modeled as overdamped particles moving over 
periodic arrangements of pinning sites
where plasticity can occur that exhibit behavior
similar to the dynamical phases that arise for random pinning arrangements.   
Such systems include superconducting vortices driven over
periodic pinning arrays and colloids moving over
periodic optical or magnetic substrates.

\subsection{Dynamic phases of superconducting vortices on periodic substrates}

Many superconducting
applications require high critical currents or strong vortex pinning, so 
considerable effort has been devoted to understanding
how to arrange pinning sites in special geometries 
that maximize the pinning effectiveness \cite{1,2,56,144a}. 
Since the ground state of the vortex lattice is triangular, a natural 
approach is to arrange the pinning sites in periodic structures
with  2D \cite{2,56,144a,145}, 
1D \cite{146}, or
quasiperodic \cite{147} ordering.
Experiments in superconductors with periodic pinning arrays show that the 
critical depinning threshold exhibits a 
series of peaks at magnetic fields at which 
the number of vortices $N$ is an integer multiple $n$ of
the number of pinning sites $N_{p}$, $n = N/N_{p}$,
while
direct imaging \cite{2,145} and simulations \cite{57,145,148}
indicate that the vortices form ordered arrangements
at the integer commensurate fillings.
Since the vortex density is controlled by
an applied magnetic field, the filling $n=1$ is called the first matching field.
At nonmatching fillings,
there can be either a disordered structure
or an ordered lattice containing well defined interstitials, vacancies, or grain boundaries.

\begin{figure}
  \includegraphics[width=\columnwidth]{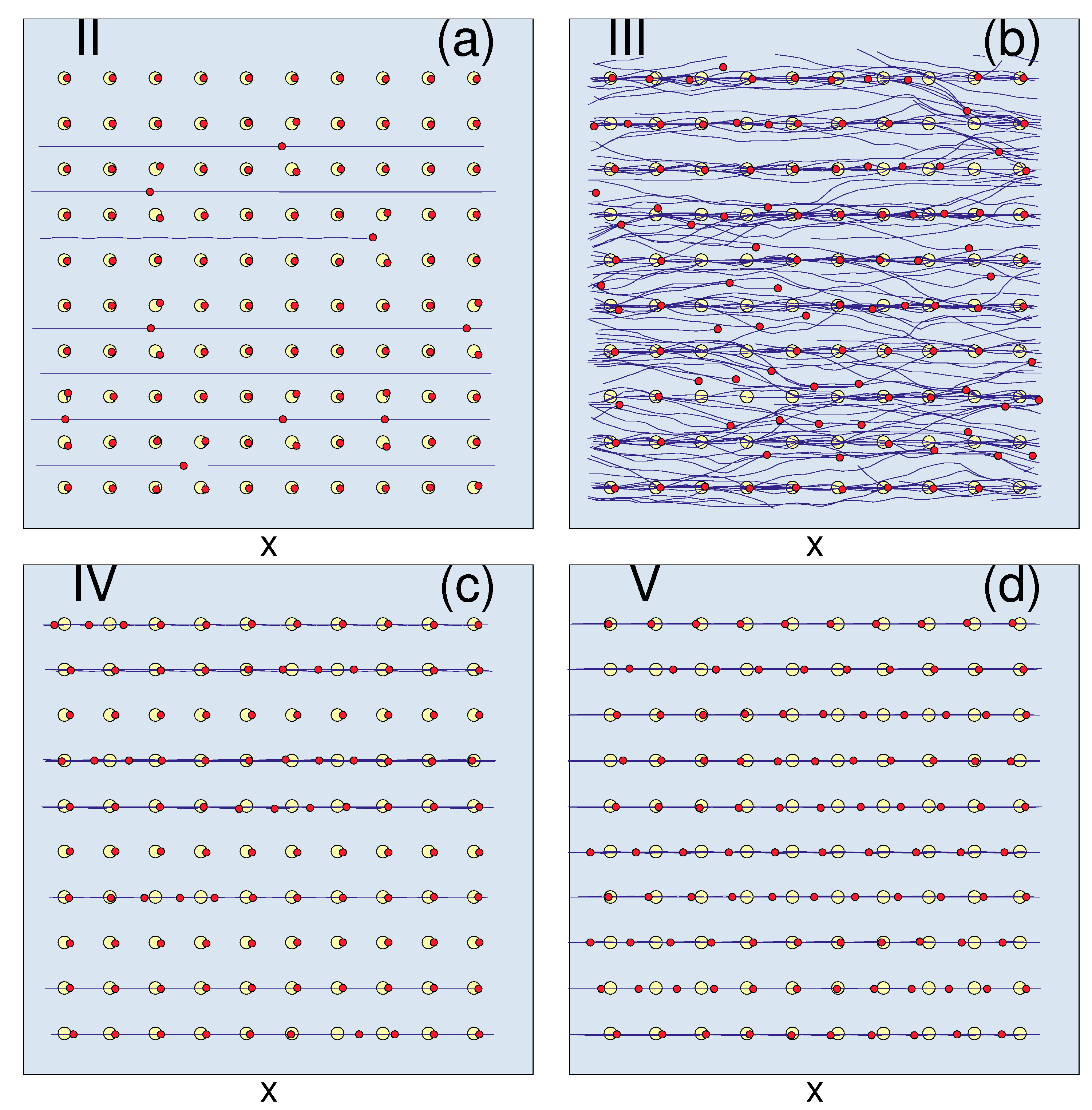}
\caption{
  Vortex locations (red dots), pinning site locations (yellow circles) and trajectories
  (blue lines) for
  simulations of superconducting vortices interacting
  with a square pinning array at a filling fraction of
  $n=1.06$
  for increasing drive.
    (a) Phase II, the 1D flow of interstitial vortices between pinned
vortices.
(b) Phase III, chaotic or turbulent flow.
(c) Phase IV, a 1D soliton type flow along the pinning sites. 
(d) Phase V, a 1D flow of all the vortices along the pinning sites.  
  Adapted with permission from C. Reichhardt, C.J. Olson, and F. Nori,
Phys. Rev. Lett. {\bf 78}, 2648 (1997). Copyright 1997 by the American Physical Society.
}
\label{fig:22}
\end{figure}

One of the key differences between vortex motion over periodic pinning arrays
and most models of atomic friction
is that the pinning sites for the vortices are localized,
so that the potential energy landscape experienced by the vortex 
resembles a muffin tin rather than an egg carton.
Additionally, the vortex interactions are much longer range than atomic
interaction potentials, so that on strong pinning landscapes two separate
vortex species can be present:
the vortices directly trapped by the pinning sites,
and the vortices located in 
the interstitial regions between pinning sites.
The interstitial vortices still feel a confining potential due to
repulsion from their neighboring vortices, some or all of which may be
trapped at the pinning sites.
For $n \lesssim 1$, just below the first matching field, all the vortices are located at 
pinning sites and there are a fixed number of vacancies, while  for $n \gtrsim 1$,
just above the first matching field, a portion of the 
vortices are trapped at interstitial sites.  
When a sample containing interstitial vortices is subjected to a driving current,
two depinning transitions should occur, with the first transition corresponding to the
depinning of the interstitial vortices, while the second is the depinning of the vortices
trapped at pinning sites.

\begin{figure}
  \includegraphics[width=\columnwidth]{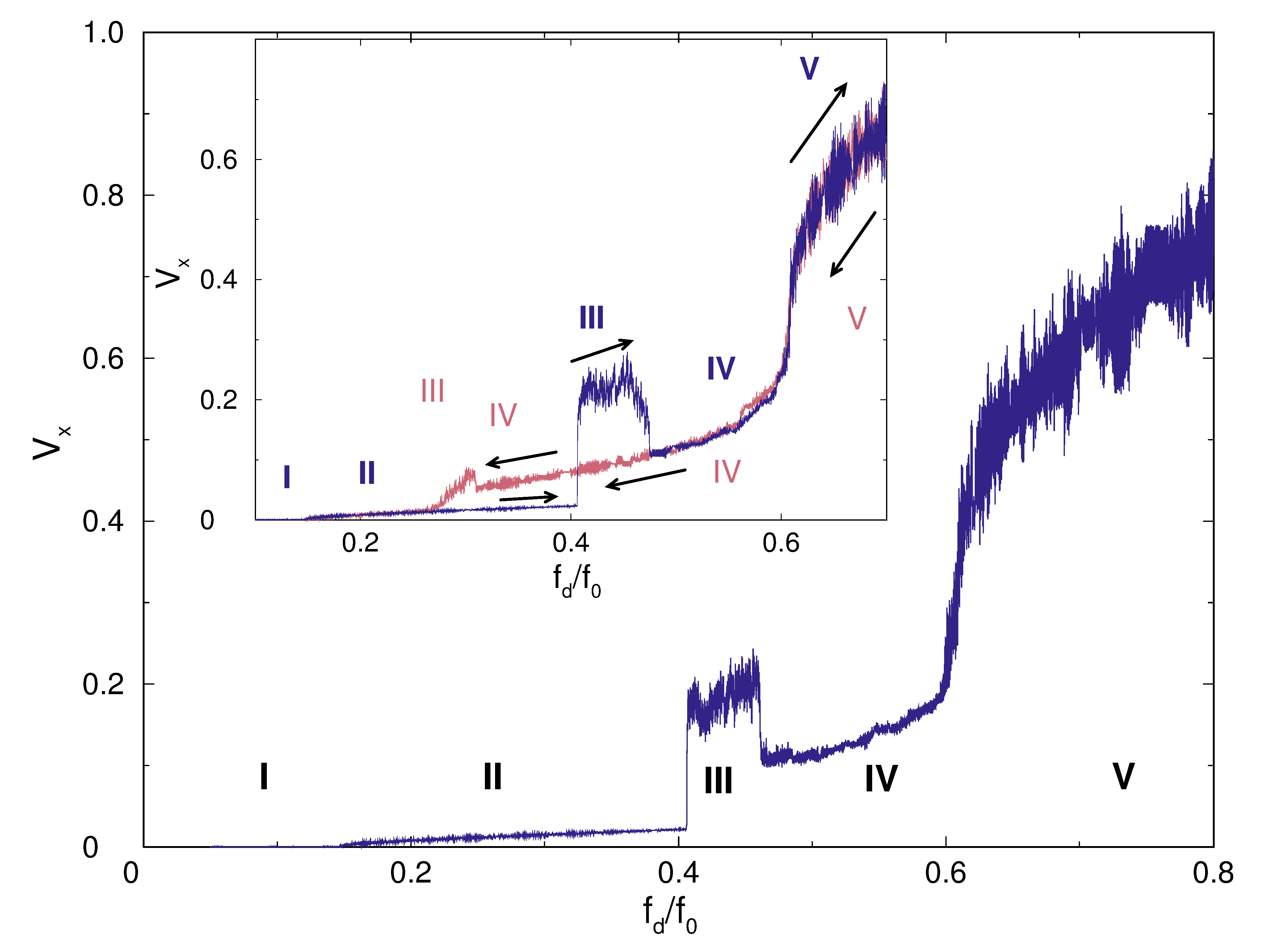}
\caption{
  The simulated velocity-force curves for the periodic pinning vortex system in
  figure~\ref{fig:22}.
  Here $V_x$ corresponds to the velocity $V$ and $f_d/f_0$ corresponds to the
  driving force $F_D$.
  The inset shows the hysteresis in the velocity-force curves when the
  driving force is swept up (blue) and down (red).
  Phase I is the pinned phase.
  The nature of the transitions between phases are as follows:
  I to II, second order;
  II to III, first order;
  III to IV, first order;
  IV to V, second order.
  Adapted with permission from C. Reichhardt, C.J. Olson, and F. Nori,
Phys. Rev. Lett. {\bf 78}, 2648 (1997). Copyright 1997 by the American Physical Society.
}
\label{fig:23}
\end{figure}

Reichhardt {\it et al.} \cite{60,149}
numerically examined the depinning dynamics of vortices 
in a 2D square pinning array.
At
$n=1.0$ there is a single continuous
non-hysteretic depinning transition at a critical value $F_c$ corresponding to
the maximum pinning force $F_p$ exerted by an individual pinning site,
indicating that the interactions between the vortices
cancel and the system responds in the same manner as in the single particle limit. 
At
$n=1.06$,
the interstitial vortices depin at a lower driving force and
follow 1D trajectories between the pinned vortices,
as illustrated in figure~\ref{fig:22}(a). 
Figure~\ref{fig:23} shows the velocity-force curve for this system,
with the pinned phase labeled phase I 
and the 1D interstitial vortex flow phase labeled
phase II.  
As  $F_{D}$ increases
there is a transition from the phase II flow to a
turbulent or chaotic flow in which
the moving interstitial vortices cause vortices at the pinning sites to depin,
producing the strongly fluctuating trajectories illustrated in figure~\ref{fig:22}(b).
This disordered pinning and depinning flow phase is
labeled phase III, and its onset coincides
with a sharp jump up in the velocity-force curve in figure~\ref{fig:23} since
a larger number of vortices are now moving.
At higher $F_{D}$ there is  another transition 
from phase III to a 1D soliton-like flow of vortices along the pinning sites,
labeled phase IV and illustrated in figure~\ref{fig:22}(c).
The transition between phases III and IV 
is associated with a sharp drop in the 
vortex velocity, as shown in figure~\ref{fig:23}, which is termed negative
differential mobility.
Flow in phase IV consists of excitations or pulses of motion that translate over the
substrate much more rapidly than the individual vortices.
At even higher drive, there is a transition to a state
called phase V in which all of the vortices
are flowing along the pinning sites, as illustrated in figure~\ref{fig:22}(d).
Since there are more vortices than pinning sites in this example,
the flow in phase V is smectic, and neighboring rows
gradually slide past each other over time.   
The inset in figure~\ref{fig:22} shows that there no hysteresis associated with
the IV-V transition,
strong hysteresis across the
III-IV and II-III transitions,
and
no hysteresis in the I-II transition.
The
I-II transition falls into the same class as the depinning of
an individual particle from a periodic substrate, with the velocity-force curve obeying
$V = (F_D- F_{c})^{\beta}$ with $\beta=0.5$.
The
II-III transition 
is a first order transition from a non-chaotic state to a chaotic state, while the
III-IV transition
is a first order transition from a chaotic state to a non-chaotic state.
The
IV-V transition is a continuous non-hysteretic transition.

\begin{figure}
  \includegraphics[width=\columnwidth]{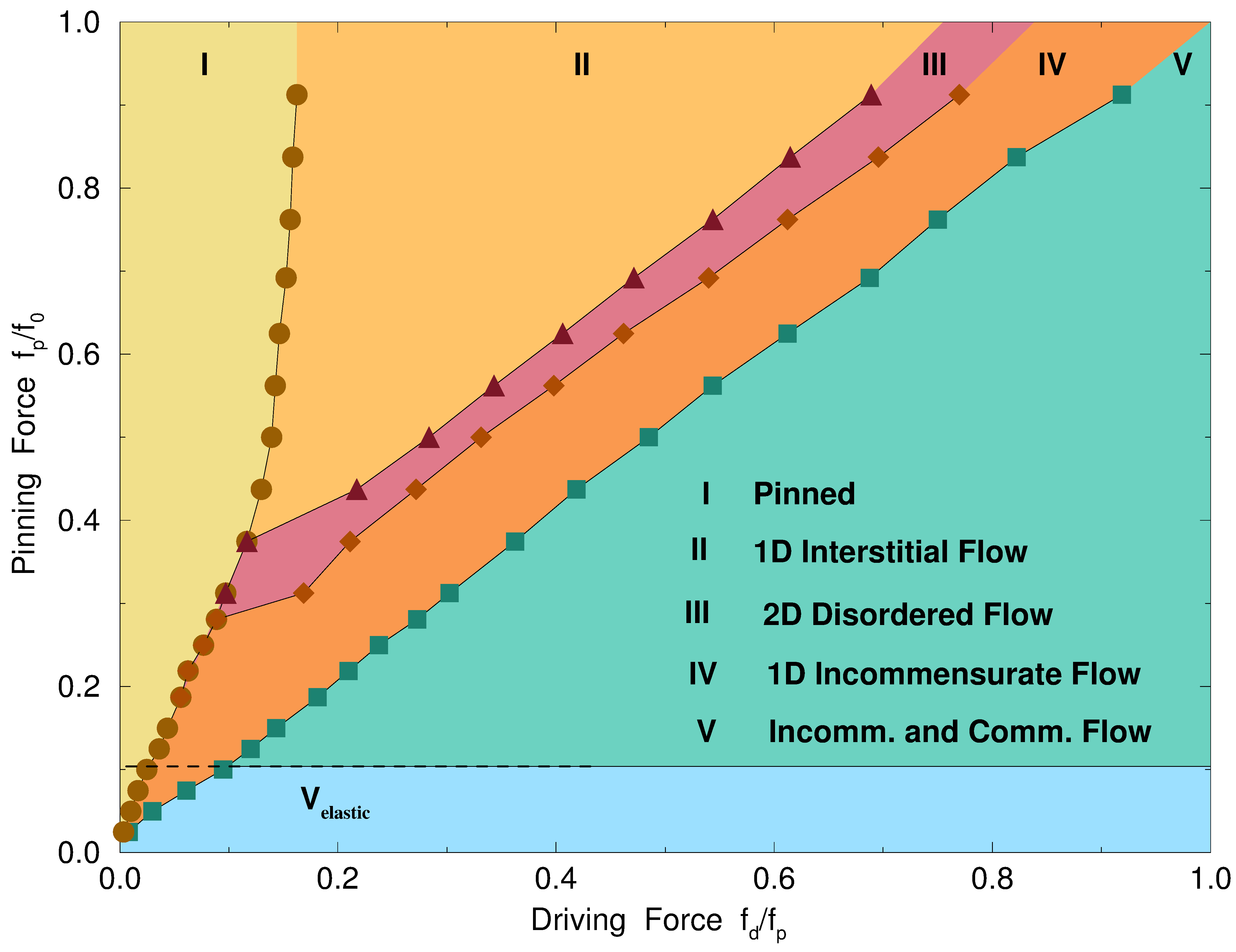}
\caption{
  Simulated dynamic phase diagram
  for vortices in periodic pinning for the same system in
  figures~\ref{fig:22} and \ref{fig:23}
  showing the
  evolution of the phases as a function of
  pinning strength $f_{p}/f_0$ vs driving force $f_d/f_p$ at $n=1.06$.
Here $f_p/f_0$ corresponds to $F_p$ and $f_d/f_p$ corresponds to $F_D$.
  Adapted with permission from C. Reichhardt, C.J. Olson, and F. Nori,
Phys. Rev. B {\bf 58}, 6534 (1998). Copyright 1998 by the American Physical Society.
}
\label{fig:24}
\end{figure}

The dynamic phase diagram constructed from
a series of simulations for the system in figure~\ref{fig:23}
is shown
in figure~\ref{fig:24}
as a function of pinning force $F_{p}$ versus
driving force $F_{D}$ at a filling of
$n=1.06$.
At low $F_{p}$, the interstitial vortices cannot be localized between the pinning sites,
and the system forms
a grain boundary state which depins elastically.
As $F_{p}$ is further increased, the interstitial vortices
become localized and phase II and phase III flows begin to appear.
The critical depinning 
threshold for the I-II transition
saturates at large $F_p$
since it is determined by the confinement of the interstitial vortices
rather than by the pinning sites themselves, and this energy scale is set by
the vortex lattice constant instead of by the pinning strength.
The III-IV and IV-V transitions
both increase linearly in $F_D$ with increasing $F_p$.
For
$n<1.0$ (not shown),
there is also a two-step depinning process in which
the vacancies in the vortex lattice depin first followed by the vortices
at the pinning sites; however,
phase III flow does not occur \cite{149}. 
Other simulations of the motion of 
vortices through periodic arrays of obstacles also show
negative differential conductivity as well as a dynamically pinned or jammed
state in which the vortices form pileups and then force their way through
the obstacle when the drive is high enough, allowing the motion to start
again \cite{150N}.  

\begin{figure}
  \includegraphics[width=\columnwidth]{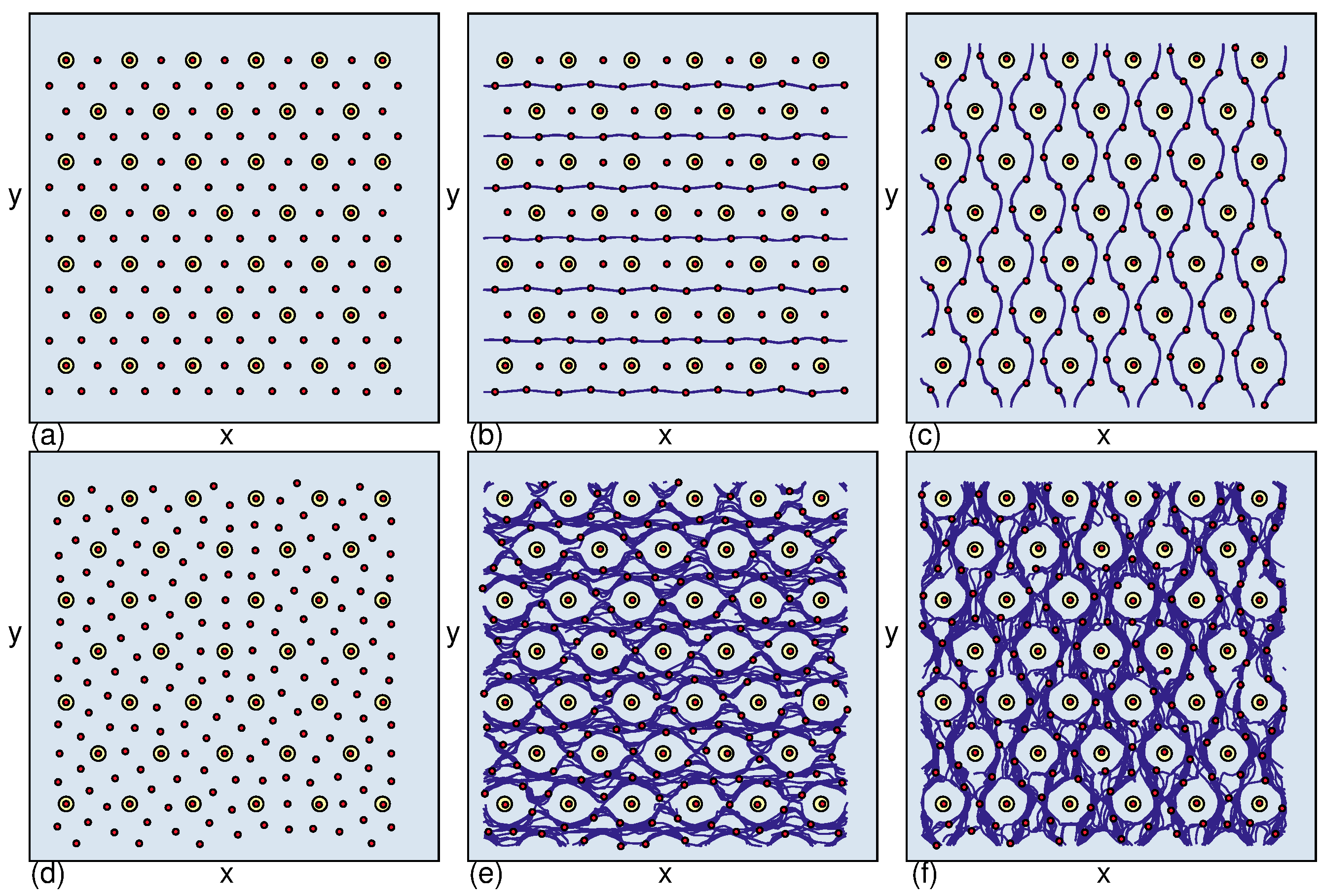}
  \caption{ Ordered and disordered plastic flow phases for
    simulations of superconducting vortices
    in a periodic triangular array of pinning sites showing
    vortex locations (red dots), pinning site locations (yellow circles), and
    trajectories (blue lines).
    (a,b,c) A filling of $n=4.0$.
  (a) In the pinned phase, the vortices form an ordered state.
  (b) For driving in the $x$-direction just above depinning,
  a portion of the vortices flow in 1D channels.
  (c) For driving in the $y$-direction,
  all the interstitial vortices flow in a sinusoidal pattern.
  (d,e,f) A filling of $n=5.0$.
(d) The frustrated pinned phase.
(e) For driving in the $x$ direction, a disordered or chaotic flow state occurs.
(f) For driving in the $y$ direction, a similar disordered flow state occurs.
  Adapted with permission from C. Reichhardt and C.J. Olson Reichhardt,
Phys. Rev. B {\bf 79}, 134501 (2009). Copyright 2009 by the American Physical Society.
}
\label{fig:25}
\end{figure}

For higher fillings and other pinning geometries, additional dynamic flow phases
occur which generically fall into two categories: a smooth non-chaotic flow of
interstitial vortices between pinned vortices and/or immobile interstitial vortices,
and chaotic flow states
\cite{149,150,151,152,153,154,155}.
The depinning into a smooth non-chaotic phase when the
pinned equilibrium vortex lattice structure is ordered 
is illustrated in figure~\ref{fig:25}(a)
for a triangular pinning array at $n=4.0$ 
where the vortices form an ordered triangular lattice
below depinning \cite{156}.
Figure~\ref{fig:25}(b) shows the flow
of the vortices at this filling just above depinning for driving in the $x$-direction. 
A portion of the interstitial vortices
flow in 1D channels between the pinning sites,
while the vortices at the pinning sites remain pinned
and a third of the interstitial vortices also remain pinned.
If the drive is instead applied along the $y$ direction,
the flow is again ordered and forms the pattern illustrated in
figure~\ref{fig:25}(c), with all the interstitial vortices flowing.
For higher drives,
additional dynamical transitions occur as the remaining vortices depin. 
In figure~\ref{fig:25}(d),
the pinned vortex phase at $n=5.0$
is disordered due to a frustration effect, and 
the depinning for driving in either the $x$ or $y$ directions
results in disordered flows such as those shown in
figure~\ref{fig:25}(e,f).

\begin{figure}
  \includegraphics[width=0.9\columnwidth]{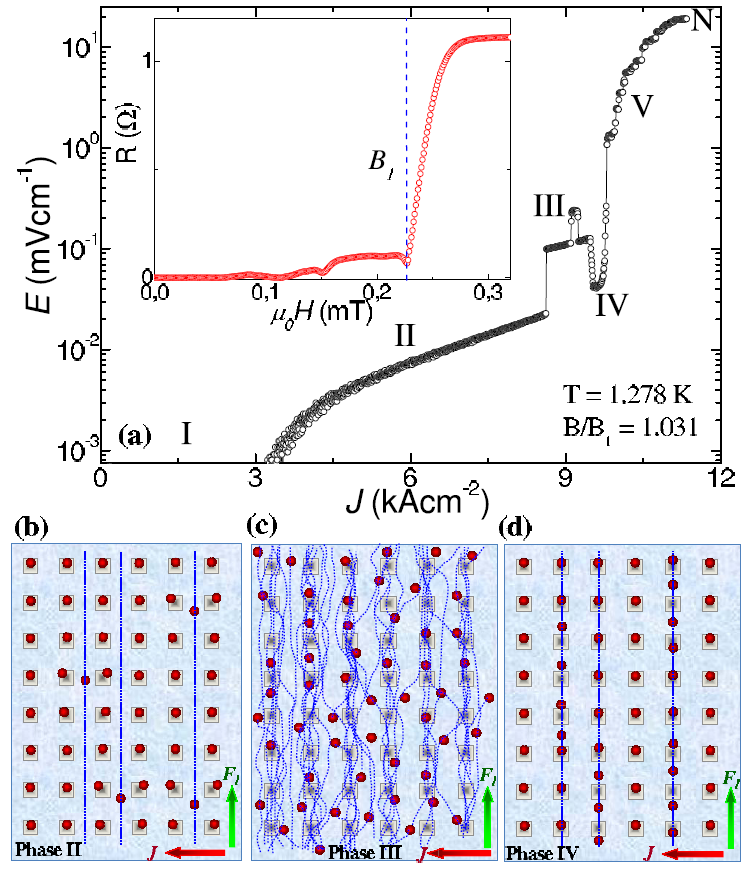}
\caption{(a) 
  Experimentally measured velocity-force curve for vortices in a square pinning
  array at a filling of $n=1.0$ as a function of voltage $E$ vs current $J$.
  Here $E$ corresponds to the velocity $V$ and $J$ corresponds to the driving
  force $F_D$.
  Labels indicate the dynamical phases. I: pinned; II: 1D interstitial motion of the type
  illustrated in panel (b); III: turbulent flow; IV: 1D soliton flow; V: motion of all the
  vortices; N: transition of the sample to the normal state.
  (b-d) Schematic images of vortex flow.  (b) 1D interstitial motion in phase II.  (c)
  Turbulent flow in phase III.  (d) 1D soliton flow in phase IV.
  Reprinted with permission from J. Gutierrez, A.V. Silhanek, J. Van de Vondel,
  W. Gillijns, and V.V. Moshchalkov,
Phys. Rev. B {\bf 80}, 140514(R) (2009). Copyright 2009 by the American Physical Society.
}
\label{fig:26}
\end{figure}

\begin{figure}
  \includegraphics[width=\columnwidth]{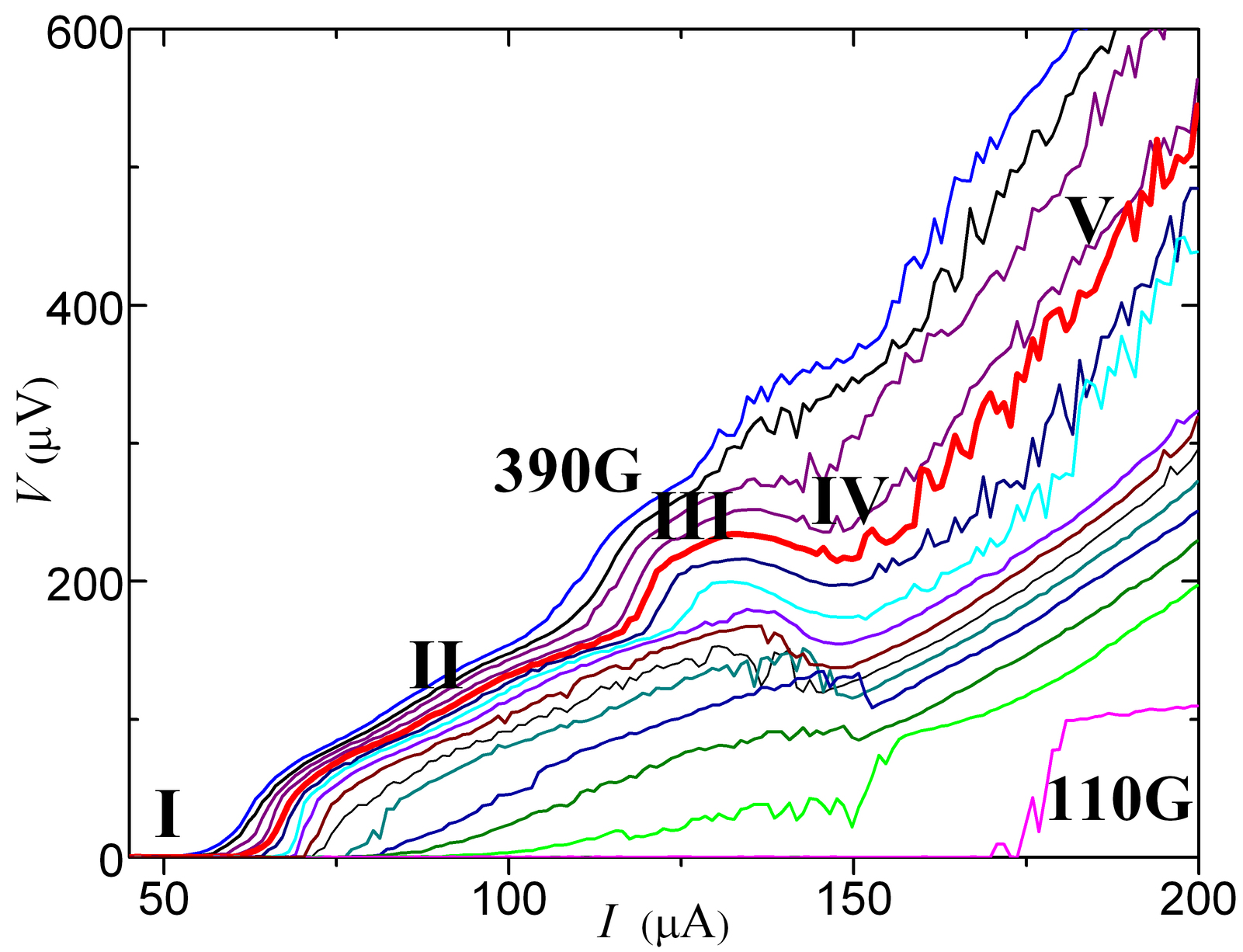}
\caption{
  Experimentally measured
  velocity-force curves for superconducting vortices in a square pinning
  array as a function of voltage $V$ vs current $I$
  for changing magnetic field or changing vortex density,
  from 110 G (lower right) to 390 G (upper right).
  $V$ corresponds to the vortex velocity and $I$ corresponds
  to the driving force $F_D$.
  Evidence of dynamical phases I through V (labeled)
  appears in the form of jumps and dips in the average vortex velocity.  
  Reprinted with permission from S. Avci, Z.L. Xiao, J. Hua, A. Imre, R. Divan, J. Pearson,
U. Welp, W.K. Kwok, and G.W. Crabtree,
Appl. Phys. Lett. {\bf 97}, 042511 (2010).  Copyright 2010, AIP Publishing LLC.
}
\label{fig:27}
\end{figure}

In experiments on superconductors with square
periodic pinning arrays near
$n=1.01$ \cite{61},
the velocity-force curves exhibit the same signatures
found in simulations of the dynamical
phases I through V,
as shown in figure~\ref{fig:26}(a), where there is
a jump up in velocity at the II-III transition and
a jump down in velocity at the III-IV transition.
Figure~\ref{fig:26}(b,c,d) shows
schematics of the  flow patterns in phases
II, III and IV. In the same set of experiments,
strong hysteresis appears across the II-III and III-IV transitions
\cite{61}.
S. Avci {\it et al.} also experimentally investigated the dynamics of vortices in periodic
pinning arrays where strong matching effects occur, and found features
in the
velocity-force curves that are
consistent with those observed for the dynamic phases I through V, as illustrated
in figure~\ref{fig:27} \cite{157}.
Other transport experiments in periodic pinning arrays have also found similar
transport signatures corresponding to dynamic phases I through V, as well as
hysteresis in the velocity-force curves \cite{158}.
 
Of the various dynamical phases observed for motion through periodic pinning
arrays, phases II and IV are locked in periodic non-chaotic orbits.  
It would be interesting to measure the Lyapunov exponents
in the different phases.
It may also be possible to achieve layering effects by constructing pinning arrays
in which every other pinning row has a different strength or size.
Another method for extracting
information about the nature of the dynamical flow
is the velocity noise signature,
since the periodicity of the pinning sites should
always induce a periodicity in the velocity as a function of time, but
the magnitude of this periodic signal will vary depending on whether the flow
is in an ordered state or a disordered state.
Simulations show that very different velocity noise spectral signatures arise depending
on the nature of the flow pattern, ranging from narrow band noise to broad
band noise \cite{150}; however, experimental noise measurements have not
yet been performed.
It would be interesting to determine whether the addition of an ac drive
to the dc drive could produce
an ordering of the chaotic flow states.
There have already been
experimental studies \cite{159} and simulations \cite{160,161}
of vortex motion in periodic pinning arrays that
reveal Shapiro step locking effects near
$n=1.0$;
it would be interesting to examine whether similar effects appear at higher fillings.
There have also been studies of dynamical transitions in the vortex flow for
systems with asymmetric pinning sites,
where different types of flow patterns emerge
for driving along different directions
with respect to the substrate asymmetry \cite{162}.    
Another open question is how the vortex dynamics
would change for periodic pinning placed in 3D systems or 3D layered systems, and whether
phases I through V would still occur, or whether additional phases would arise.
It would also be possible to study a system in which the density of pinning sites at
the top of the sample (containing the tops of the vortex lines)
is different from the bottom of the sample (containing the bottoms of the
vortex lines), and to vary
the ratio of these densities in order to determine whether a new type of
commensuration effect
can occur between the number of vortices and the two different
periodicities of the pinning array.

\subsection{Dynamic phases of colloids on periodic substrates}

\begin{figure*}
    \includegraphics[width=1.9\columnwidth]{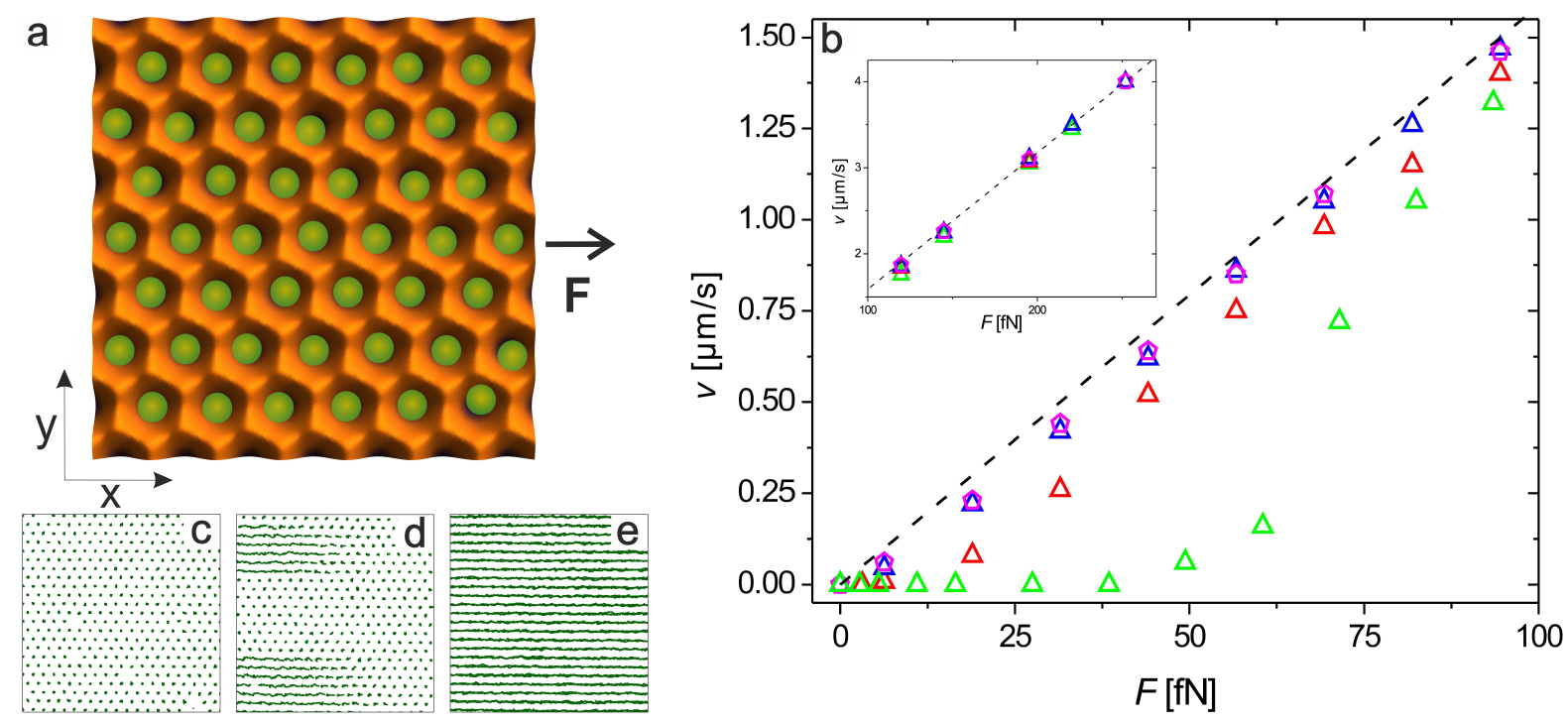}
\caption{
  (a) Schematic 
  of an experiment in which a colloidal monolayer is driven with force F
  across an energy landscape created by the interference of 
  laser beams. The potential strength and length scale of the
  substrate potential can be adjusted by varying the laser intensity and 
  the angle at which the laser beams intersect.
  Here $F$ corresponds to $F_D$.
  (b) The experimentally measured
  mean colloidal velocity $V$ of a crystalline monolayer with 
  lattice constant $a=5.7 \mu m$ vs 
  driving force $F$
  for a commensurate substrate with $n=1$ (green),
  incommensurate substrates with $n=0.83$ (red) and $n=0.71$ (blue),
  and quasiperiodic (magenta) substrates.
  The dashed line corresponds to free sliding on a flat substrate.
Inset: $V$ vs $F$ for larger $F$ where $V \propto F$.
(c,d,e) Colloid trajectories for $n=1$ commensurate conditions
at (c) $F=0$,
(d) $F=49$, 
and (e) $F=82$.
Reprinted by permission from Macmillan Publishers Ltd: Nature Materials,
T. Bohlein, J. Mikhael, and C. Bechinger,
Nature Mater. {\bf 11}, 126 (2012), copyright 2012.
}
\label{fig:28}
\end{figure*}

Colloids are an outstanding experimental system in which
to explore issues in equilibrium and nonequilibrium
statistical mechanics
due to their size scale, which allows direct access to the microscopic
degrees of freedom \cite{cref1,cref2}.
Recent advances in optics allow the creation of a variety of
substrates over which the colloids can be driven  \cite{163}. 
There have been numerous studies of
the pinned ground states and  ordering of colloids
interacting with 1D \cite{164,165,166,167} and 
2D periodic substrates \cite{168,169,170,171}.
For 2D periodic arrays,
the substrate  can have either an egg carton form
\cite{168,170,171} that does not allow interstitial particles,
or it can have a muffin tin form that allows
interstitial colloids to coexist with pinned colloids \cite{167,172}.
When individual colloids are driven through
periodic arrays at different angles with respect to the array symmetry 
direction, a variety of directional locking effects
appear in which the motion of the colloids locks to
high symmetry directions of the substrate \cite{173,174}. 
Other studies focused on the properties of the dynamic steady state
flows of colloids moving over ordered substrates \cite{175}. 
Colloids moving over periodic substrates can also be studied
for practical applications such as 
novel particle separation techniques, since in some cases one species of
colloid locks to a symmetry direction
of the substrate lattice while another species
does not, so that a mixture of different colloid species can be segregated
in the direction transverse to the drive
over time \cite{173,174,176,177,178,179}.

\subsubsection{Colloids on egg carton substrates --}

\begin{figure}
  \includegraphics[width=\columnwidth]{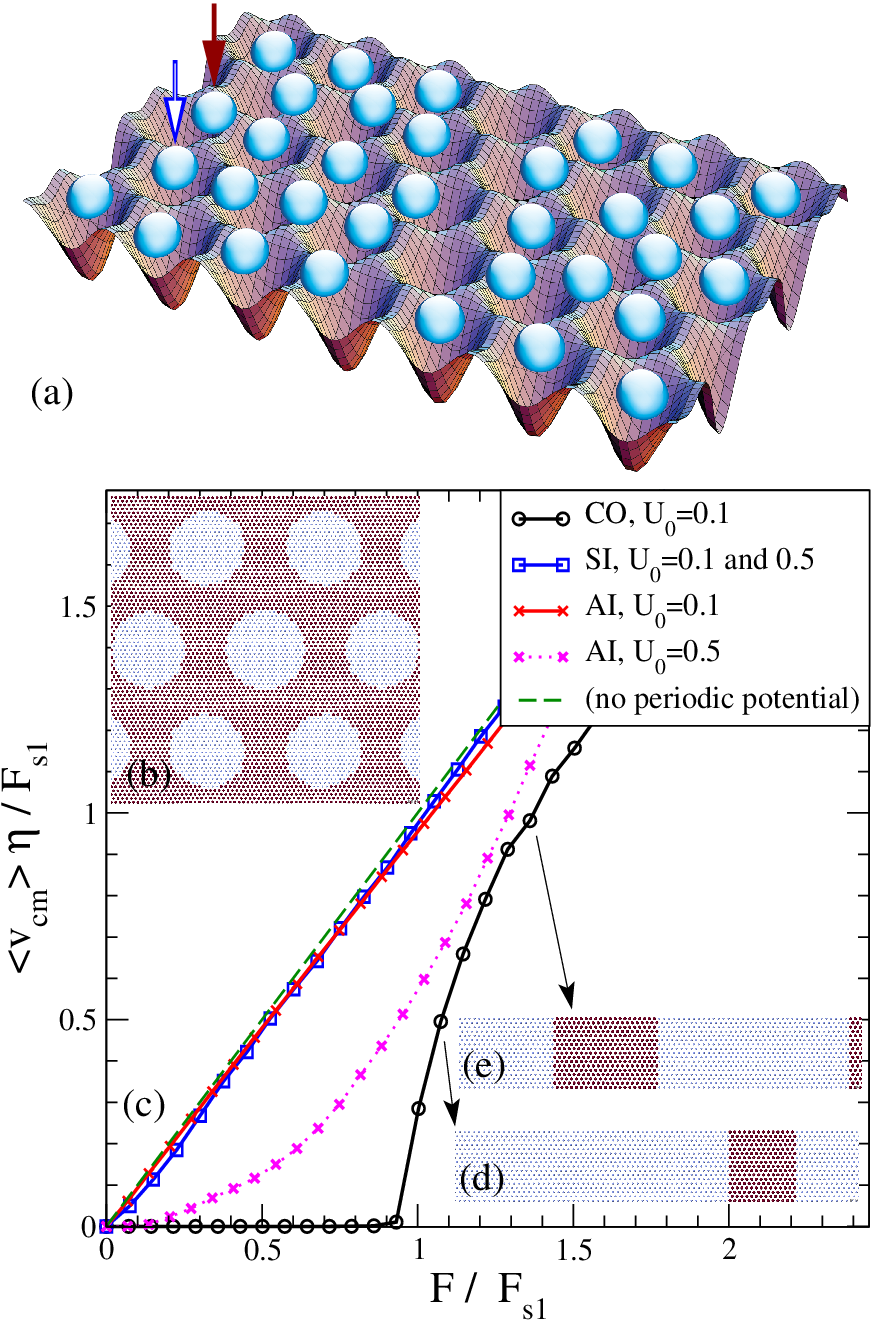}
\caption{
  (a) Schematic of a simulation of colloids interacting with a 2D triangular periodic
  substrate.  An individual colloid can sit in a minimum (white arrow) or
  on a maximum (red arrow) of the substrate
  potential.
  (b) The initial conditions at a filling of $n=0.95$.  Light dots represent colloids in
  substrate minima; dark dots indicate colloids on substrate maxima.  Here the substrate
  potential is relatively weak, so the dark areas are antisoliton regions that are floating above
  the substrate potential, while the white areas are commensurate regions in which
  the colloids sit in the substrate minima.
  (c) The velocity-force curves for fillings of
  $n=1.0$ (CO or commensurate, black),
  $n=0.95$ (AI or antisoliton-incommensurate, red and pink),
  and $n=1.05$ (SI or soliton-incommensurate, blue) showing that the commensurate case 
  has the highest depinning threshold.  $U_0$ is the substrate corrugation strength,
  corresponding to the pinning strength $F_p$, while
  $\langle v_{cm}\rangle$ corresponds to the velocity $V$ and $F/F_{s1}$ corresponds
  to the driving force $F_D$.
  (d,e) Illustration of sliding-generated solitons in the
  $n=1$ commensurate lattice at the
  points on the velocity-force curve marked with arrows.  The
  width of the soliton increases with increasing driving force.
  Reprinted with permission from A. Vanossi, N. Manini, and E. Tosatti,
  Proc. Natl. Acad. Sci. (USA) {\bf 109}, 16429 (2012).
  Copyright 2012 National Academy of Sciences, USA.
}
\label{fig:29}
\end{figure}

Bohlein {\it et al.} experimentally examined  the motion of
a monolayer of charge-stabilized colloids 
with a repulsive Yukawa colloid-colloid interaction
potential driven over a periodic substrate
\cite{7,New1}.
Figure~\ref{fig:28}(a) shows a schematic of the optically created
triangular substrate lattice, where each potential minimum traps
a single colloid and the external driving force
is applied along the $x$-direction. 
Figure~\ref{fig:28}(b) illustrates the colloid velocity versus applied
force, where the dashed line indicates
the velocity signature in the absence of a substrate.
At the $n=1$
commensurate condition,
the velocity
is roughly zero for $F_D < 45$
and the colloids are in the pinned state
shown in figure~\ref{fig:28}(c).
When $F_D = 49$
the system depins and
motion occurs in the form of sliding  kinks, 
as illustrated in figure~\ref{fig:28}(d),
while at higher drives all the colloids flow
simultaneously, as shown in figure~\ref{fig:28}(e).
At incommensurate fillings, the depinning threshold
$F_c$ is substantially reduced, but flow above depinning
still involves
running kinks (solitons) or antikinks (antisolitons).

Vanossi {\it et al.} conducted numerical simulations of the same system
and examined the velocity-force curves
at commensurate and incommensurate fillings
\cite{180}.
Figure~\ref{fig:29}(a) shows a schematic 
of their system, in which 
the colloids interact with an egg carton
corrugated substrate potential.
As illustrated in
figure~\ref{fig:29}(b),
at $n=0.95$
where there are fewer colloids than potential minima, different families of
antisolitons appear in the initial configuration.  The substrate is relatively weak
so the antisolitons are large objects containing many colloids that float
above the substrate potential.
Figure~\ref{fig:29}(c) shows velocity-force curves obtained for different
values of $n$ and different substrate strengths, with the substrate-free limit
indicated by a dashed line.
At the commensurate (CO) filling of $n=1.0$ the depinning threshold
$F_c\approx 1.0$ and the velocity-force curve has
sublinear scaling behavior;
however, the exponent $\beta$ in the
expression $V \propto (F_{D} - F_{c})^\beta$ was not analyzed. 
At the antisoliton-incommensurate (AI)
filling of $n=0.95$, depinning occurs in two steps, and 
at low drives the motion consists of antisolitons traveling in
the direction opposite to the drive,
achieved when each colloid moves one lattice constant
in the driving direction every time an antisoliton passes over it.
In the soliton-incommensurate (SI) case of $n=1.05$, the 
depinning threshold is even smaller and the solitons
move in the driving direction.
Figure~\ref{fig:29}(d,e) shows that in the CO case, the width of the moving
soliton regions
increases
as the external driving force is increased.
For both weak and strong substrates,
the solitons consist of regions in which excess particles that lack a substrate
minimum in which to sit partially float above the substrate lattice as an
incommensuration, and the
barrier for motion of these regions is small.
For weak substrates such as that shown in figure~\ref{fig:29}, the antisolitons
have the same floating incommensuration structure as the solitons, and are
similarly weakly pinned.
In contrast, for strong substrates, the antisolitons that appear at the AI filling of $n=0.95$
are composed of missing particles, and can move
only if the adjacent particles hop over the substrate
potential maximum.
As a result, the barrier for motion on strong substrates is much higher in the AI state 
than in the SI state.
Other simulation studies of the
sliding states of colloids moving over periodic substrates
show that the pinning threshold is maximum exactly
at the commensurate state, while away from commensuration, soliton flow states
arise \cite{181}.
As a function of substrate strength,
sharp transitions from a pinned state to a flowing state are observed in simulations,
leading to super-lubricity type behaviors \cite{182}.

\begin{figure}
  \includegraphics[width=\columnwidth]{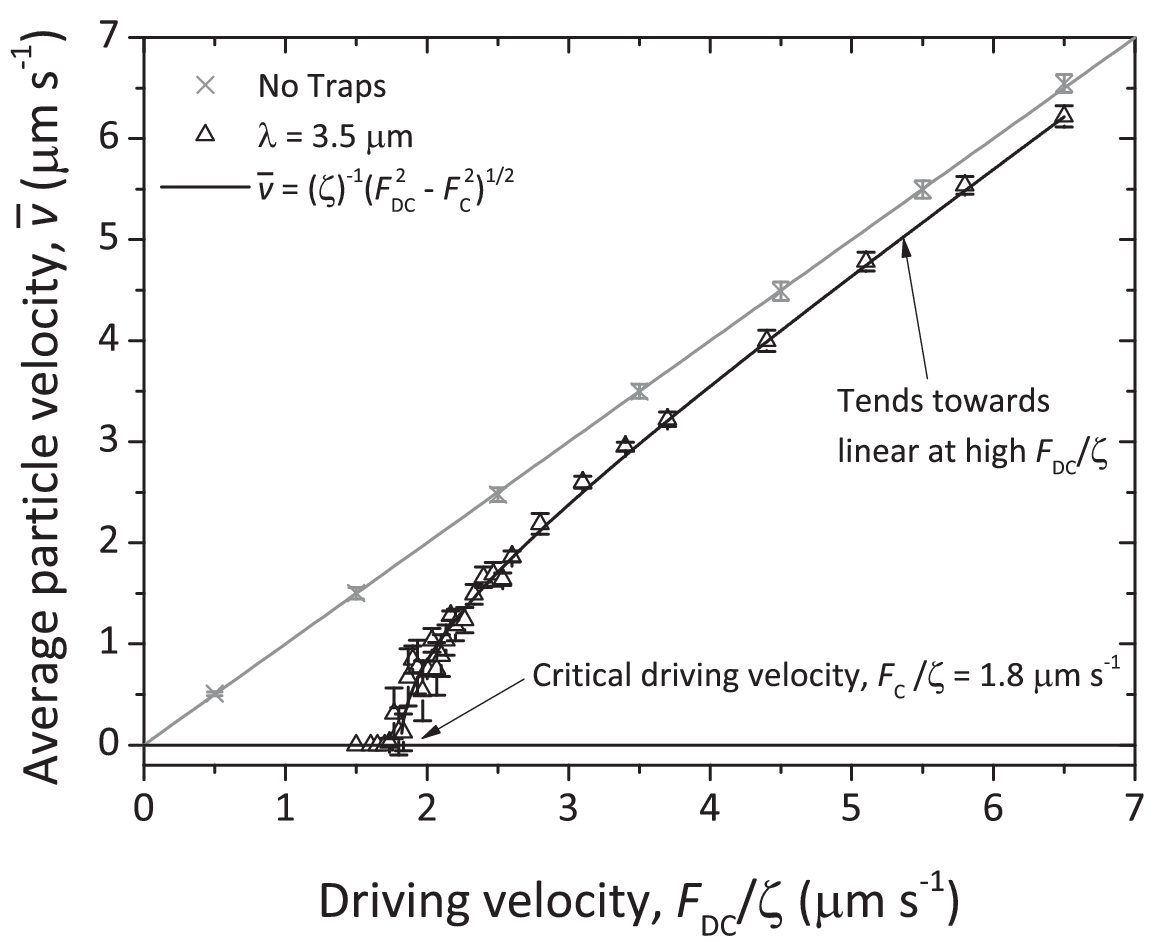}
\caption{ 
  The experimentally measured velocity-force curves for colloids moving over
  a 1D periodic substrate, showing the presence 
  of a critical depinning threshold $F_c$. At higher drive the velocity approaches the
  substrate-free limit (light line).
  Here $\bar v$ corresponds to the velocity $V$ and $F_{DC}/\zeta$ corresponds
  to the driving force $F_D$.
The solid line is a fit to $V \propto (F_{D}^2 - F^2_{c})^{1/2}$. 
  Reprinted with permission from M.P.N. Juniper, A.V. Straube, D.G.A.L. Aarts, and
  R.P.A. Dullens,
Phys. Rev. E {\bf 93}, 012608 (2016). Copyright 2016 by the American Physical Society.
}
\label{fig:30}
\end{figure}

Experimental studies of colloids moving under combined ac and dc driving
forces over 1D periodic substrates \cite{183} show phase locking steps
in the velocity-force curves
similar to those observed
for vortices and sliding CDW's under combined ac and dc drives.
Locking can occur when collective effects are present that make it possible
for a density wave or soliton excitation along the 1D colloidal chain
to act like a quasi-particle with dynamics similar to that of a single particle moving 
over a periodic substrate \cite{183}.
Simulations of this system indicate that the motion of such solitons can produce
additional steps in the velocity-force curves beyond the typical integer
Shapiro steps \cite{184}. 

Juniper {\it et al.} experimentally measured the dc velocity-force curves for 
colloids on a 1D periodic substrate and find
that the optical substrate can be effectively modeled as a sinusoidal potential for
small trap spacing,  
and that Brownian noise effects
are relevant near depinning but are absent in the  high driving limit \cite{185}.
Figure~\ref{fig:30} shows
that the average particle velocity for this system in the absence of a substrate
obeys
$V \propto F_{D}$, while when the substrate is present, a finite critical
depinning threshold appears
and the velocity approaches the clean limit at higher drives.
The solid line is a fit
to $V \propto (F^2_{D}  - F^{2}_{c})^{1/2}$;
however, near $F_c$ it is possible to fit the data to the
single particle elastic depinning
behavior of $V \propto (F_{D} - F_{c})^\beta$ with $\beta <  1.0$.

\subsubsection{Colloids on muffin tin substrates}

\begin{figure}
    \includegraphics[width=\columnwidth]{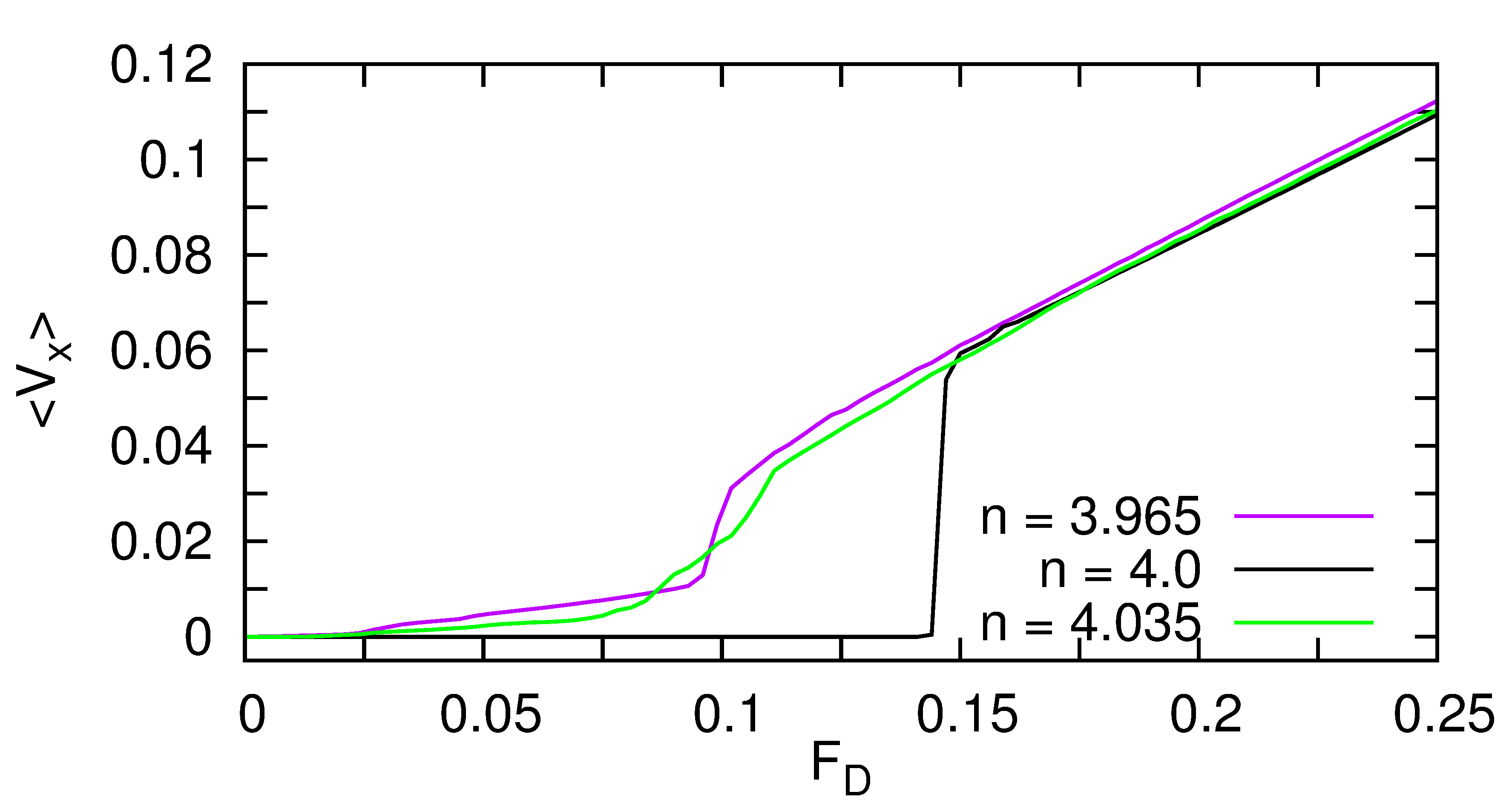}
\caption{ 
  Simulated velocity-force curves for colloids interacting with a square muffin tin
  substrate at fillings of
  $ n = 3.965$, 4.0, and $4.035$,
  showing that the depinning threshold is highest at
  the commensurate case $n = 4.0$.
  Here $\langle V_x\rangle$ corresponds to the velocity $V$.
  Adapted with permission from D. McDermott, J. Amelang, C.J. Olson Reichhardt, and
  C. Reichhardt,
Phys. Rev. E {\bf 88}, 062301 (2013). Copyright 2013 by the American Physical Society.
}
\label{fig:31}
\end{figure}

Simulation studies of colloids interacting with muffin tin type potentials reveal
the existence of a variety of dynamic phases \cite{62}.
Figure~\ref{fig:31} illustrates the velocity-force curves 
for a system with a square muffin tin pinning array 
at fillings of $n = 3.965$, 4.0, and 4.035.
At $n = 4.0$, the colloids form a commensurate ordered triangular lattice
in which the pinning sites trap $1/4$ of the colloids, and the remaining
colloids form an ordered arrangement in the interstitial region.
A two-step depinning process occurs for the $n=4.0$ filling,
with the initial depinning of the interstitial colloids occurring near $F_D=0.14$, while
the second depinning transition of the colloids in the pinning sites occurs at a drive
much higher than the range illustrated in figure~\ref{fig:31}.
At $n = 3.965$,
grain boundaries form in the pinned state that separate two different
degenerate orientations of the $n=4$ pinned state,
and the depinning transition is substantially reduced.
At depinning, the grain boundaries or antikink excitations become mobile,
while at $n=4.035$ the initial depinning is of the
kinklike grain boundaries that form.
For either incommensurate filling,
a transition occurs
at higher drives
from grain boundary-dominated dynamics to
the coherent flow of interstitial particles, while for higher drives there is a
transition to a state in which all of the colloids are flowing.

\begin{figure*}
  \includegraphics[width=1.9\columnwidth]{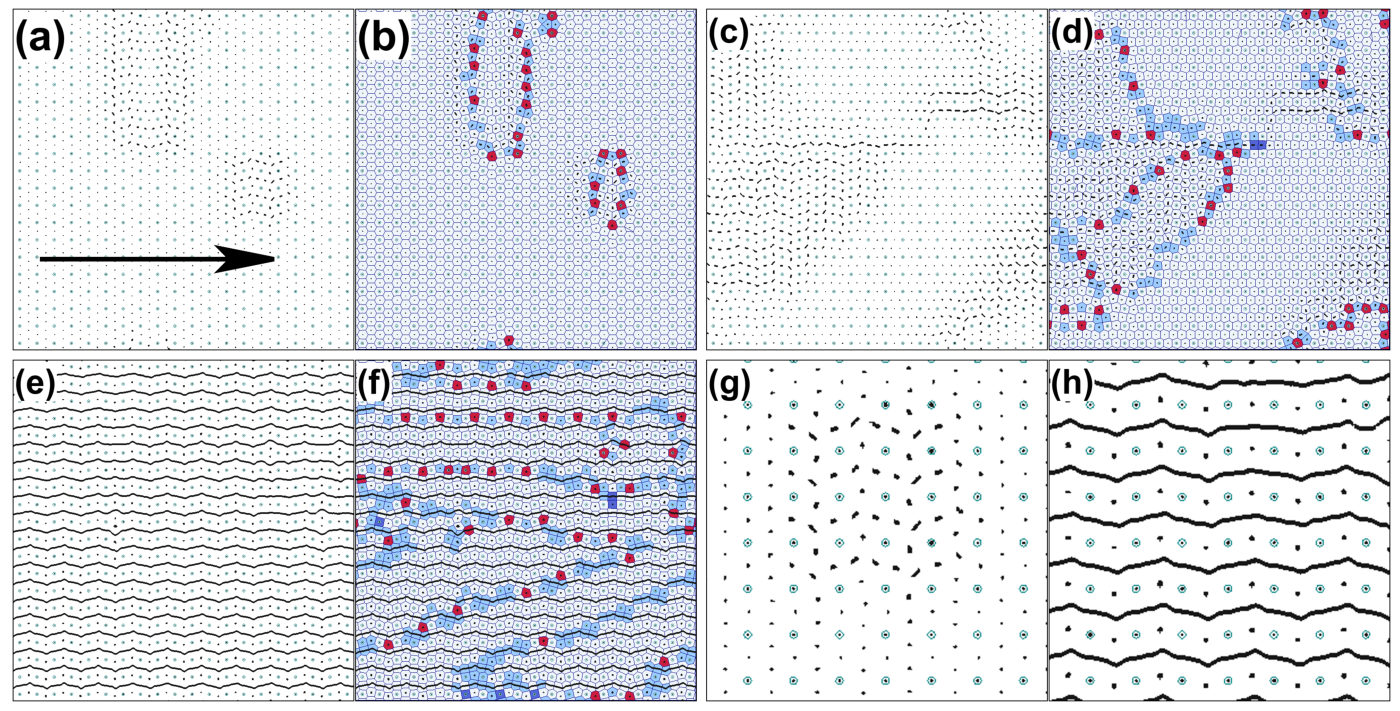}
  \caption{(a,c,e) The pinning site locations (open circles), particles (dots),
    and trajectories (lines) 
    from simulations of the system in figure~\ref{fig:31}
    for colloids driven in the $x$ direction over a square
    muffin tin pinning array at $n= 4.035$.
    The arrow indicates the driving direction.
(b,d,e) The corresponding Voronoi constructions.
    Light blue polygons indicate sixfold coordinated particles,
    while red and dark blue polygons are sevenfold and fivefold
    coordinated, respectively.
    (a,b) At $F_{D} = 0.05$, the grain boundaries depin.
    (c,d) At $F_{D} = 0.08$ there is a fluctuating grain boundary state.
    (e,f) At $F_{D} = 0.2$, interstitial colloid flow occurs.
    (g) A blowup of the trajectories from the flowing
    grain boundary state in panel (a).
    (h) A blowup of the flow of the interstitial particles in panel (e). 
  Reprinted with permission from D. McDermott, J. Amelang, C.J. Olson Reichhardt, and
  C. Reichhardt,
Phys. Rev. E {\bf 88}, 062301 (2013). Copyright 2013 by the American Physical Society.
}
\label{fig:32}
\end{figure*}

Figure~\ref{fig:32}(a) illustrates the colloid
positions, trajectories, and pinning site locations
for the $n = 4.035$ filling from
figure~\ref{fig:31},
and figure~\ref{fig:32}(b) shows 
the corresponding Voronoi construction of the colloid positions,
where light blue polygons 
indicate sixfold coordinated particles 
while the darker polygons are fivefold or sevenfold
coordinated particles.
At this filling, the system forms
two domain walls,
and  just above depinning at $F_{D} = 0.05$,
figure~\ref{fig:32}(a) shows that these  
domain walls begin moving in the direction of the drive in the form of kinks
or solitons.
The nature of the soliton 
motion is more clearly visible in
figure~\ref{fig:32}(g), which shows a blowup of the motion of a domain wall.
At $F_{D} = 0.08$, illustrated in figure~\ref{fig:32}(c,d),
the domain walls grow in size and fluctuate,
but there are still commensurate regions in which the colloids 
are immobile.
At $F_{D} = 0.2$, the flow transitions to the ordered interstitial motion shown
in figure~\ref{fig:32}(e,f),
and the domain wall  structure present at lower drives is lost.
The blowup of the ordered interstitial channel motion
in figure~\ref{fig:32}(h) shows that
the colloids at the pinning sites and a portion of the interstitial colloids remain 
pinned.
It would be interesting to examine the depinning of the domain walls in more
detail since this could provide an example of
elastic depinning of a line-like object.
In many models, domain wall depinning  is represented as
the motion of a strictly elastic system;
however, as shown in figure~\ref{fig:32}(c,d),
the domain wall can break up when it moves,
so it is possible that in many real examples of
domain wall depinning,
some plasticity or breaking of the domain wall may occur.

\subsection{Future directions}

Beyond spherical colloids, it is also possible to
consider colloidal dimers or trimers, patchy colloids, and Janus colloids,
all of which can form a variety of self-assembled equilibrium structures
\cite{glotz}.
 Simulation studies of effective colloidal dimers and trimers
 moving over a periodic substrate show that a variety of dynamical states can
 occur, such as flows in which the dimers align with the driving direction,
while
the velocity-force curves can exhibit transitions
 between ordered and sinusoidal flows \cite{186,187}.
In other numerical studies
of dimer motion on ordered
substrates, absolute negative mobility arises when a combination of
ac and dc driving forces are applied to colloidal dimers \cite{188}.
For colloids driven over quasi-1D periodic substrates, 
different types of ordered and disordered flowing states
can occur depending on the filling factor and the magnitude of the
external dc drive \cite{189,190}.
One of the next systems to examine would be
the driven motion of colloids with more complex interactions over
periodic substrates, since it may be possible to produce dynamically generated
self-assembly.
It would also be interesting to consider multiple species of colloids moving over periodic
pinning arrays to understand possible segregation or mixing effects.

Most of the studies of particles driven over periodic substrates have been performed in 2D.
One possible avenue of future research would be to study 
the dynamical phases of particles moving
through 3D pinning arrays.
Colloids represent an ideal system for such studies
since it is possible to create 3D arrays of optical traps.
In 3D, issues such as glass transitions start to become important,
and it is possible that
the depinning transitions or dynamic phases could change under
conditions at which
a glass-forming length scale becomes larger than the length scale of the periodicity
of the pinning sites.
It would also be interesting to characterize the moving phases of
colloids and other particle systems
moving over periodic pinning arrays 
using something akin to an effective temperature.
For particles moving over random disorder,
the dynamically generated fluctuations from the pinning sites
often have Boltzmann-like statistics, making it possible to apply
an effective temperature description \cite{191};
however, for particles moving over periodic arrays, the fluctuations
are much more directional and generally do not have Boltzmann-like features.
In certain cases where the pinned state is
strongly frustrated, it may be possible for
the frustration effects to survive into the moving state,
permitting the moving state for certain incommensurate fillings to be
characterized as having effective thermal fluctuations even though the commensurate
moving states do not.

\section{Dynamic Phases on Quasiperiodic Substrates}

Substrates with quasicrystalline ordering can be created
for both superconducting vortices 
and colloids.
Since quasicrystals are nonperiodic yet have long range order
\cite{192}, 
the dynamic phases of particles driven over quasiperiodic substrates
could exhibit features that are a combination of those observed in
periodic and random pinning arrays.
In the high drive limit, particles driven over random pinning
dynamically reorder into a triangular
moving solid or a moving smectic state,
while for periodic pinning arrays the particles generally form a moving smectic state,
so on a quasicrystalline pinning array, a moving solid or a moving smectic could
emerge
depending on whether the long range order or the nonperiodicity dominates.

\begin{figure}
  \includegraphics[width=\columnwidth]{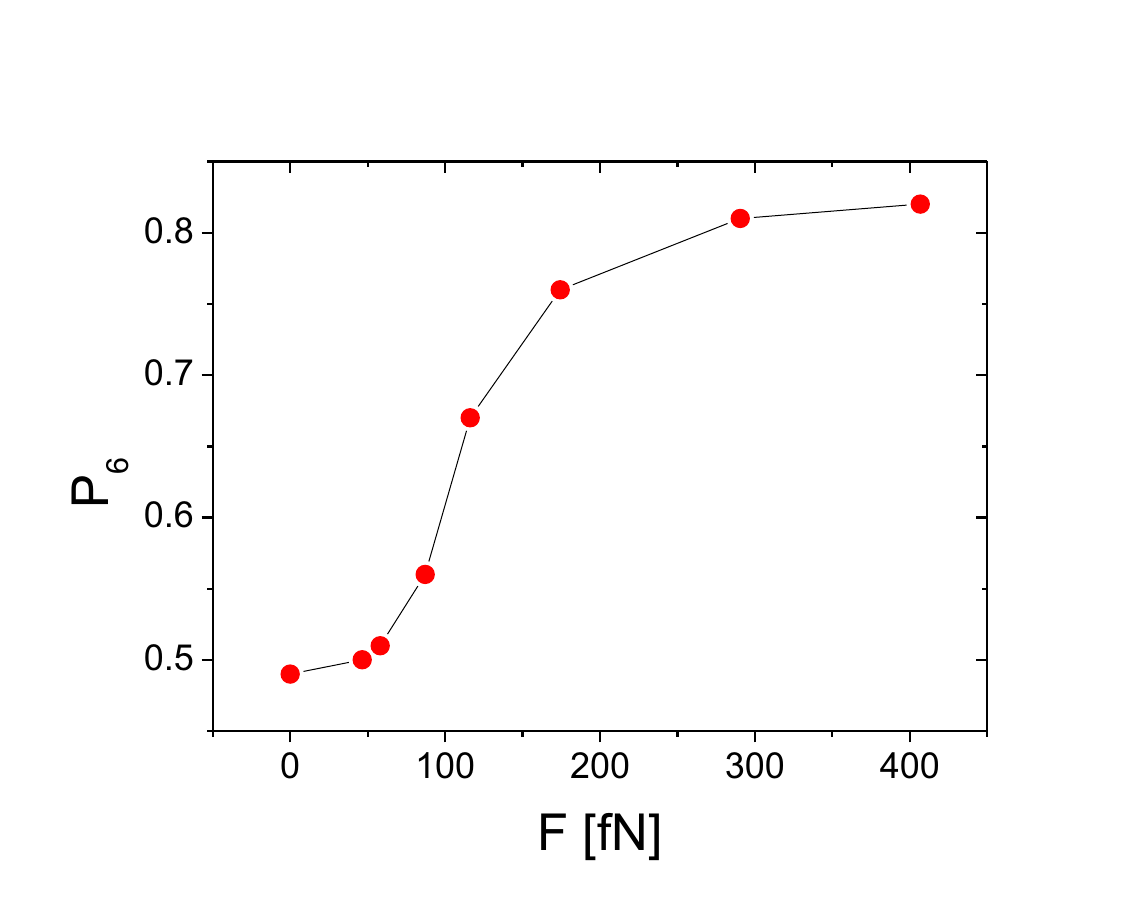}
\caption{ The fraction of sixfold coordinated colloids
  $P_6$ vs driving force $F$ for an experimental system of
  colloids driven over a quasiperiodic substrate.
  Here $F$ corresponds to the driving force $F_D$.
  Reprinted with permission from T. Bohlein and C. Bechinger,
Phys. Rev. Lett. {\bf 109}, 058301 (2012). Copyright 2012 by the American Physical Society.
}
\label{fig:33}
\end{figure}

In numerical studies of the dynamics of both superconducting
vortices and colloids moving
over a fivefold or Penrose quasiperiodic array of pinning sites
\cite{193,194},
at low drives the particles form
a pinned quasicrystal similar to that found in
the pinned state for colloids interacting with quasiperiodic
substrates in the strong trapping limit \cite{195,196}.
The system depins into a plastic flow phase, and then
at higher drives dynamically orders
into a moving Archimedean structure state in which
the particles form combinations of square and triangular rows 
that are aligned in the direction of drive \cite{193}.
Archimedean ordering was also
experimentally observed in the pinned state
for colloids on quasicrystalline substrates in the weaker substrate
limit, and was argued to arise as a compromise between the
periodic ordering favored by the colloid-colloid interactions
and the quasiperiodic ordering of the substrate \cite{195}.
If the driving direction is altered, simulations show that
the Archimedean ordering emerges only for particular orientations of the
drive with respect to the fivefold symmetry directions of the substrate, while
square moving lattices and disordered states occur
when the drive is oriented along non-symmetry directions of the substrate \cite{193}.

\begin{figure}
  \includegraphics[width=\columnwidth]{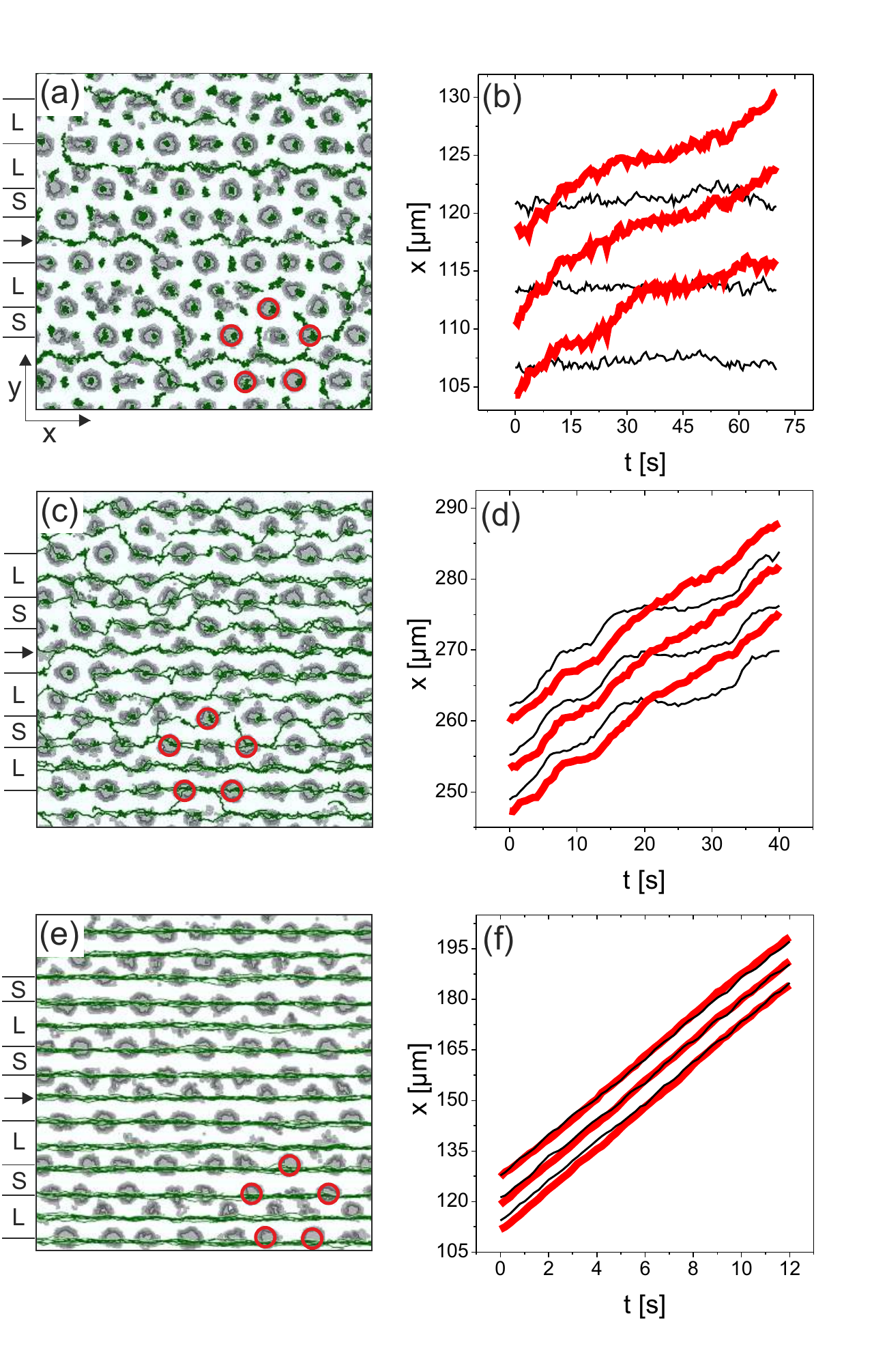}
  \caption{
    Experimental results for colloids driven
    in the $x$ direction over a quasiperiodic substrate.
  (a,c,e) The particle trajectories (green)
    and the quasiperiodic substrate (grey) at driving forces
    from figure~\ref{fig:33} of (a) $F = 47$ fN
    (c) $F = 87$ fN, and (e) $F = 291$ fN.
    The fivefold symmetry of the substrate is highlighted in red.
    Here $F$ corresponds to the driving force $F_D$.
    (b,d,f) The $x$ coordinate vs time $t$ for three sets of adjacent interstitial
    (thick red lines) and noninterstitial (thin black lines) particles.
    (b) The noninterstitial particles are pinned and the interstitial particles are moving
    at $F=47$fN.
    (d) The noninterstitial particles transiently pin and depin while the interstitial
    particles move continuously at
    $F=87$ fN.
    (f) Interstitial and noninterstitial particles move with the same velocity
    in the dynamically ordered phase at at $F=291$ fN.
  Reprinted with permission from T. Bohlein and C. Bechinger,
Phys. Rev. Lett. {\bf 109}, 058301 (2012). Copyright 2012 by the American Physical Society.
}
\label{fig:34}
\end{figure}

Bohlein and Bechinger experimentally studied
colloids driven along the $x$ axis, a symmetry direction, of
a fivefold quasiperiodic substrate,
and
observed a transition from a plastic flow state at low drives to a
dynamically ordered moving Archimedean crystal
at higher drives \cite{New1}. 
Figure~\ref{fig:33} shows a plot of $P_6$,
the fraction of sixfold coordinated colloids, versus
driving force.
At low drives, $P_{6} \approx 0.5$ and the flow consists of 
interstitial particles moving past pinned particles.
At higher
drives, $P_{6}$ increases to
a maximum value of $P_6=0.8$ in the dynamically ordered regime,
but does not reach $P_6=1.0$
despite the reordering
since the Archimedean moving lattice
structure contains a number of four-sided polygons. 
Figure~\ref{fig:34}(a) shows the particle trajectories
and the quasiperiodic substrate potential
in the plastic flow phase
at $F_D = 47$ for the system in figure~\ref{fig:33}, while
a plot of the $x$ position of adjacent interstitial and noninterstitial colloids as
a function of time in 
figure~\ref{fig:34}(b)
shows that at this drive there is a coexistence of pinned noninterstitial and moving
interstitial colloids.
Intermittent trapping of the noninterstitial colloids occurs
at $F_D=87$ fN,
as shown in figure~\ref{fig:34}(c,d), and
at $F = 291$ fN in figure~\ref{fig:34}(e,f), the system has 
entered the dynamically ordered flowing state
and the trajectories form straight lines centered on rows of pinning sites.
These experiments also demonstrated that for  driving at different angles
with respect to the substrate,
changes in $P_6$ similar to those found in simulations
occur when the colloidal motion becomes
locked to certain angles.
Moving square ordering was not observed in the experiments,
which could be due to the different scale of the pinning sites or the magnitude
of the driving force compared to the simulations.
In numerical studies of particles driven over quasiperiodic substrates by a combination
of ac and dc drives, phase locking similar to that observed for particles moving
over ordered arrays occurs; however,
for the quasiperiodic substrate
strong irrational phase locking steps are present
\cite{197}. 

Many open questions remain in
the dynamics of
driven particles on quasicrystalline arrays,
such as what phases appear on
arrays
with sevenfold, tenfold, or higher-fold ordering \cite{198}
or what effects occur
when phason dynamics are also relevant \cite{199}. In the moving state 
it would also be interesting to study the
dynamically generated fluctuations to determine
whether it is possible to
describe the behavior of the system in terms of an effective
dynamically induced temperature.

\section{Depinning and Dynamics in Charge Transport} 

\subsection{Wigner crystal depinning}

Wigner crystals \cite{3} are another system
that exhibits depinning and sliding behavior.
This electronic crystalline state is normally associated with
insulating behavior, but it
has a finite threshold for conduction under an applied drive.
2D Wigner crystal states that arise in semiconductor systems can be
pinned by charged doping sites in the material.
Experiments on 2D Wigner crystal systems
show threshold and nonlinear conduction features \cite{200,201}.
Sliding states can also occur for electrons on liquid He films, where the 
pinning is produced by dimpling effects \cite{202}.

\begin{figure}
  \includegraphics[width=\columnwidth]{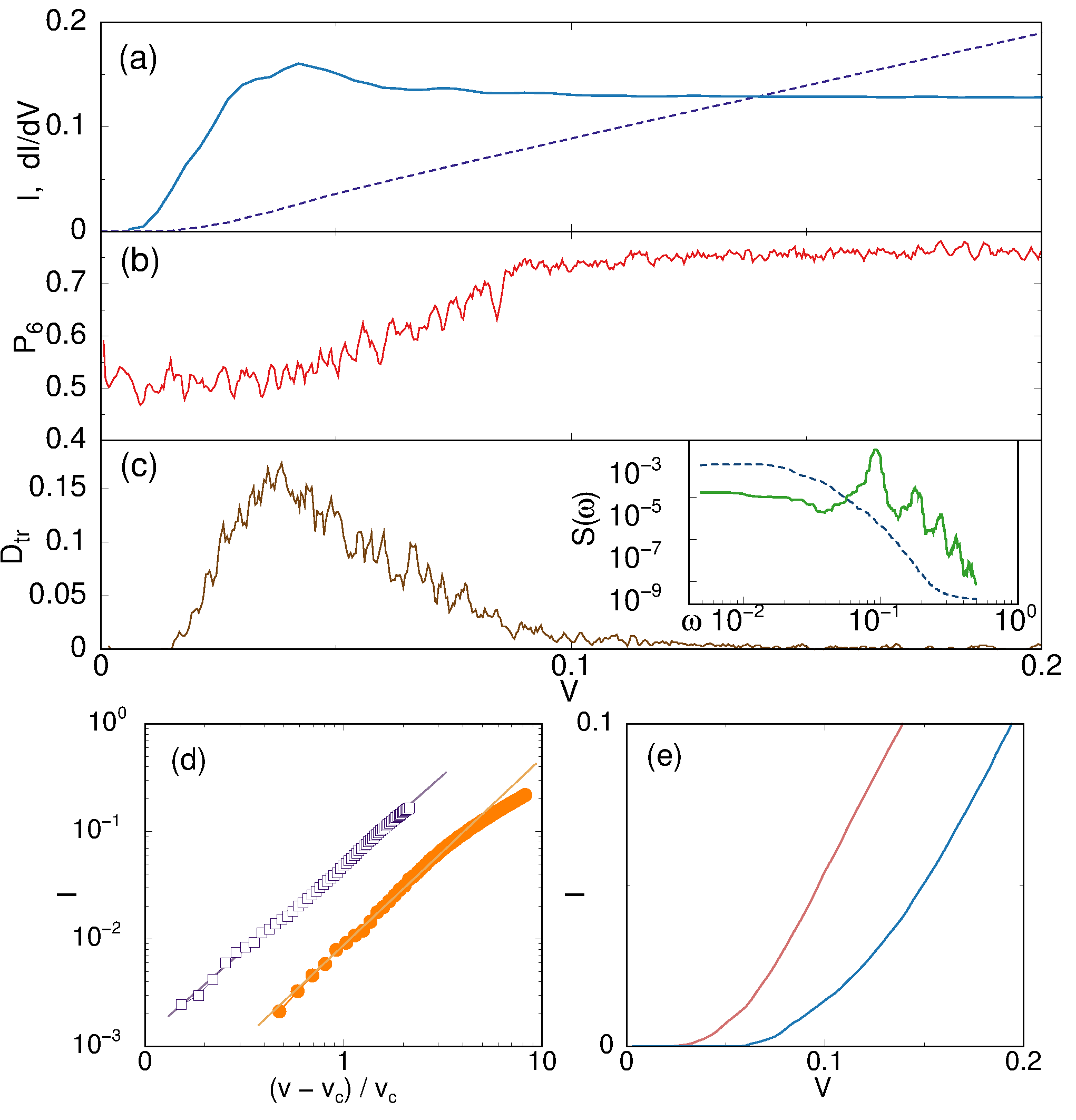}
  \caption{
    Results from simulations of a 2D Wigner crystal interacting with charged defects.
    (a) Dashed line: Current $I$ vs voltage $V$.
    $I$ corresponds to the electron velocity $V$ while the voltage $V$ corresponds
    to a driving force $F_D$.
    Solid line: $dI/dV$, corresponding to a $dV/dF_D$ curve.
  (b) The corresponding fraction of sixfold coordinated electrons
  $P_{6}$ vs voltage $V$. 
  (c) The transverse diffusion $D_{tr}$ of the electrons vs voltage $V$.
    Inset: power spectrum $S(\omega)$
    of the velocity time series for the plastic flow phase (dashed line),
  showing broad band noise, and for the moving smectic phase (solid line), showing
  narrow band noise.
  (d,e)  The scaling of the velocity-force curves,
  which obey velocity
  $V = (F_{D} - F_{c})^\beta$ with $\beta = 1.61$ and $\beta=1.71$.  
Adapted with permission from C. Reichhardt, C.J. Olson, N. Gr{\o}nbech-Jensen,
and F. Nori,
Phys. Rev. Lett. {\bf 86}, 4354 (2001). Copyright 2001 by the American Physical Society.
}
\label{fig:35}
\end{figure}

In simulations of classical electrons with long range Coulomb interactions,
Cha and Fertig observed a transition from a clean or ordered triangular Wigner
crystal state to
a disordered state as function of
the strength of the quenched disorder substrate,
and found that in the disordered regime the depinning 
is plastic \cite{38}.
In other simulations \cite{40} performed
using overdamped dynamics for the
same model of the electron crystal state, the substrate potential
produced by the charged doping sites
is represented as
a smooth pinning landscape with large-scale fluctuations.
The interaction energy in this system is given by
\begin{equation}
  U = \sum_{i\neq j} \frac{e^2}{|{\bf r}_{i} - {\bf r}_{j}|}
  - \sum_{ij}\frac{e^2}{\sqrt{(|{\bf r}_{i} - {\bf r}^{(p)}_{j}|^2 + d^2)}}
\end{equation}
where ${\bf r}_i$ is the location of electron $i$.
The first term is the electron-electron repulsion which favors
formation of a triangular lattice.
The second term represents the interaction of the electrons with impurities
at positions ${\bf r}_{j}^{(p)}$ that
are assumed to be
positively charged defects offset by a distance $d$ from the 2D plane containing
the classical electrons.
For strong disorder or small $d$, the
numerical studies of Reichhardt {\it et al.} \cite{40} indicate that the electrons 
depin into a plastic phase, followed at higher drives by a dynamical reordering
into a moving Wigner crystal phase.
Figure~\ref{fig:35}(a) shows the electron velocity, representing a
measured
current, versus driving force, representing an
applied voltage, along with the
corresponding $dV/dF_D$ curve.
The corresponding fraction $P_6$ of sixfold
coordinated electrons versus $F_D$ appears in
figure~\ref{fig:35}(b),
while
figure~\ref{fig:35}(c) shows a measure $D_{tr}$ of the fraction of electrons that
wandered a distance larger than $a_0/2$ in the direction perpendicular to the drive
during a fixed time interval, where $a_0$ is the electron lattice constant.
There is a clear finite depinning threshold, and a peak in
$dV/dF_{D}$ coincides with the drive at which $P_6$ begins to increase
and $D_{tr}$ begins to decrease.
Near $F_{D} = 0.1$, the system enters a moving smectic phase as indicated by
the saturation of $dV/dF_D$ and $P_6$.  In the moving smectic phase, $D_{tr}$ goes to zero
since the electrons flow along well-defined channels and cease to wander
in the direction transverse to the drive.
In the plastic flow phase, the velocity noise spectrum $S(\omega)$ has
a broad band signal, as shown in 
the inset of figure~\ref{fig:35}(c),
while in the moving smectic phase a narrow band 
noise feature emerges.
There is no hysteresis in the velocity-force curves obtained in these simulations,
but near the depinning threshold the electron velocity
scales as $V \propto (F_{D}- F_{c})^\beta$ with
$\beta=1.61$ to $\beta=1.71$, as shown in
figure~\ref{fig:35}(d,e).
These exponents are similar to those observed in other studies of plastic depinning.

\begin{figure}
  \includegraphics[width=\columnwidth]{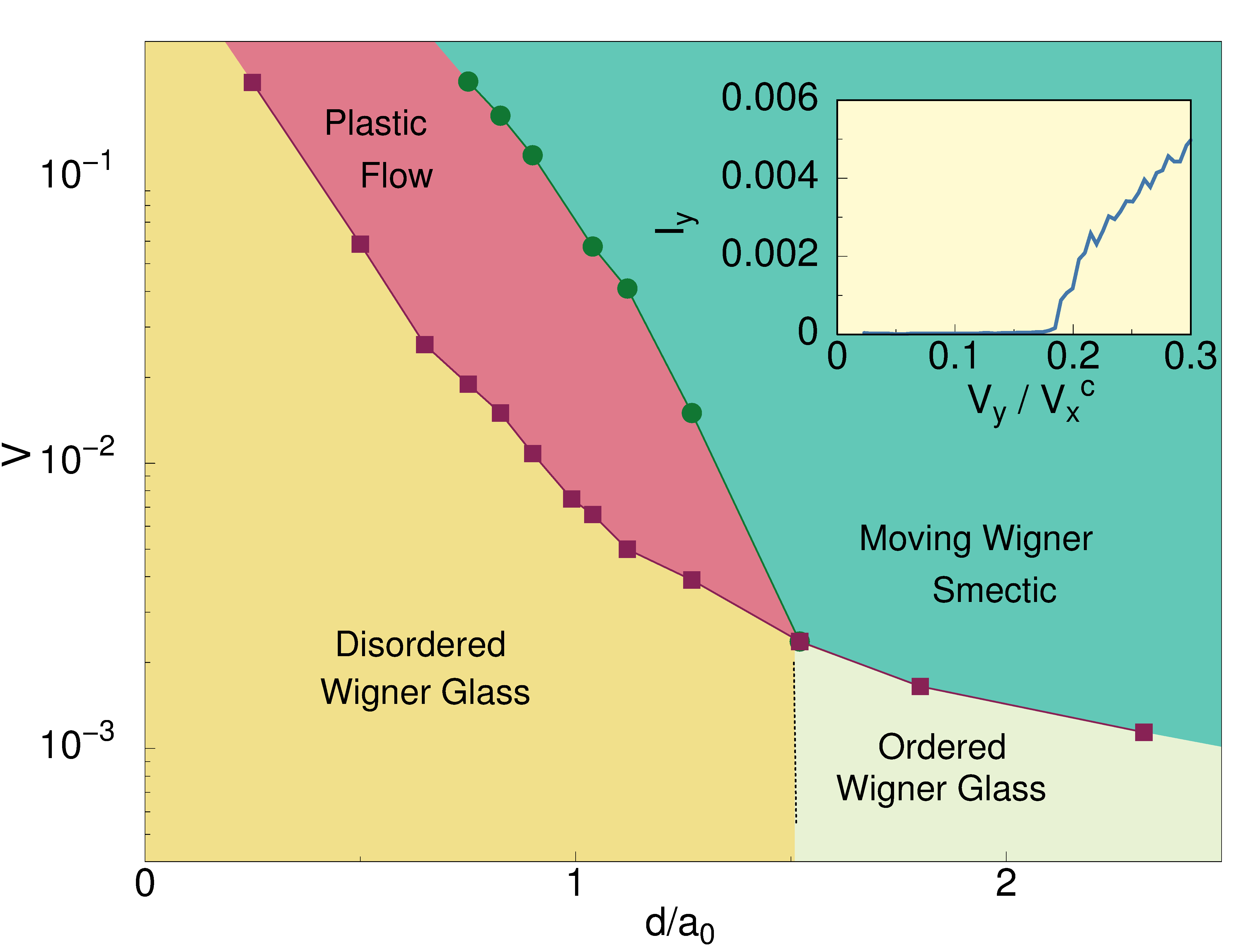}
  \caption{Dynamic phase diagram from simulations of
    the driven disordered Wigner solid as a
  function of the driving voltage $V$ vs inverse pinning strength $d/a_0$
  for the system in figure~\ref{fig:35}
  highlighting a disordered pinned phase
  (Wigner glass), an ordered pinned phase, a plastic flow phase, and a
  moving Wigner smectic phase.
  The voltage $V$ corresponds to a driving force $F_D$ and the pinning
  offset distance $d/a_0$ is inversely proportional to the pinning strength $F_p$.
  Inset: Velocity-force curve showing
  the transverse depinning of the moving smectic phase.
  The current $I_y$ corresponds to the velocity $V_y$ in the transverse direction,
  while the voltage $V_y/V_x^c$ corresponds to a transverse driving
  force $F_D^y/F_c^x$, where $F_c^x$ is the depinning threshold for longitudinal
  driving.
Adapted with permission from C. Reichhardt, C.J. Olson, N. Gr{\o}nbech-Jensen,
and F. Nori,
Phys. Rev. Lett. {\bf 86}, 4354 (2001). Copyright 2001 by the American Physical Society.
}
\label{fig:36}
\end{figure}

Figure~\ref{fig:36} shows the dynamic phase diagram for the Wigner crystal
system as a function of driving force versus
pinning offset distance $d$, where
small $d$ corresponds to high $F_p$.
For $d > 1.5$ or weak pinning, the pinned ground state is ordered and the system
depins elastically into a moving Wigner crystal state,
while for smaller $d$ or stronger pinning, plastic depinning occurs
and there is a dynamical reordering transition at higher drives into a
moving smectic phase.
The moving smectic has
a strong transverse depinning threshold that can be
observed by
applying an additional transverse dc drive to the moving smectic state,
as illustrated in the inset of figure~\ref{fig:36}.
The flowing smectic channels do not break apart at the transverse depinning
transition, which has elastic properties as indicated by the
$\beta<1$ curvature of the transverse velocity-force curve.
Due to the long-range nature of the substrate disorder potential,
the relative size of the transverse depinning threshold
$F_c^y$ compared to the longitudinal
depinning threshold $F_c^x$ is much larger than is observed in systems with short range
pinning potentials, such as superconducting vortex systems \cite{81a,82a,83a}.
Other numerical studies on the 
depinning of Wigner crystals in a constricted channel
show that the 
velocity-force curve scales as $V=(F_D-F_c)^{\beta}$ with $\beta=0.66$ 
in the elastic depinning regime and $\beta=0.95$ in the quasielastic regime
\cite{203}, while simulations using correlated pinning
give $\beta=1$ to $\beta=1.7$
in the plastic depinning regime \cite{204}. 

Transport experiments on 2D electron gases (2DEGs) show evidence of
finite depinning thresholds and hysteretic velocity-force response curves \cite{205}
as well as negative differential resistance \cite{206},
which could imply that in some
cases, a purely overdamped equation of motion as a model for a Wigner crystal
is inadequate and does not capture the true dynamics.
Alternatively, such results could
indicate that in these systems the electrons form more exotic
crystalline states such as stripes or bubbles, which can produce different dynamics as
we discuss in Section 11.
The appearance of narrow band
velocity noise in simulations suggests that 
Shapiro step phenomena could be produced by imposing an additional ac drive on
the dc drive, as previously predicted
for sliding Wigner crystal states \cite{207}; however, experiments of this
type have not been conducted.
In the case of liquid helium, the pinning is produced by
the interactions of the electrons with  the
surface of the fluid, so it
would be interesting to model this system by using 
deformable pinning sites.  

\subsection{Charge transport in metallic dot arrays}

\begin{figure}
  \includegraphics[width=0.8\columnwidth]{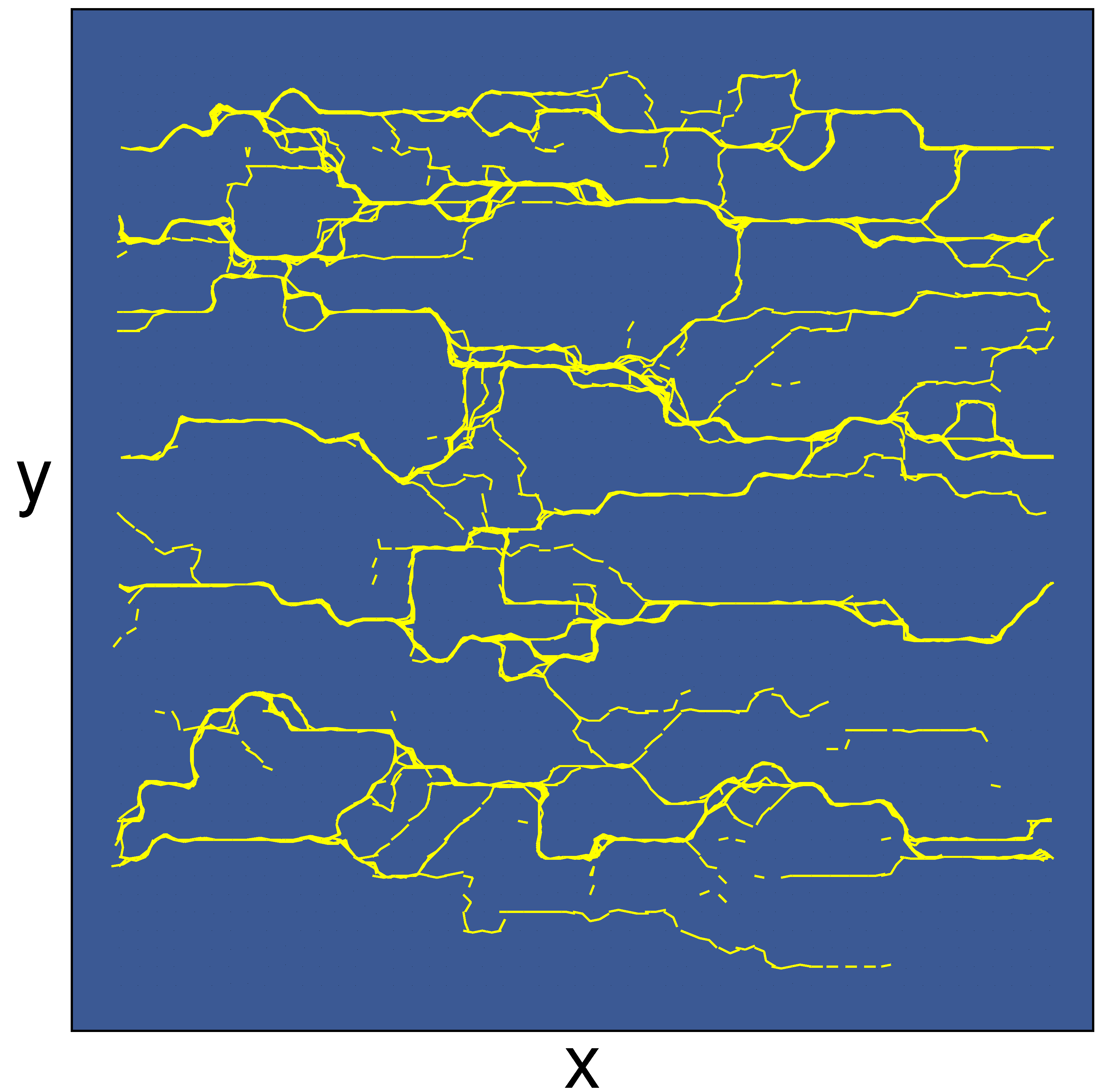}
\caption{ 
  The current paths of charge flow (yellow) from a simulation of
  a 2D metallic dot array system just above
  the plastic depinning threshold.
Adapted with permission from C. Reichhardt and C.J.O. Reichhardt,
Phys. Rev. Lett. {\bf 90}, 046802 (2003). Copyright 2003 by the American Physical Society.
}
\label{fig:37}
\end{figure}

Another system that exhibits features such as a finite
conduction threshold and nonlinear transport 
is charge transport in coupled metallic dot arrays.
Middleton and Wingreen (MW)
performed analytical and simulation studies of
both 1D and 2D models of coupled
metallic islands in which charges
hop from one island to the next across a Coulomb barrier
under a driving force
consisting of an externally applied voltage \cite{5}.
The velocity of the charges is proportional to an experimentally
measurable current.
MW observed
a finite threshold for conduction associated with critical scaling of the velocity
according to
$V = (F_D - F_{c})^{\beta}$,
with $\beta=5/3$ in analytical calculations,
$\beta=2.0$ in 2D simulations, and $\beta=1.0$ in 1D simulations.
They also found that the 
charge flows in transverse meandering channels similar to those observed for
superconducting vortices above the
plastic depinning transition.
Figure~\ref{fig:37} shows an example of such charge conduction channels
just above the plastic depinning threshold 
from simulations of a 2D metallic dot array \cite{208}.
Near depinning thresholds observed in
experiments on coupled metallic islands,
the  velocity scales as
$V \propto (F_D - F_{c})^{\beta}$ with $\beta = 1.5$ to $\beta=2.0$ \cite{209}. 
Other experiments on GaAs quantum dot arrays reveal
a scaling exponent for a single dot of $\beta = 0.5$,
consistent with single particle depinning,
while in coupled dot arrays,
$\beta =1.47$ \cite{210}.

\begin{figure}
  \includegraphics[width=\columnwidth]{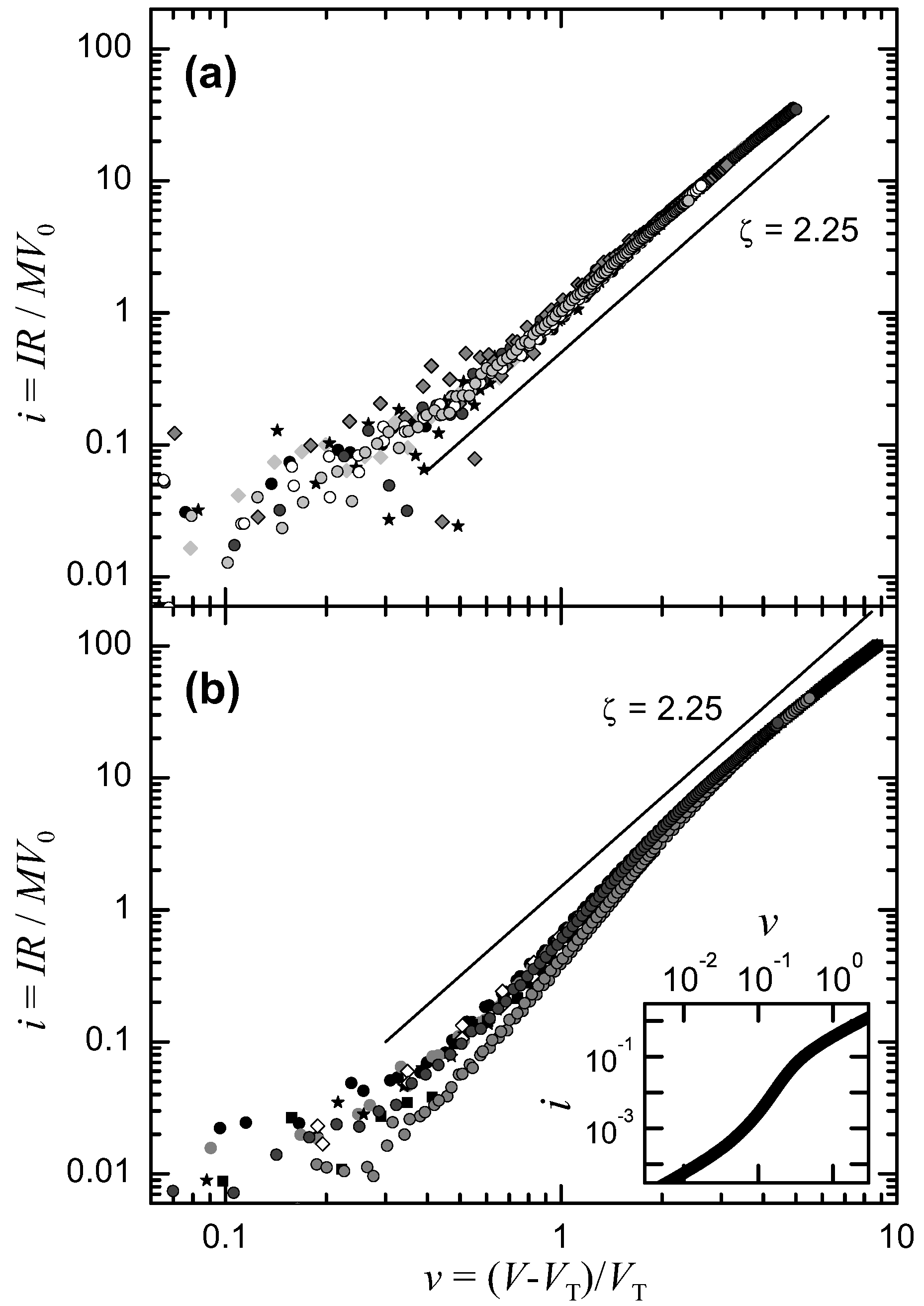}
\caption{
  (a) The scaled current $i$ vs voltage $\nu$ curves for
  experimental measurements of charge transport in
  a uniform gold nanodot array, where  $V_{T}$ is the threshold
  applied voltage at which conduction becomes finite.
  Here the current $i$ corresponds to the charge velocity $V$, while
  the voltage $\nu$ corresponds to the driving force $F_D$.
  The line indicates a fit to  
  velocity $V \propto (F_{D} - F_{c})^\beta $ with $\beta = 2.25$.
  In the figure $\zeta$ corresponds to the scaling exponent $\beta$.
 (b) Velocity-force curves for a gold dot array
  containing large scale voids.  There are two scaling regimes,
  with $\beta = 2.25$ in the higher voltage or large drive regime.
  Inset:
  The two-step transport response is also captured in  simulations.
  Reprinted with permission from R. Parthasarathy, X.-M. Lin, and H.M. Jaeger,
Phys. Rev. Lett. {\bf 87}, 186807 (2001). Copyright 2001 by the American Physical Society.
}
\label{fig:38}
\end{figure}

Parthasarathy {\it et al.} performed experimental transport studies on
monolayers of gold nanocrystals formed into arrays that have either long range
order or large scale voids \cite{211}.
Figure~\ref{fig:38}(a) shows the scaling observed in
successive velocity-force curves
obtained with ordered arrays, which
exhibit a finite depinning threshold and a scaling exponent of
$\beta = 2.25$, while figure~\ref{fig:38}(b) shows that
 an array containing large scale voids has
 two scaling regimes, with $\beta = 2.25$ at
 large drives and a lower exponent at low drives.     
 Subsequent simulations \cite{208} of uniform arrays
 give exponents of $\beta = 1.94$ for 2D arrays and
 $\beta=1.0$ for 1D arrays, while 
 systems with large voids exhibit
two scaling regimes.
 At small drives in the void system, $\beta = 1.0$ and the charges flow through a
 sparse array of static 1D channels that avoid the voids,
 placing the system in an effectively 1D regime
 for which Middleton and Wingreen predicted an exponent $\beta = 1.0$ \cite{5}. 
 At higher drives, the flow in the void system becomes fully 2D, producing
 a larger value of $\beta$.
  The simulation work in \cite{208} also
  showed a transition
  as a function of increasing drive from meandering plastic flow channels
  at low drives to 
  a coherent flow of effectively 1D chains of charges along the arrays at high drives.
  This transition is associated with a change from broad band
  velocity or conductance fluctuation noise at low drives
  to narrow band noise
  at higher drives;
however, there have been no experiments to confirm whether such a dynamical transition
to a coherent flow occurs in actual metallic dot arrays.
Additionally, it is not clear if there is universality in the scaling exponents 
for the plastic depinning. 
Several experiments on dot arrays
show discontinuous and hysteretic behavior in the velocity-force curves
\cite{210,212} which have not been captured in
simulations.
It is possible that such effects
arise from the spatial ordering
of the dot array itself,
and that they correspond to features associated with
periodic pinning arrays rather than with disordered pinning.
If so,
additional elements such as time-dependent dissipation or
local heating effects
would need to be added to the simulation
models in order to capture
these behaviors.

As advances in nanotechnology continue, greater control over the topology and
disorder of the dots should be achievable, making it possible to carefully tune
the pinning effect as well as to control longer-range correlations and
monitor local dissipation effects.
An understanding of how to control charge flows through tailored dot array geometries
would be important for certain applications of these devices.   
Future work in this field
could include studying
the driven dynamics of fully 3D coupled dot
arrays with and without layering effects \cite{213}.

\begin{figure}
    \includegraphics[width=0.8\columnwidth]{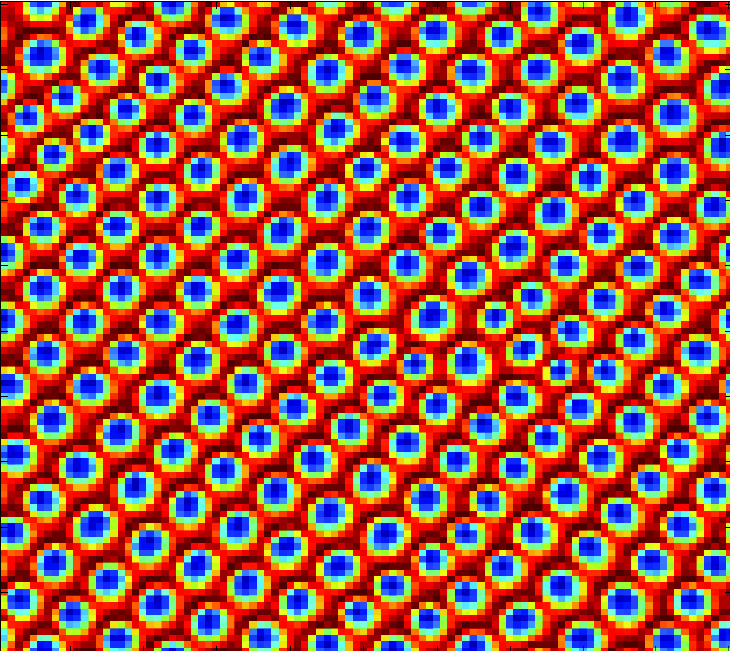}
\caption{A continuum simulation of a skyrmion lattice in the absence of a substrate.
  The skyrmions form a triangular
lattice similar to the Abrikosov lattice observed for superconducting vortices.
  Reprinted with permission from S.-Z. Lin, C. Reichhardt, C.D. Batista, and A. Saxena,
Phys. Rev. Lett. {\bf 110}, 207202 (2013). Copyright 2013 by the American Physical Society.
}
\label{fig:39}
\end{figure}

\section{Depinning and Dynamics of Skyrmions}

\begin{figure}
  \includegraphics[width=\columnwidth]{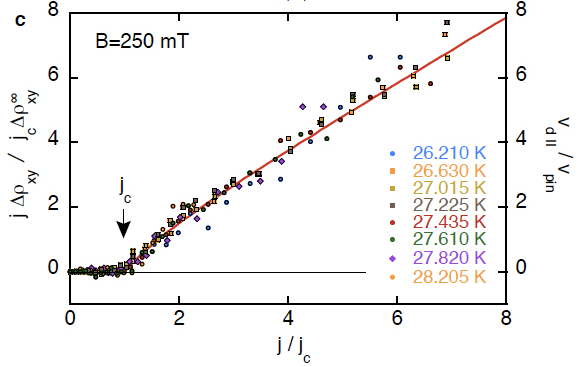}
  \caption{
    Experimental measurements obtained at different temperatures of 
    effective skyrmion velocity deduced from changes $\Delta \rho_{xy}$ in the Hall resistance
  versus driving current $j/j_c$ scaled by the critical depinning current $j_c$,
  showing a finite depinning threshold and a sliding phase.
  Here the change in the Hall resistance $\Delta \rho_{xy}$ corresponds to the
  velocity $V$, the driving current $j$ corresponds to the driving force $F_D$, and
  the critical current $j_c$ corresponds to the depinning threshold $F_c$.
  Reprinted by permission from Macmillan Publishers Ltd: Nature Physics,
  T. Schulz, R. Rita, A. Bauer, M. Halder, M. Wagner, C. Franz,
C. Pfleiderer, K. Everschor, M. Garst, and A. Rosch,
Nature Phys. {\bf 8}, 301 (2012), copyright 2012.
}
\label{fig:40}
\end{figure}

In the systems described in the preceding Sections, the equation of motion
is dominated by overdamped dynamics; however, other terms such as inertia or
Magnus terms can become relevant in certain instances.
Recently it was discovered that chiral magnets can support
skyrmion states, composed of magnetic excitations called skyrmions that
have particle-like properties and
that can form a triangular lattice similar to that observed for superconducting
vortices
\cite{214,215,216}.
Since this initial discovery, a growing number of compounds
have been identified that support skyrmions, 
including some at room
temperature \cite{216,217,218,218N}.
Skyrmions also hold promise for
a variety of applications
including magnetic storage, memory, and magnetic logic devices \cite{219}.
Skyrmions have many similarities to vortices in superconductors in that
they are both particle-like objects with repulsive interactions that favor
formation of a triangular 
lattice, as illustrated in figure~\ref{fig:39} \cite{220N}.
Additionally, both skyrmions and vortices
can be driven with an applied current 
and exhibit depinning transitions \cite{216,220,221,222,223}.
The velocity of superconducting vortices can be determined based on the
magnitude of the voltage response induced by their motion;
however, the skyrmion velocity produces no voltage drop and cannot be
measured so easily.
The motion of skyrmions does, however, produce changes in the measured Hall response,
and from these changes
it is possible to determine the depinning threshold and
construct a skyrmion velocity-force curve \cite{216,221,224}.
Skyrmion motion under an applied drive can also be
directly visualized using Lorentz microscopy \cite{222,225}. 
Figure~\ref{fig:40} shows an experimentally measured
effective skyrmion velocity-force curve
for several 
different temperatures, where the
$x$-axis has been scaled by the depinning threshold.
There is a clear finite depinning threshold along with a nonlinear feature near depinning, 
while at higher drives the velocity increases linearly with drive \cite{221}.

\begin{figure}
  \includegraphics[width=\columnwidth]{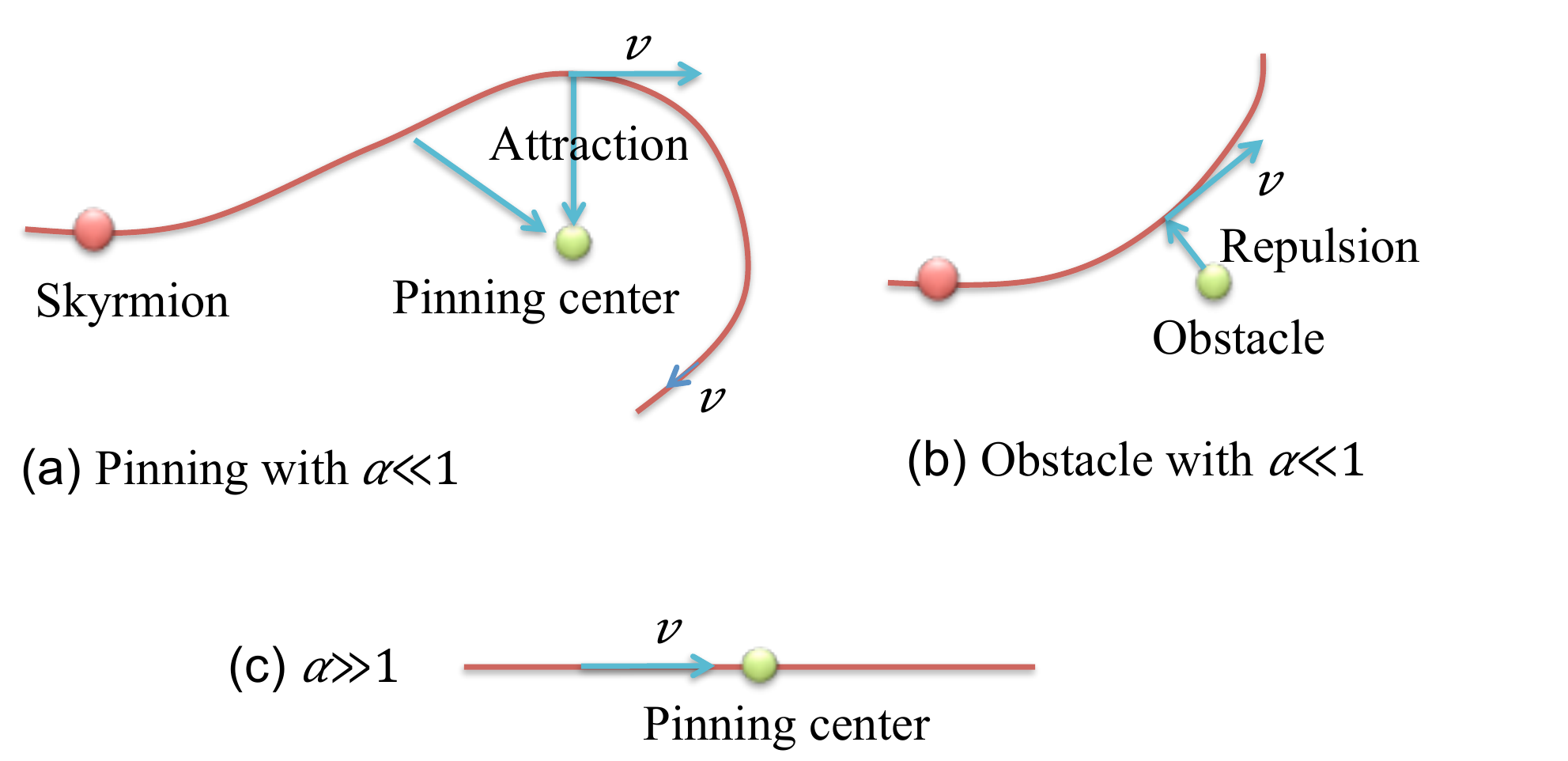}
\caption{
  (a,b) A schematic of a skyrmion system in which
  the Magnus term $\alpha_m$ dominates the dissipative term $\alpha_d$.  Here
  the ratio of the two terms is written as $\alpha=\alpha_d/\alpha_m$.
  The skyrmion is shown interacting
  with an attractive (a) and repulsive (b) pinning site.
  For the attractive pinning site,
  the Magnus term generates a velocity component 
  that is perpendicular to the attractive force of the pinning site, causing
  the skyrmion to spiral around the pinning site rather than becoming trapped.
  (c) In the overdamped limit with $\alpha \gg 1$, a particle interacting with an
  attractive pinning site travels directly to the center of the pinning site and
  becomes trapped.
  Reprinted with permission from S.-Z. Lin, C. Reichhardt, C.D. Batista, and A. Saxena,
Phys. Rev. B {\bf 87}, 214419 (2013). Copyright 2013 by the American Physical Society.
}
\label{fig:41}
\end{figure}

Due to the topological nature of skyrmions, their dynamics
is dominated by a Magnus term
which induces a velocity component that
is perpendicular to the net force acting on the skyrmion \cite{216,221,223,226,227}.
In simulation studies using a continuum model of skyrmions
based on a Landau-Lifshitz-Gilbert equation, it was argued that the
Magnus term causes skyrmions to
deflect around attractive pinning sites rather than falling into them,
producing a
reduction of the depinning threshold
compared to that observed in overdamped systems \cite{223}.
Lin {\it et al.} \cite{226}  introduced a 2D particle model for
skyrmions based on Theile's equation \cite{228}
that includes  skyrmion-skyrmion interactions and pinning effects. 
In this model, the dynamics of a skyrmion $i$
is governed by the following equation of motion: 
\begin{equation}
\alpha_{d}{\bf v}_{i} + \alpha_{m}{\hat z}\times{\bf v}_{i} = {\bf F}^{ss}_{i} + {\bf F}^{sp}_{i} + {\bf F}^{D} ,
\end{equation}
where ${\bf v}_{i}$ is the skyrmion velocity,
$\alpha_{d}$ is the dissipative term that aligns the skyrmion velocity 
with the net force,
${\alpha}_{m}$ is the Magnus term which rotates the
skyrmion velocity into the direction perpendicular to the net force,
${\bf F}_{i}^{ ss}$ is the
skyrmion-skyrmion repulsive interaction
of Bessel function form which favors triangular ordering, 
${\bf F}^{sp}_{i}$ is the pinning force, and ${\bf F}^{D}$ is the external
dc driving force.  
In most superconducting systems, $\alpha_{m}/\alpha_{d} \ll 1.0$; however, in
skyrmion systems
$\alpha_m/\alpha_d=10$ to $40$.
In most of the experimental studies performed so far, the skyrmions are
in the clean limit and form ordered triangular lattices.
More recent studies in thin films show evidence of stronger pinning that
permits a coexistence between pinned and moving skyrmions, allowing
plastic depinning to occur \cite{217,218N}.
The Magnus term can alter the depinning
transition in many ways.
For example, the Magnus term introduces a strong time dependence to the dynamics,
so that
a skyrmion entering a pinning site undergoes a spiraling motion,
illustrated schematically in figure~\ref{fig:41}(a).
This is in contrast to
the behavior of overdamped particles such as vortices, which simply travel directly
to the bottom of the pinning potential, as shown schematically in
figure~\ref{fig:41}(c) \cite{226}.
The Magnus term can also cause a skyrmion that passes through a pinning site to
undergo what is called a side jump,
in which skyrmion trajectories entering one half of the pinning site are more
strongly deflected than those entering the other half \cite{48,227}.

\begin{figure}
  \includegraphics[width=\columnwidth]{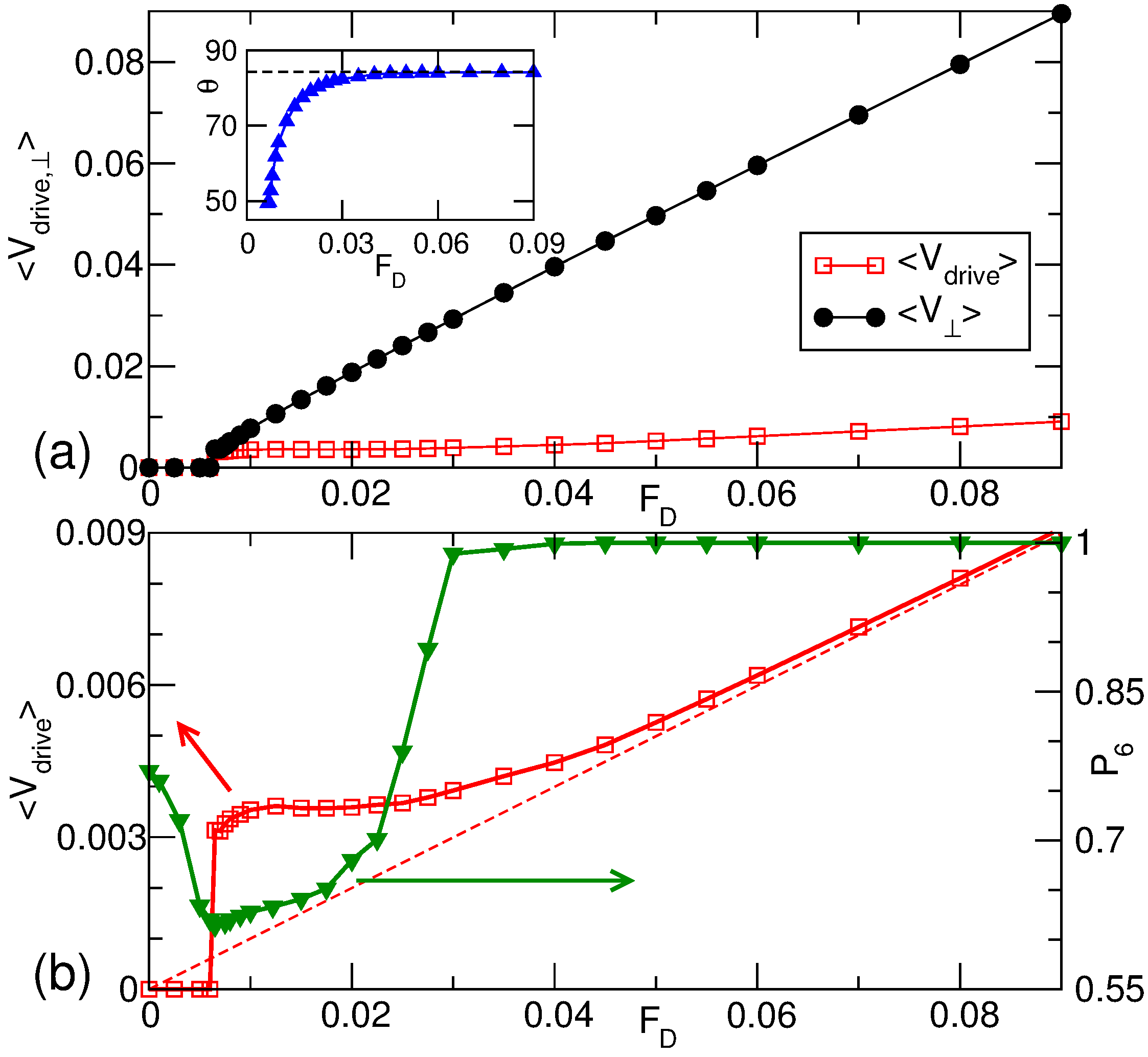}
\caption{
  Average skyrmion velocity in the direction of the drive
  $\langle V_{\rm drive}\rangle$ (open squares) and
  perpendicular
  to the drive $\langle V_{\perp}\rangle$
  (filled circles) vs driving force $F_{D}$ from
   2D particle-based simulations
   of skyrmions
   in the strong Magnus regime  $\alpha_{m}/\alpha_{d} = 10$
   interacting with randomly
   placed pinning sites.  
   The inset shows the Hall angle $\theta$ vs $F_{D}$;
   the dotted line indicates the value of the Hall angle in a clean, substrate-free system.
   (b) The corresponding $\langle V_{\rm drive}\rangle$
   (open squares) and the fraction of sixfold coordinated
   particles $P_{6}$ (filled triangles)
   vs $F_{D}$, where the dashed line indicates the results from a clean system.
   A dynamical reordering transition
   occurs at $F_{D} = 0.03$.
   Here $\langle V_{\rm drive}\rangle$ corresponds to the velocity $V$ or $V_x$ in
   the driving direction, and $\langle V_{\perp}\rangle$ corresponds to the
   velocity $V_y$ perpendicular to the driving direction.
  Adapted with permission from C. Reichhardt, D. Ray, and C.J. Olson Reichhardt,
Phys. Rev. Lett. {\bf 114}, 217202 (2015). Copyright 2015 by the American Physical Society.
}
\label{fig:42}
\end{figure}

The spiraling motion of 
skyrmions and the capture probabilities for different
types of pinning sites have been studied 
in various continuum and particle based models
\cite{227,229} and observed in experiments \cite{30}. 
Reichhardt {\it et al.}
\cite{48} conducted particle-based simulations of
skyrmions interacting with random pinning
under the Magnus-dominated condition $\alpha_{m}/\alpha_{d} = 10$. 
Figure~\ref{fig:42}(a)
shows the skyrmion velocity in the direction of drive, $V$ or $V_x$, 
and perpendicular to the drive,
$V_y$,
versus driving force $F_{D}$ in the strong pinning limit where the depinning is plastic. 
There is a
finite depinning threshold and $V_x \ll V_y$
due to the Magnus term,
which produces a Hall angle in the substrate-free limit of
$\theta = \tan^{-1} (\alpha_{m}/\alpha_{d}) = 84.25$, meaning that
the skyrmions move nearly perpendicular to the direction of the net
force acting on them.
The inset of figure~\ref{fig:42}(a)
shows the measured Hall angle
$\theta = \tan^{-1}(V_y/V_x)$ versus $F_{D}$,
where the dashed line indicates the value of $\theta$ in the clean limit.
The Hall angle has a strong drive dependence
due to the side jump effect
induced by the pinning, with the size of the side jumps
decreasing with increasing $F_{D}$.
The changing Hall angle indicates that the net direction of
the skyrmion flow is changing as a function of drive.
In figure~\ref{fig:42}(b),
the corresponding fraction of sixfold coordinated skyrmions $P_6$
versus $F_{D}$ indicates that
the system is the most disordered just above the depinning transition, and that a
dynamical ordering transition to a nearly defect-free state occurs at high drive,
as shown by the increase in  $P_{6}$ to  $P_6=1.0$ near $F_{D} = 0.3$. 
Figure~\ref{fig:42}(b) also contains a plot of $V_x$ versus $F_D$
where the dashed line indicates the behavior expected in the clean limit.
The depinning transition is discontinuous, and
the skyrmion velocities are {\it larger}
than the clean limit values due to an
overshoot or pinning-induced speed-up effect.
Such behavior never occurs in overdamped inertia-free systems.
Speed-up effects have also been observed for the depinning of single
skyrmions from pinning sites,
and are generated when the Magnus term
rotates the velocity component induced by the
pinning into the velocity component induced by the drive \cite{227,230}.

\begin{figure}
  \includegraphics[width=\columnwidth]{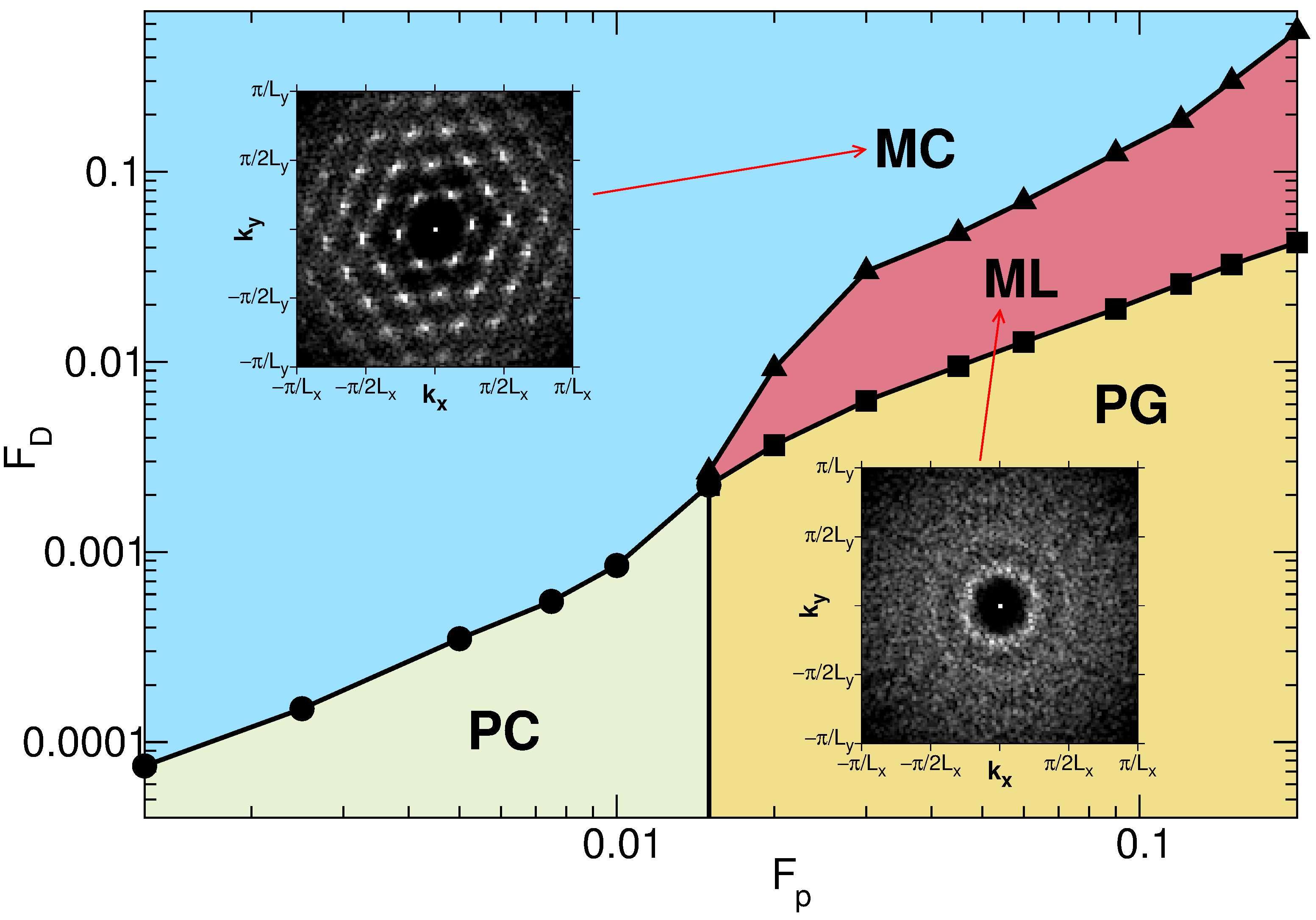}
\caption{
  Simulated dynamic phase diagram for the skyrmion system from figure~\ref{fig:42} as a
  function of driving force $F_D$ vs
  pinning force $F_p$.  PC: pinned crystal phase; PG: pinned glass phase;
  ML: moving liquid phase; MC: moving crystal phase.  Insets show the structure
  factor $S({\bf k})$ calculated in the moving crystal (upper left) and moving liquid
  (lower right) phases.
  Adapted with permission from C. Reichhardt, D. Ray, and C.J. Olson Reichhardt,
Phys. Rev. Lett. {\bf 114}, 217202 (2015). Copyright 2015 by the American Physical Society.
}
\label{fig:43}
\end{figure}

Figure~\ref{fig:43} illustrates the dynamical  phase diagram
as a function of driving force $F_{D}$ versus
pinning strength $F_{p}$ for the system in figure~\ref{fig:42}. 
At low $F_{p}$ and low $F_D$, there is a pinned skyrmion
crystal that depins elastically with increasing drive
into a moving skyrmion crystal.
As the pinning strength
increases, there is a critical value $F^c_{p}$ at which topological defects proliferate
in the pinned phase and the system depins plastically into a liquid like state, as indicated
by the plot of $S({\bf k})$ in the lower inset of figure~\ref{fig:42}.
At higher drives, dynamical reordering occurs
into a moving crystal phase free of dislocations,
as shown by the six-fold ordering of
$S({\bf k})$ plotted in the upper inset of figure~\ref{fig:42}.
A difference between the dynamically ordered skyrmion phases and the dynamically
ordered vortex phases is that the skyrmions
order into an isotropic moving crystal rather than
into a moving smectic state, since the
shaking temperature induced in the presence of a strong Magnus term
is distinct from the shaking temperature
produced in
the overdamped limit.
In the moving phase in superconducting vortex systems, the
pinning
generates vortex velocity fluctuations that are stronger in the direction of drive than
perpendicular to the drive,
while for skyrmion systems, the
Magnus term
causes the velocity fluctuations induced by the pinning to be more isotropic.

The study of the collective dynamics of skyrmions
is still in its infancy, and potentially represents an entirely 
new field of Magnus-dominated collective dynamical phases. 
Open questions include whether the Magnus
term changes the scaling behaviors at elastic or plastic depinning transitions
compared to the overdamped limit.
The plastic flow itself
could have different properties when the Magnus term dominates than 
in overdamped systems, and the velocity noise fluctuations
in the plastic flow and/or the moving ordered phases
could be different from those observed in overdamped systems.
In 3D systems,
skyrmions form line-like objects similar to vortex lines,
so it would also be interesting to study the 3D dynamical plastic
flow of skyrmions when the Magnus term is important.
Another direction of study is to consider
skyrmions moving through periodic pinning arrays,
which could be created with various nanostructuring techniques.
Simulations of individual
skyrmions moving
over 2D periodic \cite{231} and 1D periodic \cite{232} substrates
show threshold behavior, directional 
locking, and Shapiro steps that have features distinct from the overdamped case,
so it would interesting to explore 
the behavior of collectively interacting skyrmions
moving on periodic pinning arrays.
The Magnus term could affect 
thermal creep, the velocity-force curves \cite{233},
and  avalanche behaviors.
Some studies have provided evidence that inertial effects may 
be important
for skyrmion dynamics \cite{230},
and there is evidence that 3D point-like skyrmion objects could be stabilized in 
certain systems,
which would open up a new class
of the dynamics of 3D point-like objects \cite{124}.

\section{Jamming and Pinning}

\begin{figure}
  \includegraphics[width=\columnwidth]{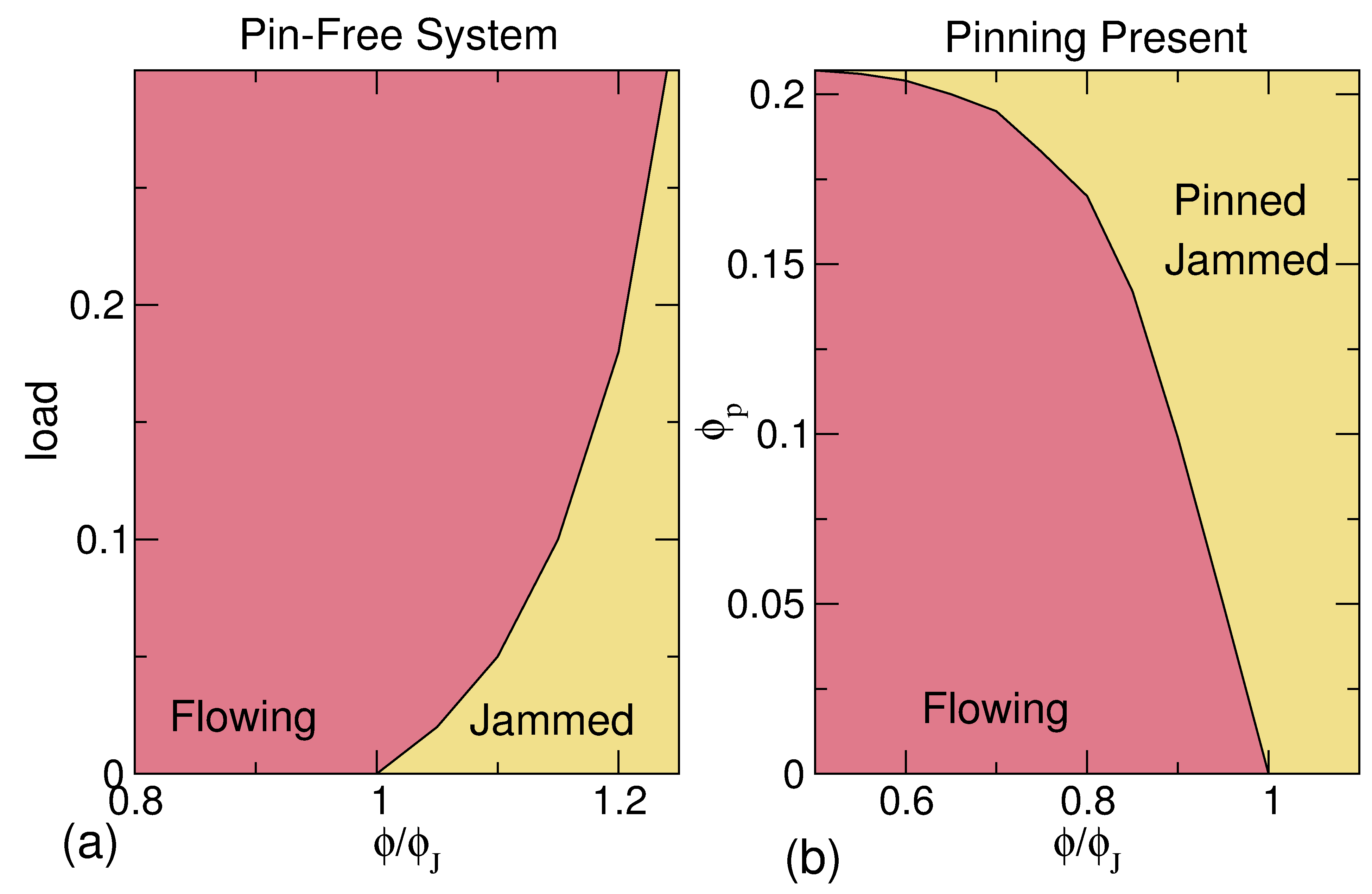}
\caption{ Schematic jamming phase diagrams.
  (a) In a pin-free system, as a function of load or shear vs particle
  density $\phi/\phi_J$,
  jamming occurs for sufficiently large load
and sufficiently high $\phi/\phi_J$, where $\phi_J$ is
the density called point J at which the system jams in the absence of load.
Jamming occurs only for $\phi/\phi_J \geq 1$.
(b) When 
pinning is present, the particle density 
at which jamming occurs under zero load drops below $\phi/\phi_J=1.0$ as
the pinning density $\phi_p$ increases.
}
\label{fig:44}
\end{figure}

The jamming of loose assemblies of particles with 
only contact interactions, 
such as grains, bubbles, or emulsions, 
has received 
considerable attention
over the last 15 years 
since Liu and Nagel proposed a jamming phase diagram in which,
at high enough particle density, the particles 
come into contact and the system  transitions from a fluid-like state
to a rigid jammed state
\cite{234,235,236,237}.
The density at which the system begins to support
a finite shear is called point J.
An appealing feature of this picture
is that the physics that describes point J may also describe 
other phenomena such as the onset of yielding under load
or the onset of rigidity as the temperature is lowered. 
One difference between jamming and depinning is that the jamming 
transition is controlled solely by the particle-particle interactions
and does not involve quenched disorder.
Jamming is typically discussed 
in the context of assemblies of particle-like objects that
interact with each other through a short range steric repulsion, making it possible
to  define a unique particle density at which all the particles have
just barely come into contact with each other.
In contrast,
in most
of the systems in which depinning transitions have been studied,
even the short-range particle-particle interactions that have been
considered are longer range than steric,
so that in the absence of pinning the system can act
like a crystal or solid down to very low particle densities.
Although it is possible for systems containing particles which do not
interact with each other to exhibit a depinning transition, such systems
cannot exhibit a jamming transition, which is a truly collective effect.
These points suggest that depinning and jamming transitions represent
quite distinct phenomena; however, there are many cases in which
depinning transitions exhibit properties similar to those associated with
a jamming transition.
For example, in regimes in which there are fewer pinning sites than particles,
the entire system can remain pinned even when a driving force is applied
due to particle-particle interactions that confine and trap the
interstitial particles which are
not directly trapped by pinning sites.  This resembles a localized jamming
effect in which the interstitial particles
between the pinning sites act like a solid.
In systems where the particle-particle
interactions are short-ranged and have a well-defined cutoff distance,
if a random substrate potential is introduced, both jamming and pinning can occur.
Examples of such systems include
sterically stabilized colloids, bubbles, or grains moving over a
rough landscape, and in these systems the interplay between jamming behavior
and depinning transitions can be explored.

In figure~\ref{fig:44} we illustrate
an example of
how jamming and pinning can be connected. 
One of the pairs of axes on the Liu-Nagel phase diagram is particle
density $\phi/\phi_J$ versus external load \cite{234,236}, 
as shown schematically in figure~\ref{fig:44}(a) for a pin-free system. 
Below the critical jamming density $\phi_{J}$,
the system is in a liquidlike state and 
can flow under any applied load, while for
$\phi/\phi_{J} > 1.0$, a finite  load must be applied before the system
begins to flow, 
and the magnitude of this load increases
with increasing $\phi/\phi_J$.
Adding pinning to the jamming system produces the phase diagram
shown schematically in
figure~\ref{fig:44}(b)
as a
function of pinning site density $\phi_p$ versus particle
density $\phi/\phi_J$ at zero load.
When $\phi_p=0$, there are no pins and the system jams only
when $\phi/\phi_J \geq 1.0$; however, for finite pinning density,
the onset of jamming drops to particle densities
$\phi/\phi_J<1.0$,
and a state emerges that is both jammed and pinned.
This suggests that a new axis could be added to the jamming
phase diagram proposed by Liu and Nagel showing the interplay between
jamming and pinning.

\begin{figure}
  \includegraphics[width=\columnwidth]{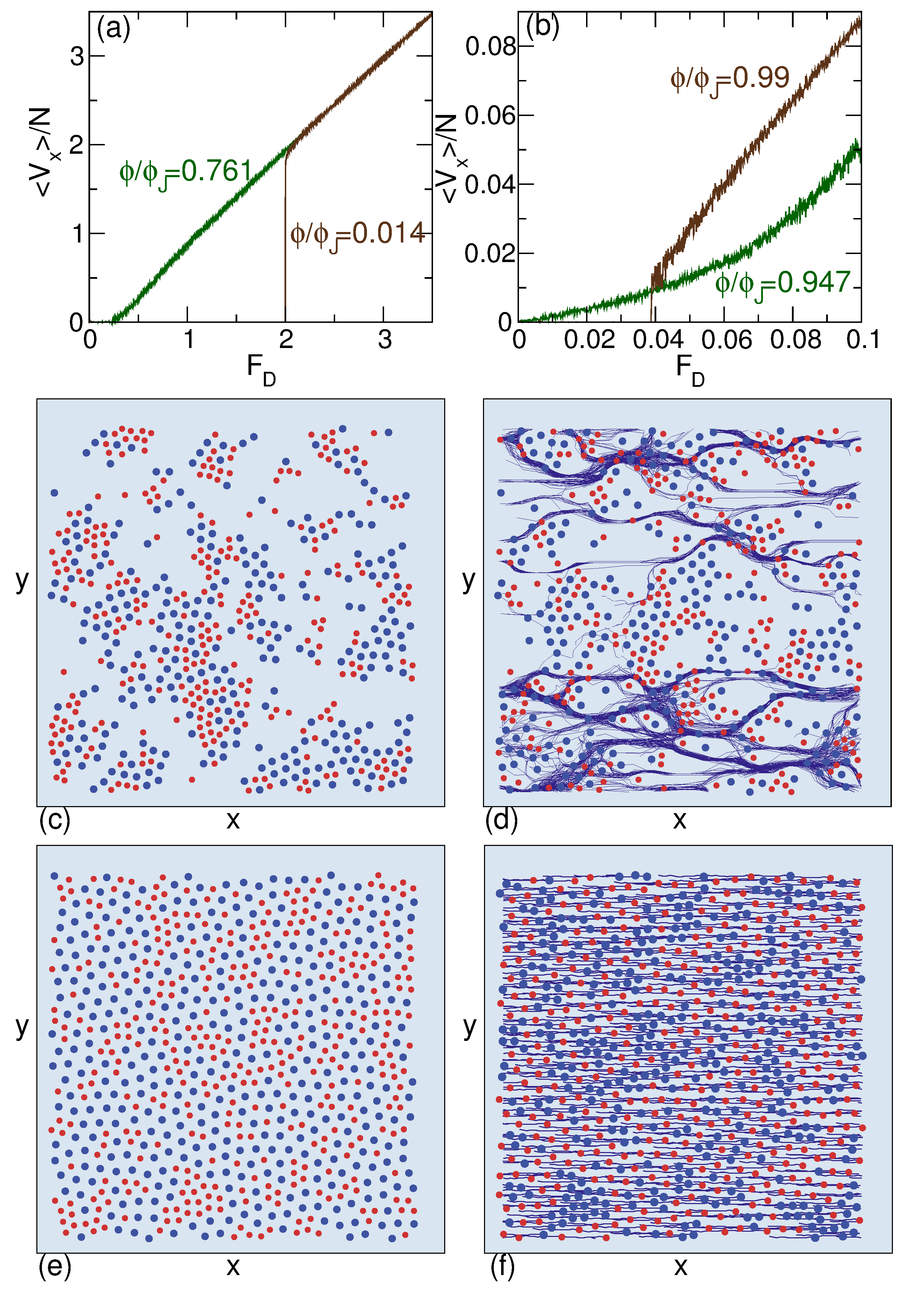}
\caption{
(a,b) Velocity-force curves from simulations of a 
  system of binary hard disks at an area coverage
  or disk density of $\phi/\phi_J$,
  where $\phi_{J}$ is the pin-free jamming density.
  Here $\langle V_x\rangle$ corresponds to the velocity $V$ and $F_D$ is the driving force.
  (a) Samples with a pinning density of $\phi_p/\phi_J=0.415$.
  At a disk density of $\phi/\phi_{J} = 0.014$ (brown),
  there are more pins than disks and the system is in the single particle
pinning limit, while at $\phi/\phi_{J} = 0.761$ (green)
a pinned jammed configuration appears.
(b) Samples with sparse pinning at a density of $\phi_p/\phi_J=0.09$.
There is no depinning threshold
at $\phi/\phi_{J} = 0.947$ (green),
while at $\phi/\phi_{J} = 0.99$ (brown)  a finite depinning
threshold appears.
(c,d) The pinned (c) and flowing (d) configurations for the system in 
panel (a) at $\phi/\phi_{J} = 0.761$, showing
plastic depinning. 
The two sizes of disks are indicated by two sizes and colors of dots, while
lines show the disk trajectories in the flowing state.
(e,f) The pinned (e) and flowing (f) configurations for the system in 
panel (b) at $\phi/\phi_{J} = 0.99$, showing elastic depinning.
  Adapted with permission from C.J. Olson Reichhardt, E. Groopman, Z. Nussinov, and
  C. Reichhardt,
Phys. Rev. E {\bf 86}, 061301 (2012). Copyright 2012 by the American Physical Society.
}
\label{fig:45}
\end{figure}

Reichhardt {\it et al.} considered a model of a 
bidisperse assembly of 2D hard disks interacting with
a random pinning array and subjected to a uniform driving force \cite{238}.
This system has a
well-defined pin-free jamming density corresponding to
an areal disk coverage of
$\phi=\phi_{J} =  0.844$.  
If only a single pinning site is present in the system,
the depinning threshold is finite
only when $\phi/\phi_J \geq 1$ and the system jams into a solid state
that can be held in place by a single obstacle.
Depending on the strength of the single pinning site, the depinning
transition
may be either plastic or elastic.
At disk densities $\phi/\phi_J < 1$, 
the system may or may not exhibit a finite depinning threshold
depending on the ratio of the number of pinning sites to 
the number of disks.
Figure~\ref{fig:45}(a) shows the disk velocity versus 
applied drive for a sample in which the pinning strength is
$F_{p} = 2.0$ and the pinning density is $\phi_p/\phi_J=0.415$.
At a disk density of $\phi/\phi_J=0.014$, each disk can be
pinned independently since
there are many more pinning sites than disks,
and the depinning threshold is $F_c = F_p$.
At a higher disk density of $\phi/\phi_J=0.761$,
the depinning threshold remains finite even though there are more
disks than pinning sites, indicating that the disks trapped directly
at the pinning sites are able to block or jam the flow of the interstitial
disks and prevent them from moving.
Figure~\ref{fig:45}(c) illustrates
the pinned state for the $\phi/\phi_J=0.761$ sample from 
figure~\ref{fig:45}(a), showing that the pinned state is strongly clustered.
The local disk density is equal to $\phi_J$ within each cluster and
there are large disk-free regions between clusters.
Under an applied drive,
the system depins plastically
as illustrated in figure~\ref{fig:45}(d). 
If the density of pinning sites is reduced, 
the depinning threshold vanishes but the velocity-force curve
can still show nonlinearity, as shown 
in figure~\ref{fig:45}(b)
for samples in which there are only a small number of pinning sites present,
$\phi_{p}/\phi_{J} = 0.09$.
Here, the depinning threshold is zero
when the disk density $\phi/\phi_{J} = 0.947$ 
but 
the velocity-force curve is nonlinear, indicating that plastic
flow is occurring.
In contrast, at $\phi/\phi_{J} = 0.99$, 
the depinning threshold is finite
and the system forms a uniform pinned solid,
shown in figure~\ref{fig:45}(e), which depins elastically into
the uniform moving state illustrated
in figure~\ref{fig:45}(f). 
These results show that elastic jammed solids can in some 
cases show a finite depinning threshold, while liquidlike states
do not.

\begin{figure}
  \includegraphics[width=0.8\columnwidth]{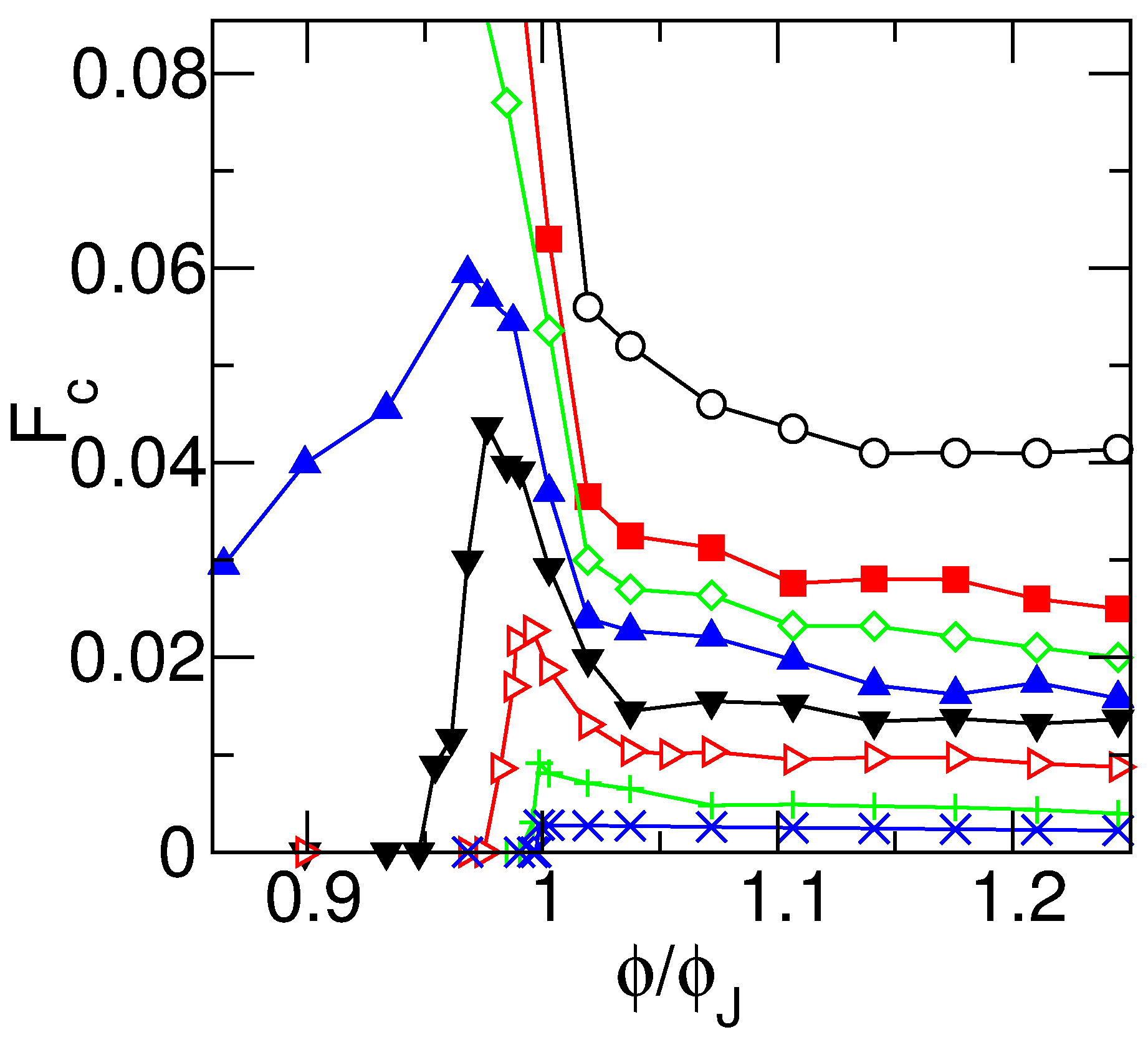}
  \caption{
    The depinning threshold $F_{c}$ vs
    disk density $\phi/\phi_J$ for
    simulations of the system in Figure~\ref{fig:45}
at pinning densities of
$\phi_{p}/\phi_{J} = 0.828$, 0.415, 0.277, 0.138, 0.09267, 0.0346, 0.00692, and 
$0.00138$,
from top to bottom.
For $\phi/\phi_J$ below the peak in $F_c$, the disks depin
plastically, while for $\phi/\phi_J$ above the peak in $F_c$, the disks depin elastically.
  Adapted with permission from C.J. Olson Reichhardt, E. Groopman, Z. Nussinov, and
  C. Reichhardt,
Phys. Rev. E {\bf 86}, 061301 (2012). Copyright 2012 by the American Physical Society.
}
\label{fig:46}
\end{figure}

Figure~\ref{fig:46} shows the evolution of the depinning 
threshold $F_{c}$ as a function of $\phi/\phi_{J}$ for varied
pinning densities ranging from 
$\phi_{p}/\phi_{J} =  0.828$ to $\phi_p/\phi_J=0.00138$.
For the low pinning densities, the depinning threshold $F_{c} = 0$. 
There is a peak in $F_{c}$ at $\phi^*$ just below
$\phi/\phi_{J} = 1.0$.
For disk densities $\phi/\phi_J<\phi^*$, the disks depin plasticity, 
while for $\phi/\phi_{J} > \phi^*$, the disks
depin elastically.
For $\phi_{p}/\phi_{J} > 0.3$, the depinning threshold is always finite, and
$F_c$ is
much higher
at particle densities 
$\phi/\phi_J<1$ below the jamming density
than
for $\phi/\phi_J>1$, above the jamming density.
Other numerical studies also show
that the introduction of a small number of pinning sites 
can reduce the disk density at which jamming occurs \cite{239}.

Relatively little work has been done on connecting jamming and depinning, leaving
the field wide open with a range of topics for further study.
Since critical properties are associated with point J, it would be
interesting to understand whether the pinning-induced jamming that occurs
at densities below point J preserves these critical properties or
destroys them.
Another open question is whether dynamical ordering transitions are
possible in pinned jamming systems when the jamming density has been
depressed below point J, since in principle the system could organize
to a state in which the particles do not interact with each other.

\section{Dynamics of Driven Systems with Competing Interactions}

In the systems discussed in the preceding Sections,
the  particle-particle interactions are purely repulsive,
so that in monodisperse 2D samples in
the clean limit, the ground states have triangular ordering.
A natural question to ask is 
whether
collections of particles with different 
types of clean limit ground states can exhibit similar
nonequilibrium dynamical phases, or if 
entirely new types of phases can arise. 
In one such class of system, the interactions between the particles includes
a competition between a long-range repulsion and a short-range attraction.
In the absence of quenched disorder, systems of this type are known to
form a rich variety of 
crystalline, bubble, stripe, void crystal, 
and densely packed triangular states \cite{240}. 
In soft matter systems, patterns of this type can arise in 
colloidal assemblies, phase separating mixtures, and binary fluids
\cite{240}. 
In hard condensed matter systems, pattern 
formation is relevant for stripes in quantum Hall systems \cite{241}, 
charge ordering in cuperates \cite{242,243}, 
pattern formation in systems with Jahn-Teller interactions \cite{244},
flux states in type-I superconductors \cite{245}, and vortices in 
in multiband superconductors \cite{246,247,248,249,250}. 
In quantum Hall systems, the existence of stripe crystals 
instead of Wigner crystals has been 
proposed to explain the strongly anisotropic transport found 
at certain magnetic fields \cite{251}, as well as 
narrow band noise fluctuations \cite{253} and the hysteresis 
in certain transport measurements \cite{252}. 

\begin{figure}
  \includegraphics[width=\columnwidth]{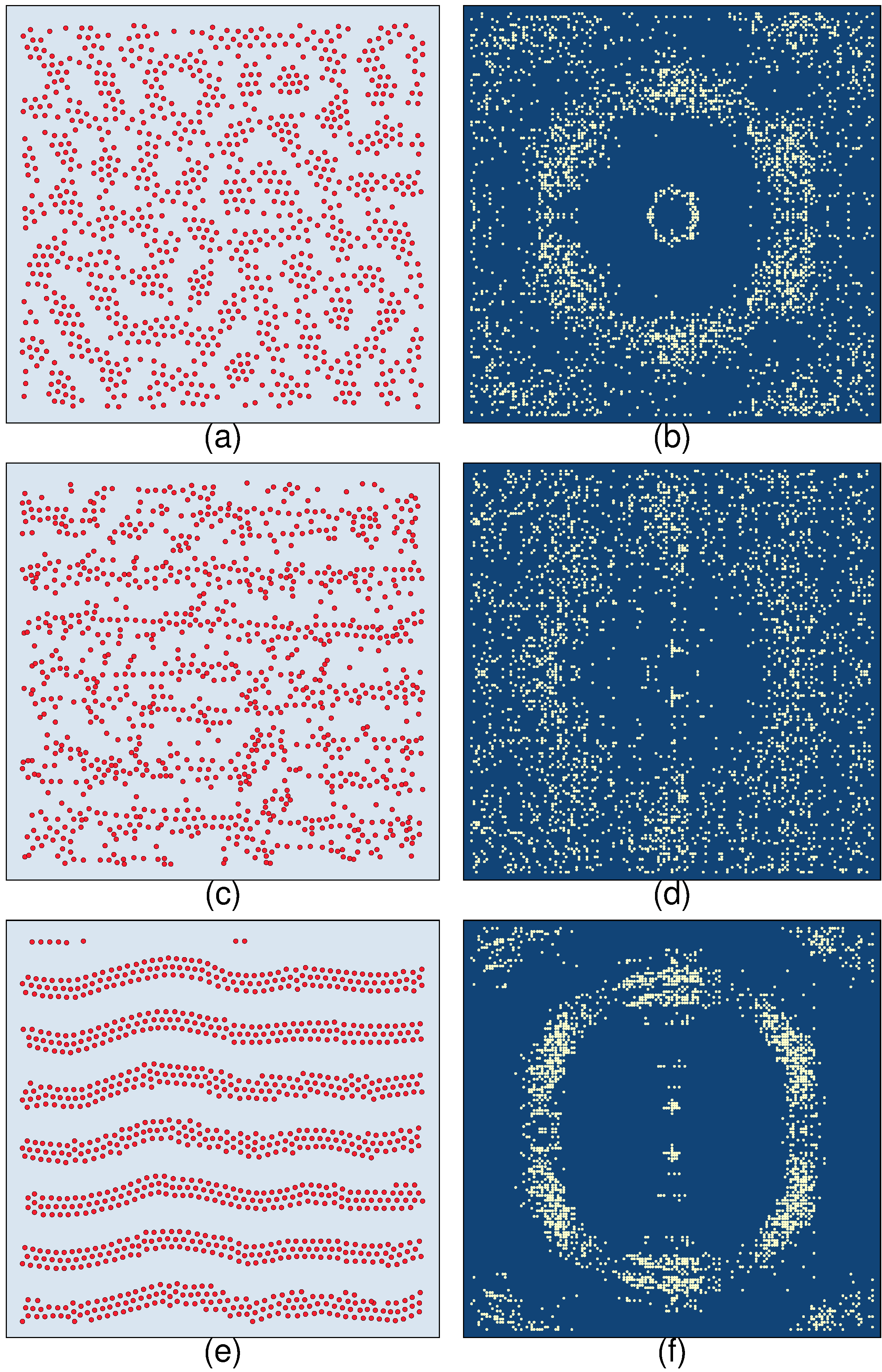}
  \caption{
    Simulations of a pattern-forming system in which the particles have a
    competing long-range repulsion and short-range attraction and are
    driven over a random pinning array.
(a,c,e) The particle positions (dots) and 
(b,d,f) the
    corresponding structure factor $S({\bf k})$.
(a,b) In the pinned state the particles form disordered clumps. 
(c,d) Just above depinning, the structure is further disordered. 
(e,f) The dynamically reordered stripe phase at high drives. 
  Adapted with permission from C. Reichhardt, C.J. Olson Reichhardt, I. Martin, and
  A.R. Bishop,
Phys. Rev. Lett. {\bf 90}, 026401 (2003). Copyright 2003 by the American Physical Society.
}
\label{fig:47}
\end{figure}

\subsection{Pairwise competing interactions}

As an example of a pattern forming system,
Reichhardt {\it et al.} \cite{8,254} considered a model in which
the particle-particle interaction contains
a long-range Coulomb repulsion combined with a short-range attractive term.
The interaction energy
for particles located a distance $R_{ij}$ apart has the following form: 
\begin{equation} 
U(R_{ij}) = \frac{1}{R_{ij}}  - e^{-BR_{ij}} .
\end{equation}
A clump crystal forms at low particle densities, with
the particles forming a triangular lattice within each clump 
and the clumps themselves
forming a larger scale triangular
lattice due to the long-range repulsive interaction term.
At intermediate particle densities, a stripe phase
forms, and a void crystal appears at higher particle densities \cite{254}.
If weak pinning is added to the stripe phase,
the stripe ordering can be preserved.
In the presence of strong pinning, however,
the stripes break up and the system forms
an amorphous phase that still exhibits two length scales \cite{8}, as
illustrated in 
figure~\ref{fig:47}(a).
The corresponding structure factor $S({\bf k})$ in figure~\ref{fig:47}(b) has
two rings since
disordered structures appear at
the two characteristic length scales of
the distance between
particles within individual clumps
and the average spacing between clumps.
If a driving force is applied along the positive $x$ direction, 
there is a finite depinning threshold.
Just above depinning, the system
becomes even more disordered
as shown in figure~\ref{fig:47}(c), and 
some particles form 1D  
channels along the drive direction, producing faint smectic-like peaks
in $S({\bf k})$ in figure~\ref{fig:47}(d).
At higher drives, the system 
dynamically orders into a moving stripe phase that resembles
the static stripe state in a sample without disorder, except that
the stripes are
aligned with the driving direction 
as illustrated in figure~\ref{fig:47}(e,f).

\begin{figure}
  \includegraphics[width=\columnwidth]{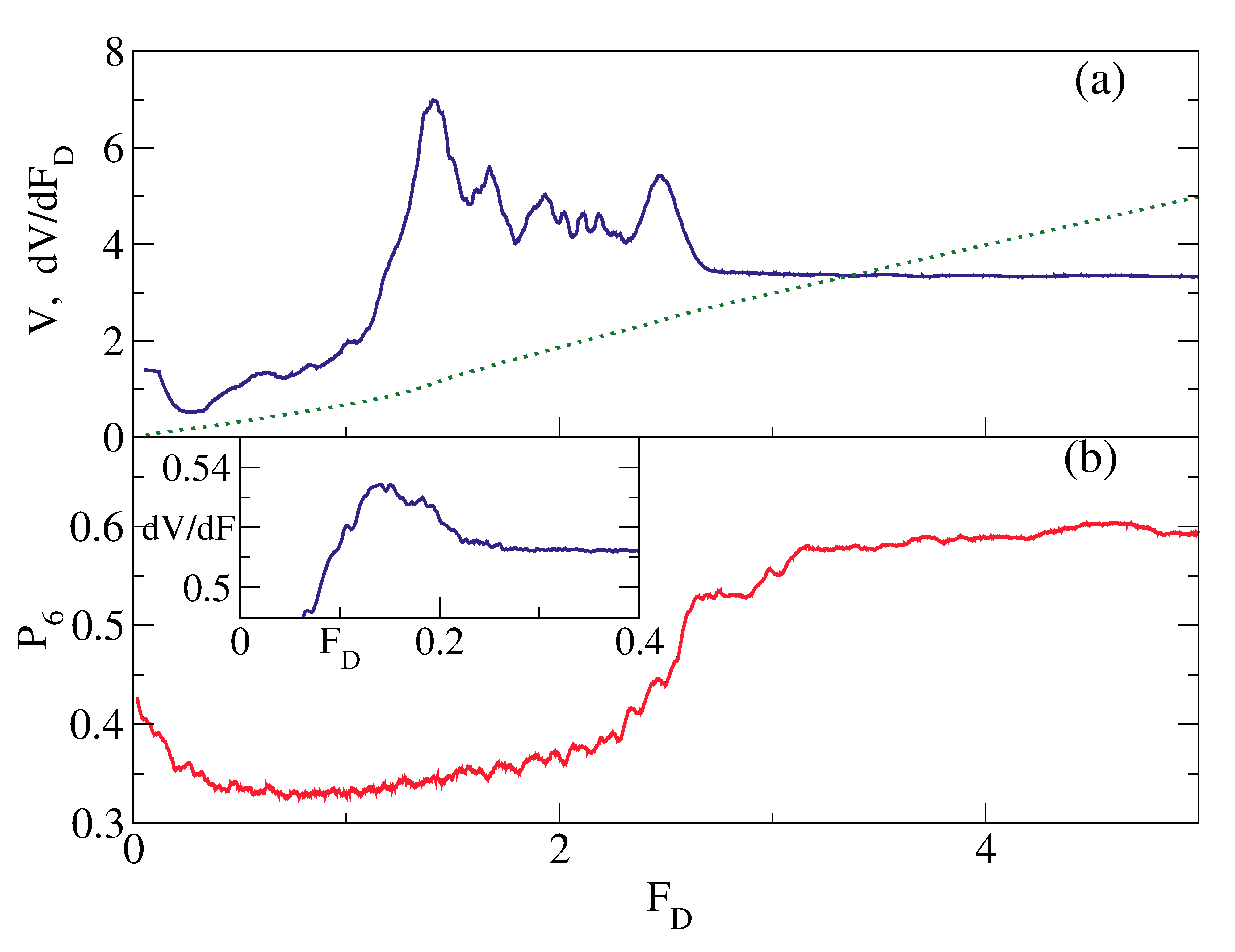}
\caption{
  (a) Velocity-force curve (dashed line) for
  simulations of the stripe forming system in 
figure~\ref{fig:47}                
and the corresponding 
$dV/dF_D$ curve (solid line), which shows a two peak feature.
(b) The corresponding fraction of six-fold particles $P_6$
vs $F_D$.  Inset: $dV/dF_D$ vs $F_D$ for a sample with weak pinning has
only a single peak.
  Adapted with permission from C. Reichhardt, C.J. Olson Reichhardt, I. Martin, and
  A.R. Bishop,
Phys. Rev. Lett. {\bf 90}, 026401 (2003). Copyright 2003 by the American Physical Society.
}
\label{fig:48}
\end{figure}

Figure~\ref{fig:48}(a) shows
the velocity-force curve for the stripe state on a strong random substrate
along with $dV/dF_D$.
There are two peaks in $dV/dF_{D}$ 
rather than the single peak typically found for
systems with purely repulsive particle-particle interactions. 
The double peak feature arises due to the two 
length scales in the system. 
At the first peak
in $dV/dF_D$, there is a dynamical transition to clump ordering,
while at the second peak in $dV/dF_D$, the system dynamically
orders into the aligned stripe state illustrated in figure~\ref{fig:47}(e).
A plot of 
the fraction of six-fold coordinated particles 
$P_6$ versus $F_{D}$
in figure~\ref{fig:48}(b)
shows that $P_6$ begins to increase at the first peak 
in $dV/dF_D$ but remains at a low value since,
even though clumps are beginning to form, the particles 
within each clump are still disordered. 
Near the second peak  in $dV/dF_D$, there is
a jump up in $P_6$ that occurs when the stripes form, since
the particles within the stripes have
a considerable amount
of sixfold ordering, as illustrated in figure~\ref{fig:47}(e,f). 
The inset of figure~\ref{fig:48}(b) shows that when the pinning is weak, 
there is only one peak in $dV/dF_D$. 
At this weaker pinning, the stripes do not break up when placed on the
disordered substrate; however,
some plastic flow occurs within the stripes at drives above depinning,
while at high enough dives the particles within the stripes order. 
These results show that in 
systems with multiple length scales, 
multiple dynamical ordering transitions can occur at
well-separated drives.

\begin{figure}
  \includegraphics[width=\columnwidth]{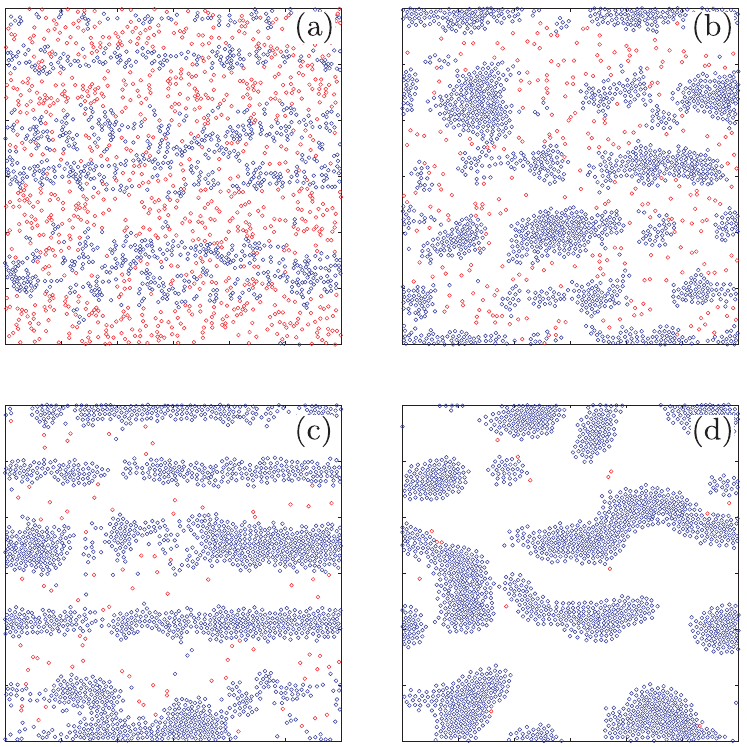}
\caption{
  The particle positions for
  simulations of a system in which the particles
  interact with a short-range repulsion and a long-range attraction
  and are driven over a random pinning array. Red 
  particles are pinned and blue particles are moving.
  The driving force increases from (a) to (d).
(a) The disordered flow phase just above depinning.  (b) Depinned clusters. 
(c) Moving stripes. (d) Cluster coarsening regime.
Reprinted with permission from H.J. Zhao, V.R. Misko, and F.M. Peeters,
Phys. Rev. E {\bf 88}, 022914 (2013). Copyright 2013 by the American Physical Society.
}
\label{fig:49}
\end{figure}

Pattern forming systems exhibit
strong hysteresis in the velocity-force curves, with the
reordered stripe state persisting on the decreasing drive sweep
down to drives that are much lower than the drive at which
dynamical reordering into the stripe state occurs for an increasing drive sweep.
Further studies show that
hysteresis also occurs as a function of particle density 
for systems of driven clump crystals, stripes, and void crystals 
whenever two length scales come into play \cite{255}. 
At very low densities where the particle-particle interactions are
dominated by the long-range repulsive term,
there is no hysteresis, and 
similarly at very high densities where the system forms a 
dense triangular lattice that is also dominated by the long-range repulsive 
term,
there is no hysteresis. 
Simulations also show that anisotropic transport in the 
stripe phase appears only at low pinning strengths
for which the stripes remain ordered above the depinning
transition \cite{256}. 
These results suggest that the appearance of hysteresis 
in the quantum Hall systems could be due to the formation of
stripe or bubble states rather than Wigner crystals.  
Other simulations
of colloidal systems with competing short range attraction and 
longer range repulsion
show dynamical ordering transitions to aligned stripe phases 
\cite{257}.  
Experiments on 2DEGs 
in regimes where stripes are believed to form 
show evidence of dynamical ordering of 
stripes along the direction of drive 
associated with a memory of the drive direction 
in the dynamically ordered stripe state \cite{258}. 

\begin{figure}
  \includegraphics[width=\columnwidth]{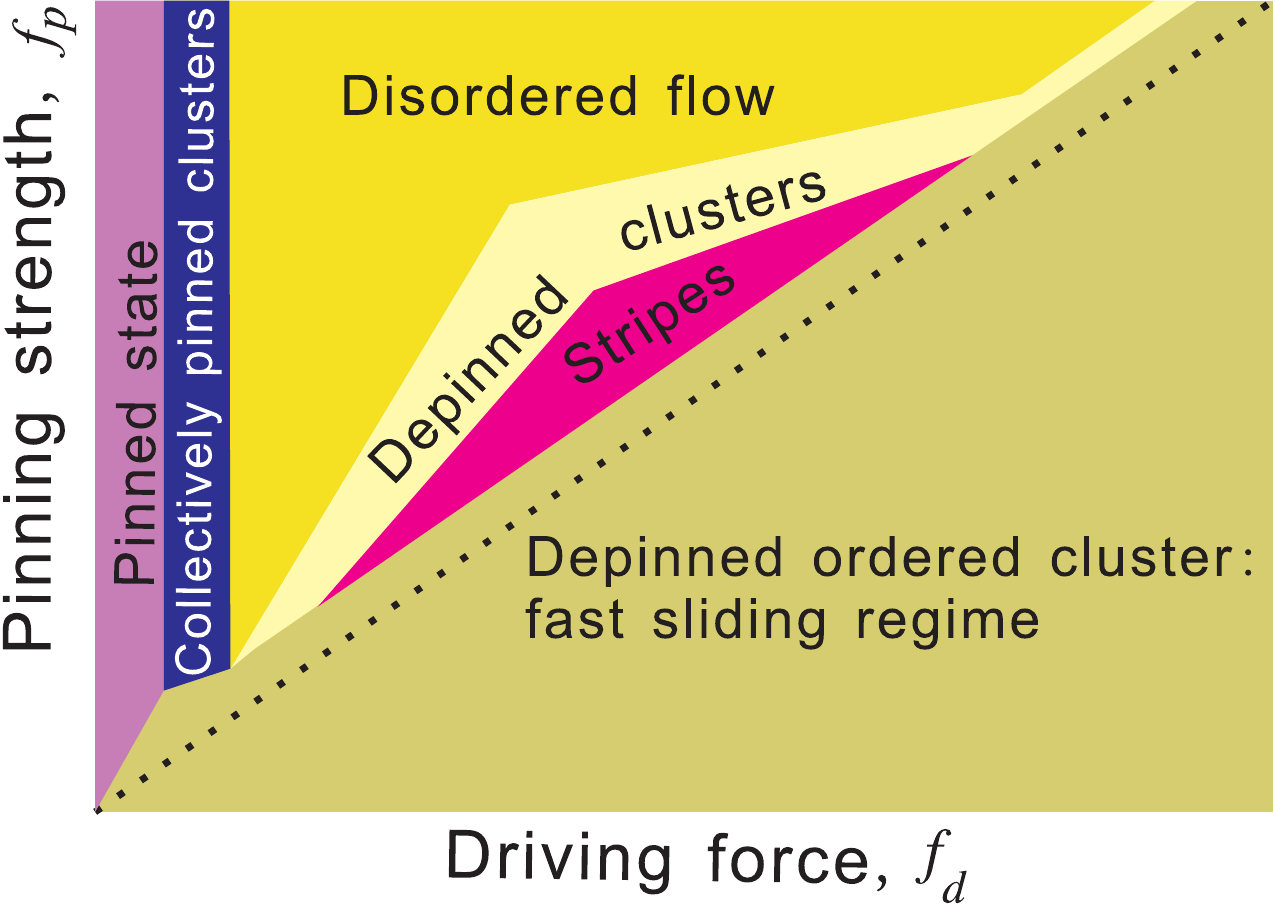}
\caption{
  Schematic dynamic phase diagram for the
  pattern forming system in 
  figure~\ref{fig:49} as a function of pinning strength
  $f_p$ vs driving force $f_d$ showing
a pinned phase, collectively pinned clusters, 
a disordered phase, depinned clusters, stripes, 
and a rapidly sliding cluster regime.
Here $f_p$ corresponds to the pinning strength $F_p$ and $f_d$ corresponds
to the driving force $F_D$.
Reprinted with permission from H.J. Zhao, V.R. Misko, and F.M. Peeters,
Phys. Rev. E {\bf 88}, 022914 (2013). Copyright 2013 by the American Physical Society.
}
\label{fig:50}
\end{figure}

Zhao {\it et al.} \cite{259} studied the dynamical 
flow and ordering in pattern forming systems where the particles have a 
short range repulsion and a longer range attraction. 
In the absence of pinning, the system forms clump or labyrinth states.
For strong disorder, a uniform disordered state 
appears as shown in figure~\ref{fig:49}(a).
Under an applied drive, the particle flow just above depinning is plastic and
occurs in a riverlike pattern. 
At higher drives, as illustrated in 
figure~\ref{fig:49}(b), there is a regime in which
depinned clumps flow through a background of pinned particles, 
while at even higher drives, the clumps dynamically align into a
stripe-like state as shown in figure~\ref{fig:49}(c).
At the highest drives, these clumps coarsen as
shown in figure~\ref{fig:49}(d). 
A schematic dynamic phase diagram for this system appears in
figure~\ref{fig:50}  as a function of pinning force $F_p$ versus applied drive
$F_D$.
The dashed line denotes the drive at which
the clumps 
dynamically order. 
For weak pinning, the pinned clumps depin elastically 
into a moving clump phase; 
however, for stronger pinning, the system passes through a
pinned phase, a disordered plastic phase, a depinned cluster phase,
a stripe phase, and finally a sliding phase as a function of 
increasing drive.
The transport curves for this system have two peaks in $dV/dF_D$ as well
as hysteresis.
Other studies of systems with
long range attraction and short range repulsion 
also reveal a similar set of dynamical phases as well as
the double peak feature in 
the $dV/dF_D$ curves, and show that when periodic rather than random pinning is 
introduced, the stripe phases 
generally become more ordered \cite{260}. 
These results underscore the fact
that the double peak feature 
in $dV/dF_D$ is associated with multiple ordering 
transitions. The appearance of hysteresis is also similar to 
that observed in 3D layered superconducting vortex systems, where
the layering provides the additional length 
scale required to generate the hysteresis. 
Simulation studies of a 2D system with long-range repulsion 
and short range attractions also reveal
a peak effect similar to that observed in the 3D vortex systems
\cite{ourstripe}. 

\begin{figure}
  \includegraphics[width=\columnwidth]{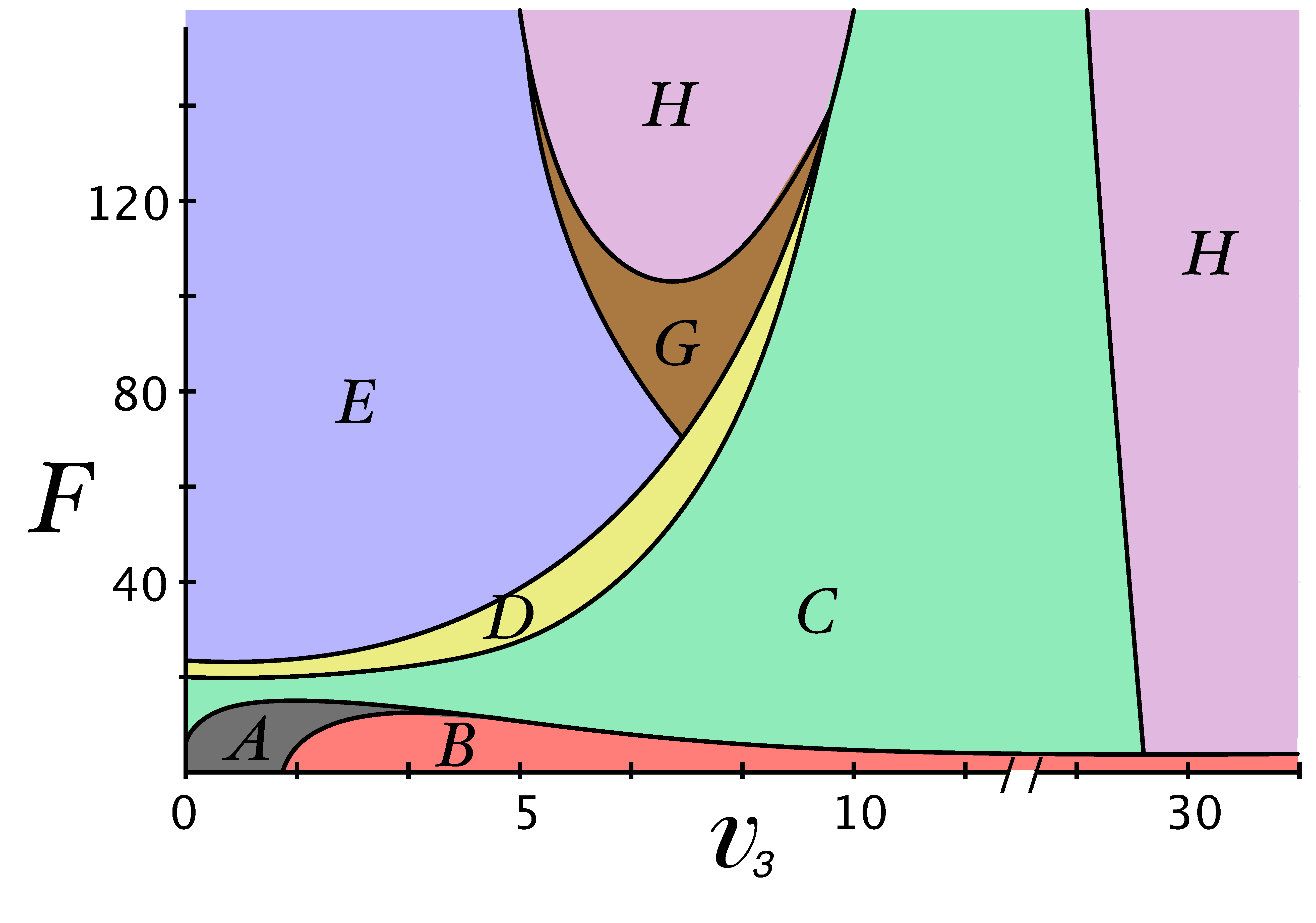}
\caption{
Schematic dynamic phase diagram 
as a function of driving force F vs 
the strength of the three-body interaction term
$v_3$ for a system with a two-body repulsive
force and a three-body interaction term that favors square ordering. 
A: pinned triangular lattice, B: pinned square lattice, 
C: plastic flow, D: anisotropic triangular lattice, 
E: moving triangular lattice, H: moving square lattice, 
G: coexistence between moving square and moving triangular lattice.
Here $F$ corresponds to the driving force $F_D$.
Reprinted with permission from A. Sengupta, S. Sengupta, and G.I. Menon,
Phys. Rev. B {\bf 81}, 144521 (2010). Copyright 2010 by the American Physical Society.
}
\label{fig:51}
\end{figure}

Stripes and other types of patterns can also form 
in systems with purely repulsive interactions provided that there are 
multiple length scales in the interaction potential. 
In general, such patterns occur whenever the Fourier transform of the 
particle-particle interaction potential 
has a negative peak in ${\bf k}$-space \cite{261}, 
so that even a simple two-step repulsive shoulder interaction potential
can produce a remarkable number of 
patterned mesophases in 2D \cite{262,263}. 
There is also evidence that purely repulsive interactions with two length scales
can produce stripelike vortex patterns 
in certain types of superconductors \cite{264}. 
Up until now,
studies of pattern forming systems driven over 
random substrates have involved systems in which
the competing repulsive and attractive interactions are 
explicit in the real space form
of the particle-particle interaction potential.
It would be interesting to 
study whether systems with purely repulsive interactions that have
two or multiple length scales
would also exhibit the same general dynamical 
ordering and hysteresis. 
The ordering and dynamics 
of pattern forming systems on periodic substrates, about which little is known,
would
be another interesting direction for future study.

\subsection{Non-pairwise competing interactions} 

\begin{figure}
  \includegraphics[width=\columnwidth]{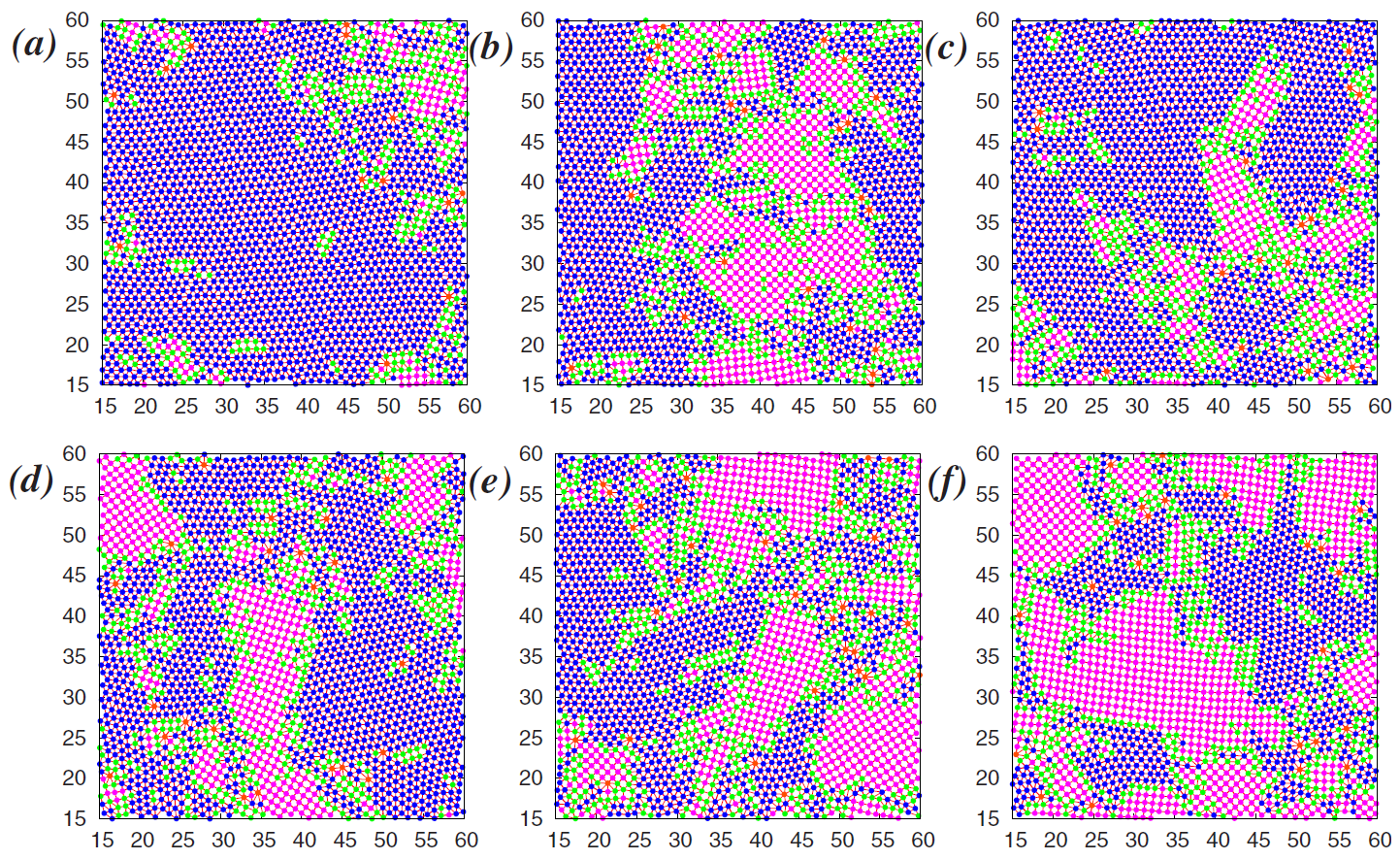}
  \caption{
   Snapshots of
the dynamical evolution of the moving triangular lattice phase 
(phase E)
into the coexistence regime (phase G) 
for the three-body competing interaction
system in figure~\ref{fig:51} for increasing driving forces of
 $F =$ (a) 90, (b) 95, (c) 100, (d) 105, (e) 110, and (f) 115. 
The particles are colored
according  to their coordination number $z_i$ with
$z_i = 4$ (magenta), 5 (green),  6 (blue) and 7 (orange).
Here $F$ corresponds to the driving force $F_D$.
Reprinted with permission from A. Sengupta, S. Sengupta, and G.I. Menon,
Phys. Rev. B {\bf 81}, 144521 (2010). Copyright 2010 by the American Physical Society.
}
\label{fig:52}
\end{figure}

Non-pairwise interactions or three-body effects can also lead to the formation
of patterns
\cite{265}.
Interactions of this type
can arise for particles coupled 
to a deformable substrate or
particles that are hydrodynamically coupled.
Sengupta {\it et al.} \cite{266,267} 
considered a system of repulsively interacting particles 
containing an additional three-body interaction term 
with fourfold symmetry that favors a square lattice ground state,
so that a 
transition from a triangular lattice to a square lattice 
can occur as a function of the strength $v_3$ of the three-body term. 
A schematic dynamic phase diagram
in figure~\ref{fig:51} 
of this system 
in the presence of pinning
as a function of driving force $F_D$ versus
$v_3$ shows that a rich variety of dynamical 
flow phases occur, including
a pinned triangular lattice at small $v_3$, 
a pinned square lattice at higher $v_3$, a plastic flow phase, 
an anisotropic hexatic phase, a moving triangular lattice, 
a coexistence regime, and a flowing square lattice
regime. 
Figure~\ref{fig:52} shows the evolution of the triangular lattice 
into a regime of coexisting square and triangular lattices
as a function of increasing drive.
The system is initially mostly triangular, but regions of square ordering
form and grow as the drive increases.

\subsection{Phase field models}

\begin{figure}
  \includegraphics[width=\columnwidth]{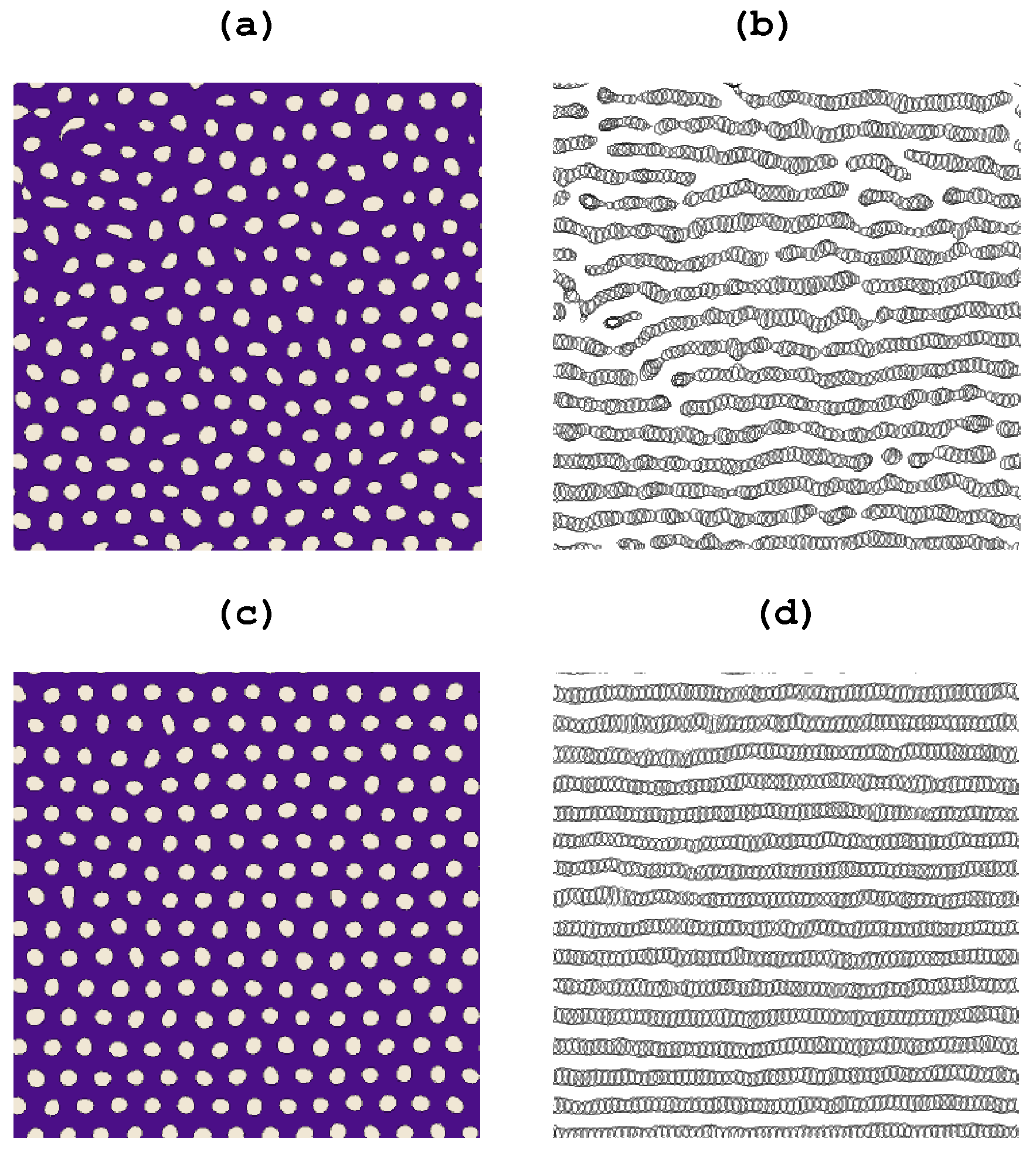}
  \caption{
    Phase field simulation studies of a system that forms a triangular
    lattice in the clean ground state driven over random disorder.
(a,c) Snapshots of the phase field 
$\Psi({\bf x})$  and (b,d) the trajectories of the peaks in
$\Psi({\bf x})$ in the moving state.
(a,b) The moving liquidlike state at a drive of $f=0.2$.
    (c,d) The moving smectic state at a drive of $f=0.8$.
    Here $f$ corresponds to the driving force $F_D$.
  Reprinted with permission from E. Granato, J.A.P. Ramos, C.V. Achim, J. Lehikoinen,
S.C. Ying, T. Ala-Nissila, and K.R. Elder,
Phys. Rev. E {\bf 84}, 031102 (2011). Copyright 2011 by the American Physical Society.
}
\label{fig:53}
\end{figure}

\begin{figure}
  \includegraphics[width=\columnwidth]{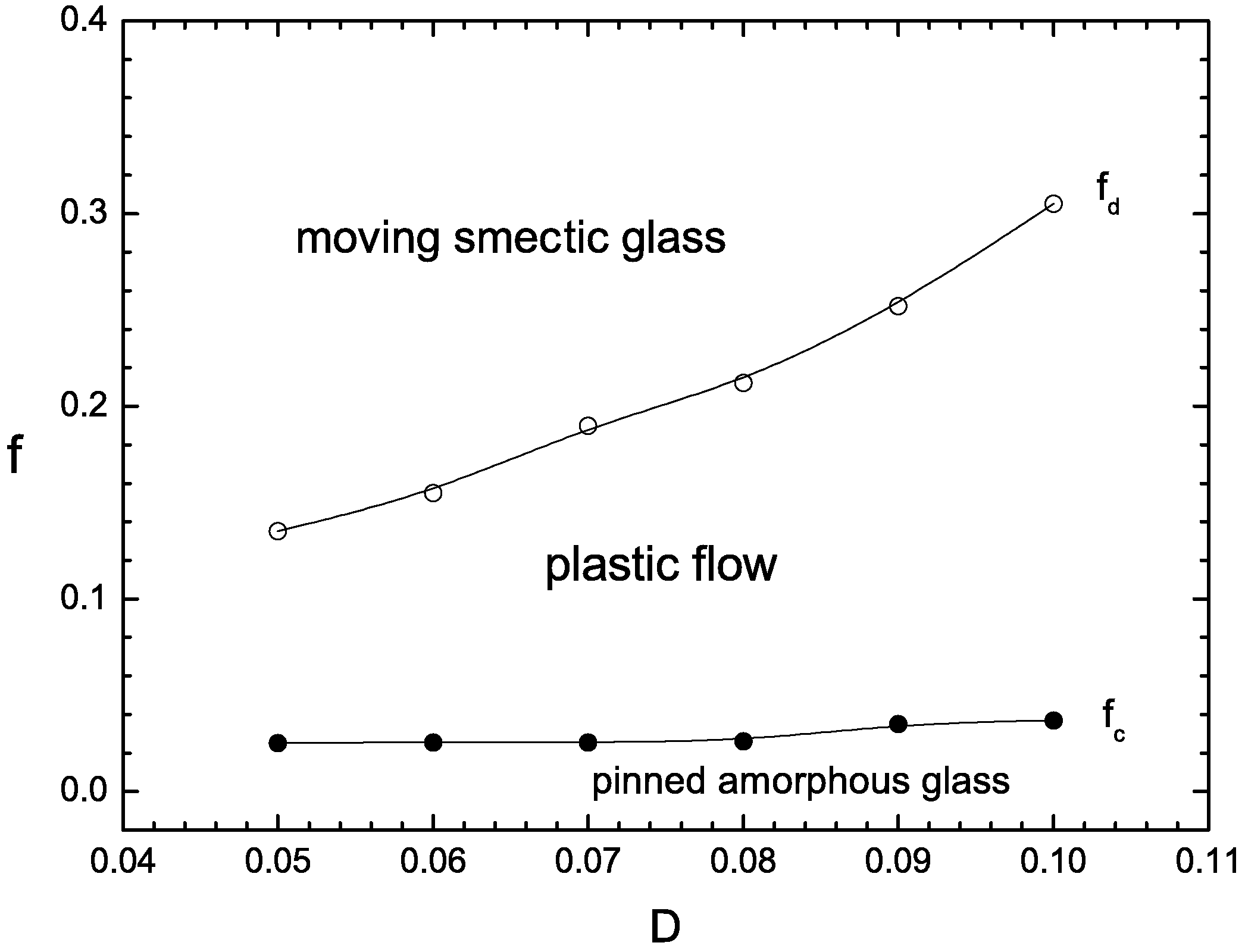}
\caption{
The simulated dynamic phase diagram for the phase field model 
from figure~\ref{fig:53} as a function of 
external drive $f$ versus disorder strength $D$
showing a pinned phase, plastic flow phase, and moving smectic phase.
Here $f$ corresponds to the driving force $F_D$ and $D$ corresponds
to the pinning strength $F_p$.
Solid circles are the depinning threshold $f_c$ which corresponds to $F_c$, while
open circles indicate the dynamical reordering transition.
  Reprinted with permission from E. Granato, J.A.P. Ramos, C.V. Achim, J. Lehikoinen,
S.C. Ying, T. Ala-Nissila, and K.R. Elder,
Phys. Rev. E {\bf 84}, 031102 (2011). Copyright 2011 by the American Physical Society.
}
\label{fig:54}
\end{figure}

Pattern forming systems can also be studied using phase field models
\cite{268}, 
which describe the particle density $\Psi({\bf x})$ on a diffuse time 
scale rather than a real microscopic time scale. 
Such methods are computationally efficient and have been used to study
longitudinal and transverse depinning  as well as the
scaling of velocity-force curves for pattern forming
systems on periodic substrates \cite{269}.
Granato {\it et al.} \cite{270}
used a phase field model to study a system that forms a triangular
lattice in the clean ground state.
Nonlinear velocity-force curves appear when the system is
driven over random disorder, 
along with a transition from a pinned state 
to the plastic flow phase shown in 
figure~\ref{fig:53}(a,b) and a transition at higher drives to
the smectic flow state
illustrated in figure~\ref{fig:53}(c,d).
The dynamical phase diagram
in figure~\ref{fig:54} highlights the different flow phases
as a function of driving force $F_D$ versus pinning strength $F_p$.
Within the moving smectic phase
Granato {\it et al.} observe a finite transverse
depinning threshold; however, this threshold decreases with increasing
system size, suggesting that it may vanish in the thermodynamic limit.
It should be possible to use similar phase field models for
systems in which the ground state consists of stripes or other patterns
to study 2D and 3D pattern forming systems 
driven over periodic or random substrates.

\subsection{Driven binary systems} 

\begin{figure}
  \includegraphics[width=\columnwidth]{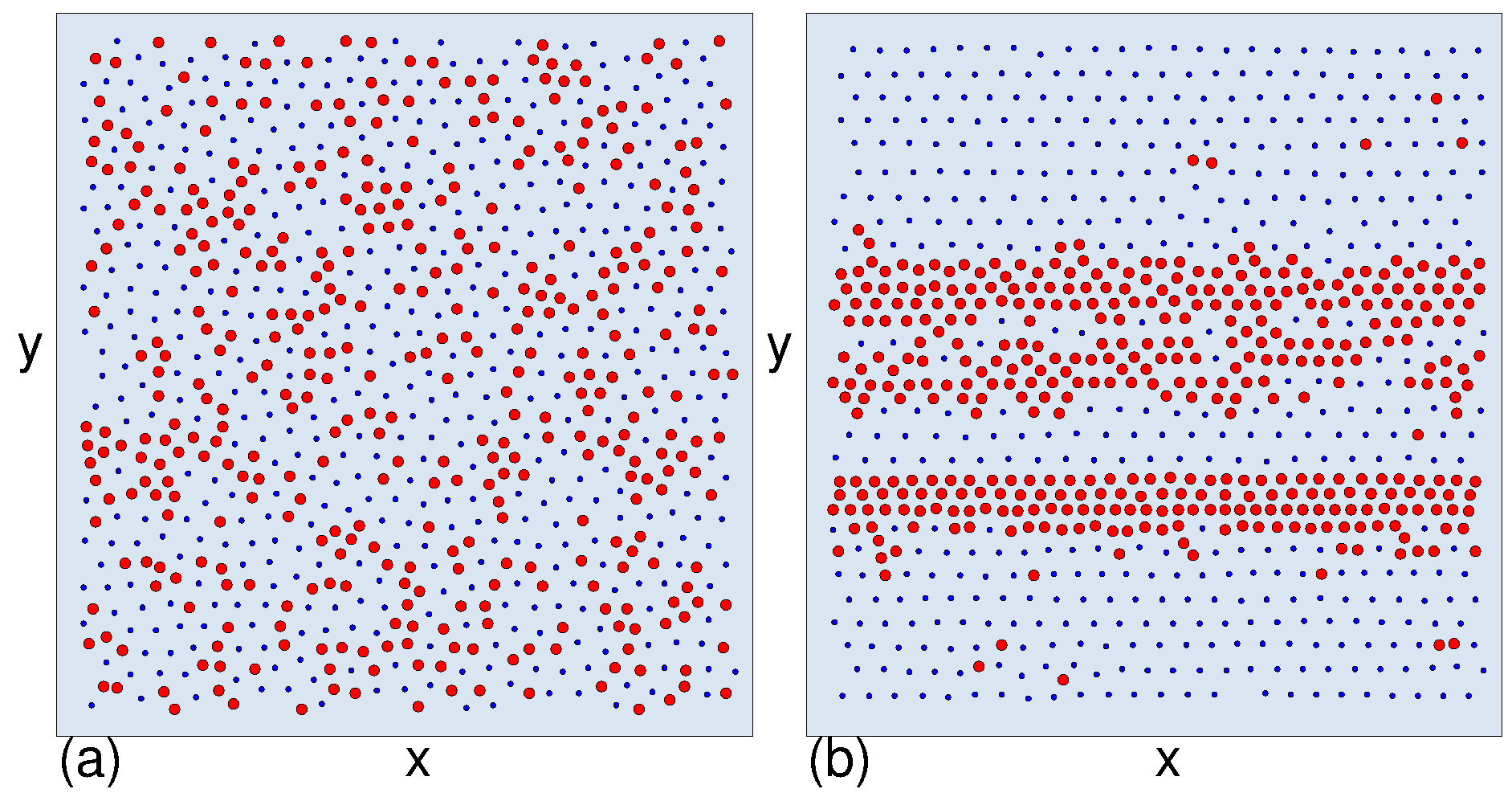}
\caption{ Simulation of a binary assembly of Yukawa interacting particles driven over
  random disorder.  The two different particle species have different effective
  charges. The smaller red particles are more strongly charged and the larger
  blue particles are more weakly charged.
  (a) The pinned mixed phase.
  (b) The strongly driven dynamically ordered phase forms
  stripelike domains, with triangular ordering of the particles inside each domain.
  Adapted with permission from C. Reichhardt and C.J. Olson Reichhardt, 
Phys. Rev. E {\bf 75}, 040402(R) (2007). Copyright 2007 by the American Physical Society.
}
\label{fig:55}
\end{figure}

In addition to systems with competing interactions, 
there are also examples of pattern forming binary systems
that can demix over time. 
It would be interesting to study the dynamics of such 
systems when they are driven over random disorder. 
For example, in the case of binary colloidal systems,
2D simulations of Yukawa interacting particles that have different effective charges
show that in the pinned state, the two species of particles
are mixed and form a disordered structure, 
but that at higher drives, the system can order into
a segregated  state \cite{271}. 
Figure~\ref{fig:55}(a) shows an example of the pinned
mixed phase for this system, while in figure~\ref{fig:55}(b),
for strong driving the system dynamically orders and phase separates into stripes of
triangularly ordered particles.

\section{Absorbing Phase Transitions and Reversibility At Depinning}

The nature of plastic depinning is still an open question, and although there is 
considerable evidence to suggest that in many cases plastic depinning 
has critical properties, there is still a lack
of knowledge of the associated universal exponents or even
the identity of the proper order parameters \cite{41}. 
A wide range of critical scaling exponents have been reported
for elastic depinning,
so it is not even clear whether there is a single unique behavior associated with the
elastic regime \cite{272}.
Both elastic and plastic depinning have been characterized using the statistics of
avalanches that occur near depinning, where
critical behavior is associated with a power law distribution of avalanche sizes,
\begin{equation}
P(s) \propto s^{-\alpha} .
\end{equation}
In many cases, however, it is difficult to identify the exact critical points
or depinning thresholds,
and experiments and simulations examining avalanche behavior at depinning
have produced a wide variety of exponents, and, in some cases, no exponent at all
when the avalanche sizes are not power-law distributed.
The form of the avalanche size distribution depends strongly on
the ratio of the number of pinning sites to the number of particles,
the temperature, the dissipation mechanism,
and how the system is driven  \cite{273,274,275,276}.
Other methods that have been used to characterize the nature of plastic depinning
include a Horton river structure analysis of the changes in the flow channel
morphologies 
\cite{277,278} and changes in the fractal dimension of the system  \cite{279,280}.

Recently a new method to identify and characterize nonequilibrium
phase transitions was developed in
studies of  periodically sheared 
dilute colloidal systems  where thermal effects are negligible \cite{281}.
A stroboscopic technique reveals
that as a function of shear amplitude or colloid density, there is a  transition
from a reversible flow where the particles return to the
same position at the end of each shear cycle 
to an irreversible state where the particles no longer return to the
same position and undergo anisotropic diffusion.
Additional experiments and computational studies show
that the system always begins in an irreversible state, but
over time it settles into either
a reversible state or a steadily fluctuating irreversible state.
The time scale $\tau$ to reach either state 
diverges at the transition point
as a power law with
$\tau \propto |\sigma - \sigma_{c}|^{-\nu}$,
where $ \sigma$ is the strain amplitude,
$\nu= 1.33$ in 2D, and $\nu=1.1$ in 3D  \cite{282}.
The reversible state is termed
a  random organized  state since the particle positions appear to be random
even though the particles undergo no net diffusion.
This system 
exhibits characteristics consistent with a
nonequilibrium absorbing phase transition, in which
a
driven system becomes trapped or absorbed in
a non-fluctuating or non-chaotic state 
\cite{25}.

\begin{figure}
  \includegraphics[width=\columnwidth]{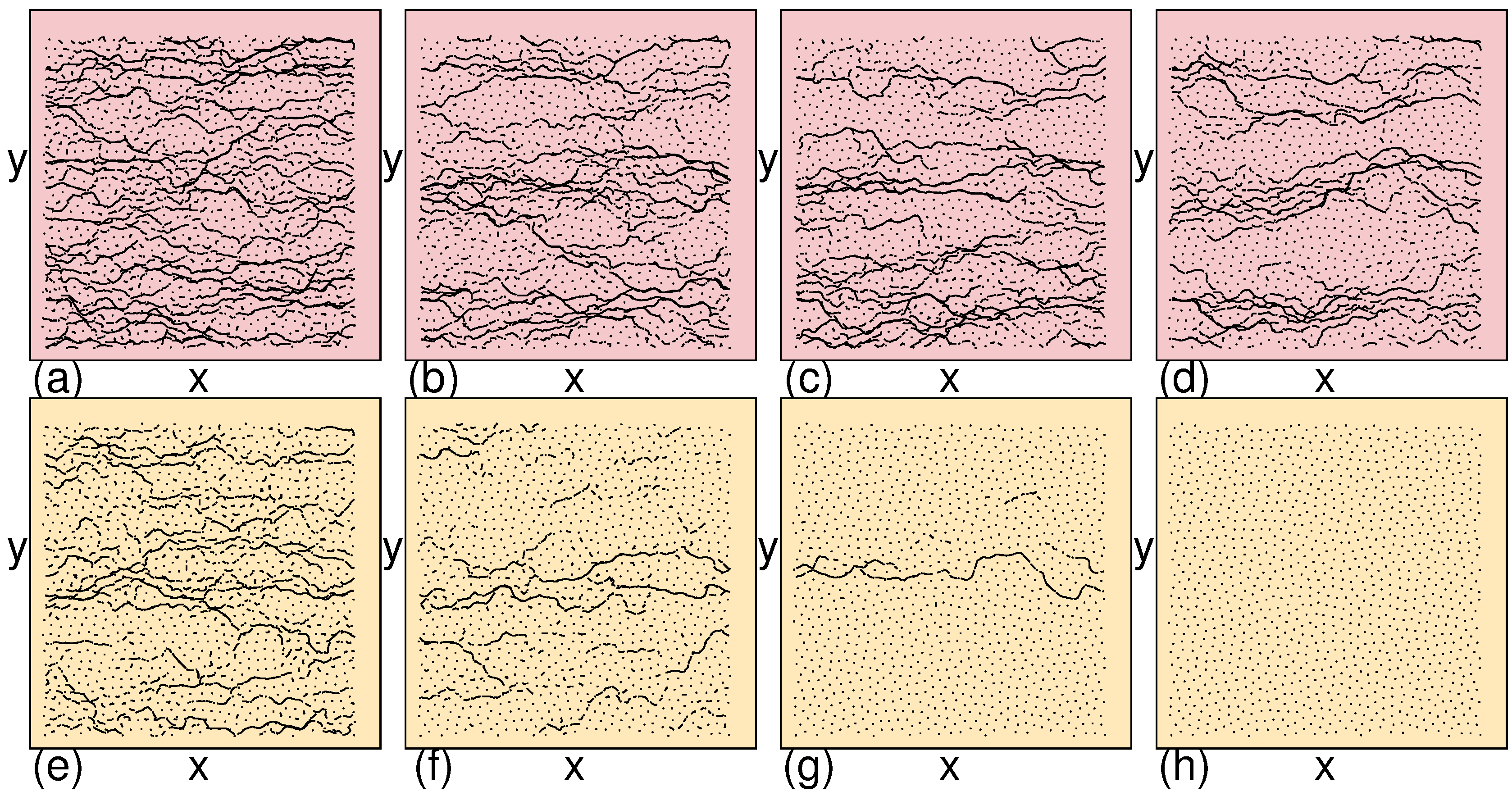}
\caption{
  Colloid positions (dots) and trajectories (lines)
  for simulations of
  a system with random pinning subjected to the sudden application of a driving force
  of fixed direction and fixed magnitude $F_D$.
  (a-d) Time evolution of the particle trajectories 
  as the system settles into a steady fluctuating state for a pinning force of
$F_{p}/F^{c}_{p} = 0.93$, where $F^{c}_{p}$ is the pinning force at the critical point. 
  (e-h) The same for a sample with $F_{p}/F^{c}_{p} = 1.05$,
  which settles  into a pinned state.  
  Adapted with permission from C. Reichhardt and C.J. Olson Reichhardt, 
Phys. Rev. Lett. {\bf 103}, 168301 (2009). Copyright 2009 by the American Physical Society.
}
\label{fig:56}
\end{figure}

\begin{figure}
  \includegraphics[width=\columnwidth]{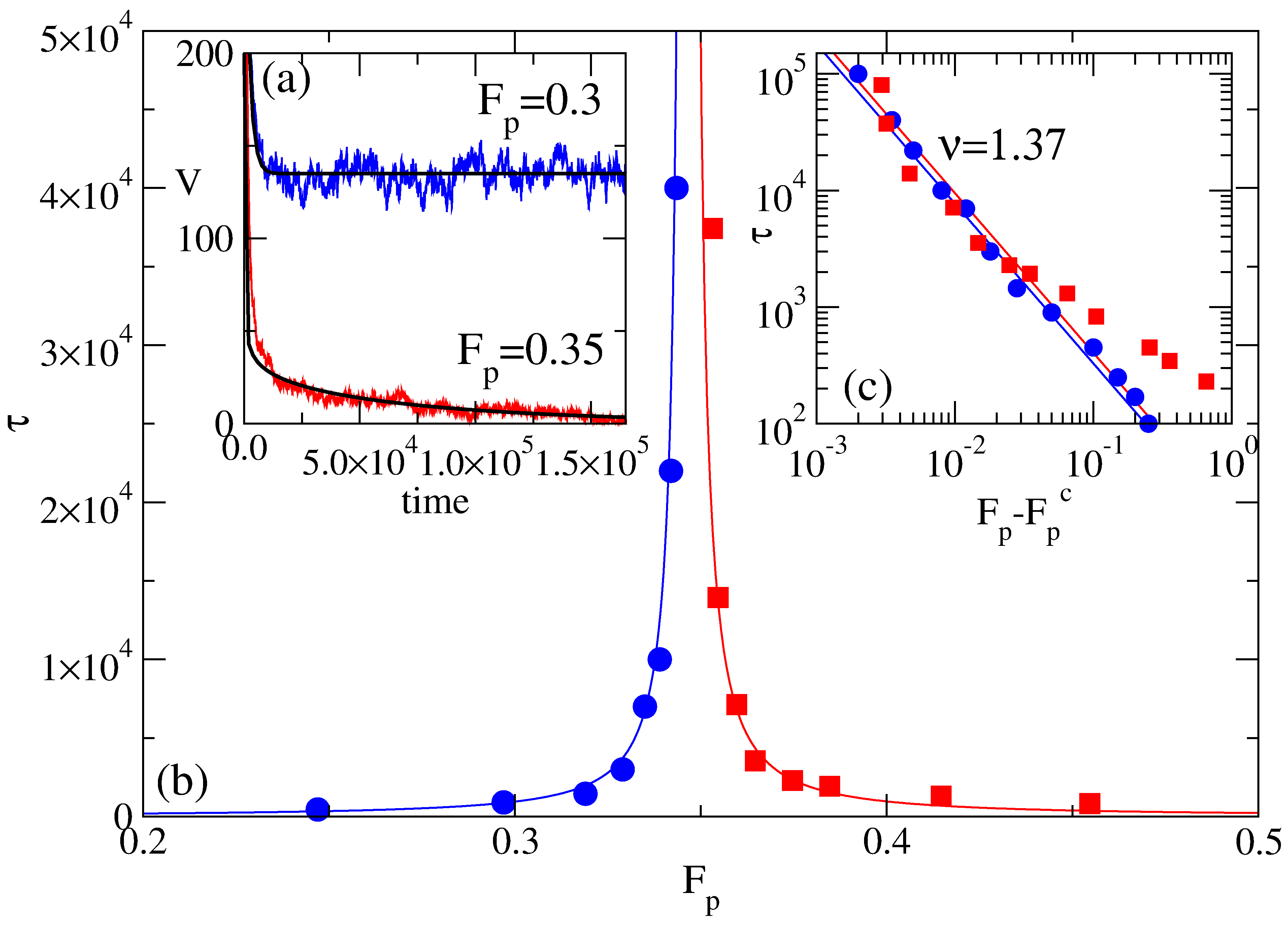}
\caption{ 
  (a) The simulated colloid velocity $V$ vs time for the system in
  figure~\ref{fig:56} at  $F_{p}/F^{c}_{p} = 0.86$ (blue) where the system settles into 
  a steady fluctuating state
  and at $F_{p}/F^{c}_{p} = 1.008$ (red) where the system settles into a pinned state. 
  (b)  The time scale $\tau$ required to reach either state vs pinning force $F_p$.
  Here $\tau$ diverges at the critical point $F^{c}_{p}$. 
  Blue dots are systems with $F_p/F^{c}_p<1$ and red squares are systems
  with $F_p/F^c_p>1$.
(c) Power law fits of $\tau \propto |F_{p} - F^c_{p}|^{-\nu}$ with $\nu = 1.37 \pm 0.06$.  
  Adapted with permission from C. Reichhardt and C.J. Olson Reichhardt, 
Phys. Rev. Lett. {\bf 103}, 168301 (2009). Copyright 2009 by the American Physical Society.
}
\label{fig:57}
\end{figure}

\begin{figure}
  \includegraphics[width=\columnwidth]{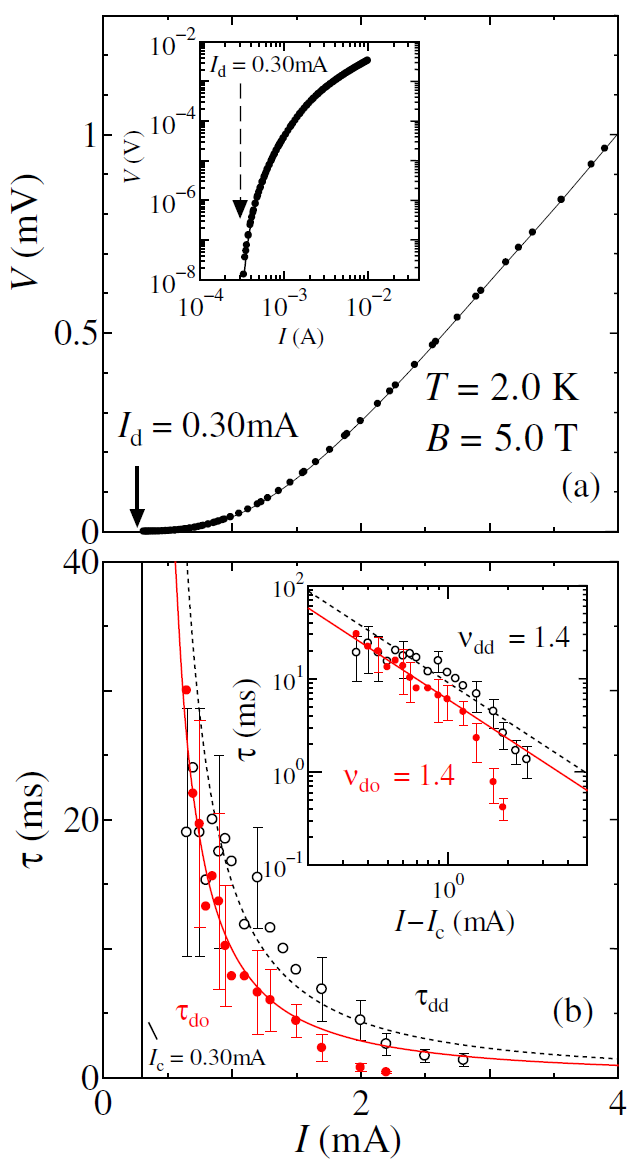}
\caption{
  (a) Voltage $V$ vs driving current $I$  from experiments on superconducting
  vortices subjected to a suddenly applied current.
  The voltage $V$ corresponds to the vortex velocity $V$, the
  driving current $I$ corresponds
  to the driving force $F_D$, and the critical depinning current $I_d$
  corresponds to the critical depinning threshold
  $F_c$.  Inset: The same data on a log-log scale.
  (b) The time scale $\tau$
  required to reach a disordered state from an ordered initial condition
  ($\tau_{dd}$, black) or an ordered state from a disordered initial condition
  ($\tau_{do}$, red)
  vs driving current $I$.
  There is a power law divergence in $\tau$ near a critical drive $I_{c}$,
  which corresponds to a critical driving threshold $F_D^c$.
  Inset: $\tau$ vs $(I-I_c)$
can be fit to a power law with exponent $\nu = 1.4$.  
 Reprinted with permission from S. Okuma and A. Motohashi,
 New J. Phys. {\bf 14}, 123021 (2012).  Open Access (CC-BY) article,
 http://creativecommons.org/licenses/by/3.0/
}
\label{fig:58}
\end{figure}

In numerical examinations of the plastic depinning transition for colloids under an
external drive in the presence of random pinning
of varied strength $F_p$,
Reichhardt {\it et al.} found that the sudden application of a driving
force of fixed direction and fixed magnitude $F_D$
produces plastic rearrangements of the colloids which
subsequently settle into either a pinned
state or a fluctuating flowing state
\cite{283}.  The time scale $\tau$ required for the system
to reach either of these states diverges near a critical pinning force $F_p^c$.
Figure~\ref{fig:56}(a-d) shows the time evolution of the particle trajectories 
at $F_{p}/F^{c}_{p}  = 0.93$, where the fixed drive $F_D$ is larger than the
depinning threshold of the system.
The number of trajectories gradually decreases with time but settles eventually into
a steady flowing state.
In contrast, in figure~\ref{fig:56}(e-h) at $F_{p}/F^{c}_{p}=1.05$,
the fixed drive $F_D$ is smaller than the depinning threshold and
the number of trajectories continues
to decrease until the system reaches a pinned state and
can no longer evolve.
In analogy with the reversible-irreversible transition,
the pinned state corresponds to a reversible or absorbed state
where the fluctuations are lost,
while the fluctuating flowing state corresponds to the irreversible state.    
Figure~\ref{fig:57}(a) shows the colloid velocity
$V$ versus time in two samples: one with
$F_{p}/F^{c} _{p} = 0.86$ which settles into a steady fluctuating state, and
one with $F_{p}/F^{c}_{p} = 1.008$ which settles into a pinned state. 
The solid lines are fits to the time-dependent colloidal velocity of the form
$V(t) = (V^{0} - V^{s})\exp(-t/\tau)/t^{\alpha} + V^{s}$, where
$V^{0}$ is the initial velocity and $V^{s}$ is the steady state velocity.
At the critical point $F_p^c$,
$V(t)$  has a power law form
and  $\tau \rightarrow \infty$.
Based on this measure, $\tau$ can be identified on either side of the transition,
as shown in figure~\ref{fig:57}(b), and used to determine the
critical value of $F_{p}$ at which $\tau$ diverges.
In figure~\ref{fig:57}(c), a log-log plot of the
time scale $\tau$ versus $F_{p} - F_{p}^{c}$
can be fit to a power law form with exponent $\nu = 1.37$, close to the value
measured in the sheared colloidal system.
At the critical point,
the critical exponent for decay of the velocity is  $\alpha =  0.5$. 
The exponents obtained in both the sheared colloidal system
and the depinning system are close to the 
values predicted for conserved directed percolation (CDP), which are
$\nu = 1.29$ in 2D and $\nu=1.12$ in 3D.
Ordinary directed percolation (DP) has exponents $\nu = 1.295$ in 2D and
$\nu=1.105 $ in 3D \cite{25}.
Similarly, $\alpha = 0.5$ is close to the values
$\alpha=0.45$ in DP and $\alpha=0.5$ in CDP.
Experiments on nonequilibrium absorbing phase transitions between different
turbulent states also give the value
$\alpha = 1/2$ \cite{27}.

\begin{figure}
  \includegraphics[width=\columnwidth]{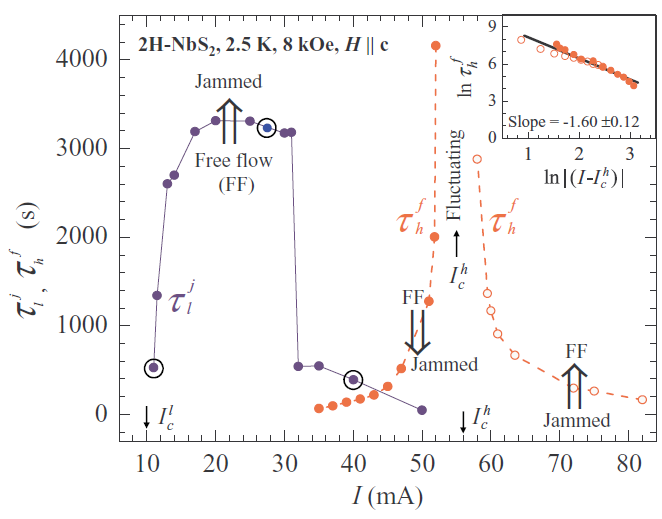}
  \caption{
    Experimental measurements of transient time scales $\tau$ vs driving current $I$ for
    superconducting vortex states near a depinning
    threshold.
    Blue symbols: Vortices prepared in an ordered state depin above a critical current
    $I_c^l$ and flow during a time scale $\tau_l^j$ before suddenly dropping into
    a disordered pinned state.
    Orange symbols: Vortices in the disordered pinned state
    exhibit a diverging transient time scale $\tau_h^f$ near the depinning
    current $I_c^h$; for $I<I_c^h$ (filled circles), the steady state is a pinned phase,
    while for $I>I_c^h$ (open circles), the steady state is flowing.
    Here $\tau_h^f$ corresponds to the time scale $\tau$ measured near a
    depinning transition, the current $I$ corresponds to the driving force $F_D$, and
    the critical current $I_c^h$ corresponds to the critical driving
    threshold $F_D^c$.
    Inset: The divergence of $\tau$ can be fit to the form
    $\tau  \propto |(F_D-F_D^c)|^{-\nu}$ with $\nu=1.6$.
  Reprinted with permission from G. Shaw, P. Mandal, S.S. Banerjee, A. Niazi, A.K. Rastogi,
  A.K. Sood, S. Ramakrishnan, and A.K. Grover,
Phys. Rev. B {\bf 85}, 174517 (2012). Copyright 2012 by the American Physical Society.
}
\label{fig:59}
\end{figure}

Okuma {\it et al.} \cite{283a} experimentally studied
the plastic
flow of
superconducting vortices in a Corbino disk geometry under the sudden application 
of a dc driving current and measured the transient voltage
response, which is proportional to the vortex velocity. 
Figure~\ref{fig:58}(a) shows the velocity-force curve for this system, indicating that
there is a clearly defined finite depinning threshold $F_c$.
After the vortices are prepared in an initially ordered or initially disordered state,
the time scale $\tau$ required to reach a steady state 
can be measured.
Based on changes in $\tau$ it is possible to 
identify a critical driving threshold $F_D^c$ at which
$\tau$ diverges as a power law
that can be fit with an exponent of $\nu=1.4$,
as shown in
figure~\ref{fig:58}(b).
For $F_D>F_D^c$, the flowing vortices become more ordered when the drive is applied,
while for $F_D<F_D^c$, they become less ordered when the drive is applied.
These results show that critical
behavior in the form of
a diverging time scale can occur near a dynamical ordering transition as well as
near a depinning transition.
Further studies of the same system at different magnetic fields
where the  ordering of the pinned state changes also show
power law divergences in the transient time scale with exponents $\nu = 1.3$ to
$\nu=1$ \cite{284}.
Shaw {\it et al.} \cite{285}
experimentally examined the lifetimes of transient
states near a superconducting vortex depinning threshold.
When the system is prepared in an ordered state, there is a lower
critical driving threshold above which the vortices flow
transiently during a time scale $\tau_l^j$ 
before jumping abruptly to a pinned state.
This sudden jump was argued
to be a jamming phenomenon associated with the disordering of the vortex lattice.
Figure~\ref{fig:59} shows how the transient time scale $\tau_l^j$
changes as a function of driving force $F_D$.
The disordered vortex system exhibits a
higher critical driving threshold $F_D^c$ near which
there is another transient behavior
associated with the time scale $\tau$
required  for the system to settle into a pinned phase for
$F_D<F_D^c$
or into a moving steady state for
$F_D>F_D^c$,
as shown in figure~\ref{fig:59}.
The inset of figure~\ref{fig:59} indicates that
$\tau$
diverges as a power law at
$F_D^c$
according to $\tau \propto |F_D - F_D^{c}|^{-\nu}$ with $\nu = 1.6$.  

The behavior of periodically driven superconducting vortex systems near
depinning transitions has also been examined in
numerical \cite{286,287,288} and experimental 
\cite{288,289} studies.
The experiments of  Okuma {\it et al.} on a periodically sheared vortex system 
show diverging time scales for the system to settle into a reversible or irreversible state
with  $\nu = 1.3$ \cite{290}. 
The consistency of the critical exponents in these different studies
provides evidence that examining the
transient time scales might be the most promising way to
characterize nonequilibrium phase transitions in systems with pinning.

Criticality at reversible-irreversible transitions is currently an open topic
with many new developments.
For example,
it was recently proposed that just at the transition,
the spatial distribution of the particles is not truly random but is an example
of a hyperuniform distribution \cite{291}.
Such a distribution has particular scaling features
in $S({\bf k})$, indicating that the
large scale density fluctuations associated with a truly
random distribution are lost \cite{292}.
Several studies of systems of particles with contact interactions
have produced evidence for hyperuniformity at the critical point
\cite{293,295}.
It would be interesting to determine whether hyperuniformity also arises
near reversible-irreversible transitions in systems
with longer range interactions, such as superconducting vortices,
charge-stabilized colloids, or 
dislocations.
It would also be interesting to apply the concepts of reversible-irreversible transitions
to systems that undergo elastic depinning.
In this case, although the
system may always organize to a reversible state,
there could be different types of reversible states, and
there could still be a diverging time scale at transitions between these states.
There is some work showing how 
avalanche statistics can be related to a diverging time scale
near the reversible-irreversible transition in sheared materials \cite{ott5}, 
and it would  be interesting to make a similar connection between
avalanches and depinning transitions.

\section{Other Systems and Future Directions}

\subsection{Single driven particles}
It is possible to construct a
velocity-force curve for
a single probe particle driven
through an assembly of other particles \cite{297,298,299} or a combination of other 
particles and pinning sites \cite{300,301}.
Such a system can exhibit a finite threshold for motion
as well as nonlinear velocity-force curves with
$V \propto (F_D- F_{c})^{\beta}$. 
Simulations by
Hastings {\it et al.}
\cite{297} for a colloid driven through a disordered
assembly of other colloids
showed both a finite depinning threshold
and a power-law velocity-force curve with exponent $\beta = 1.5$
associated with
plastic rearrangements of the particles surrounding the probe particle at the depinning
transition.
In experiments
on a magnetic colloid driven through a 3D assembly of
monodisperse colloids, Habdas {\it et al.} \cite{298}
found $\beta = 1.5$ below the 3D glass transition and $\beta=2.5$ near 
the glass transition.
It is not known whether the depinning process in this class of system
is the same as that
in collectively driven particle systems or
whether it is possible for dynamical transitions to occur at higher drives.  
The local probe technique has been used in a variety of soft matter systems
as a way to explore the rheological properties and
is termed active microrheology; the interested reader is referred to a recent
review on the topic \cite{302}.    

\subsection{Vortices in Bose-Einstein condensates}

Particle-like vortices also occur in
Bose-Einstein condensates (BECs) \cite{303}
and there have already been theoretical \cite{304} and experimental 
\cite{305} studies of BEC vortices in a co-rotating optical trap array
showing
commensurate and incommensurate pinning transitions.
The BEC vortices have much less damping in their dynamics and much
stronger hydrodynamic interaction effects than superconducting vortices,
making them ideal for studying new types of
nonequilibrium dynamics, depinning, and dynamic phases that are not accessible using
superconducting vortices.
Experimentally it would be possible to introduce a
co-rotating optical trap into the BEC and determine, by
slowing the rotation  of the optical trap,
when the vortex lattice decouples from the optical
traps and
whether the lattice breaks up or forms coherent patterns.
There has already been some computational work on
dynamic phases of BEC vortices in optical trap arrays \cite{306}. 
It is also possible to create binary BECs that can contain
multiple species of vortices  which
form  more complicated structures \cite{307}.
It is not known how such vortex systems would depin
or what their dynamics would be.
For example, if multiple vortex species
enter a moving phase, it is possible that dynamical phase
separation of the different vortex species could occur into moving bands, and
such structure formation
would be associated with hysteresis in a velocity-force curve.

\subsection{Time dependent traps}  

The ability to pin BEC vortices, colloids, and even
superconducting vortices with optical traps opens the possibility of
studying the effect of a time-dependent substrate on the particle dynamics.
For example, the pinning sites could
be periodically flashed on and off or moved around in a correlated manner. 
Although it may seem
that introducing time dependence to the pinning sites would only reduce the effectiveness
of the pinning,
it may be possible to introduce some type of dynamical feedback that could actually
result in an
increase in the pinning effectiveness.
There has already been some work
on colloids and vortices on time dependent substrates
that shows interesting interference effects in the transport curves \cite{308,309}. 
In general,
driven particle dynamics on a time dependent
landscape is an open area of research
that could be relevant to new nanoscale devices as well as to biological systems. 

\subsection{Flowing granular matter}
In a simulation model of granular bed erosion
in which particles with a short-range steric repulsive interaction
are driven over a random substrate
by a constant force representing a viscous fluid,
Yan {\it et al.} \cite{Wyart} find that above a threshold
drive $F_c$ for motion, the plastic flow of the particles through channels
is associated with a velocity scaling of
$V \propto (F_D-F_c)^{\beta}$ with $\beta=1.0$, an exponent
considerably lower than that found for plastic depinning in other systems.
In addition, near $F_c$ 
the transient time
scale $\tau$ required for the system to settle into a steady state diverges as
$\tau = |F_D-F_c|^{-\nu}$  with $\nu=2.5$, which is considerably larger than
the values of $\nu=1.3$ to $\nu=1.5$ observed in the colloidal and vortex
depinning systems described in Section 12.
These unusual exponents could be an indication of
strong finite size effects, where systems of extremely
large size would be required to obtain the correct exponent,
they could arise since the flow in this model becomes
effectively 1D rather than 2D, 
or they could indicate
that plastic flow in granular systems with short-range steric repulsive interactions
falls into a different universality class than that of
particles that have much longer range repulsive interactions,
such as vortices,
charge-stabilized colloids, or Wigner crystals.
It would be interesting to
consider a system with short-range
interactions and gradually increase the interaction range to determine how this
modifies the scaling properties of the dynamics.
 
\subsection{Active matter} 

In active matter systems,
the particles are self-driven or self-propelled.
Examples of active matter include
swimming bacteria, pedestrian flow, vehicular traffic, molecular motors, 
flocks of birds, herds of animals, and crawling cells \cite{310}.
Recently,
the creation of artificial swimmers such as self-driven colloids has
attracted growing attention as a means
for understanding collective nonequilibrium behaviors \cite{311}.  
In most active matter studies, the particles move over a smooth substrate
or interact with a fixed number of walls.
In many real systems such as 
swimming bacteria, the active particles
also interact with some type of disordered substrate, so the 
dynamics of active mater moving over random or periodic substrates
will be a growing area of research.
Although increasing the activity level of the particles might be expected to produce
effects similar to increasing the temperature, it was shown that
increased activity can produce new types of behaviors such 
as self-clustering \cite{311,312}. 
Studies of the mobility of active particles 
in disordered media under an external drive
show that the mobility of the particles is nonmonotonic
as a function of increasing
activity, and that higher activity levels can cause the particles
to become more strongly pinned \cite{312N}.
Other studies of
flocking models in disordered media
indicate that there can be an optimal noise level
that produces coherent flow \cite{314}.  
It is also possible to create active solid systems \cite{315},
and it would be interesting to understand whether plastic or elastic depinning can occur in 
such systems, and if so, in what ways the depinning differs
from the depinning of non-active systems. 

\subsection{Dusty plasmas}

Dusty plasmas,  which are composed of
solid particles that become charged when
they are placed in a plasma,
can be modeled as repulsively interacting
Yukawa particles that form a triangular lattice in the strongly
coupled limit 
\cite{316,317}.  
The major difference between dusty plasma systems and colloidal systems is that 
in dusty plasmas the particle dynamics are
underdamped.
Due to the size scale, the speed of sound in a dusty plasma can be 
extremely low, making it possible to create shock waves, solitons,
and other nonlinear wave phenomena with relatively small perturbations.
In principle it should be possible to create dusty plasma systems in which
the particles interact with some form of
disordered or periodic landscape, and then to drive the particles
across this landscape. 
Studies of this type 
could open a new class of depinning dynamics in which inertia plays
a dominant role, 
and it would be interesting to see how the generation of shock waves
might modify the depinning transition.
It is also possible to create charged colloidal systems in air
which obey underdamped dynamics, while a study of
charged metal balls in air 
has demonstrated commensurate-incommensurate pinning transitions \cite{318}.  

\subsection{Nonequilibrium fluctuation relations}

Over the last 30 years, various nonequilibrium fluctuation and work relations 
have been derived  \cite{319,320},
including those of Jarzynski \cite{321} and Crooks \cite{322},
and these relations have been applied
to a variety of systems including a colloid dragged through fluid \cite{323}
and biological systems \cite{324}.
It would be interesting to test whether there are any conditions under which such
relations hold in the nonequilibrium flowing phases of systems with pinning.
There has already been some 
initial work in this direction for colloids moving over random \cite{325} 
and periodic \cite{326} substrates.

\subsection{Topological states}

Topological states have been receiving growing attention in
quantum systems \cite{327}, and similar ideas are beginning to be applied
to the dynamics of classical systems as well,  
such as to the motion of solitons in classical mechanical
insulators \cite{328} and to edge mobility in 
coupled gyroscope arrays \cite{329}.  
It is possible that some of these ideas could also be applied
to systems that exhibit depinning phenomena.
For example,
the mobility at depinning could be limited to or enhanced along
the edges of pinning array, or it may be possible to design
pinning arrays in which only certain topologically restricted modes of motion can occur.

\subsection{Curved geometries}

In the systems described up to this point,
depinning occurs on flat surfaces.
There is, however, growing interest in studying
particle dynamics and ordering
in systems with curved geometries,
such as dislocation patterns and motion on surfaces with positive
or negative curvature \cite{330,331,332}. 
It would be interesting to study elastic or plastic depinning
phenomena on such curved geometries 
and compare them to the dynamics observed in flat geometries.
In curved geometries, dynamically ordered states could correlate with
certain types of pattern formation of moving topological defects.  

\subsection{Deformable substrates} 

In the studies described so far, the pinning sites or substrates 
remain fixed in space and only the particles can move. Another class 
of system that is wide open for study is one in which
the pinning or substrate itself is deformable or can even exhibit plasticity. 
Such effects could come into play for
particles moving over deformable membranes or elastic sheets,
where the particle could create indentations in the surface that would introduce
modified pairwise interactions between the particles. 
Another possibility is that near the depinning threshold,
a wrinkling transition could occur in which
the system could become strongly buckled into the
third dimension, and that once the particles depin, the 
sheet could straighten again.
It could also be possible to
create systems in which the pinning sites can
move past each other and exhibit plasticity.
For example, the pinning sites could be modeled as
obeying
their own equation of motion with dynamics that differ from those of the driven particles.
Effects of this type could occur for
electron crystals pinned by 
charged doping sites, where the doping sites gradually electromigrate over time. 
It would also be possible to
drive colloids through obstacle arrays
in which the obstacles gradually change position.  
Such systems could exhibit strong memory effects
that can change the transport over time, and
may offer a new approach for
creating novel memrister type devices \cite{333}.

\subsection{Quantum effects} 
One of the major areas of condensed matter physics
is focused on equilibrium quantum phase transitions 
that are driven by quantum fluctuations \cite{334}. 
In principle there could also be nonequilibrium quantum phase transitions, in analogy
with classical nonequilibrium phase transitions, and 
driven quantum systems that exhibit a depinning threshold or 
nonlinear transport at low temperatures
could be ideal candidates in which to observe such effects.
It might be expected that adding dynamical fluctuations
to quantum fluctuations would destroy the quantum effects;
however,  recent work shows that features of quantum critical states
can survive in the presence of nonequilibrium noise \cite{335}.
It may also be be possible that the application of a drive could 
in some cases suppress thermal noise or
other dynamical noise while preserving the quantum fluctuations.
Open issues include 
what the proper methods are to measure such nonequilibrium quantum transitions,
how they would differ from their classical analogs, whether there are quantum
analogs to dynamical ordering transitions, and whether new types of quantum phenomena
exist only in the driven phase but not in the equilibrium state.

\ack
We thank the following people for discussions:
Clemens Bechinger,
Shobo Bhattacharya,
Alan Bishop,
Karen Dahmen,
Roel Dullens,
Thierry Giamarchi,
John Goree,
David Grier,
Matt Hastings,
Boldizar Jank{\' o},
Wai Kwok,
Andras Lib{\' a}l,
Shi-Zeng Lin,
Andrea Liu,
Cristina Marchetti,
Jose Mart{\' \i}n,
Danielle McDermott,
Victor Moshchalkov,
Satoshi Okuma,
David Pine,
Dipanjan Ray,
Ido Regev,
Ivan Schuller,
Alejandro Silhanek,
Gabe Spalding,
Pietro Tierno,
Eric Weeks,
and Zhili Xiao.
This work was carried out under the auspices of the 
NNSA of the 
U.S. DoE
at 
LANL
under Contract No.
DE-AC52-06NA25396.

\end{document}